\documentclass[manuscript,nonacm]{acmart}

\setcopyright{none}
\usepackage{booktabs}
\usepackage{multirow}
\usepackage{enumitem}
\usepackage{xspace}
\usepackage{graphicx}
\usepackage{hyperref}
\usepackage{microtype}
\usepackage{xcolor}

\definecolor{keyborder}{RGB}{66,133,244}
\definecolor{keybg}{RGB}{240,248,255}

\newsavebox{\keyboxcontent}
\newenvironment{keybox}{%
  \par\smallskip\noindent
  \begin{lrbox}{\keyboxcontent}
  \begin{minipage}{\dimexpr\linewidth-2\fboxsep-2\fboxrule-1.5pt\relax}
  \raggedright
  \emergencystretch=1em
  {\color{keyborder}\faLightbulbO}\ \textbf{\textcolor{keyborder}{Observation.}}\space
}{%
  \end{minipage}
  \end{lrbox}
  \fcolorbox{keyborder}{keybg}{\usebox{\keyboxcontent}}%
  \par\smallskip
}

\newcommand{\inlinefigurewithcaption}[4]{%
  \begin{center}
    \refstepcounter{figure}%
    \includegraphics[#1]{#2}\par
    \smallskip
    {\small Fig.~\thefigure.~#3\par}%
    \label{#4}%
  \end{center}
}


\makeatletter
\renewcommand{\@mkauthorsaddresses}{}
\makeatother

\usepackage{fontawesome}

\begin{document}

\title{Beyond Storage: State as a Runtime Control Problem in Parallel and Distributed Systems}

\author{Shuhao Zhang}
\authornote{Corresponding author.}
\authornote{Supported by New Generation Artificial Intelligence-National Science and Technology Major Project (No. 2025ZD0123804).}
\orcid{0000-0002-9927-6925}
\affiliation{%
  \institution{Huazhong University of Science and Technology}
  \city{Wuhan}
  \country{China}}
\email{shuhaozhang@hust.edu.cn}

\author{Haoran Peng}
\orcid{0009-0003-6745-1762}
\affiliation{%
  \institution{Huazhong University of Science and Technology}
  \city{Wuhan}
  \country{China}}
\email{q1293132751@163.com}

\author{Chun Ho Ma}
\orcid{0009-0008-1636-0381}
\affiliation{%
  \institution{Huazhong University of Science and Technology}
  \city{Wuhan}
  \country{China}}
\email{u202490042@hust.edu.cn}

\author{Yancan Mao}
\orcid{0000-0002-7824-5978}
\affiliation{%
  \institution{Huazhong University of Science and Technology}
  \city{Wuhan}
  \country{China}}
\email{maoyancan@u.nus.edu}

\author{Yufeng Du}
\orcid{0009-0009-8919-0018}
\affiliation{%
  \institution{Huazhong University of Science and Technology}
  \city{Wuhan}
  \country{China}}
\email{yufeng050601@gmail.com}

\author{Shifeng Liu}
\orcid{0009-0001-1457-5839}
\affiliation{%
  \institution{Huazhong University of Science and Technology}
  \city{Wuhan}
  \country{China}}
\email{u202315663@hust.edu.cn}

\author{Ruijie Qiu}
\orcid{0009-0004-0659-3510}
\affiliation{%
  \institution{Huazhong University of Science and Technology}
  \city{Wuhan}
  \country{China}}
\email{Ying01020@outlook.com}

\author{Xiaofei Liao}
\orcid{0000-0002-2960-4982}
\affiliation{%
  \institution{Huazhong University of Science and Technology}
  \city{Wuhan}
  \country{China}}
\email{xfliao@hust.edu.cn}

\author{Hai Jin}
\orcid{0000-0002-3934-7605}
\affiliation{%
  \institution{Huazhong University of Science and Technology}
  \city{Wuhan}
  \country{China}}
\email{hjin@hust.edu.cn}

\begin{abstract}
Shared state increasingly shapes both performance and failure behavior in streaming, serving, retrieval, and continual-learning systems. Existing studies, however, often isolate access control, hardware-aware execution, memory management, and long-horizon updates. The review organizes this literature around three coupled dimensions: state access and scheduling, state-aware execution, and state evolution and reuse. Across these dimensions, the literature is synthesized through a common scaffold: state object, control surface, coupling path, evaluation boundary, and unresolved contract. This comparison identifies recurring anti-patterns and informs a contract-oriented blueprint and disturbance-aware evaluation agenda. Taken together, the evidence characterizes state management as a runtime control problem.
\end{abstract}

\maketitle
\renewcommand{\shortauthors}{Zhang et al.}

\section{Introduction}

Systems have moved away from largely stateless throughput engines toward long-running services shaped by evolving state. Foundational streaming systems already treated state and time as runtime concerns~\cite{akidau2013millwheel,murray2013naiad,zaharia2013dstreams,akidau2015dataflow}; newer LLM serving and retrieval systems expose similar pressure through KV caches, scheduling metadata, vector memories, and retained update traces~\cite{kwon2023pagedattention,agrawal2024sarathi,zheng2023sglang,zhou2025ferret,zhang2026flowrag}. Across these settings, the central question is whether a runtime can keep shared, long-lived, performance-critical state governable without sacrificing service stability.

That shift makes \emph{state management} in parallel and distributed systems harder to reason about. Traditional decompositions examine operator placement, memory layout, or model adaptation in isolation, but deployed systems do not preserve those boundaries: access hotspots reshape queues and locality, kernel-level gains disappear once data movement or quality constraints dominate, and aggressive memory updates can destabilize downstream retention or retrieval. What matters, then, is not optimization within one phase, but control over a lifecycle in which a local gain often comes back as a later constraint.

We use \emph{state management} to refer to the mechanisms that organize, observe, access, update, retain, and reuse shared state across execution boundaries. The underlying object may be a shared hash table, a compression dictionary, a KV cache, or retriever memory. Whatever its form, the runtime has to make the object visible, account for its access cost, tie it to execution policy, and preserve its future reuse value.

We organize the literature around three dimensions that are analytically distinct but operationally coupled: access decisions reshape execution, execution choices change update cost, and evolution policies alter future access conditions.

\begin{figure*}[t]
\centering
\includegraphics{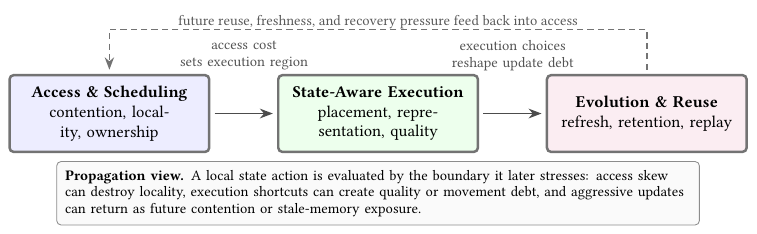}
\caption{A propagation-oriented view of state management. Access, execution, and evolution are coupled through feedback rather than separable pipeline stages.}
\label{fig:propagation}
\end{figure*}

\begin{enumerate}[leftmargin=*]
\item \textbf{State access and scheduling}: how systems expose contention, model access cost, and schedule work around hotspots, topology, and recovery requirements.
\item \textbf{State-aware execution optimization}: how systems turn state layout and execution coupling into stable end-to-end gains under hardware, energy, latency, and quality constraints.
\item \textbf{State evolution and reuse}: how systems continuously write, retain, organize, and reuse state for dynamic learning and inference.
\end{enumerate}
Figure~\ref{fig:propagation} summarizes this organizing perspective.

\begin{keybox}
\textbf{Access, execution, and evolution form one control loop over long-lived state.} A paper advances state management when it makes a decision in that loop observable, analyzable, or enforceable.
\end{keybox}

The survey makes four contributions. It first introduces a vocabulary for comparing streaming, approximate execution, serving, retrieval, and memory-centric intelligent services. It then develops a three-axis taxonomy around access, execution, and evolution, identifies recurring anti-patterns together with a contract-oriented blueprint for stateful runtimes, and proposes disturbance-aware evaluation boundaries for mechanisms whose local gains may not survive composition. The emphasis is comparative rather than exhaustive.

Section~\ref{sec:foundations} defines the survey scope and analytical model. Section~\ref{sec:core-dimensions} develops the three dimensions of state management. Section~\ref{sec:comparative-synthesis} compares mechanisms across domains and identifies when they transfer beyond one workload family. Section~\ref{sec:design-implications} distills design principles, anti-patterns, and a contract-oriented blueprint. Section~\ref{sec:evaluation-outlook} gives an evaluation vocabulary and research agenda, and Section~\ref{sec:conclusion} concludes.

\section{Foundations: Scope, Model, and Taxonomy}
\label{sec:foundations}

We focus on systems in which state is both shared and performance-critical: stream processing engines, transactional streaming systems, heterogeneous analytics runtimes, approximate execution frameworks with quality constraints, and dynamic AI services such as continual learning and RAG. We exclude static storage engines, model-centric training papers that do not expose runtime state-management mechanisms, and applications in which state is incidental.

Our source material includes representative systems papers, journal articles, and recent intelligent-service systems papers. Instead of assembling the corpus around a single lineage, we organize it around recurring control problems. Shared-state access and scheduling, hardware-conscious execution, and evolving memory services anchor the set~\cite{zhang2024survey,zhang2020hardware,mao2023morphstream,wu2023sentistream,zhang2026flowrag}; streaming, serving, recovery, approximation, and dynamic retrieval systems then test whether those mechanisms transfer across domains~\cite{akidau2013millwheel,murray2013naiad,zaharia2013dstreams,akidau2015dataflow,kwon2023pagedattention,agrawal2024sarathi,zheng2023sglang}. Papers enter the analytical core when they externalize a governable runtime seam over state; otherwise, we treat them as contextual.

We use three inclusion criteria:
\begin{itemize}[leftmargin=*]
\item \textbf{State visibility.} The system must surface state as an explicit runtime concern rather than a hidden storage detail.
\item \textbf{Exposed mechanism.} The paper must expose a systems mechanism such as scheduling, cost modeling, update control, migration, retention, or reuse orchestration.
\item \textbf{End-to-end connection.} The evaluation must connect that mechanism to end-to-end properties such as throughput, tail latency, recovery time, quality degradation, or long-horizon reuse quality.
\end{itemize}
These criteria favor papers that make control decisions legible. In the retrieval and learning lines, model-centric papers enter the analytical core only when they expose managed memory objects, planner-visible controls, or maintenance boundaries.

\subsection{Survey Protocol and Evidence Posture}

The review protocol is comparative and mechanism-centered rather than narrowly bibliometric. We seed the corpus from canonical systems lines and adjacent surveys, then trace forward and backward citation paths around shared-state access, state-aware execution, and state evolution. The goal is to cover mechanism families, not to maximize paper counts. When a stable venue version exists, we use it rather than a moving preprint. Recent papers enter the analytical core only when they expose a new state object, runtime control seam, or evaluation boundary. We also distinguish foundational papers, transfer-oriented systems, and frontier systems, so local speedups, transfer demonstrations, and long-horizon governance claims are not all read as the same kind of evidence.

The protocol also separates \emph{coverage} from \emph{representativeness}. Coverage asks whether the survey includes the mechanism families required by its taxonomy: visibility and progress, ownership transfer, hardware-conscious placement, approximation control, serving-memory lifecycle management, retrieval maintenance, and retention governance. Representativeness asks a different question, namely whether the selected papers expose enough design variation to support meaningful comparison. For example, a single KV-cache allocator would not justify claims about serving-memory lifecycle control, so the survey compares paging, prefix reuse, phase separation, tenant multiplexing, disaggregation, and restoration mechanisms under a common latency and memory boundary. By the same logic, one vector-index paper would not justify claims about retrieval evolution; the survey instead combines retriever invocation, mutable ANN maintenance, publication safety, freshness control, and placement.

This distinction matters because the paper is not meant to function as an encyclopedic bibliography. A paper can be historically important yet remain contextual if it does not expose a runtime control surface over state; conversely, a recent system can enter the analytical core if it makes a previously hidden lifecycle boundary measurable. For borderline cases, we ask whether at least one field of the comparison tuple changes. A storage-engine paper, for instance, may remain a boundary case when it assumes stable object identities and clear invalidation semantics, but it becomes analytically relevant once compaction, admission, eviction, or tiering is made visible to a scheduler or service-level controller. This rule keeps the scope broad enough to connect streaming, serving, retrieval, and retention without turning the survey into a list of systems that merely store data.

The evidence classes used later in the paper follow from this rule. \emph{Foundational} papers introduce state objects or control boundaries that later systems inherit, such as progress frontiers, checkpoint barriers, ownership metadata, bounded-error summaries, or reusable memory stores. \emph{Transfer-oriented} papers show that a mechanism remains useful under a related systems constraint: a placement policy survives recovery, a cache policy survives tenant churn, or a retrieval policy survives update drift. \emph{Frontier} papers expose a control boundary that is not yet mature but appears likely to matter as workloads change, such as disaggregated serving memory, planner-visible retrieval freshness, or budget-elastic retention. This classification helps avoid a familiar survey mistake: treating an early abstraction paper, a mature runtime system, and a speculative frontier system as if they made interchangeable claims. The comparison is strongest when it states which class a paper belongs to and what evidence would be needed to move the mechanism into a stronger class.

Omitted measurements also delimit the available evidence. When a paper reports strong local performance but does not measure migration, recovery, freshness, restoration, or quality debt that its mechanism could plausibly affect, the result should not be extrapolated beyond that boundary. For example, an LLM serving paper that improves goodput through aggressive reclamation may still leave open whether the reclaimed state harms prefix reuse under follow-up bursts. A retrieval paper that improves recall-latency trade-offs on a static index may likewise leave open whether deletion repair and publication lag alter the service boundary. The survey's unresolved-contract field records these limits explicitly.

\subsection{Positioning Relative to Existing Surveys}

Several prior surveys already cover important slices of this design space, including catalogs of stream-processing design choices~\cite{hirzel2014catalog,cugola2012processing}, transactional stream processing~\cite{zhang2024survey}, hardware-conscious stream processing~\cite{zhang2020hardware}, and continual learning~\cite{delange2021survey}. These are valuable adjacent references, but they adopt narrower units of analysis than the present manuscript. Our comparison unit is not one application class or one execution substrate; it is the runtime control problem exposed over state. That choice lets us place migration, checkpointing, KV-cache lifecycle control, vector-index maintenance, retrieval freshness governance, and retention budgeting into the same analytical frame without flattening them into a chronology of subfields.

This manuscript does not replace specialized surveys. It asks which systems abstractions recur once state becomes the primary object of runtime control across streaming, serving, retrieval, and long-horizon memory services. Adjacent storage-engine and cache-management surveys provide boundary cases: they expose controls such as compaction, admission, eviction, and placement~\cite{luo2020lsm,ali2011survey}, but usually assume stabler object models and clearer invalidation semantics.

\subsection{Broader Literature Landscape}

Several systems lines expose the same runtime seams under different names. Together they motivate the three-axis decomposition used in this survey: access and scheduling, execution optimization, and evolution and reuse.

\subsubsection{Stream and Dataflow Systems}

Stream and dataflow systems provide the historical root for the access-and-scheduling axis. Aurora and Borealis made operator state and adaptation explicit inside continuous-query plans, while later systems showed that progress tracking, incremental maintenance, elasticity, and recovery all reshape the same state boundary rather than living in isolated subsystems~\cite{abadi2003aurora,abadi2005borealis,carbone2015flink,carbone2015asynchronous,chandramouli2014trill,armbrust2018structured,murray2013differential,gulisano2012streamcloud}. Performance and correctness depend on when state is exposed, who owns it, and how safely it can move under disturbance.

\subsubsection{Transactional, Replicated, and Memory-Resident Systems}

Transactional, replicated, and memory-resident data systems supply the correctness boundary that the rest of the survey reuses. Systems such as H-Store, Spanner, Calvin, RAMCloud, and FaRM treat state as a jointly governed object spanning consistency, placement, logging, and remote access~\cite{kallman2008hstore,corbett2012spanner,thomson2012calvin,ousterhout2011ramcloud,dragojevic2014farm}; snapshot and replication results clarify which transitions may be migrated, replayed, or shared without violating service guarantees~\cite{chandy1985snapshot,ongaro2014raft}. This lineage explains why later mechanisms such as ownership transfer, shard exposure, and short-lived memory reuse must be evaluated against explicit authority and replay boundaries, not only against throughput.

\subsubsection{Serving and Memory-Governance Systems}

Serving and memory-governance systems motivate the execution-optimization axis. Clipper and Clockwork already treated caching, batching, predictability, and dispatch as runtime controls~\cite{crankshaw2017clipper,gujarati2020clockwork}; modern LLM-serving systems make the state objects explicit through KV pages, adapter state, model residency, and prefill/decode separation~\cite{kwon2023pagedattention,sheng2023slora,zheng2023sglang,agrawal2024sarathi,zhong2024distserve,patel2024splitwise,yu2025prism,zhang2025jenga}. Tiered and disaggregated memory systems reach the same abstraction from another direction: extra capacity helps only when the runtime decides which state must stay near compute and which movement or maintenance debt can be deferred safely~\cite{hu2024aceso,zhong2024unimem,jalalian2024extmem,raybuck2021hemem,cho2024coaxial,zhao2025equilibria,peng2024scalacache,wang2025finemem,chang2024gmt,li2024streamcache,qiu2025hotrap}. Stable execution gains require governed placement, representation, and movement, not faster kernels alone.

\subsubsection{Retrieval, Retention, and Approximation Systems}

Retrieval, long-horizon memory, continual retention, and approximation-aware systems motivate the evolution-and-reuse axis. Retrieval systems show that reusable memory is valuable only when refresh, lookup, and maintenance remain coordinated~\cite{guu2020realm,lewis2020rag,borgeaud2022retro,izacard2022atlas,asai2024selfrag,wang2024hipporag,sarthi2024raptor}. Continual-learning and bounded-retention systems expose the same problem under update pressure, where memory budgets, replay choices, and future reuse value must be co-governed~\cite{kirkpatrick2017ewc,lopez2017gem,rebuffi2017icarl,chaudhry2019agem,rolnick2019experience,aljundi2019mir,buzzega2020der,hayes2020remind,prabhu2020gdumb,delange2021survey}. Approximate analytics adds a complementary constraint: once summaries or samples stand in for full state, the runtime must reason jointly about error, update cost, and future reuse~\cite{agarwal2013blinkdb,goiri2015approxhadoop,wang2018approxjoin,hu2018streamapprox,chen2018incapprox}. Together these lines make state a lifecycle control problem: retention, compression, and refresh choices change future freshness, utility, and service stability.

\subsubsection{Adjacent Traditions and Analytical Boundary}

Several adjacent traditions fix the analytical boundary. General-purpose dataflow and storage systems show how large services externalize state into programmable, replicated layers with explicit update semantics~\cite{isard2007dryad,chambers2010flumejava,murray2011ciel,chang2008bigtable,decandia2007dynamo,lakshman2010cassandra,baker2011megastore,peng2010percolator}. CRDT-style replicated objects, dynamic graph systems, vector indexes, and low-level attention kernels add further boundary cases: merge legality, mutable structure, index maintenance, and memory-traffic shape become runtime-visible once the state object persists long enough to constrain later control~\cite{shapiro2011crdt,gao2026grace,karpukhin2020dpr,khattab2020colbert,jegou2011pq,malkov2020hnsw,dao2024flashattention2}. These cases ground the comparison in recurring runtime control problems rather than unrelated applications.

\subsection{System Model: Where State Lives in a Runtime}

The literature landscape motivates a common runtime abstraction: before comparing mechanisms, we need to specify where state lives and when it becomes manageable. Prior work on stream processing provides a methodological template~\cite{fernandez2010seep,fernandez2013osm,mao2025spacker,carbone2015flink}: first describe computation, communication, deployment, and execution boundaries, then define which internal objects must be exposed as managed state. The modeling style is general: state becomes meaningful only after the runtime boundary around computation, communication, scheduling, and recovery has been made explicit.

We view a parallel or distributed system as a runtime that executes computational units over data items, requests, or events. These units may be streaming operators, transactions, serverless functions, tasks or actors, LLM workers, retrievers, or agentic stages; they communicate through shuffles, RPCs, remote memory transfers, object-store reads, collective communication, KV-cache transfers, or vector-index lookups. A scheduler assigns work, decides placement and batching, reacts to load changes, and may trigger migration, checkpointing, eviction, refresh, or recovery. State is any runtime object read, written, moved, retained, or reused across these execution boundaries.

\smallskip\noindent\textbf{Formal sketch.}
A runtime can be abstracted as a tuple \(\mathcal{R}=(\mathcal{W},\,\mathcal{P},\,\Sigma,\,\Gamma)\) where
\(\mathcal{W}\) is a set of computational workers,
\(\mathcal{P}\) is a set of communication paths,
\(\Sigma\) is a set of state objects, and
\(\Gamma\) is a scheduler mapping work to resources.
An object \(s\in\Sigma\) qualifies as \emph{managed state} iff it satisfies three conditions:
\begin{enumerate}[leftmargin=*,nosep]
\item \(\textsc{Lifetime}(s)\gg\text{one invocation}\): \(s\) persists beyond a single function call, request step, or operator firing;
\item \(\textsc{Influence}(s)\neq\varnothing\): there exists at least one future scheduling, placement, recovery, quality, or reuse decision whose outcome depends on the value or location of \(s\);
\item \(\textsc{Controllable}(s)\): the runtime exposes a primitive \(\pi\) such that applying \(\pi\) to \(s\) changes its placement, representation, ownership, or retention status.
\end{enumerate}
This definition covers both persistent state (database indexes, checkpoint logs) and semi-persistent state (KV-cache pages, retriever-update queues) as long as all three conditions hold.

As summarized in Table~\ref{tab:state-system-model}, the concrete state object changes across runtime architectures, but its relationship to the runtime is stable. State constrains computation, communication, scheduling, memory, recovery, and long-horizon service quality because workers must read or update it, moving work often requires moving or reconstructing it, and future decisions depend on its size, ownership, hotness, freshness, and reuse potential.

\begin{table}[!htbp]
\centering
\caption{State in representative runtime architectures.}
\label{tab:state-system-model}
\small
\setlength{\tabcolsep}{4pt}
\renewcommand{\arraystretch}{1.10}
\resizebox{\textwidth}{!}{%
\begin{tabular}{@{}p{3.1cm}p{3.1cm}p{4.2cm}p{6.2cm}@{}}
\toprule
\textbf{System class} &
\textbf{Computational unit} &
\textbf{Communication path} &
\textbf{Representative state objects} \\
\midrule
Stream processing &
Operators and tasks &
Streams, shuffles, backpressure channels &
Keyed state, windows, operator state, watermarks, checkpoint metadata \\
Transactional dataflow and databases &
Transactions, execution fragments, workers &
Dependency exchange, log replication, remote reads &
Records, indexes, versions, locks, dependency graphs, logs, recovery snapshots \\
Distributed task/actor runtimes &
Tasks, actors, executors &
RPCs, object-store transfers, lineage replay &
Objects, actor-local state, lineage metadata, placement state, scheduler queues \\
LLM serving &
Prefill and decode workers &
KV transfer, model-parallel communication, request queues &
KV caches, page tables, prefix trees, adapter state, model-residency metadata \\
RAG and agentic workflows &
Retrievers, planners, tools, generators &
Vector-index queries, tool calls, memory updates &
Vector indexes, document memories, workflow context, tool traces, retained memories \\
Edge and approximate analytics &
CPU/GPU operators, sensor tasks &
Device transfer, edge--cloud transfer &
Summaries, sketches, compression dictionaries, quality-control metadata \\
\bottomrule
\end{tabular}%
}
\end{table}

This model also clarifies why state has multiple coupled forms: \emph{logical state} captures the semantic object; \emph{physical state} captures representation and placement; \emph{metadata state} records ownership, routing, and progress; \emph{control state} summarizes observations such as hotness or freshness; and \emph{evolution state} records updates, compaction, retention, refresh, eviction, or reuse. A change in one form often forces movement, metadata updates, scheduler decisions, and new recovery constraints in the others. The model therefore does not force systems into one implementation template; it provides a vocabulary for deciding when an object should be treated as managed state and which lifecycle control problem dominates: access and scheduling, execution optimization, or evolution and reuse.

\subsection{A Unifying View of Stateful Computing Systems}

\subsubsection{What Counts as State?}

In this survey, \emph{state} includes any persistent or semi-persistent runtime object whose value influences future execution beyond a single operator invocation. It spans windows, shared indexes, KV caches, retriever memory, and retained samples. Such state is systems-hard because it is shared across tasks, requests, or time windows and temporally extended enough for today's update decisions to reshape tomorrow's scheduling, quality, or recovery space.

\subsubsection{The Propagation Perspective}

We organize \emph{state management} in parallel and distributed systems using a propagation perspective. A request first encounters \emph{state access}: it must find, read, or synchronize on shared state. The resulting data path then shapes \emph{state-aware execution}: work is mapped to cores, sockets, accelerators, or approximate operators under latency and quality goals. The outputs in turn contribute to \emph{state evolution}: some state is updated, some retained, and some exposed for future reuse. Failures in any one stage propagate. Uncontrolled contention in access can destroy locality and distort execution modeling; mis-modeled execution trade-offs can make update policies unsustainable; over-aggressive updates can increase future access pressure and destabilize inference. The literature repeatedly shows that these seams cross traditional layer boundaries~\cite{zhang2019briskstream,zhang2020concurrent,zeng2024pecj,zhou2025ferret,li2026streamfp,zhang2026flowrag}, so the propagation perspective provides a more faithful abstraction for long-running stateful systems.

\smallskip\noindent\textbf{Propagation structure.}
Let the runtime state at logical time \(t\) be denoted \(S_t\).
Define three stage operators---access (\(\mathcal{A}\)), execution (\(\mathcal{E}\)), and evolution (\(\mathcal{V}\))---each producing the next state snapshot:
\[
S_{t+1}= \mathcal{V}\bigl(\mathcal{E}\bigl(\mathcal{A}(S_t;\;W_t,\Gamma_t);\;\mathcal{H}\bigr);\;\mathcal{U}_t\bigr),
\]
where \(W_t\) is the arriving workload, \(\Gamma_t\) captures scheduling decisions, \(\mathcal{H}\) denotes hardware constraints, and \(\mathcal{U}_t\) captures update and retention policies.
The core claim is that these stages are \emph{not} separable:
\[
\frac{\partial\,\mathcal{E}}{\partial\,\mathcal{A}}\neq 0,\qquad
\frac{\partial\,\mathcal{V}}{\partial\,\mathcal{E}}\neq 0,\qquad
\frac{\partial\,\mathcal{A}}{\partial\,\mathcal{V}}\neq 0.
\]
In words, a local perturbation in one stage propagates and reappears as a constraint in the other two.
State management is therefore a coupled control loop, not a pipeline of independent optimizations.

\begin{keybox}
\textbf{Local state decisions do not stay local.} Access contention, execution placement, and update policy propagate through the state lifecycle and reappear as constraints on the other two axes, so each mechanism must be read together with the downstream boundary it may quietly reshape.
\end{keybox}

\subsection{Taxonomy of State Management Problems}

Table~\ref{tab:taxonomy} summarizes the taxonomy used throughout the survey. It decomposes runtime control problems rather than application communities: a serving stack, for example, can simultaneously expose access skew, execution-level memory pressure, and evolution-level reclamation debt. The taxonomy is intentionally operational and nonexclusive; a stream engine and a RAG service may both fall under evolution and reuse if the dominant challenge is continuous updates, while a compression engine and a transactional stream processor may both fall under execution optimization if the central problem is coordinating stateful execution with hardware constraints.

\begin{table*}[t]
\small
\caption{Taxonomy of state-management problems.}
\label{tab:taxonomy}
\begin{tabular}{@{}p{0.18\textwidth}p{0.37\textwidth}p{0.37\textwidth}@{}}
\toprule
Dimension & Core question & Representative concerns \\
\midrule
Access and scheduling & How is shared state exposed, costed, and scheduled under concurrency? & hotspot diagnosis, conflict propagation, locality, topology, recovery, migration, prefix-aware sharing \\

Execution optimization & How does state interact with hardware and service objectives? & placement, compression, KV-cache paging, prefill/decode coupling, approximation, latency-energy-quality trade-offs \\

Evolution and reuse & How is state updated, retained, and reused over time? & online updates, memory budgets, cache growth and reclamation, sample selection, retriever drift, knowledge maintenance \\
\bottomrule
\end{tabular}
\end{table*}

\subsubsection{A Reusable Analysis Tuple}

We apply the same five-field tuple whenever a system line is summarized: what is the \emph{state object}; what \emph{control surface} does the runtime expose; what \emph{coupling path} makes that surface important; what \emph{evaluation boundary} is optimized; and what \emph{unresolved contract} remains. The tuple keeps the comparison stable across domains and prevents the survey from becoming disconnected implementation summaries.

\smallskip\noindent\textbf{Comparison unit.}
Each paper under review is normalized to the same five-field comparison tuple
\[
\mathcal{Q}=(\sigma,\;\kappa,\;\chi,\;\beta,\;\gamma)
\]
where \(\sigma\) is the \emph{state object} governed by the mechanism,
\(\kappa\) is the \emph{control surface} the runtime exposes over \(\sigma\),
\(\chi\) is the \emph{coupling path} connecting local decisions on \(\sigma\) to system-wide behavior,
\(\beta\) is the \emph{evaluation boundary} (the service-level property optimized or bounded), and
\(\gamma\) is the \emph{unresolved contract} left after the mechanism is applied.
Two papers are directly comparable when their tuples differ in at most one field; they address fundamentally different control problems when they differ in \(\sigma\) or \(\beta\).

The tuple is meant to be reusable rather than merely descriptive. When reading a new paper, the first pass identifies the concrete state object and its lifetime: an object that exists only inside one kernel invocation does not create the same control problem as a KV page, mutable shard, progress frontier, replay buffer, or retriever memory that survives across requests or failures. The second pass identifies the control surface and coupling path: a mechanism is relevant only if the runtime can act on the object and if that action changes an end-to-end property through contention, locality, memory pressure, quality loss, replay, or future reuse. The final pass asks whether the evaluation boundary matches the coupling path. A scheduler that claims to manage lifecycle state but reports only steady-state throughput leaves a different gap from one that tests burst, drift, or recovery but omits ownership semantics.

This procedure also clarifies how future work can extend the survey without rewriting the taxonomy. A new system can be inserted by changing one tuple field at a time. If the state object is familiar but the control surface is new, the contribution is a mechanism refinement; if the control surface is familiar but the evaluation boundary changes, the contribution is a transfer result; if the unresolved contract changes, the contribution is architectural. This discipline is especially useful for fast-moving LLM and retrieval systems, where papers often introduce new names for related cache, memory, or planner objects. Normalizing them through the tuple makes it harder to over-credit a local optimization and easier to see when a paper has actually exposed a new runtime boundary.

\begin{table*}[t]
\small
\caption{Reusable analysis tuple for extending the survey.}
\label{tab:scaffold}
\begin{tabular}{@{}p{0.18\textwidth}p{0.37\textwidth}p{0.37\textwidth}@{}}
\toprule
Question & What to extract from each paper & Typical answers in this survey \\
\midrule
State object & Which runtime object persists across tasks, requests, or time? & windows, indexes, logs, dictionaries, KV caches, retriever memories, coresets \\

Control surface & What decision does the system make over that object? & placement, batching, migration, paging, admission, retention, reuse orchestration \\

Coupling path & Why does this state decision propagate to end-to-end behavior? & contention, locality, memory pressure, phase asymmetry, drift, quality-loss amplification \\

Evaluation boundary & Which system-level property is optimized or bounded? & throughput, p99 latency, recovery time, energy, bounded error, forgetting, reuse quality \\

Unresolved contract & What is still missing after the paper's contribution? & weak observability, no closed loop, limited portability, incomplete lifecycle control, no long-horizon evaluation \\
\bottomrule
\end{tabular}
\end{table*}

\subsubsection{Representative System Classes}

The state-management lens cuts across at least five recurring system classes: high-throughput streaming and transactional dataflow, hardware-conscious analytics, LLM serving and memory-bound inference, continual learning and adaptive services, and memory-centric retrieval or agentic inference. They matter because they expose structurally different failure modes, yet the same runtime ideas often reappear under different names:
\begin{itemize}[leftmargin=*]
\item \textbf{Streaming and transactional dataflow} fail by contention amplification, migration cost, or slow recovery when state ownership and progress semantics are not co-designed.
\item \textbf{Hardware-conscious analytics} fail by misplacing state relative to device topology or by violating quality boundaries when using approximate methods.
\item \textbf{LLM serving systems} fail when KV-cache growth, fragmentation, or prefix-sharing policy turns memory into the primary throughput bottleneck.
\item \textbf{Continual-learning systems} fail by exhausting retention budgets or forgetting useful history when admission and replay policies remain static.
\item \textbf{Memory-centric retrieval and agentic systems} fail by update drift, unstable retrieval, or uncontrolled growth in reusable state.
\end{itemize}

\section{Core Dimensions of State Management}
\label{sec:core-dimensions}

The recurring control problems can be grouped into three coupled dimensions: state access and scheduling, state-aware execution, and state evolution and reuse. A local gain in one dimension often returns as a downstream constraint on the other two. We keep the discussion grounded with three running examples: (A) a streaming fraud detector under hotspot skew, (B) a multi-tenant LLM assistant under KV-memory pressure, and (C) a RAG knowledge service under corpus drift.

\subsection{State Access and Scheduling}

\subsubsection{From Invisible Contention to Observable State Access}

In running example A, a fraud-detection pipeline suddenly receives a skewed burst after a campaign launch. The model logic has not changed, but a small set of hot keys now serializes queue and lock paths. The practical question is therefore not ``can the operator run faster,'' but ``can the runtime detect and reshape shared access before the skew turns into replay and migration debt.'' As multicore stream and event systems scaled, this issue became hard to ignore: operator logic was often not the main bottleneck, because hidden contention in shared access paths could dominate end-to-end behavior. Studies of multicore stream processing and complex event processing made that point concrete. They showed that shared queues, synchronization patterns, fine-grained operator interactions, and sub-computation sharing can all limit scalability when the runtime cannot see where interference is forming~\cite{zhang2017revisiting,zhang2017mqo}. State management begins with observability, but access metrics are useful only if they preserve enough structural context for the runtime to decide whether the right response is repartitioning, migration, admission throttling, or a change in execution grain.

Earlier stream systems exposed the same lesson at different boundaries. Aurora and Borealis made operator graphs and adaptation policies runtime concerns~\cite{abadi2003aurora,abadi2005borealis}; MillWheel, Naiad, and Dataflow made time and completion visible~\cite{akidau2013millwheel,murray2013naiad,akidau2015dataflow}; Trill, Differential Dataflow, Flink, and StreamCloud connected progress tracking, incremental maintenance, elasticity, and fault handling to the same state boundary~\cite{chandramouli2014trill,murray2013differential,carbone2015flink,gulisano2012streamcloud}. Together they define a ladder of control surfaces: adaptation makes placement explicit, progress semantics determine when state may advance, and ownership transfer determines where it may move.

\subsubsection{Cost Modeling Under Locality and Topology}

Returning to Example~A, once the fraud-detection runtime has identified a hotspot, it must estimate the cost of repartitioning or migrating the hot keys under the current NUMA topology, checkpoint state, and downstream dependencies. That is a cost-modeling problem rather than a detection problem alone. Once access patterns become visible, the harder task is to price them correctly under skew and topology. Topology-aware and concurrent stateful streaming systems suggest that hotspot placement, NUMA distance, synchronization design, state partitioning, and task assignment have to be considered together; locality decisions that ignore any one of these factors can wipe out the gains from parallelization~\cite{zhang2019briskstream,zhang2020concurrent}.

Taken together, these results point to a broader systems lesson: locality does not come from state layout alone, but from its interaction with scheduling and workload structure. Similar patterns appear in Spacker's migration substrate~\cite{mao2025spacker}, interval-join systems with shared indexes and ordering constraints~\cite{zhang2023openmldb}, deployment planners such as MaveriQ~\cite{liakopoulos2025maveriq}, and intra-window join studies in which execution style, join method, and partitioning reshape skew and latency trade-offs~\cite{zhang2021intrawindowjoin}. Across these works, it is more accurate to think in terms of an access-cost surface shaped by contention and hardware placement than in terms of a static data structure.

Partitioning and synchronization choices also shape later recovery behavior by determining checkpoint grain, replay skew, and migration debt under disturbance. Topology-aware placement, concurrency control, and state-transfer design should therefore be considered together~\cite{zhang2019briskstream,zhang2020concurrent,carbone2015asynchronous,zhao2024recovery,mao2025spacker}. Access management cannot stop at hotspot detection: a runtime that cannot account for the recovery consequences of a placement decision addresses only part of the state-management problem.

\subsubsection{Runtime Control, Recovery, and Stateful Governance}

Example~A again: once the fraud detector's access-cost model decides that migration is warranted, the runtime must execute the move without losing in-flight transactions or stalling downstream operators. That execution requires runtime control and stateful governance: the runtime must know who owns a state shard, where replay may safely resume, and when the old owner can be reclaimed. Effective access management therefore extends to runtime control. MorphStream and related transactional-stream systems treat execution as a scheduling problem over stateful dependencies; Flink's asynchronous snapshots and fast parallel recovery show that checkpoint barriers, in-flight records, replay metadata, and access control belong to the same runtime path rather than to separate steady-state and failure-time subsystems~\cite{mao2023morphstream,zhao2025morphstream,carbone2015asynchronous,zhao2024recovery}. Earlier elasticity work and later migration substrates reach the same conclusion from another angle: SEEP, operator-state management, Megaphone, and Spacker all show that operator ownership, transfer granularity, and recovery semantics are inseparable parts of one state-governance problem~\cite{fernandez2010seep,fernandez2013osm,nasir2019megaphone,mao2025spacker}.

Across these systems, the natural abstraction is an ownership-transfer contract: (i) fence or freeze updates, (ii) transfer a consistent shard snapshot plus progress metadata, (iii) hand off authority with idempotent replay boundaries, and (iv) reopen updates under a new owner. Megaphone, Spacker, Flink's asynchronous barriers, and fast-recovery designs each expose part of this contract~\cite{nasir2019megaphone,mao2025spacker,carbone2015asynchronous,zhao2024recovery}. Adjacent transactional and replicated systems make the missing fields clearer: Spanner, Calvin, Raft, and RAMCloud show that a reusable handoff protocol needs current authority, replay boundary, and reclamation condition for the old owner~\cite{corbett2012spanner,thomson2012calvin,ongaro2014raft,ousterhout2011ramcloud}. Most systems still encode these fields implicitly in separate migration and recovery paths.

Comparable governance patterns appear in workflow schedulers, checkpointing systems, large-model execution stacks, and serving-oriented schedulers, all of which make warm state, snapshot timing, KV residency, or cross-workload handoff visible to the scheduler~\cite{mahgoub2021sonic,mahgoub2022orion,zhao2025rtsfaas,wan2025bytecheckpoint,huang2025flowcheck,lian2025universal,lin2025weipipe,liu2025mario,hong2025sola,oh2024exegpt,wang2025sirius}. In these systems, recovery, rebalancing, serving admission, and utilization control can be viewed as closely coupled scheduling decisions over shared state. When migration, checkpointing, serving admission, and rescaling rely on separate ownership models, they can incur repeated coordination costs and brittle recovery semantics. Access governance would therefore benefit from a unified path linking observation, cost modeling, scheduling, recovery, and subsequent observation.

\subsubsection{Open Challenges in Access Management}

Despite the progress, the access dimension leaves five recurring control gaps:
\begin{itemize}[leftmargin=*]
\item \textbf{Conflict observability.} Most systems still rely on coarse metrics and do not expose state conflicts, ownership pressure, and migration debt as first-class runtime objects.
\item \textbf{Explicit handoff contracts.} Recovery, migration, and rescaling often maintain separate notions of authority, replay boundary, and reclamation condition.
\item \textbf{Lifecycle-closed serving control.} KV admission, eviction, transfer, and restoration are usually optimized separately even though they share one latency and memory boundary.
\item \textbf{Semantic shared state.} Intelligent-service runtimes increasingly create vector indexes, retriever memories, prefix-shared KV caches, and workflow context whose conflict patterns are semantic, dynamic, and cross-request.
\item \textbf{Compositional controllers.} Local access policies rarely compose cleanly with topology changes, disaggregation, colocated training, or memory middleware.
\end{itemize} These gaps push access modeling beyond local contention management toward ownership and lifecycle contracts over dynamic state objects.

\subsection{State-Aware Execution Optimization}

\subsubsection{Why Faster Kernels Do Not Guarantee Better Services}

Running example B highlights the same point in serving form: a multi-tenant assistant with mixed prompt lengths may have excellent single-kernel speed, yet still miss SLOs once KV residency, prefill/decode phase asymmetry, and adapter multiplexing collide. The runtime wins only when it governs where short-lived state lives and when that state moves. A common limitation is to equate stateful execution with faster kernels or higher operator throughput. Once state interacts with heterogeneous hardware, data movement and coordination often dominate raw compute speed, while latency, energy, and quality objectives can overturn microbenchmark gains.

Integrated CPU-GPU stream systems make this point concrete. Fine-grained window processing on integrated architectures showed that topology and data path design directly alter the feasible performance envelope, while co-running studies on the same class of hardware reached a similar conclusion from the broader angle of shared-resource interference~\cite{zhang2020finestream,zhang2016corun,zhang2016elastic}. These systems emphasize that execution gains emerge only when state organization, data movement, and device mapping are optimized together.

\subsubsection{Energy, Compression, and Stateful Data Paths}

Example~B surfaces a variant of this point: when the multi-tenant assistant offloads certain KV pages to CPU DRAM to relieve GPU memory pressure, the decision is not merely ``which pages to evict'' but which representation (full precision, quantized, or compressed) makes later restoration affordable without violating the p99~SLO. The system is navigating a stateful data path, not choosing a single compression level.

The CStream line deepens this theme by examining stateful compression on edge and asymmetric multicore devices~\cite{zeng2023cstreamicde,zeng2024cstream}. Compression dictionaries are state: their access pattern, update granularity, and task decomposition shape energy, latency, and compression ratio. The observation generalizes beyond compression. Whenever intermediate state crosses operators or requests, device selection, data movement, and representation choice become one coupled decision rather than independent knobs.

\subsubsection{KV-Cache Management as a State-Execution Problem}

Large language model serving makes this point unusually concrete. In these systems, the KV cache is not a minor implementation artifact. It is the dominant short-lived runtime state that grows with prompt and output length, mediates prefix reuse, and constrains feasible batch size. Once the runtime hits memory pressure, almost every higher-level serving policy becomes a decision about KV ownership, residency, transfer cost, or reclamation timing.

The serving literature can be read through six control surfaces: orchestration, memory layout, representation, tiering, disaggregation, and structured sharing. Orchestration systems such as Clipper, Clockwork, Nexus, Orca, InferLine, and INFaaS show that queue state, batching opportunity, provisioning state, and stage imbalance are themselves runtime-managed controls~\cite{crankshaw2017clipper,gujarati2020clockwork,shen2019nexus,yu2022orca,crankshaw2020inferline,romero2021infaas}. vLLM and Sarathi-Serve then make memory layout and phase asymmetry explicit by treating KV memory as a paged state object whose footprint evolves differently in prefill and decode~\cite{kwon2023pagedattention,agrawal2024sarathi}. DistServe and Splitwise extend the same logic across worker boundaries: once prefill and decode are disaggregated, the control problem shifts from local placement alone to queue isolation, inter-stage transfer, and goodput-aware admission~\cite{zhong2024distserve,patel2024splitwise}.

Recent systems refine this lifecycle by changing representation, tiering, or sharing semantics. QServe, Oaken, Keyformer, Q-Hitter, and VQ-LLM treat the cache as a representation-control problem, trading fidelity, memory footprint, and future reuse value inside one execution boundary~\cite{lin2025qserve,kim2025oaken,adnan2024keyformer,dong2024qhitter,liu2025vqllm}. FlexGen, Neo, AQA, LServe, LoongServe, and LeanAttention expose complementary tiering and long-context surfaces in which bytes may be reduced, moved, or avoided through sparse and tiled access~\cite{sheng2023flexgen,jiang2025neo,kumar2025aqa,yang2025lserve,loongserve2024,sanovar2024leanattention}. Punica, AlpaServe, SGLang, and FastGen add multi-tenant, model-parallel, structured-program, and long-prompt variants, showing that adapter sharing, partition multiplexing, and phase transition control all belong to the same short-lived state lifecycle~\cite{chen2024punica,li2023alpaserve,zheng2023sglang,holmes2024fastgen}.

These mechanisms differ mainly in which boundary they protect first: queue stability, memory efficiency, or phase-local goodput. Their shared lesson is stronger than any single serving optimization: KV-cache management is a state-management problem because the cache persists across token steps, affects future scheduling, exposes reuse across requests, and demands explicit allocation, sharing, reclamation, placement, and transfer policies. Closely related work on compressed stream processing without decompression~\cite{zhang2023compressstreamdb} and data-aware adaptive compression~\cite{yu2025adaptivecompression} extends this view to database-style execution. Here, state-aware execution means choosing representation, operator placement, and access path jointly. The system's real unit of optimization becomes a stateful data path rather than an isolated operator.

\subsubsection{Approximation and Quality Boundaries}

Example~B sharpens this lesson further: if the serving runtime applies token-level KV eviction to fit more concurrent requests, it must reason not only about memory savings but also about the quality floor at which downstream answer utility degrades faster than throughput improves. The same quality-boundary logic that governs approximate stream joins governs KV-budget management.

Approximate execution adds output quality to the control boundary. PECJ~\cite{zeng2024pecj} addresses disorder in stream window joins by compensating for missing or late data using explicit error modeling. LibAMM~\cite{zeng2024libamm} shows that algorithm choice, dataset properties, and memory behavior jointly determine whether approximate matrix multiplication yields a stable efficiency-accuracy trade-off. FreeSAM~\cite{tang2025freesam} and LEAP~\cite{xu2024leap} reach similar conclusions for joins and video databases: approximation works only when quantity, quality, and system cost are modeled together. When intermediate or retained state influences correctness, the runtime needs an \emph{execution boundary}, not a single performance target.

\subsubsection{Toward Closed-Loop Execution Control}

Existing systems provide the ingredients for online execution control: topology-aware mapping~\cite{zhang2020finestream}, energy-aware decomposition~\cite{zeng2024cstream}, and quality-aware compensation~\cite{zeng2024pecj}. The missing piece is an integrated runtime that combines state observability with online execution control across heterogeneous processors and service objectives. The need is acute in inference stacks, where accelerator choice, memory transfer, approximation, and retrieval quality interact.

\begin{keybox}
\textbf{Execution gains emerge only when state organization, data movement, and device mapping are optimized together.} The unit of optimization is a \emph{stateful data path}, not an isolated operator or kernel. Stable improvements require representation, placement, and quality boundaries to be controlled together.
\end{keybox}

\subsection{State Evolution and Reuse}

\subsubsection{From Updates to Long-Horizon Memory}

Running example C makes the evolution problem tangible: a RAG service ingesting daily document churn can keep answering queries, but quality silently drifts unless refresh cadence, shard exposure, and compaction debt are co-managed. ``Update more'' and ``serve faster'' become conflicting objectives unless the runtime prices their future interaction. The third major dimension of \emph{state management} in parallel and distributed systems therefore concerns what happens after execution produces new information. In many older data systems, state updates were treated as maintenance; in dynamic AI services, updates are themselves central because the system must decide what to absorb, what to retain, what to discard, and how to expose retained state for future reuse.

Online adaptation and clustering systems make this shift visible. SentiStream~\cite{wu2023sentistream} frames model update as part of a coupled co-training loop, while MOStream~\cite{wang2024mostream} and empirical stream-clustering studies~\cite{wang2023dscstudy} expose summarization, windowing, refinement, and outlier control as interacting maintenance decisions. Online update is therefore not one mechanism but a coupled control problem.

\subsubsection{Memory Budgets, Retention, and Sample Selection}

Example~C illustrates the pressure directly: if the RAG service's vector index grows by 50k documents per day while its memory budget remains fixed, the runtime must decide which older embeddings to compress, demote, or evict---and those decisions shape which queries can still retrieve relevant context tomorrow. Retention is not a background garbage-collection task; it is a forward-looking service-quality control. Once updates become continuous, retention becomes equally important: FERRET~\cite{zhou2025ferret}, StreamFP~\cite{li2026streamfp}, and CANDOR-Bench~\cite{wang2026candor} show that update rate, memory budget, pipeline structure, and vector-index churn must be coordinated so the runtime can preserve future reuse value while controlling immediate update cost.

\subsubsection{Structured Memory and Dynamic Retrieval}

Retrieval-augmented and knowledge-enhanced systems make reuse explicit by turning external knowledge into a managed runtime substrate. The relevant substrate has four layers: retriever-facing memory, retrieval policy, index structure, and maintenance. FlowRAG shows that reuse quality decays as corpora shift without controlled retriever adaptation~\cite{zhang2026flowrag}; KELDAR and Global Planning show that planner-visible knowledge structures can become runtime-managed memory objects~\cite{li2024keldar,li2025global}.

Recent memory-centric systems begin to make the middleware layer itself explicit. Neuromem decomposes memory into insertion, consolidation, retrieval, and integration stages, making latency cost and accuracy decay visible as lifecycle properties~\cite{zhang2026neuromem}. SAGE points in a workflow-native direction by exposing pluggable vector-index layers, asynchronous update queues, declarative window operators, and dual-stream joins as modular stateful services~\cite{liu2026sage}. Placed alongside FlowRAG and Self-RAG, these lines suggest that long-horizon reasoning will eventually require memory middleware that externalizes both retrieval policy and memory-maintenance policy~\cite{zhang2026flowrag,asai2024selfrag}. The missing layer is a policy kernel that can decide when to consolidate, refresh, defer, or expose memory objects across heterogeneous reasoning workflows.

Modern retrieval pipelines expose concrete control surfaces. REALM, RAG, RETRO, and Atlas make document memory explicit~\cite{guu2020realm,lewis2020rag,borgeaud2022retro,izacard2022atlas}; Self-RAG, RAPTOR, Global Planning, and KELDAR add retrieval-policy and planning layers~\cite{asai2024selfrag,sarthi2024raptor,li2025global,li2024keldar}; DPR, ColBERT, PQ, HNSW, and HippoRAG expose vector-state representation, traversal, compression, and graph-structure choices~\cite{karpukhin2020dpr,khattab2020colbert,jegou2011pq,malkov2020hnsw,wang2024hipporag}. CANDOR-Bench and FlowRAG then make the maintenance boundary concrete: insertion pressure, deletion churn, rebuild cadence, and freshness policy can move an index off its recall-latency operating point before the retriever architecture changes~\cite{wang2026candor,zhang2026flowrag}.

The comparison separates four roles: memory systems make reuse visible, planning systems make traversal policy visible, vector-index systems make representation and maintenance debt visible, and dynamic-memory systems make freshness timing visible. What remains underdeveloped is middleware closure: few systems expose one policy kernel that governs retrieval policy, index maintenance, freshness exposure, and cross-workflow reuse together. KV caches are a useful boundary case because future decoding cost depends on whether short-lived state is retained, shared, compacted, or discarded at the right moment.

\subsubsection{Bridging Evolution with Access and Execution}

All three running examples converge here. In Example~A, aggressive hot-key migration (an access decision) reshapes future checkpoint cost (an evolution concern). In Example~B, token-level KV eviction (an execution decision) reduces future prefix-reuse value (an evolution loss). In Example~C, deferred index compaction (an evolution deferral) concentrates future query traffic onto a shrinking set of fresh shards (an access hotspot). Each case shows that one-axis optimization silently shifts debt onto the other two. A central implication is that evolution cannot be managed independently: continuous updates create future access hotspots, retention decisions alter memory footprint and execution cost, and retrieval quality affects downstream approximation choices. Memory evolution should be an online control loop coupled to access and execution.

\subsection{Quantitative Propagation Trace: A Worked Example}
\label{sec:propagation-trace}

To make the propagation claim falsifiable rather than merely aspirational, we trace one concrete decision through all three axes using Example~B (multi-tenant LLM serving) with representative numbers drawn from the vLLM~\cite{kwon2023pagedattention} and Sarathi-Serve~\cite{agrawal2024sarathi} literature. Under bursty multi-tenant load, an admission decision that looks locally beneficial can trigger KV eviction, destroy prefix reuse, increase TTFT, and create restoration debt that feeds back into future access pressure. In five-field terms, the state object is shared KV-cache state; the control surface is admission plus eviction; the coupling path is prefix-reuse destruction; the evaluation boundary is latency and goodput; and the unresolved contract is a lifecycle-closed controller over admit, expose/transfer, evict/reclaim, and subsequent re-materialization costs. The example shows why eviction debt belongs on the future-reuse ledger, not only on the instantaneous-throughput ledger.

Consider a multi-tenant LLM serving cluster with continuous batching, mixed prompt lengths, shared prefix reuse, and a fixed memory budget per decode pool. The runtime must decide whether to admit a new request whose prompt is long enough to threaten the current KV working set. At first glance, admitting the request looks locally beneficial: utilization stays high, and the prefill stage can begin immediately. The decision also changes the future legality and value of shared prefix state.

The coupling path unfolds in five steps. First, admission pressure rises when a long-prompt request arrives while the decode pool is already populated with shorter ongoing sessions. Second, local memory slack shrinks: admitting the request widens the active prefill footprint and may force reclamation of KV pages belonging to lower-priority or seemingly colder sessions. Third, reuse structure is damaged because some reclaimed pages would have supported prefix reuse or cheaper continuation for later requests from the same conversation family or structured workflow. Fourth, restoration debt appears when the next burst or follow-up call must repopulate state that was previously available cheaply, competing with live decode work. Fifth, the original gain can reverse: an admission decision that improved instantaneous occupancy can reduce goodput or inflate p99 latency after the coupling path fully unfolds.

\begin{keybox}
\textbf{Lifecycle debt is temporal before it is numerical.} The runtime can spend future reuse value to improve present occupancy, and that trade is invisible unless admission, reclamation, and restoration are evaluated on the same boundary.
\end{keybox}

The example isolates four serving-lifecycle boundaries. Admission-versus-restoration tradeoffs unfold over time, not at one instant. Prefix reuse is also a legality question: which shared state may be reclaimed? Burst-after-warmup and tenant-switch scenarios are necessary to expose lifecycle debt. The governed object is not one allocator event but a trajectory across admission, reclamation, and later restoration.

\section{Comparative Synthesis Across Domains}
\label{sec:comparative-synthesis}

The preceding sections established a mechanism inventory, but coverage alone does not show which lessons transfer. Normalizing the surveyed work to one comparison unit---state object, control surface, coupling path, evaluation boundary, and unresolved contract---separates transferable mechanisms from domain-local optimizations that look strong in isolation but do not compose across workloads and disturbance regimes.

\subsection{Comparative Criteria and Integration Principles}

Using that tuple, we group work by recurring control problem rather than by venue order or application label. A mature comparison contrasts multiple mechanisms under one service boundary and names the contract still missing. A paper that makes a seam visible is not yet equivalent to one that shows the seam remains governable under disturbance, multi-tenancy, and lifecycle interaction.

\subsection{Cross-Domain Comparative Synthesis}

\begin{figure*}[t]
\centering
\includegraphics[width=0.96\textwidth]{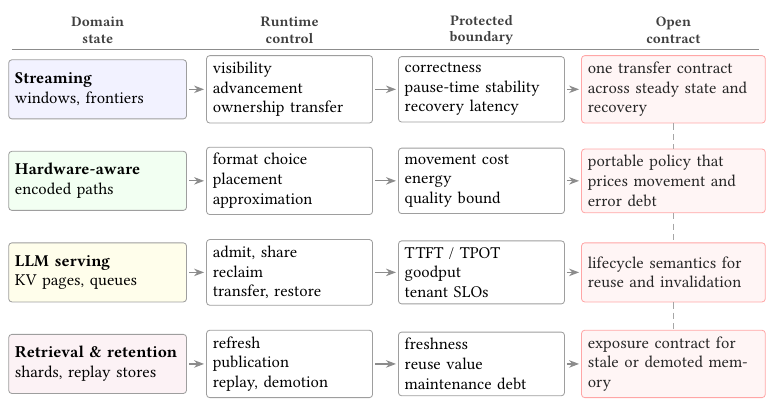}
\caption{Cross-domain comparison. The visual index summarizes the recurring seams discussed in the comparative synthesis.}
\label{fig:cross-domain-comparison}
\end{figure*}

Across the surveyed areas, four seams dominate: streaming centers on visibility and ownership, hardware-conscious execution on movement and quality boundaries, serving on short-lived lifecycle control, and retrieval plus retention on maintenance debt over longer horizons. Figure~\ref{fig:cross-domain-comparison} and Table~\ref{tab:comparison-guide} summarize these seams as a visual index and a compact reading key. The domains do not need to share terminology for the same control pattern to recur: each must identify a state object, expose a control surface, protect a service boundary, and name the contract that prevents local gains from composing cleanly.

\begin{table*}[t]
\small
\setlength{\tabcolsep}{3.5pt}
\caption{Compact cross-domain reading key.}
\label{tab:comparison-guide}
\begin{tabular}{@{}p{0.17\textwidth}p{0.17\textwidth}p{0.18\textwidth}p{0.17\textwidth}p{0.22\textwidth}@{}}
\toprule
Domain cluster & Dominant state object & Mature control surfaces & Boundary usually protected first & Persistent open seam \\
\midrule
Streaming and transactional dataflow & windows, frontiers, dependency state, movable shards & visibility, advancement, migration, replay, ownership transfer & correctness plus pause-time stability & one ownership-transfer contract across steady state, elasticity, and recovery \\

Hardware-conscious execution & encoded paths, compressed intermediates, tiered residency state & format choice, placement, decompression path, approximation steering & movement cost and quality boundary & portable control that prices movement and error debt together \\

LLM serving and short-lived lifecycle control & KV pages, phase queues, residency layers & admission, paging, sharing, disaggregation, transfer, restoration & TTFT, TPOT, and goodput under tight SLOs & lifecycle closure over reuse legality, isolation, restoration, ownership \\

Retrieval and retention governance & mutable shards, freshness metadata, replay stores, demotable memory objects & invocation, refresh, publication, replay, compression, demotion, budget shrink & freshness, reuse value, maintenance stability & policy-visible exposure contract for stale or demoted memory \\
\bottomrule
\end{tabular}
\end{table*}

\subsubsection{Streaming and Transactional Dataflow Systems}

Streaming and transactional dataflow systems are the clearest historical example of state management becoming a runtime control problem rather than a storage afterthought. Their dominant state objects are windows, progress metadata, dependency state, and movable operator shards; their mature controls govern visibility, advancement, placement, and replay~\cite{akidau2013millwheel,murray2013naiad,chandramouli2014trill,carbone2015flink,mao2023morphstream}. Progress semantics answer when a transition is safe, while ownership-transfer mechanisms answer where it may occur; strong systems need both.

The tradeoff is between finer-grained control and stronger correctness obligations. Topology-aware placement, elasticity, and fast recovery reduce pause time or contention only when the runtime can preserve progress fences, dual-ownership bounds, and replay legality across disturbance~\cite{zhang2019briskstream,nasir2019megaphone,zhao2024recovery,mao2025spacker}. The more durable target is a reusable ownership-transfer contract that spans steady state, reconfiguration, and recovery under one disturbance-aware policy, not another isolated migration heuristic.

\subsubsection{Hardware-Conscious and Approximation-Aware Stateful Execution}

Hardware-conscious execution shows that stateful efficiency depends on movement and representation choices as much as arithmetic intensity. Here the dominant objects are encoded data paths, compressed intermediates, and tiered-residency state; the mature controls govern format, placement, decompression path, and approximation budget~\cite{zhang2020finestream,zeng2024cstream,zhang2023compressstreamdb,yu2025adaptivecompression}. Approximation-aware systems add the boundary model that makes this domain comparable to the rest of the survey. Once a runtime saves work by sampling, compressing, or summarizing state, it must also govern update debt, reuse value, and error stability over time~\cite{agarwal2013blinkdb,goiri2015approxhadoop,hu2018streamapprox,chen2018incapprox,zeng2024pecj,zeng2024libamm}. Read together, these papers expose three recurring families: state-path shaping, budgeted approximation, and online boundary steering. Quality budgets become scheduling inputs, not post-hoc validation checks.

Portability across hardware and disturbance regimes remains weak. Capacity expansion, offload, and phase specialization help only when the runtime can also account for movement debt, metadata locality, and ownership overhead across devices and tiers~\cite{ren2020hmg,zhang2023g10,liu2025hermes,na2025flexinfer,kim2025paise,jang2024smartinfinity}. In comparative terms, this domain contributes a clear rule: state representation is a dynamic control variable, and its error budget belongs inside the protected service boundary.

\subsubsection{LLM Serving and Short-Lived Memory Lifecycles}

LLM serving is the clearest recent case where state management effectively \emph{is} the runtime. KV pages, prefix-sharing metadata, phase-skewed queues, and multiplexed adapter or model-residency layers often define the service envelope more directly than the model graph alone~\cite{kwon2023pagedattention,agrawal2024sarathi,li2023alpaserve,chen2024punica}. The local control problem is therefore not generic batching, but how a runtime governs short-lived state through the same canonical lifecycle used elsewhere in the survey: admit, place, mutate, expose or transfer, compact, and evict or reclaim. In serving, these stages take domain-specific forms: structural reuse and adapter-scoped sharing clarify when state may be exposed to a new request path; disaggregation and offload mechanisms govern transfer; and restoration debt appears as a downstream consequence of earlier evict or reclaim decisions. The literature is most comparable when grouped into a few control seams over one lifecycle: SLO-facing orchestration, KV allocation and sharing, prefill/decode phase control, and tenant or residency governance~\cite{crankshaw2017clipper,gujarati2020clockwork,kwon2023pagedattention,agrawal2024sarathi,zhong2024distserve,patel2024splitwise,chen2024punica,sheng2023slora,zhang2025jenga,yu2025prism}. Later predictive, structural-reuse, and temporal-lifecycle systems sharpen the same point from different angles: short-lived serving memory should be treated as a governed lifecycle rather than as a passive buffer pool~\cite{windserve2025,guo2025gllm,gong2025pastfuture,mo2025hetis,su2025seesaw,jiang2025thunderserve,gim2024promptcache,pan2025marconi,yu2025iccache,yao2025cacheblend,gim2025pie,yu2025pensieve,qin2025mooncake,wu2026fastserve,sun2024llumnix,agarwal2026symphony,lou2026hydraserve,fu2024serverlessllm}.

The first seam is predictive control. Systems such as WindServe, GLLM, Past-Future, HETIS, Seesaw, and ThunderServe do not merely predict load; they predict different future debts and bind those predictions to different actuation points~\cite{windserve2025,guo2025gllm,gong2025pastfuture,mo2025hetis,su2025seesaw,jiang2025thunderserve}. Some forecasts protect the prefill/decode balance, some anticipate future KV peaks, and others expose heterogeneous placement pressure before it becomes queue amplification. The systems lesson is therefore narrower and stronger than ``prediction helps.'' A prediction is useful only when the runtime can say which later cost it is avoiding, which action surface can still change that cost, and when a wrong prediction will be charged back to the same SLO boundary.

The second seam is reuse legality. Structural-reuse systems such as PromptCache, Marconi, ICCache, CacheBlend, and PIE make prompt segments, chunks, branches, or adapter-scoped state visible enough that the runtime can decide whether reuse is legal rather than merely cheap~\cite{gim2024promptcache,pan2025marconi,yu2025iccache,yao2025cacheblend,gim2025pie}. This differs from footprint reduction. Compression, paging, and offload mechanisms change the representation or location of state that has already been admitted as relevant; structural reuse decides whether the state object may be shared in the first place. Confusing these two control questions hides the isolation and invalidation contract behind memory-efficiency numbers.

The third seam is temporal lifecycle control. Pensieve, Mooncake, and Symphony operate at the session-restoration or disaggregated-cache boundary by preserving reuse value across turns and devices~\cite{yu2025pensieve,qin2025mooncake,agarwal2026symphony}. FastServe and LLMnix operate during active contention, where preemption, migration, and priority redistribute latency debt over live request state~\cite{wu2026fastserve,sun2024llumnix}. HydraServe and ServerlessLLM operate earlier at the bootstrap-residency boundary, where loading and tiered initialization determine which state can become reusable later~\cite{lou2026hydraserve,fu2024serverlessllm}. Grouped this way, the serving literature still fits the survey's canonical lifecycle, but with serving-specific emphasis on reuse legality inside expose/transfer and on restoration debt after evict/reclaim, rather than on a separate domain-exclusive stage list.

This distinction also clarifies why local serving metrics are often misleading. Allocator efficiency can improve while restoration cost grows; prefill/decode separation can improve goodput while transfer ordering becomes the next bottleneck; prefix reuse can reduce compute while weakening tenant isolation or invalidation semantics. A mature serving runtime therefore needs to state which reuse is legal, which state may be reclaimed, when restoration debt is charged, and whether the SLO boundary is request-local, session-local, or tenant-wide. The strongest serving papers expose at least part of this contract; the open problem is to make the whole lifecycle portable across batching, disaggregation, adapter multiplexing, and speculative execution.

The unresolved issue is contractual. Current systems expose local interference more clearly than they reconcile prediction, reuse legality, tenant isolation, and restoration under one service boundary. Practical runtimes still lack portable semantics for when KV fragments may be shared, transferred, restored, or invalidated across continuous batching, adapter multiplexing, speculative branches, and disaggregated decode paths.

\subsubsection{Retrieval Memory and Vector-Index State Layers}

Retrieval systems are easiest to misread as a sequence of better retrievers, but the more durable systems lesson is about memory governance. Once retrieval moved from static lexical indexes to dense and hybrid memory substrates, invocation, refresh, exposure, and maintenance became coupled runtime decisions rather than offline preprocessing details~\cite{karpukhin2020dpr,khattab2020colbert,guu2020realm,lewis2020rag}. The local control problem is therefore how to keep a living memory layer queryable without hiding freshness debt, rebuild debt, or exposure risk. The literature is most comparable as a trigger hierarchy: invocation policy, index maintenance, safe publication, and placement or movement policy~\cite{guu2020realm,lewis2020rag,borgeaud2022retro,izacard2022atlas,asai2024selfrag,jegou2011pq,malkov2020hnsw,singh2021freshdiskann,xu2023spfresh,xu2025place,yue2025marco,wang2021milvus,guo2022manu,chen2024singlestorev,zhang2024rummy,quinn2025accelerating}. Planner-mediated and continuous-maintenance systems make the missing coupling explicit: one controller can preserve short-run latency by querying stale memory while another silently accumulates rebuild debt~\cite{zhang2026flowrag,wang2026candor,liu2025heterrag}.

The trigger hierarchy is worth making explicit. Request-time invocation decides whether external memory should be consulted at all; index-time maintenance decides whether insertion, deletion, repair, or compaction has pushed the memory substrate away from its intended recall-latency operating point; publication-time exposure decides whether a partially refreshed shard may serve internal traffic, restricted traffic, or general requests; placement-time residency decides whether movement is justified by downstream quality gain rather than lookup throughput alone. These triggers often live in different modules, but the user-facing answer quality depends on their ordering. A retriever can appear accurate in a static benchmark while its publication lag, deletion debt, or rebuild backlog makes the live memory layer unsafe to expose.

Retrieval systems still lack a compositional trigger policy that orders stale-memory tolerance, rebuild triggers, shard-exposure legality, and movement under one service boundary. Without that contract, reported gains remain hard to compare across request-time quality, maintenance stability, exposure safety, and placement efficiency. Retention systems arrive at the same abstraction from a different timescale. Retrieval asks when partially stale memory is still safe to expose; retention asks when compressed, demoted, or budget-constrained memory is still safe to preserve and replay. In both cases, the systems question is when the runtime can still treat memory as legally useful.

\subsubsection{Continual Learning and Retention Governance}

Continual-learning systems matter here not mainly because they report forgetting curves, but because they provide a runtime language for bounded retention. Once memory is scarce, the main question is which historical state deserves protection, admission, replay, compression, demotion, or eviction as future updates arrive~\cite{kirkpatrick2017ewc,rebuffi2017icarl,lopez2017gem,aljundi2019mir,buzzega2020der,prabhu2020gdumb,hayes2020remind}. The reusable comparison is therefore between lifecycle controls over one bounded store rather than between named replay heuristics alone.

Protection methods decide what must not drift; admission methods decide what enters the scarce retained set; replay methods decide when stored state should re-enter the update path; compression methods decide whether a cheaper representation preserves enough future utility; and systems such as Ferret and StreamFP make budget elasticity and future-use ranking explicit runtime concerns~\cite{kirkpatrick2017ewc,rebuffi2017icarl,lopez2017gem,chaudhry2019agem,prabhu2020gdumb,hayes2020remind,zhou2025ferret,li2026streamfp}. This same logic extends to broader long-horizon reuse settings such as federated adaptation and tiered training-state storage~\cite{kumar2020quiver,khan2025flstore,zhan2025assyllm}. The missing abstraction is a retention-store contract that composes protect, admit, replay, compress, demote, and budget-shrink decisions under one service boundary, rather than treating end-task accuracy as the only judge~\cite{delange2021survey}.

The retrieval-retention comparison exposes a useful asymmetry. Retrieval systems usually know which memory object is queryable now but struggle to price the future cost of stale exposure and rebuild debt. Retention systems usually know the memory budget now but struggle to price the future value of a protected, compressed, or demoted item. In both cases, the runtime needs a demotion rule: when does memory remain fully usable, when does it become restricted, when is it only useful for replay or reconstruction, and when may it be dropped? Making those states explicit would let freshness, recall, forgetting, and budget pressure be evaluated as different views of the same lifecycle problem rather than as unrelated metrics.

\subsubsection{Comparative Mechanism Matrices}

The matrix makes the comparison unit explicit. It does not rank systems; it shows how the same five fields expose recurring seams across different state lifetimes and evaluation boundaries, and it keeps the comparison from collapsing into a list of domain-specific optimizations.

\begin{table*}[t]
\small
\setlength{\tabcolsep}{3.5pt}
\caption{Representative mechanism matrix for cross-domain comparison.}
\label{tab:representative-mechanism-matrix}
\begin{tabular}{@{}p{0.17\textwidth}p{0.18\textwidth}p{0.18\textwidth}p{0.16\textwidth}p{0.15\textwidth}@{}}
\toprule
Cluster & State object & Control surface & Evaluation boundary & Remaining gap \\
\midrule
Streaming visibility and progress~\cite{akidau2013millwheel,murray2013naiad,akidau2015dataflow} & windows, frontiers, progress metadata & visibility, advancement, buffering, replay timing & correctness plus latency & progress and movement remain weakly unified \\

Streaming migration and recovery~\cite{carbone2015asynchronous,nasir2019megaphone,zhao2024recovery,mao2025spacker} & movable shards, replay logs, ownership metadata & migration trigger, transfer grain, replay scheduling, barrier alignment & pause time, recovery time, overhead & ownership-transfer legality across steady state and disturbance \\

Hardware-conscious execution~\cite{zhang2020finestream,zeng2024cstream,zhang2023compressstreamdb} & encoded paths, local and remote representations & format choice, placement, decompression path, execution mapping & speed, energy, latency, movement cost & portability across devices and tiers \\

Approximation control~\cite{agarwal2013blinkdb,hu2018streamapprox,zeng2024pecj,zeng2024libamm} & samples, summaries, compensation state & sample sizing, compensation, approximation activation, quality steering & latency/error or cost/quality & service-level quality contracts \\

Serving paging and phase control~\cite{kwon2023pagedattention,agrawal2024sarathi,zhong2024distserve,patel2024splitwise} & KV pages, phase queues, transfer-ready state & paging, admission, prefill/decode separation, transfer scheduling & TTFT, TPOT, goodput, p99 latency & lifecycle closure over admit, expose/transfer, evict/reclaim, and restoration debt \\

Serving tenant and model multiplexing~\cite{chen2024punica,sheng2023slora,zhang2025jenga,yu2025prism} & adapters, heterogeneous caches, residency windows & batching, eviction, co-placement, reactivation, pooled ownership & throughput, SLO stability, GPU memory efficiency & lease and invalidation semantics across request, model, adapter, node boundaries \\

Retrieval index maintenance~\cite{singh2021freshdiskann,wang2021milvus,guo2022manu,xu2023spfresh,xu2025place} & mutable ANN shards, segment metadata, deletion and repair state & local repair, compaction, sealing, publication, routing & recall, freshness, maintenance cost, latency & cross-shard exposure contracts and unified triggers \\

Planner-mediated retrieval~\cite{asai2024selfrag,zhang2026flowrag,wang2026candor} & evolving retriever memory, freshness and confidence state & query invocation, refresh, rebuild trigger, exposure control & answer quality, freshness, lifecycle cost & explicit policy for stale memory \\
\bottomrule
\end{tabular}
\end{table*}

\begin{table*}[t]
\small
\setlength{\tabcolsep}{3.5pt}
\caption{Representative mechanism matrix for cross-domain comparison (continued).}
\begin{tabular}{@{}p{0.17\textwidth}p{0.18\textwidth}p{0.18\textwidth}p{0.16\textwidth}p{0.15\textwidth}@{}}
\toprule
Cluster & State object & Control surface & Evaluation boundary & Remaining gap \\
\midrule

Retention governance~\cite{kirkpatrick2017ewc,aljundi2019mir,prabhu2020gdumb,hayes2020remind,zhou2025ferret,li2026streamfp} & protected parameters, exemplar stores, replay buffers, budget traces & protect, admit, replay, compress, demote, budget shrink & forgetting, reuse value, store cost, disturbance response & retention-store contract with precedence \\
\bottomrule
\end{tabular}
\end{table*}

\subsection{Analytical Maturity}

Across access, execution, and evolution, local gains often disappear once effects propagate through the state lifecycle. Modern runtimes increasingly schedule around mutable state objects, but they still lack stable inter-layer contracts for observability, lifecycle coordination, policy composition, and long-horizon debt accounting~\cite{zhang2019briskstream,zeng2024cstream,kwon2023pagedattention,agrawal2024sarathi,zhou2025ferret,zhang2026flowrag}. Evaluation should therefore treat state management as closed-loop control: every experiment should name its state object, control surface, and service boundary before reporting gains, then test burst, skew, topology, or semantic disturbance so stationary replay does not hide lifecycle debt~\cite{kwon2023pagedattention,agrawal2024sarathi,mao2023morphstream,zhang2019briskstream}. The next stage of progress is architectural: observability linked to service drift, state-centric scheduling, recovery-aware lifecycle control, and explicit semantics for composing overlapping policies~\cite{zhang2019briskstream,kwon2023pagedattention,agrawal2024sarathi}.

\section{Design Implications for Stateful Runtimes}
\label{sec:design-implications}

The comparative synthesis points to a bounded set of design implications. It does not imply one settled architecture; it identifies constraints that stateful mechanisms must satisfy if they are to remain composable under disturbance: make the boundary observable, schedule around state rather than passive work, close the loop between action and accounting, and model the service boundary explicitly.

\subsection{A Design Space for Stateful Runtime Architectures}

\subsubsection{State Granularity: Record, Key, Segment, and Session}

A first architectural choice is the granularity at which state is represented and controlled. Record, key, segment/page, and session/conversation granularities trade flexibility, locality, and metadata overhead differently~\cite{zhang2019briskstream,kwon2023pagedattention,zheng2023sglang}. The earlier migration, paging, and retrieval examples all exposed this same choice indirectly: fine-grained objects improve local placement and reuse decisions, while coarser objects simplify ownership and recovery semantics at the cost of wasted movement or weaker selectivity.

\subsubsection{Control-Plane Coupling: Embedded, Sidecar, or Shared Service}

A second choice is where control logic lives. Embedded control is fast but fragmented, sidecar-style control is modular but delayed, and shared services are consistent but can bottleneck if their APIs are too fine-grained~\cite{zhang2020concurrent,kwon2023pagedattention,zhang2026flowrag}. This axis follows directly from the surveyed control surfaces: stream schedulers often embed legality into the operator path, serving systems increasingly split decision and actuation across allocators or disaggregated pools, and retrieval middleware starts to externalize policy into shared services that outlive any single request path.

\subsubsection{Lifecycle Management: Admit, Place, Mutate, Expose/Transfer, Compact, Evict/Reclaim}

A robust stateful runtime should make six lifecycle stages explicit: admit, place, mutate, expose or transfer, compact, and evict or reclaim. If any stage is implicit, local optimizations tend to create lifecycle debt under disturbance~\cite{sheng2023slora,zhang2025jenga,yu2025prism}. That lesson is visible across domains: migration and replay expose admit, place, and expose/transfer semantics in streaming; KV management exposes admit, expose/transfer, and evict/reclaim semantics in serving; and dynamic retrieval and retention expose mutate, compact, and evict/reclaim semantics once freshness or budget pressure is no longer negligible. Domain-specific actions such as sharing, refresh, demotion, or restoration are best read as specializations of these canonical stages rather than as parallel lifecycle vocabularies.

\subsubsection{Service-Boundary Alignment: Local Wins vs End-to-End Wins}

A final architectural axis is alignment between local optimization targets and the user-facing boundary. Without that contract, queue-local improvements can hurt p99 latency, retrieval-local freshness can hide publication debt, and retention-local compression can damage future reuse value~\cite{zeng2024pecj,gujarati2020clockwork,asai2024selfrag}. The design question is therefore not only whether a local controller improves its own metric, but whether the metric is the boundary that should absorb the action's future cost.

\subsection{Failure Modes and Anti-Patterns}

Many failures are repeated contract omissions rather than isolated bugs. Table~\ref{tab:antipattern-repairs} summarizes recurring anti-patterns and the repairs that make them reviewable. Read as missing contracts, they expose hidden invalid actions, conflicting objectives, and deferred debt; the repair names the state object, legal action, protected boundary, and charged cost.

\begin{table}[t]
\small
\caption{Recurring anti-patterns and the contract repairs that make them reviewable.}
\label{tab:antipattern-repairs}
\begin{tabular}{@{}p{0.22\textwidth}p{0.26\textwidth}p{0.18\textwidth}p{0.26\textwidth}@{}}
\toprule
Anti-pattern & Missing contract & Primary blueprint layer & Expected design effect \\
\midrule
Telemetry-action mismatch & signals must name a legal actuation target and its delay or confidence semantics & observation and disturbance detection & dashboards become control inputs rather than passive monitoring \\

Policy Layering Without Conflict Semantics & hard constraints, soft objectives, and tie-breakers must be ordered explicitly & boundary model and decision kernel & overlapping controllers stop oscillating under burst or skew \\

One-shot benchmarking of non-stationary mechanisms & disturbance phases and re-stabilization windows must be part of the contract & observation plus safety and recovery envelope & reported gains remain meaningful outside stationary replay \\

Ignoring Write Amplification in Read-Optimized Designs & mutation, compaction, deletion repair, and maintenance debt must be charged to the same boundary as read gains & state catalog plus actuation interfaces & deferred rebuild or compaction cost becomes visible during design review \\

Treating State Ownership as Static & ownership transfer and no-eviction rules must survive migration, recovery, and restoration & actuation interfaces plus safety and recovery envelope & scale-out and fault handling share one transferable state model \\
\bottomrule
\end{tabular}
\end{table}

\subsection{A Contract-Oriented Blueprint for Stateful Runtimes}

The literature supports a compact contract view with five layers: typed state catalog, observation and disturbance detection, boundary model and decision kernel, actuation interfaces, and safety and recovery envelope. Systems that made these layers explicit made state more governable; systems that left them implicit moved debt into disturbance handling, recovery, or later reuse.

Stronger systems approximate a typed state catalog containing object name, granularity, mutability class, lifecycle stage, and quality sensitivity. Reliable migration, paging, refresh, or replay requires these fields to be explicit enough for multiple controllers to reason about the same object. The observation layer then combines steady metrics with disturbance indicators such as hotspot flips, phase imbalance, retrieval drift, and fragmentation spikes, so the runtime can distinguish ordinary load variation from a change that threatens the service boundary.

The decision kernel is where the state catalog and observation streams become boundary-aware choices. Whether heuristic, model-based, or learned, the comparative requirement is the same: decisions should remain analyzable and their precedence over competing objectives should be explicit. Actuation interfaces are most comparable when they are explicit and bounded; the surveyed mechanisms repeatedly distinguish between observation, decision, and actuation only when the actuation surface is narrow enough to describe. Finally, a stateful system remains analytically incomplete without a safety envelope: invariants that must hold during and after control actions. These guarantees depend on the fault model, whether crash-stop, crash-recovery with replay, network partition, or silent degradation~\cite{carbone2015asynchronous,chandy1985snapshot,mao2023morphstream,zhao2024recovery,corbett2012spanner,ongaro2014raft,wang2026candor,zhang2026flowrag}. In retrieval and retention settings, the key invariants are freshness exposure and retention demotion.

The blueprint is not intended to prescribe a single control-plane architecture. Its role is to make design reviews sharper. A runtime with an embedded scheduler can satisfy the blueprint if it names the objects it controls, exposes the signals that trigger action, and states the invariants that survive migration or recovery. A sidecar controller can also satisfy it, but only if the delay between observation and actuation is part of the contract. A shared memory service can satisfy it when its APIs expose not just allocation and release, but also lease duration, invalidation rules, restoration cost, and the boundary to which deferred debt is charged. The same review therefore applies to stream processors, serving stacks, and retrieval middleware even when their implementations look unrelated.

The blueprint also helps separate mechanism novelty from integration maturity. A new eviction policy, approximation rule, or retrieval trigger may be technically clever, but it remains a local policy until the system states how it composes with neighboring controllers. Integration maturity increases when the mechanism has a typed state object, a measurable trigger, a bounded action surface, an explicit precedence relation, and a recovery or safety condition. This is why the survey repeatedly emphasizes contracts over algorithms. Algorithms decide what to do under a local objective; contracts decide when an action is legal, which boundary absorbs its cost, and how later controllers observe the debt it created.

\subsection{Cross-Domain Case Studies}

The three running scenarios expose the same rule from different directions. Multi-tenant LLM serving under burst arrival shows how admit, expose/transfer, evict/reclaim, and restoration debt share one p99-facing boundary. Stateful stream reconfiguration shows that migration, progress, and replay need one transfer protocol. Retrieval under corpus drift shows that freshness, rebuild debt, and exposure legality must be co-governed. Local gains are not stable unless the runtime also names the debt they push into neighboring phases.

\section{Evaluation and Research Outlook}
\label{sec:evaluation-outlook}

The remaining question is evidentiary: how should the field recognize progress once state-management mechanisms interact across layers and over time? We use a disturbance-oriented vocabulary for judging stability, composability, and long-horizon reuse quality, because stationary runs rarely expose the debt created by access reshaping, memory reclamation, delayed publication, or retention under shrinking budgets.

\subsection{Evaluation Dimensions for Future Surveys and Systems}

An enduring weakness is evaluation fragmentation. Access papers often report throughput or contention, execution papers report speedup or energy, and evolution papers report quality or forgetting; each often leaves one adjacent boundary weakly audited, so local gains look stronger than they remain under disturbance or lifecycle debt. Four dimensions should be tracked:
\begin{itemize}[leftmargin=*]
\item \textbf{Steady-state efficiency:} throughput, amortized cost per useful result, and locality-sensitive resource use.
\item \textbf{Tail behavior:} percentile latency, burst amplification, and recovery delay.
\item \textbf{State quality:} bounded approximation error, retrieval quality under drift, and retention value under memory budgets.
\item \textbf{Operational sustainability:} migration cost, update overhead, and how quickly a control loop re-stabilizes after workload or hardware changes.
\end{itemize}
Systems that score well on only one dimension deliver local improvements rather than full state-management solutions.

These dimensions form an evidence ladder. A strong paper makes its state object and control boundary explicit, demonstrates local gains, shows that they survive disturbance, and tests whether they remain meaningful when composed with neighboring controllers or maintenance tasks. The benchmark unit should shift from one-shot runs to multi-phase disturbance traces: warmup, burst, reconfiguration, drift, and long-horizon maintenance belong in one connected experiment.

\begin{table}[t]
\small
\caption{Representative disturbance-oriented evaluation matrix.}
\label{tab:disturbance-evaluation-matrix}
\begin{tabular}{@{}p{0.18\textwidth}p{0.24\textwidth}p{0.22\textwidth}p{0.28\textwidth}@{}}
\toprule
Cluster & Recommended disturbance tests & Common blind spot & Research extension \\
\midrule
State access under skew & hotspot flips, burst tenant switch, rebalance under live joins & separating locality and recovery experiments & connect placement choices to future replay and migration debt \\

Serving lifecycle control & mixed prompt lengths, burst after warmup, reclaim-and-restore stress, tenant churn & allocator-only or phase-only evaluation & measure admit, expose/transfer, evict/reclaim, and restoration debt on one SLO boundary \\

Retrieval under drift & continuous updates, deletion bursts, stale-data injection, compaction lag & one-shot index assumption & add exposure-safety and maintenance-debt metrics \\

Retention under budget elasticity & abrupt budget contraction, asynchronous updates, background compaction & reporting accuracy alone & define retention-store metrics and disturbance protocols \\

Cross-domain middleware & concurrent serving, retrieval, and update pressure & layer-by-layer optimization & compare policy-composition and precedence models \\
\bottomrule
\end{tabular}
\end{table}

\subsection{Boundary Conditions and Threats to Transfer}

The comparison above is deliberately mechanism-centered, but the mechanisms do not transfer automatically. A scheduler that works for keyed stream partitions may fail for KV-cache pages because the ownership interval is shorter, the reuse value is more speculative, and the protected boundary is tail latency rather than exactly-once progress. Conversely, a serving allocator that reacts well to phase imbalance may offer little guidance for retrieval maintenance, where the costly action is not immediate eviction but publication, repair, or rebuild under freshness constraints. The transferable unit is therefore not the policy itself, but the contract it implies: which object is governed, which actions are legal, which service boundary absorbs the cost, and which future debt is recorded.

Three threats are especially easy to miss when reading across domains. First, state identity is often less stable than the paper's terminology suggests. A stream shard, a KV page, an ANN segment, and an exemplar buffer can all be called ``state,'' but their mutability, ownership, and invalidation semantics differ sharply. Second, disturbance horizons differ. Stream recovery may unfold over seconds, serving reclamation over milliseconds, retriever rebuilds over minutes or hours, and continual-retention effects over many update rounds; an evaluation that is long enough for one domain can still be too short for another. Third, quality boundaries are not interchangeable. Bounded approximation error, answer freshness, p99 latency, and forgetting are all service properties, but they tolerate different kinds of temporary debt. These differences explain why the survey emphasizes tuples and contracts rather than mechanism names.

Transfer claims are most convincing when they include at least one explicit non-transfer case. A mechanism that generalizes from stream migration to KV transfer, for example, should say which field of the tuple changes when progress fences become tenant SLOs, when shard ownership becomes page residency, or when replay legality becomes restoration priority. Likewise, a retrieval-maintenance policy that borrows from retention should identify whether it is transferring an admission rule, a demotion rule, or only an evaluation pattern. This negative boundary is not a weakness in the contribution; it is what makes the contribution reviewable. Without it, cross-domain synthesis can overstate similarity by reusing words such as locality, freshness, or memory while hiding the different legality conditions under which those words operate.

For future systems papers, the practical implication is to state the transfer claim narrowly. A contribution is strongest when it names the state class it governs, reports the disturbance horizon over which the result holds, and exposes at least one adjacent boundary where the mechanism could fail. For example, a KV-cache policy should report not only memory savings and decode latency, but also whether reclamation harms later prefix reuse or restoration cost; a retrieval-maintenance policy should report not only recall and query latency, but also whether delayed compaction changes exposure safety or update backlog. Such reporting does not make every paper larger; it makes the claim reviewable. It also prevents the field from treating local throughput, local accuracy, and local freshness as interchangeable evidence for state-management progress.

\subsection{An Integrated Research Agenda}

The design implications above distill into four pressure points: cross-domain observability contracts, state-centric scheduling on heterogeneous hardware, memory middleware for dynamic AI services, and end-to-end evaluation beyond local speedups~\cite{zhang2019briskstream,zhang2020finestream,kwon2023pagedattention,sheng2023flexgen,cong2026mico,zhang2026neuromem,liu2026sage}. Access mechanisms remain brittle when visibility and ownership semantics are implicit; execution mechanisms lose composability when movement cost, representation, and service boundary are optimized separately; evolution mechanisms remain unstable when refresh, retention, and exposure decisions hide future debt. The common research target is a shared control layer that makes state identity, movement cost, exposure legality, and lifecycle debt composable under one service boundary.

\section{Conclusion}
\label{sec:conclusion}

Efficient state management is best understood as one coupled control loop over access, execution, and evolution. Comparing streaming, serving, retrieval, and retention through the same five-field tuple shows that the recurring bottleneck is not the absence of clever local mechanisms, but the absence of contracts that keep those mechanisms compatible once state is shared across time, hardware, and service phases. Effective systems name the state object, bind local actions to a measurable service boundary, and charge deferred debt to the same boundary that received the immediate gain. Progress will therefore depend on runtimes that keep state identity, observability, and policy precedence stable under disturbance.

\clearpage

\appendix
\makeatletter
\gdef\shorttitle{Appendix}
\gdef\shortauthors{Zhang et al.}
\makeatother
\section*{Appendix}

This appendix provides supporting material for the main paper, including literature-boundary context, design checklists, case walkthroughs, and additional domain figures. Its bibliography includes the works cited directly in the appendix.

\section{Literature Boundary and Historical Context}

This appendix material clarifies which adjacent traditions contributed reusable state-management mechanisms and which mainly provide background or contrast.

\subsection{Streaming and Dataflow Context}

A broader historical reading helps explain why visibility, ownership, and disturbance-aware control emerged as recurring seams. Early systems such as Aurora and Borealis already made continuous query plans, adaptation paths, and stateful operators explicit runtime artifacts rather than passive storage structures~\cite{abadi2003aurora,abadi2005borealis}. Later systems such as MillWheel, Naiad, and the Dataflow model shifted the comparison unit from raw operator throughput to controlled advancement over time, frontiers, and externally visible progress~\cite{akidau2013millwheel,murray2013naiad,akidau2015dataflow}. That shift matters because it reveals an early version of the survey's main thesis: performance-relevant state becomes governable only after the runtime exposes when it is safe to publish, buffer, replay, or move it.

The same line then broadened toward elasticity, migration, and recovery. Trill and Differential Dataflow show that partially materialized state can still be updated incrementally when timestamps and frontier metadata make legality explicit~\cite{chandramouli2014trill,murray2013differential}. Flink, StreamCloud, Megaphone, and later migration/recovery work add the deployment-facing extension: once operator state is movable, the runtime must govern not just update cost but also transfer legality, replay scope, and disturbance-time pause behavior~\cite{carbone2015flink,gulisano2012streamcloud,nasir2019megaphone,zhao2024recovery,mao2025spacker}. The broader landscape therefore helps justify why the survey treats visibility and ownership as coupled rather than separate concerns.

\subsection{Replicated, Transactional, and Memory-Resident Systems as Boundary References}

Classic transactional or replicated systems are not treated as a full co-equal cluster, but they remain important boundary references because they make legality contracts more explicit than many newer runtime papers do.

Spanner clarifies leaseholder continuity and writer legitimacy across distributed transactions~\cite{corbett2012spanner}. Calvin clarifies how deterministic ordering can turn replay from conflict resolution into continued execution~\cite{thomson2012calvin}. Raft and related replicated-log systems clarify term-scoped authority and old-owner retirement semantics~\cite{ongaro2014raft}. FaRM and RAMCloud contribute the perspective that reconstruction speed and serving ownership after failure are themselves runtime-managed state problems rather than mere recovery afterthoughts~\cite{dragojevic2014farm,ousterhout2011ramcloud}. These systems are not inserted into the survey to broaden venue coverage; they help specify what many modern stateful runtimes still leave implicit.

\subsection{Serving, Retrieval, and Memory-Governance Systems in Broader Perspective}

The main text intentionally centers the serving literature on short-lived lifecycle control. A broader perspective shows why this concentration happened.

Earlier serving systems such as Clipper, Clockwork, Nexus, InferLine, and INFaaS primarily exposed orchestration, provisioning, predictability, and admission as control surfaces around model invocation~\cite{crankshaw2017clipper,gujarati2020clockwork,shen2019nexus,crankshaw2020inferline,romero2021infaas}. Later LLM-oriented systems internalized more of the memory substrate itself: vLLM and related paging systems govern KV state directly; Sarathi, DistServe, Splitwise, and FastGen govern phase asymmetry; Punica, S-LoRA, Jenga, and Prism govern tenant multiplexing, adapter residency, heterogeneous cache layouts, and whole-model activation windows~\cite{kwon2023pagedattention,agrawal2024sarathi,zhong2024distserve,patel2024splitwise,holmes2024fastgen,chen2024punica,sheng2023slora,zhang2025jenga,yu2025prism}. This trajectory is exactly why the survey reads serving as a state-governance problem rather than as a mere systems-for-LLMs topic.

Retrieval and memory-governance systems underwent a parallel shift. Dense retrieval and late-interaction systems such as DPR and ColBERT originally foregrounded query-time quality versus cost over mostly offline-built memory substrates~\cite{karpukhin2020dpr,khattab2020colbert}.

Dynamic ANN systems, vector stores, and planner-mediated retrievers later made freshness, repair, shard exposure, and long-horizon memory evolution first-class runtime concerns~\cite{singh2021freshdiskann,wang2021milvus,guo2022manu,asai2024selfrag,zhang2026flowrag,wang2026candor}. The larger landscape therefore supports the claim that retrieval quality should increasingly be read as a lifecycle property rather than a static model property.

\subsection{Retention, Approximation, and Middleware as Adjacent Traditions}

Approximation and bounded-retention lines also matter because they make evaluation boundaries explicit in ways many mainstream runtime papers still do not. BlinkDB, ApproxHadoop, ApproxJoin, StreamApprox, and IncApprox show that speedup claims over sampled or summarized state only become comparable after the runtime states which error boundary is being protected~\cite{agarwal2013blinkdb,goiri2015approxhadoop,wang2018approxjoin,hu2018streamapprox,chen2018incapprox}.

PECJ and LibAMM extend that logic toward online control and explicit quality compensation~\cite{zeng2024pecj,zeng2024libamm}. Continual-retention lines such as EWC, GEM, MIR, GDumb, REMIND, and Ferret similarly show that bounded memory is not just a model problem; it creates operational questions over protection, admission, replay, compression, and budget shrinkage~\cite{kirkpatrick2017ewc,lopez2017gem,aljundi2019mir,prabhu2020gdumb,hayes2020remind,zhou2025ferret}.

Finally, recent memory-middleware proposals such as Neuromem and SAGE are useful not because they already solve the full problem, but because they expose lifecycle decomposition and workflow-visible actuation as first-class abstractions~\cite{zhang2026neuromem,liu2026sage}.

Together, these adjacent lines help motivate contract-oriented blueprint and middleware pressure-point arguments rather than one more domain-specific taxonomy.

\section{Design Checklist and Contract Skeleton}

This appendix material keeps the design-space, anti-pattern, blueprint, and boundary evidence behind the design conclusions. The organization follows recurring principles, architecture axes, contract repairs, and concrete examples.

\subsection{Design Checklist}

The checklist turns the comparative synthesis into concrete design questions. The table names recurring questions, and the following subsections unpack them into principles, axes, and repair patterns.

\begin{table}[H]
\small
\centering
\caption{Appendix guide for translating synthesis into runtime design choices.}
\label{tab:supp-design-guide}
\begin{tabular}{@{}p{0.25\textwidth}p{0.43\textwidth}p{0.22\textwidth}@{}}
\toprule
Recurring synthesis result & Design question & Concrete artifact to look for \\
\midrule
Observability gaps dominate failure analysis & Which signal names the state object, lifecycle stage, delay semantics, and legal action? & typed telemetry and disturbance indicators \\
Local speedups disappear under composition & Which boundary is hard, which objective is soft, and who wins on conflict? & precedence-aware decision kernel \\
Lifecycle debt accumulates across updates and movement & Where are admit, place, mutate, expose/transfer, compact, and evict/reclaim made explicit? & lifecycle pipeline with bounded actuation interfaces \\
Recovery reuses the same state objects as steady state & Which invariants survive migration, replay, rebuild, or restore? & safety envelope and ownership-transfer rules \\
\bottomrule
\end{tabular}
\end{table}

\subsection{Contract Principles}

Observation before optimization means the relevant state boundary must already be named in telemetry. State as a first-class scheduling object means runtimes should schedule around mutable state rather than passive tasks. Closed-loop rather than open-loop policies means action must feed back into accounting. Stable gains require boundary modeling means every gain must declare what is held fixed, what debt is deferred, and what disturbance invalidates the claim.

\subsection{Architecture Axes}

The architecture axes are state granularity, control-plane coupling, lifecycle management, and service-boundary alignment. Granularity ranges from record to session. Control logic may be embedded, sidecar-based, or shared. Lifecycle management should make admit, place, mutate, expose or transfer, compact, and evict or reclaim explicit. Boundary alignment asks whether local wins survive end-to-end service constraints.

\subsection{Anti-Pattern Repairs}

Telemetry-action mismatch is repaired by naming a legal control action for each signal. Policy Layering Without Conflict Semantics is repaired by ordering hard constraints, soft objectives, and tie-breakers. One-shot benchmarking is repaired by adding warmup, disturbance, adaptation, and re-stabilization. Ignoring Write Amplification in Read-Optimized Designs is repaired by charging mutation and repair debt to the same service boundary as read gains. Treating State Ownership as Static is repaired by transferable ownership and lease semantics.

\begin{figure*}[t]
\centering
\includegraphics[width=0.96\textwidth]{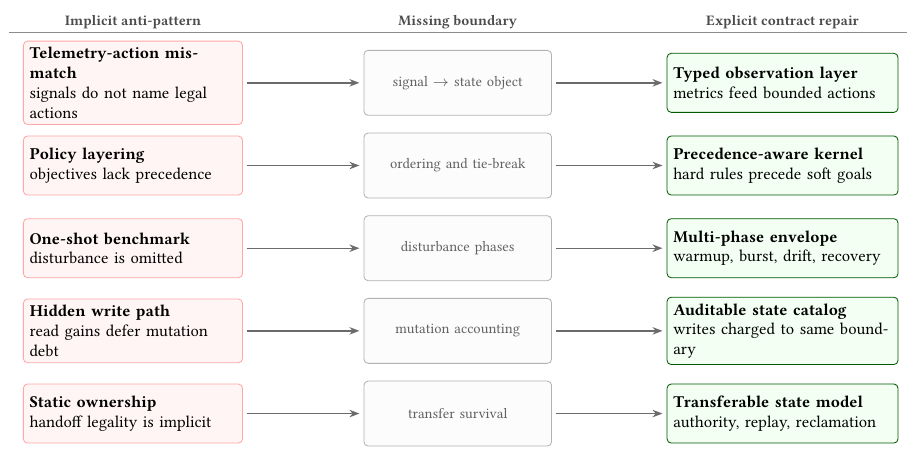}
\caption{Anti-pattern to contract repair map.}
\label{fig:supp-antipattern-map}
\end{figure*}

\subsection{Blueprint Contract Skeleton}

The blueprint is a conceptual reference model, not a monolithic runtime. Its five layers are typed state catalog, observation and disturbance detection, boundary model and decision kernel, actuation interfaces, and safety and recovery envelope.

The main paper includes the compact anti-pattern repair table. The appendix keeps the abstract policy skeleton and additional evidence anchors for readers who want the contract view in a more operational form.

\begin{table*}[t]
\small
\caption{Abstract policy skeleton for unified state governance.}
\label{tab:supp-policy-skeleton}
\begin{tabular}{@{}p{0.09\textwidth}p{0.83\textwidth}@{}}
\toprule
Step & Policy skeleton \\
\midrule
1 & \texttt{observe\_signals()} and update telemetry confidence for contention, queueing, local freshness, exposure freshness, and retention pressure. \\
2 & \texttt{identify\_state\_objects()} whose boundary is currently active: hot operator shards, KV pages, retrieval indexes, replay buffers, or demotion candidates. \\
3 & \texttt{check\_hard\_constraints()} for SLA, memory limit, recovery envelope, shard-exposure safety, and quality floor before any optimization action. \\
4 & \texttt{rank\_actions()} over admit, place, mutate, expose/transfer, compact, and evict/reclaim; domain-specific actions such as migrate, refresh, or demote are evaluated as specialized instances of those lifecycle stages using expected utility and maintenance debt. \\
5 & \texttt{apply\_bounded\_action()} with explicit rollback or retry semantics when telemetry is delayed or conflicting. \\
6 & \texttt{account\_outcome()} by charging the decision against tail latency, quality drift, effective budget loss, and long-horizon reuse value. \\
\bottomrule
\end{tabular}
\end{table*}

\subsection{Boundary Case Bridges}

These case bridges highlight three recurring boundary failures. Multi-tenant LLM serving under burst arrival shows how admit, expose/transfer, evict/reclaim, and restoration debt share one p99-facing boundary. Stateful stream reconfiguration shows that migration, progress, and replay need one transfer protocol. Retrieval under corpus drift shows that freshness, rebuild debt, and exposure legality must be co-governed.

\subsection{Hardware-Conscious Evidence Anchors}

The hardware-conscious cluster is most useful when read through three debts that the main paper could only summarize briefly. The first is \emph{representation debt}: approximate execution and compressed-state systems such as Anda, TurboAttention, QServe, ATOM, VAttention, FlashInfer, SampleAttention, QuantLLM, and related low-bit cache work all show that encoding decisions reshape memory traffic, kernel regularity, and quality loss simultaneously~\cite{fang2025anda,kang2025turboattention,lin2025qserve,zhao2024atom,prabhu2025vattention,ye2025flashinfer,zhu2024sampleattention,xia2024quantllm,dong2024qhitter}. Even systems that make representation look cheaper, such as FuseMax-style lines, do so by shifting complexity into calibration, selection, or error accounting rather than eliminating it~\cite{nayak2024fusemax,kwon2020fvm,jha2025hycache}.

The second is \emph{movement debt}. Systems such as Chimera, HeterInfer, LIA, HETIS, Mercury, WaferLLM, DRAMCache, Tiertune, MoE-lightweight offloading, PhoenixOS, JIT-Serve, MLP offload, Colloid, BeyondHotness, Optimus, Tigon, and MemStrata all show that placement helps only when the runtime can jointly price compute skew, cache residency, interconnect bandwidth, and cross-device movement~\cite{huang2025chimera,chen2025heteroinfer,kim2025lia,mo2025hetis,mercury2025rms,he2025waferllm,hong2024dramcache,lee2025tiertune,cao2025moelightning,wei2025phoenixos,zhang2026jitserve,maurya2025mlpoffload,vuppalapati2024colloid,liu2025beyondhotness,feng2025optimus,huang2025tigon,zhong2024memstrata}. Storage- and I/O-adjacent systems such as AutoScratch, PFSCK, PolarDB, CXL-based designs, Toleo, and VM-control make the same point from another boundary: once state crosses devices or tiers, metadata handling, legality, and transport cost become first-order control variables rather than implementation details~\cite{fu2024autoscratch,domingo2021pfsck,cao2020polardb,ji2025cxl,dong2024toleo,zheng2024vmcu,tabatabai2024fbmm}.

The third is \emph{resilience and repair debt}. Save, ChecknRun, CXL memory-protection work, Concealing, TurboAttention-style repair-aware approximation, SampleAttention, and PFSCK show that movement and representation cannot be priced independently from failure handling, degradation tolerance, or hidden repair cost~\cite{zheng2025save,eisenman2022checknrun,zhao2025equilibria,jin2024concealing,kang2025turboattention,zhu2024sampleattention,domingo2021pfsck}. Heterogeneity-aware execution and server-placement lines such as GMLake, Flex-MoE, Kamath-style overlap, Clap, DecDec, Coruscant, Zico, MEPipe, Clone, PowerInfer, PIMBA, QFactory, and MoC reinforce the same lesson from scheduling: overlap, offload, or recomputation only help if the control overhead and recovery debt stay below the service debt they are meant to avoid~\cite{guo2024gmlake,cao2025moelightning,kamath2025podattention,park2025clap,park2025decdec,joo2025coruscant,zico21,sun2025mepipe,tian2025clone,song2024powerinfer,kim2025pimba,zhang2025qfactory,cai2025moc}.

Dynamic input pruning with cache-aware masking, near-core decompression, and CXL-centered inference expose different actuators, yet each still ties placement and movement to an explicit runtime choice over state: what may be skipped, decompressed, kept near compute, or shifted into the memory hierarchy~\cite{federici2025dip,gerogiannis2025deca,gu2025cent}. Placement, movement, and state transformation therefore remain one coupled control problem even when the local actuator changes.

Grouped this way, the hardware cluster reads less like a long annex and more like a precise extension of the main claim. The exact mechanisms differ, but the comparative result is stable: once state crosses a hardware boundary, representation, movement, and resilience have to be priced together or the local optimization stops being trustworthy.

\section{Cross-Domain Case Walkthroughs}

These walkthroughs preserve the same five-field structure while making the deployment-shaped reasoning more concrete.

\subsection{Multi-Tenant LLM Serving Under Burst Arrival}

\paragraph{State object.}
The governed objects are paged KV state, prefix-sharing metadata, tenant-specific adapters, model-residency windows, and the restoration debt inherited by subsequent requests.

\paragraph{Control surface.}
The runtime controls admission, batch shaping, paging, prefill/decode separation, expose/transfer decisions, evict/reclaim decisions, and tenant-sensitive placement~\cite{kwon2023pagedattention,agrawal2024sarathi,zhong2024distserve,patel2024splitwise,chen2024punica,nian2026cacheflow}.

\paragraph{Coupling path.}
A locally beneficial admission may force evict/reclaim decisions that destroy future reuse, causing later re-materialization traffic and queue amplification. Once phase disaggregation is enabled, the next bottleneck may shift from raw memory capacity to expose/transfer ordering and the downstream cost of restoring discarded state.

\paragraph{Evaluation boundary.}
The meaningful boundary is TTFT, TPOT, p99 latency, and goodput over a multi-phase window rather than isolated allocator efficiency.

\paragraph{Remaining gap.}
The unresolved problem is still a unified contract that ranks admit, expose/transfer, evict/reclaim, offload, tenant isolation, and restoration debt against the same service boundary instead of letting each sub-policy optimize its own phase locally.

\subsection{Stateful Stream Processing Under Reconfiguration}

\paragraph{State object.}
The runtime governs keyed windows, operator shards, progress metadata, checkpoint boundaries, replay buffers, and migration ownership metadata.

\paragraph{Control surface.}
The active controls are migration grain, trigger policy, checkpoint cadence, replay scheduling, and ownership-transfer timing~\cite{akidau2013millwheel,carbone2015flink,nasir2019megaphone,zhao2024recovery,mao2025spacker}.

\paragraph{Coupling path.}
Reconfiguration changes locality and dual-ownership windows; recovery policy changes replay pressure and convergence cost. These effects feed back into future contention and scheduling stability rather than remaining isolated maintenance events.

\paragraph{Evaluation boundary.}
The right boundary combines pause time, throughput stability, correctness, and post-failure recovery behavior. Measuring migration and recovery separately hides the control loop that actually governs the state object.

\paragraph{Remaining gap.}
The missing piece is a disturbance-invariant ownership protocol that unifies visibility, transfer legality, and replay semantics across scale-out and recovery rather than re-deriving them per subsystem.

\subsection{Retrieval Memory Under Continuous Corpus Drift}

\paragraph{State object.}
The relevant objects are mutable ANN shards, segment-sealing metadata, shard-exposure state, planner-visible freshness signals, and long-horizon graph or hierarchy memory.

\paragraph{Control surface.}
The runtime controls incremental indexing, local repair, compaction, rebuild trigger, publication, stale-memory eviction, and planner-mediated retrieval invocation~\cite{singh2021freshdiskann,wang2021milvus,guo2022manu,asai2024selfrag,zhang2026flowrag,wang2026candor}.

\paragraph{Coupling path.}
Aggressive refresh improves freshness in the short term but can increase maintenance debt, publication lag, and answer instability if exposure, rebuild, and planner behavior are not coordinated.

\paragraph{Evaluation boundary.}
The relevant boundary is answer quality under drift plus freshness lag, query latency, maintenance cost, and exposure safety.

\paragraph{Remaining gap.}
The missing abstraction is an exposure-aware accounting layer that can combine recall drift, rebuild debt, planner confidence, and publication legality into one explicit trigger policy.

\section{Domain-Specific Figures and Evidence Anchors}

This appendix material gathers domain evidence under the same analytical frame as the main survey. It is organized around the pressure regions where additional mechanism detail most improves comparative clarity.

\subsection{Access-Scheduling Design Space}

The access-scheduling design-space visual matters because state access becomes a control problem when the runtime must decide what is visible, when a transition is legal, who owns the object now, and how disturbance changes the cost of the next action. The figure is a compact reminder that observability, locality, ownership transfer, and recovery are not separate subtopics but one coupled design space.

\begin{figure*}[t]
\centering
\includegraphics[width=0.94\textwidth]{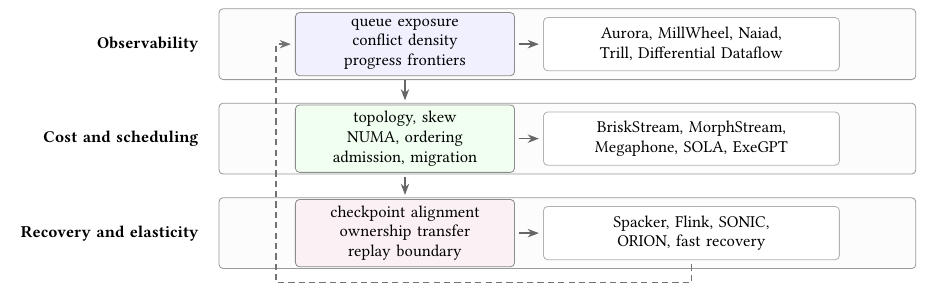}
\caption{Access-scheduling design space for locality, legality, transfer, and recovery.}
\label{fig:supp-access-designspace}
\end{figure*}

The comparative lesson behind the figure should remain explicit. Visibility mechanisms tell the runtime when a state transition is safe to expose; locality mechanisms tell it which placement keeps the next access affordable; ownership-transfer mechanisms tell it when relocation may occur without dual writers or replay ambiguity; recovery mechanisms tell it how those same promises survive disturbance. These are not four optional implementation concerns. They are the four places where access control becomes system-wide behavior rather than local bookkeeping.

The broader access-scheduling lineage is equally important for keeping the survey's boundary crisp. Classical consistency and concurrency systems such as TicToc, Percolator, and RACE show that timestamping, metadata layout, and remote coordination are already control surfaces over state, even before modern serving runtimes made the same pattern explicit~\cite{yu2016tictoc,peng2010percolator,zuo2021race}. Systems such as CheckFreq, ChecknRun, PipeFill, and the newer runtime-monitoring lines show that the access problem is not just placement, but also when the runtime may safely observe, checkpoint, and resume without invalidating later actions~\cite{mohan2021checkfreq,eisenman2022checknrun,arfeen2024pipefill,zhou2024neomem}. Tooling for dynamic reconfiguration and transfer, including Caerus, MORPH-style recovery, and queue-aware stream control, matters because it keeps the state-ownership question visible instead of treating movement as a mere implementation detail~\cite{zhang2021caerus,mao2023morphstream,fu2024autocheck,gao2025deck}.

The same lineage also includes systems that turn placement pressure into explicit runtime policy rather than hidden implementation cost. Resource-multiplexing and scheduler-centric systems such as G10, resource multiplexing, PFS-style reconfiguration, and workload-sensitive monitoring show that queueing, hot-state exposure, and placement are inseparable once the runtime manages shared state across stages or tenants~\cite{jiang2024megascale,he2025resourcemultiplexing,domingo2021pfsck,jiang2025shadowviews}. Transactional dataflow and distributed-storage systems such as Flash-style state flow, Cloud-style buffering, and multi-stage transfer control reinforce the same point: the access problem is always also a visibility problem and a handoff problem~\cite{fu2022sfs,fu2024autocheck,gao2025deck,li2020pegasus}. That is why the survey keeps access and scheduling as a first-class axis rather than folding it into generic runtime management.

Additional boundary references preserve the historical path. Transaction-processing systems such as TicToc, Percolator, and TiDB-style concurrency control show that control surfaces over mutable state have long existed in distributed data systems, even if they were not described in the same vocabulary as modern state management~\cite{yu2016tictoc,peng2010percolator}. Stream-processing systems such as Corun, Elastic stream systems, and BriskStream-style monitoring show that scheduling, state exposure, and reconfiguration were already intertwined before serving and retrieval brought those issues to the foreground~\cite{zhang2016corun,zhang2016elastic,zhang2019briskstream}. More recent monitoring and admission lines such as CheckFreq, ChecknRun, and PipeFill show that the runtime often needs an explicit contract for when observations can be trusted after disturbance~\cite{mohan2021checkfreq,eisenman2022checknrun,arfeen2024pipefill}.

The tail of the access line also includes a few systems that are easy to overlook because they sit between streaming, scheduling, and memory management. More recent scheduling and observability papers such as MOC, Optimum-style optimization, FS-MoE runtime control, and tree-aware transfer paths show that access control is still the main lever whenever a runtime has to reason about hidden contention and state movement together~\cite{cai2025moc,feng2025optimus,fsmoe25,wang2025wlbllm}. Likewise, Tigon, VM-control, and the updated memory-aware runtime lines keep the same lesson visible: once state crosses a device or tenancy boundary, legality and placement become one problem~\cite{huang2025tigon,zheng2024vmcu,tabatabai2024fbmm}. Even storage-to-compute bridges such as Snowflake-style disaggregation and mixed-access control belong in this lineage because they expose how latency, placement, and handoff interact~\cite{vuppalapati2020snowflake,zhou2024neomem}.

Two compact boundary references keep the access story complete. DRTMH-style transaction scheduling and the state-machine tradition remind us that legality over mutable state was a control problem long before modern runtimes named it that way~\cite{ren2015drtmh,schneider1990state}. EndMyth-style stream repair work shows the same issue from the failure side: access, recovery, and trust in observations are inseparable once the system can replay or resume~\cite{zamanian2017endmyth}.

\subsection{Serving Lifecycle Control}

The main paper now carries the core predictive, structural-reuse, and temporal-lifecycle comparison. The appendix keeps the supporting bridge references and the lifecycle figure so readers can trace how the broader serving literature maps onto the same control loop without repeating the main argument.

Additional serving papers refine the same boundary rather than forming a separate taxonomy. SplitZip, CacheFlow, and ThunderServe make transfer cost and cloud heterogeneity explicit; ChandGPT-style reuse systems, Jenga, Prism, and Aegaeon make multi-model or multi-tenant residency explicit; Spinfer, FlexGen, and related offload designs make memory pressure and phase placement explicit; HeterInfer, MoE-Lightning, and resource multiplexing make the hardware/topology asymmetry explicit~\cite{guo2026splitzip,nian2026cacheflow,jiang2025thunderserve,gao2025weaver,zhang2025jenga,yu2025prism,fan2025spinfer,sheng2023flexgen,chen2025heteroinfer,cao2025moelightning,he2025resourcemultiplexing}. Together, they keep the lifecycle argument from shrinking into a single KV-cache story.

Bridge references connect older and newer state-governance lines. PagedAttention, Orca, and AlpaServe explain why batching and scheduling were already state-sensitive before the latest disaggregation wave; SlimPipe, What-If Stragglers, and Pumped trace-based systems explain why scheduling debt becomes visible only after phase or hardware asymmetry is modeled; output-length prediction and cache-aware routing systems show why admission and future-memory forecasting belong in the same discussion as restoration and transfer~\cite{kwon2023pagedattention,yu2022orca,li2023alpaserve,liu2025slimpipe,lin2025whatifstragglers,gong2025pastfuture,jiang2025shadowviews}. That bridge keeps the survey's lineage coherent across older and newer serving lines.

CacheSlide and SpecInfer-style speculative reuse make legality and verification a precondition for cross-position or tree-structured reuse; CachedAttention and HCache expose restoration cost across conversation turns; and MELL, Libra, fairness-aware serving, LayerKV, and Prefill-Only show that GPU residency, request partitioning, fairness, layer-wise allocation, and prefill admission determine which cached state remains reusable later~\cite{cacheslide2025,miao2024specinfer,gao2025cachedattention,gao2025hcache,liu2025mell,ruan2026libra,sheng2024fairnessserving,xiong2024layerkv,du2025prefillonly}. These systems make reuse, restoration, contention, and initialization part of the same serving control problem rather than separate local optimizations.

Three more serving families sharpen the boundary. Application-centric serving such as Parrot shows that workflow-level state and prompt DAGs can dominate placement decisions even before the cache story starts~\cite{lin2024parrot}. Tree-shaped speculative and group-wise systems such as FastTree and InstAttention show that reuse and scheduling depend on structure in the request stream, not just on raw throughput targets~\cite{pan2025fasttree,pan2025instattention}. Decentralized and overlay-based systems such as PlanetServe and GLLM-adjacent coordination lines show that cache reuse, routing, and trust become intertwined once the scheduler is no longer centralized~\cite{fang2026planetserve}. The same cluster also includes systems such as Infinity-style eviction, Duplex-style cross-turn reuse, and world-load balancing variants, which are best understood as refinements of lifecycle control rather than new top-level subfields~\cite{infinigen2024,yun2024duplex,wang2025wlbllm}.

Two final serving references are useful as compact examples of the same lifecycle logic. Prefill-optimized training-serving crossover systems such as MEPipe show that the serving boundary is often governed by how pipeline slack is distributed, not just by the cache itself~\cite{sun2025mepipe}. Copy-aware or clone-style state movement lines show the same thing from a different angle: once request state is shareable, the runtime has to decide whether cloning, relocating, or reusing is cheaper than recomputing~\cite{tian2025clone}.

The remaining serving papers that have not yet been mentioned directly still fit the same lifecycle view. Multi-model and adapter-aware systems such as Aegaeon, DLoRA, Chameleon, and Toppings expose tenant-locality and residency pressure; cache and reuse systems such as PromptCache, CacheWild, and Droidspeak expose the boundary between prompt overlap and safe reuse; inference-control systems such as Alise, Atom, and Prism-like lineage expose the boundary between active scheduling and future debt; and cloud-disaggregation systems such as Mooncake and PRFaaS expose the difference between local phase control and cross-cluster ownership~\cite{xiang2025aegaeon,wu2024dlora,iliakopoulou2024chameleon,li2025toppings,gim2024promptcache,wang2025kvcachewild,liu2026droidspeak,zhao2024alise,zhao2024atom,yu2025prism,qin2026prfaas}. Those distinctions keep the lifecycle account tied to concrete control boundaries.

Across these systems, the serving cluster still fits the survey's canonical short-lived lifecycle: \emph{admit -> place -> mutate -> expose/transfer -> compact -> evict/reclaim}. The family-level differences matter because serving papers attach to different parts of that loop: structural reuse clarifies what may be exposed safely, disaggregation governs transfer, and restoration debt appears after evict/reclaim decisions have already been made.

\begin{figure*}[t]
\centering
\includegraphics[width=0.94\textwidth]{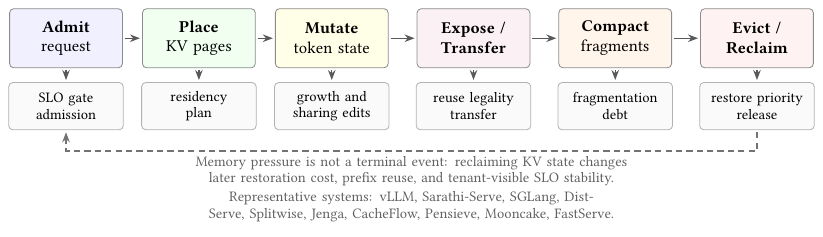}
\caption{KV-cache lifecycle.}
\label{fig:supp-kvcache-lifecycle}
\end{figure*}

The lifecycle figure makes the control seams concrete. Admit decides whether a request is allowed to create new short-lived state at all. Place decides where that state lives and therefore which queues or buses it will burden next. Mutate covers decode-time growth, prefix edits, and other in-place evolution of live serving state. Expose/transfer covers whether prefix, adapter, or session-local state may be reused safely and when it may move across workers or phases. Compact covers fragmentation control and structural cleanup. Evict/reclaim covers what can be released without hiding future restoration debt and how that debt is later charged back to the original SLO.

The following evidence is especially useful in this subsection:
\begin{itemize}[leftmargin=*]
\item predictive-serving families grouped by the debt they forecast;
\item legality conditions for prefix, branch, and multi-call reuse;
\item temporal-lifecycle policies over expose/transfer, evict/reclaim, warm-start, and fairness debt;
\item clearer distinctions between allocator realism, transfer scheduling, and ownership semantics.
\end{itemize}

\subsection{Retrieval and Retention Governance}

The unifying point is that retrieval and retention both expose a \emph{longer-lived memory layer} whose value decays gradually and whose maintenance debt is easy to hide if the runtime only reports steady-state quality. In retrieval systems, the state object is not just embeddings or an ANN graph in the abstract, but mutable shards, publication metadata, freshness signals, deletion debt, and planner-visible exposure state~\cite{singh2021freshdiskann,wang2021milvus,guo2022manu,asai2024selfrag,zhang2026flowrag,wang2026candor}. In retention systems, the state object is not just ``memory'' generically, but a bounded store whose protected parameters, exemplars, replay entries, compressed features, and effective budget must be co-governed over time~\cite{kirkpatrick2017ewc,rebuffi2017icarl,aljundi2019mir,prabhu2020gdumb,hayes2020remind,zhou2025ferret,li2026streamfp}.

Retrieval systems expose at least four trigger layers: request-time invocation, index-time maintenance, publication-time exposure, and placement-time residency. Self-RAG changes whether external memory should be consulted at all; FlowRAG changes when retriever state should be refreshed before reuse quality decays too far; CANDOR-Bench and FreshDiskANN expose when update churn, deletion debt, or local repair have already pushed the index away from its intended operating region; Milvus and Manu expose when partially refreshed state may be published safely under concurrent updates; Rummy and composable-memory search expose when movement or placement is justified by downstream quality gain instead of raw lookup throughput~\cite{asai2024selfrag,zhang2026flowrag,wang2026candor,singh2021freshdiskann,wang2021milvus,guo2022manu,zhang2024rummy,quinn2025accelerating}. The comparison shows that retrieval quality is a lifecycle property, not merely an encoder property.

Retention systems expose an analogous trigger hierarchy over one bounded store. EWC protects already-committed parameter state; iCaRL, GDumb, and StreamFP govern what enters the scarce retained set; GEM, A-GEM, MIR, and DER govern when stored state should re-enter the update path; REMIND governs whether retained signal is compressed and later reconstructed; Ferret governs what happens when the budget itself contracts under live update pressure~\cite{kirkpatrick2017ewc,rebuffi2017icarl,prabhu2020gdumb,li2026streamfp,lopez2017gem,chaudhry2019agem,aljundi2019mir,buzzega2020der,hayes2020remind,zhou2025ferret}. Read this way, the continual-learning literature matters to the survey not because it supplies another application area, but because it provides one of the clearest examples of how protect, admit, replay, compress, and budget-shrink decisions compose over time.

The comparative lesson across these two clusters should remain explicit. Retrieval runtimes lack a stable rule for when stale-but-local memory remains queryable; retention runtimes lack a stable rule for when compressed or demoted memory remains good enough to preserve. In both cases, the missing abstraction is a policy-visible exposure contract that turns hidden maintenance debt into an explicit control surface.

Several remaining papers fit this same long-lived-memory picture. Dynamic ANN and index-maintenance lines such as HNSW, DiskANN, Milvus, and graph- or shard-centric systems make clear that freshness, shard publication, and rebuild debt are all part of one operational budget~\cite{malkov2020hnsw,singh2021freshdiskann,wang2021milvus,guo2022manu}. Retrieval-augmented and memory-augmented systems such as DPR, ColBERT, REALM, RAG, ATLAS, HippoRAG, and Raptor show the same phenomenon at the model boundary: the runtime must manage what gets exposed, refreshed, abstracted, and reused rather than treating memory as a static auxiliary index~\cite{karpukhin2020dpr,khattab2020colbert,guu2020realm,lewis2020rag,izacard2022atlas,wang2024hipporag,sarthi2024raptor}. Those references preserve the lifecycle logic behind the retrieval chapter.

Retention has a similar tail of boundary papers. Experience replay, memory budget, and continual-update lines such as MIR, GDumb, REMIND, Ferret, and StreamFP establish that protect/admit/replay/demote decisions are all budgeted governance acts over one bounded store~\cite{aljundi2019mir,prabhu2020gdumb,hayes2020remind,zhou2025ferret,li2026streamfp}. The resulting comparison connects long-horizon retention with short-lived serving reuse without treating the two as the same problem.

\begin{figure*}[t]
\centering
\includegraphics[width=0.94\textwidth]{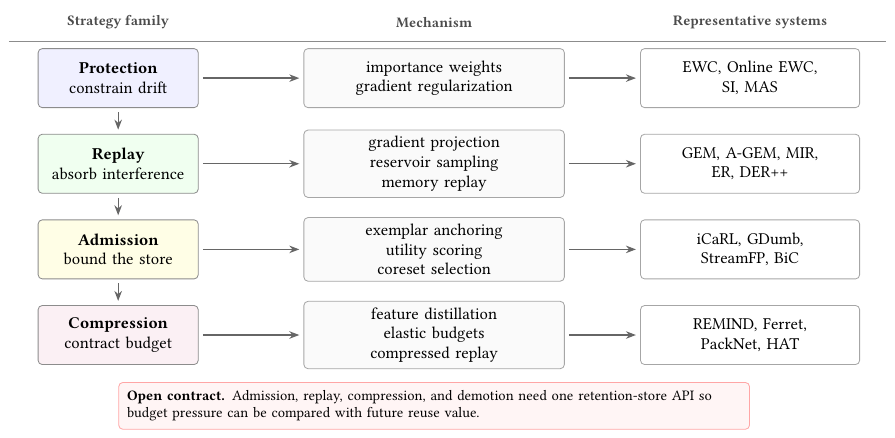}
\caption{Retention-governance taxonomy.}
\label{fig:supp-retention-taxonomy}
\end{figure*}

The taxonomy figure prevents a recurring misreading: retention is not one replay heuristic with a few engineering variants. It is a store-level governance problem over multiple actions with different temporal meanings. Protect decisions stabilize already-committed state; admission decisions determine future coverage; replay decisions determine when retained state becomes active again; compression and demotion decisions decide how much of that future value survives under tighter budgets; budget-shrink rules decide which guarantees fail first when the store contract is stressed.

\subsection{Long-Horizon Evolution Anchors}

The longest-horizon evidence keeps the survey from collapsing ``evolution'' into a narrow continual-learning story. SuperNeurons and related memory-reuse systems show that activation liveness, recomputation, and offload policy already form a managed state loop in deep-learning training, not just in inference-side serving~\cite{wang2018superneurons}. Bubble-filling and recomputation-aware training systems such as Obscura show the same thing from another angle: when state is expensive to keep live, the runtime must decide which debt to shift into pipeline slack and which debt to pay immediately~\cite{huang2025obscura}. That is the same governance pattern that later appears in retention stores, but with a different temporal horizon and a different type of reuse value.

\subsection{Blueprint and Fault-Model Boundary Conditions}

The first piece that should remain visible here is the distinction between \emph{typed state objects}, \emph{observation contracts}, \emph{decision precedence}, \emph{actuation scope}, and \emph{safety-envelope invariants}. The blueprint is not a proposal for one monolithic runtime; its role is to make recurring hidden assumptions explicit enough that neighboring mechanisms can compose without silently violating one another's boundaries.

One implication is that recovery should not be treated as a separate semantic regime. If steady-state control reasons over typed KV pages, refreshable vector shards, replay buffers, or demotable retention objects, then recovery must restore and validate the same objects under the same exposure and action boundaries. Otherwise a system effectively runs one state model in steady state and another during failure. That inconsistency matters because it explains why fault-model notes are part of the mechanism, not optional decoration.

The fault-model distinction also deserves fuller treatment here. A safety claim that holds under crash-stop may fail under crash-recovery with replay, may fail differently under network partition, and may fail silently under degradation modes such as stale telemetry, compaction lag, or sudden budget contraction~\cite{chandy1985snapshot,corbett2012spanner,ongaro2014raft,mao2023morphstream,zhao2024recovery,wang2026candor,zhang2026flowrag}. These are not small implementation variants. They determine whether a controller may expose partially refreshed retrieval state, whether ownership may transfer across disaggregated serving pools, and whether a retention tier may demote memory before replaying it. The blueprint's value lies in precisely these boundary conditions.

Two invariant families deserve to be carried here explicitly. The first is \emph{freshness exposure invariants}: a retrieval shard or index segment should specify when it is queryable for internal warmup, restricted traffic, or general external service after rebuild, deletion repair, or compaction. The second is \emph{retention demotion invariants}: a retention object should specify whether it is protected, compressible, summarizable, demotable, or droppable when effective budget shrinks. These invariants preserve the blueprint's operational meaning.

\subsection{Foundations Figures and Boundary Material}

The inclusion rule is not venue prestige or application relevance. A paper is central when it turns state into an explicit runtime-managed object and exposes a reusable control surface over that state. That is why transactional streaming, migration, KV-cache lifecycle control, retrieval maintenance, and retention governance sit in the core, while many application-facing LLM, agent, or prediction papers stay outside it even when they are technically strong.

That definition also has a formal lineage. The state-machine tradition matters not because the survey is about formal methods, but because it shows that a system only becomes governable when the runtime can name, observe, and transform its state explicitly~\cite{schneider1990state}. That conceptual boundary is the reason the survey keeps the access, serving, retrieval, and retention lines together instead of splitting them into unrelated application silos.

\inlinefigurewithcaption
  {width=0.9\linewidth}
  {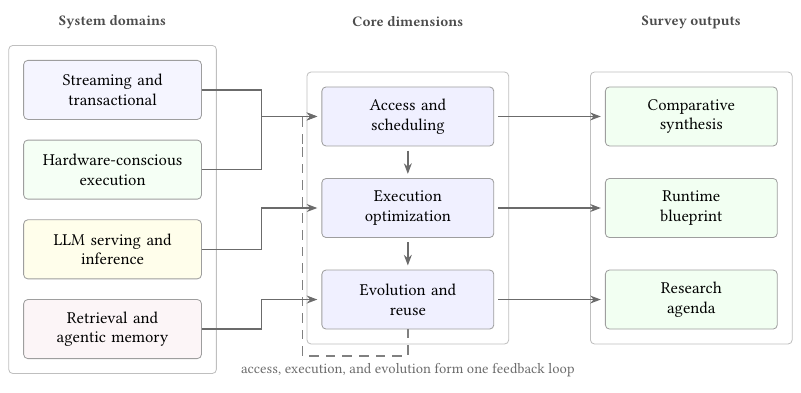}
  {Poster-style survey overview. It serves as a compact map of the survey's scope and argumentative flow.}
  {fig:supp-poster-overview}

The core-versus-context distinction can also be stated operationally. Core material changes the survey's comparative result by sharpening the state object, control surface, coupling path, evaluation boundary, or remaining gap. Context material instead clarifies neighboring traditions, records boundary-setting decisions, or preserves mappings between triage labels and the five-field frame. Keeping that distinction explicit prevents the survey from drifting back into a broad but weakly comparative literature tour.

Two overview visuals also belong conceptually with this foundations rationale. The first is the poster-style overview, which summarizes the survey spine and clarifies how the three axes and cross-domain synthesis fit together. The second is the running-example triptych, which explains why the paper reuses one access case, one serving case, and one retrieval case across sections instead of multiplying disconnected examples.

\inlinefigurewithcaption
  {width=0.96\linewidth}
  {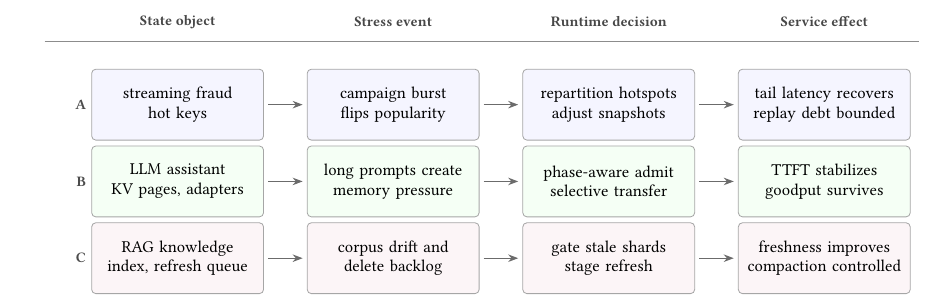}
  {Running examples that recur across access, execution, and evolution.}
  {fig:supp-running-examples}

\bibliographystyle{ACM-Reference-Format}
\bibliography{refs}


\begin{thebibliography}{287}


\ifx \showCODEN    \undefined \def \showCODEN     #1{\unskip}     \fi
\ifx \showISBNx    \undefined \def \showISBNx     #1{\unskip}     \fi
\ifx \showISBNxiii \undefined \def \showISBNxiii  #1{\unskip}     \fi
\ifx \showISSN     \undefined \def \showISSN      #1{\unskip}     \fi
\ifx \showLCCN     \undefined \def \showLCCN      #1{\unskip}     \fi
\ifx \shownote     \undefined \def \shownote      #1{#1}          \fi
\ifx \showarticletitle \undefined \def \showarticletitle #1{#1}   \fi
\ifx \showURL      \undefined \def \showURL       {\relax}        \fi
\providecommand\bibfield[2]{#2}
\providecommand\bibinfo[2]{#2}
\providecommand\natexlab[1]{#1}
\providecommand\showeprint[2][]{arXiv:#2}

\bibitem[Abadi et~al\mbox{.}(2005)]%
        {abadi2005borealis}
\bibfield{author}{\bibinfo{person}{Daniel~J. Abadi}, \bibinfo{person}{Yanif
  Ahmad}, \bibinfo{person}{Magdalena Balazinska}, \bibinfo{person}{Ugur
  Cetintemel}, \bibinfo{person}{Mitch Cherniack}, \bibinfo{person}{Jeong-Hyon
  Hwang}, \bibinfo{person}{Wolfgang Lindner}, \bibinfo{person}{Anurag Maskey},
  \bibinfo{person}{Alex Rasin}, \bibinfo{person}{Esther Ryvkina},
  \bibinfo{person}{Nesime Tatbul}, \bibinfo{person}{Ying Xing}, {and}
  \bibinfo{person}{Stan Zdonik}.} \bibinfo{year}{2005}\natexlab{}.
\newblock \showarticletitle{The Design of the Borealis Stream Processing
  Engine}. In \bibinfo{booktitle}{\emph{Proceedings of the Second Biennial
  Conference on Innovative Data Systems Research (CIDR)}}.
  \bibinfo{pages}{277--289}.
\newblock
\urldef\tempurl%
\url{https://www.cidrdb.org/cidr2005/papers/P23.pdf}
\showURL{%
\tempurl}


\bibitem[Abadi et~al\mbox{.}(2003)]%
        {abadi2003aurora}
\bibfield{author}{\bibinfo{person}{Daniel~J. Abadi}, \bibinfo{person}{Don
  Carney}, \bibinfo{person}{Ugur \c{C}etintemel}, \bibinfo{person}{Mitch
  Cherniack}, \bibinfo{person}{Christian Convey}, \bibinfo{person}{Sangdon
  Lee}, \bibinfo{person}{Michael Stonebraker}, \bibinfo{person}{Nesime Tatbul},
  {and} \bibinfo{person}{Stan Zdonik}.} \bibinfo{year}{2003}\natexlab{}.
\newblock \showarticletitle{Aurora: a new model and architecture for data
  stream management}.
\newblock \bibinfo{journal}{\emph{The VLDB Journal}} \bibinfo{volume}{12},
  \bibinfo{number}{2} (\bibinfo{date}{Aug.} \bibinfo{year}{2003}),
  \bibinfo{pages}{120–139}.
\newblock
\showISSN{1066-8888}
\href{https://doi.org/10.1007/s00778-003-0095-z}{doi:\nolinkurl{10.1007/s00778-003-0095-z}}


\bibitem[Adnan et~al\mbox{.}(2024)]%
        {adnan2024keyformer}
\bibfield{author}{\bibinfo{person}{Muhammad Adnan}, \bibinfo{person}{Akhil
  Arunkumar}, \bibinfo{person}{Gaurav Jain}, \bibinfo{person}{Prashant~J.
  Nair}, \bibinfo{person}{Ilya Soloveychik}, {and} \bibinfo{person}{Purushotham
  Kamath}.} \bibinfo{year}{2024}\natexlab{}.
\newblock \showarticletitle{Keyformer: KV Cache reduction through key tokens
  selection for Efficient Generative Inference}. In
  \bibinfo{booktitle}{\emph{Proceedings of Machine Learning and Systems}},
  \bibfield{editor}{\bibinfo{person}{P.~Gibbons},
  \bibinfo{person}{G.~Pekhimenko}, {and} \bibinfo{person}{C.~De Sa}} (Eds.),
  Vol.~\bibinfo{volume}{6}. \bibinfo{pages}{114--127}.
\newblock
\urldef\tempurl%
\url{https://proceedings.mlsys.org/paper_files/paper/2024/file/48fecef47b19fe501d27d338b6d52582-Paper-Conference.pdf}
\showURL{%
\tempurl}


\bibitem[Agarwal et~al\mbox{.}(2026)]%
        {agarwal2026symphony}
\bibfield{author}{\bibinfo{person}{Saurabh Agarwal}, \bibinfo{person}{Bodun
  Hu}, \bibinfo{person}{Anyong Mao}, \bibinfo{person}{Aditya Akella}, {and}
  \bibinfo{person}{Shivaram Venkataraman}.} \bibinfo{year}{2026}\natexlab{}.
\newblock \showarticletitle{{SYMPHONY}: Enabling {Compute-Memory}
  Disaggregation in {LLM} Serving Systems}. In \bibinfo{booktitle}{\emph{23rd
  USENIX Symposium on Networked Systems Design and Implementation (NSDI 26)}}.
  \bibinfo{publisher}{USENIX Association}, \bibinfo{address}{Renton, WA},
  \bibinfo{pages}{2027--2041}.
\newblock
\showISBNx{978-1-939133-54-0}
\urldef\tempurl%
\url{https://www.usenix.org/conference/nsdi26/presentation/agarwal}
\showURL{%
\tempurl}


\bibitem[Agarwal et~al\mbox{.}(2013)]%
        {agarwal2013blinkdb}
\bibfield{author}{\bibinfo{person}{Sameer Agarwal}, \bibinfo{person}{Barzan
  Mozafari}, \bibinfo{person}{Aurojit Panda}, \bibinfo{person}{Henry Milner},
  \bibinfo{person}{Samuel Madden}, {and} \bibinfo{person}{Ion Stoica}.}
  \bibinfo{year}{2013}\natexlab{}.
\newblock \showarticletitle{BlinkDB: queries with bounded errors and bounded
  response times on very large data}. In \bibinfo{booktitle}{\emph{Proceedings
  of the 8th ACM European Conference on Computer Systems}} (Prague, Czech
  Republic) \emph{(\bibinfo{series}{EuroSys '13})}.
  \bibinfo{publisher}{Association for Computing Machinery},
  \bibinfo{address}{New York, NY, USA}, \bibinfo{pages}{29–42}.
\newblock
\showISBNx{9781450319942}
\href{https://doi.org/10.1145/2465351.2465355}{doi:\nolinkurl{10.1145/2465351.2465355}}


\bibitem[Agrawal et~al\mbox{.}(2024)]%
        {agrawal2024sarathi}
\bibfield{author}{\bibinfo{person}{Amey Agrawal}, \bibinfo{person}{Nitin
  Kedia}, \bibinfo{person}{Ashish Panwar}, \bibinfo{person}{Jayashree Mohan},
  \bibinfo{person}{Nipun Kwatra}, \bibinfo{person}{Bhargav~S. Gulavani},
  \bibinfo{person}{Alexey Tumanov}, {and} \bibinfo{person}{Ramachandran
  Ramjee}.} \bibinfo{year}{2024}\natexlab{}.
\newblock \showarticletitle{Taming throughput-latency tradeoff in LLM inference
  with sarathi-serve}. In \bibinfo{booktitle}{\emph{Proceedings of the 18th
  USENIX Conference on Operating Systems Design and Implementation}} (Santa
  Clara, CA, USA) \emph{(\bibinfo{series}{OSDI'24})}.
  \bibinfo{publisher}{USENIX Association}, \bibinfo{address}{USA}, Article
  \bibinfo{articleno}{7}, \bibinfo{numpages}{18}~pages.
\newblock
\showISBNx{978-1-939133-40-3}


\bibitem[Akidau et~al\mbox{.}(2013)]%
        {akidau2013millwheel}
\bibfield{author}{\bibinfo{person}{Tyler Akidau}, \bibinfo{person}{Alex
  Balikov}, \bibinfo{person}{Kaya Bekiro\u{g}lu}, \bibinfo{person}{Slava
  Chernyak}, \bibinfo{person}{Josh Haberman}, \bibinfo{person}{Reuven Lax},
  \bibinfo{person}{Sam McVeety}, \bibinfo{person}{Daniel Mills},
  \bibinfo{person}{Paul Nordstrom}, {and} \bibinfo{person}{Sam Whittle}.}
  \bibinfo{year}{2013}\natexlab{}.
\newblock \showarticletitle{MillWheel: fault-tolerant stream processing at
  internet scale}.
\newblock \bibinfo{journal}{\emph{Proc. VLDB Endow.}} \bibinfo{volume}{6},
  \bibinfo{number}{11} (\bibinfo{date}{Aug.} \bibinfo{year}{2013}),
  \bibinfo{pages}{1033–1044}.
\newblock
\showISSN{2150-8097}
\href{https://doi.org/10.14778/2536222.2536229}{doi:\nolinkurl{10.14778/2536222.2536229}}


\bibitem[Akidau et~al\mbox{.}(2015)]%
        {akidau2015dataflow}
\bibfield{author}{\bibinfo{person}{Tyler Akidau}, \bibinfo{person}{Robert
  Bradshaw}, \bibinfo{person}{Craig Chambers}, \bibinfo{person}{Slava
  Chernyak}, \bibinfo{person}{Rafael~J. Fern\'{a}ndez-Moctezuma},
  \bibinfo{person}{Reuven Lax}, \bibinfo{person}{Sam McVeety},
  \bibinfo{person}{Daniel Mills}, \bibinfo{person}{Frances Perry},
  \bibinfo{person}{Eric Schmidt}, {and} \bibinfo{person}{Sam Whittle}.}
  \bibinfo{year}{2015}\natexlab{}.
\newblock \showarticletitle{The dataflow model: a practical approach to
  balancing correctness, latency, and cost in massive-scale, unbounded,
  out-of-order data processing}.
\newblock \bibinfo{journal}{\emph{Proc. VLDB Endow.}} \bibinfo{volume}{8},
  \bibinfo{number}{12} (\bibinfo{date}{Aug.} \bibinfo{year}{2015}),
  \bibinfo{pages}{1792–1803}.
\newblock
\showISSN{2150-8097}
\href{https://doi.org/10.14778/2824032.2824076}{doi:\nolinkurl{10.14778/2824032.2824076}}


\bibitem[Ali et~al\mbox{.}(2011)]%
        {ali2011survey}
\bibfield{author}{\bibinfo{person}{Waleed Ali}, \bibinfo{person}{Siti~Mariyam
  Shamsuddin}, {and} \bibinfo{person}{Abdul~Samad Ismail}.}
  \bibinfo{year}{2011}\natexlab{}.
\newblock \showarticletitle{A Survey of Web Caching and Prefetching A Survey of
  Web Caching and Prefetching}.
\newblock \bibinfo{journal}{\emph{International Journal of Advances in Soft
  Computing and its Applications}}  \bibinfo{volume}{3} (\bibinfo{date}{03}
  \bibinfo{year}{2011}).
\newblock


\bibitem[Aljundi et~al\mbox{.}(2019)]%
        {aljundi2019mir}
\bibfield{author}{\bibinfo{person}{Rahaf Aljundi}, \bibinfo{person}{Eugene
  Belilovsky}, \bibinfo{person}{Tinne Tuytelaars}, \bibinfo{person}{Laurent
  Charlin}, \bibinfo{person}{Massimo Caccia}, \bibinfo{person}{Min Lin}, {and}
  \bibinfo{person}{Lucas Page-Caccia}.} \bibinfo{year}{2019}\natexlab{}.
\newblock \showarticletitle{Online Continual Learning with Maximal Interfered
  Retrieval}. In \bibinfo{booktitle}{\emph{Advances in Neural Information
  Processing Systems}}, \bibfield{editor}{\bibinfo{person}{H.~Wallach},
  \bibinfo{person}{H.~Larochelle}, \bibinfo{person}{A.~Beygelzimer},
  \bibinfo{person}{F.~d\textquotesingle Alch\'{e}-Buc},
  \bibinfo{person}{E.~Fox}, {and} \bibinfo{person}{R.~Garnett}} (Eds.),
  Vol.~\bibinfo{volume}{32}. \bibinfo{publisher}{Curran Associates, Inc.}
\newblock
\urldef\tempurl%
\url{https://proceedings.neurips.cc/paper_files/paper/2019/file/15825aee15eb335cc13f9b559f166ee8-Paper.pdf}
\showURL{%
\tempurl}


\bibitem[Arfeen et~al\mbox{.}(2025)]%
        {arfeen2024pipefill}
\bibfield{author}{\bibinfo{person}{Daiyaan Arfeen}, \bibinfo{person}{Zhen
  Zhang}, \bibinfo{person}{Xinwei Fu}, \bibinfo{person}{Gregory Ganger}, {and}
  \bibinfo{person}{Yida Wang}.} \bibinfo{year}{2025}\natexlab{}.
\newblock \showarticletitle{PipeFill: Using GPUs During Bubbles in
  Pipeline-parallel LLM Training}. In \bibinfo{booktitle}{\emph{Proceedings of
  Machine Learning and Systems}},
  \bibfield{editor}{\bibinfo{person}{M.~Zaharia}, \bibinfo{person}{G.~Joshi},
  {and} \bibinfo{person}{Y.~Lin}} (Eds.), Vol.~\bibinfo{volume}{7}.
  \bibinfo{publisher}{MLSys}.
\newblock
\urldef\tempurl%
\url{https://proceedings.mlsys.org/paper_files/paper/2025/file/53d3f45797970d323bd8a0d379c525aa-Paper-Conference.pdf}
\showURL{%
\tempurl}


\bibitem[Armbrust et~al\mbox{.}(2018)]%
        {armbrust2018structured}
\bibfield{author}{\bibinfo{person}{Michael Armbrust},
  \bibinfo{person}{Tathagata Das}, \bibinfo{person}{Joseph Torres},
  \bibinfo{person}{Burak Yavuz}, \bibinfo{person}{Shixiong Zhu},
  \bibinfo{person}{Reynold Xin}, \bibinfo{person}{Ali Ghodsi},
  \bibinfo{person}{Ion Stoica}, {and} \bibinfo{person}{Matei Zaharia}.}
  \bibinfo{year}{2018}\natexlab{}.
\newblock \showarticletitle{Structured Streaming: A Declarative API for
  Real-Time Applications in Apache Spark}. In
  \bibinfo{booktitle}{\emph{Proceedings of the 2018 International Conference on
  Management of Data}} (Houston, TX, USA) \emph{(\bibinfo{series}{SIGMOD
  '18})}. \bibinfo{publisher}{Association for Computing Machinery},
  \bibinfo{address}{New York, NY, USA}, \bibinfo{pages}{601–613}.
\newblock
\showISBNx{9781450347037}
\href{https://doi.org/10.1145/3183713.3190664}{doi:\nolinkurl{10.1145/3183713.3190664}}


\bibitem[Asai et~al\mbox{.}(2024)]%
        {asai2024selfrag}
\bibfield{author}{\bibinfo{person}{Akari Asai}, \bibinfo{person}{Zeqiu Wu},
  \bibinfo{person}{Yizhong Wang}, \bibinfo{person}{Avi Sil}, {and}
  \bibinfo{person}{Hannaneh Hajishirzi}.} \bibinfo{year}{2024}\natexlab{}.
\newblock \showarticletitle{Self-RAG: Learning to Retrieve, Generate, and
  Critique through Self-Reflection}. In \bibinfo{booktitle}{\emph{International
  Conference on Learning Representations}},
  \bibfield{editor}{\bibinfo{person}{B.~Kim}, \bibinfo{person}{Y.~Yue},
  \bibinfo{person}{S.~Chaudhuri}, \bibinfo{person}{K.~Fragkiadaki},
  \bibinfo{person}{M.~Khan}, {and} \bibinfo{person}{Y.~Sun}} (Eds.),
  Vol.~\bibinfo{volume}{2024}. \bibinfo{pages}{9112--9141}.
\newblock
\urldef\tempurl%
\url{https://proceedings.iclr.cc/paper_files/paper/2024/file/25f7be9694d7b32d5cc670927b8091e1-Paper-Conference.pdf}
\showURL{%
\tempurl}


\bibitem[Baker et~al\mbox{.}(2011)]%
        {baker2011megastore}
\bibfield{author}{\bibinfo{person}{Jason Baker}, \bibinfo{person}{Chris Bond},
  \bibinfo{person}{James~C. Corbett}, \bibinfo{person}{JJ Furman},
  \bibinfo{person}{Andrey Khorlin}, \bibinfo{person}{James Larson},
  \bibinfo{person}{Jean-Michel Leon}, \bibinfo{person}{Yawei Li},
  \bibinfo{person}{Alexander Lloyd}, {and} \bibinfo{person}{Vadim Yushprakh}.}
  \bibinfo{year}{2011}\natexlab{}.
\newblock \showarticletitle{Megastore: Providing Scalable, Highly Available
  Storage for Interactive Services}. In \bibinfo{booktitle}{\emph{Proceedings
  of the Conference on Innovative Data system Research (CIDR)}}.
  \bibinfo{pages}{223--234}.
\newblock
\urldef\tempurl%
\url{http://www.cidrdb.org/cidr2011/Papers/CIDR11_Paper32.pdf}
\showURL{%
\tempurl}


\bibitem[Borgeaud et~al\mbox{.}(2022)]%
        {borgeaud2022retro}
\bibfield{author}{\bibinfo{person}{Sebastian Borgeaud}, \bibinfo{person}{Arthur
  Mensch}, \bibinfo{person}{Jordan Hoffmann}, \bibinfo{person}{Trevor Cai},
  \bibinfo{person}{Eliza Rutherford}, \bibinfo{person}{Katie Millican},
  \bibinfo{person}{George~Bm Van Den~Driessche}, \bibinfo{person}{Jean-Baptiste
  Lespiau}, \bibinfo{person}{Bogdan Damoc}, \bibinfo{person}{Aidan Clark},
  \bibinfo{person}{Diego De~Las~Casas}, \bibinfo{person}{Aurelia Guy},
  \bibinfo{person}{Jacob Menick}, \bibinfo{person}{Roman Ring},
  \bibinfo{person}{Tom Hennigan}, \bibinfo{person}{Saffron Huang},
  \bibinfo{person}{Loren Maggiore}, \bibinfo{person}{Chris Jones},
  \bibinfo{person}{Albin Cassirer}, \bibinfo{person}{Andy Brock},
  \bibinfo{person}{Michela Paganini}, \bibinfo{person}{Geoffrey Irving},
  \bibinfo{person}{Oriol Vinyals}, \bibinfo{person}{Simon Osindero},
  \bibinfo{person}{Karen Simonyan}, \bibinfo{person}{Jack Rae},
  \bibinfo{person}{Erich Elsen}, {and} \bibinfo{person}{Laurent Sifre}.}
  \bibinfo{year}{2022}\natexlab{}.
\newblock \showarticletitle{Improving Language Models by Retrieving from
  Trillions of Tokens}. In \bibinfo{booktitle}{\emph{Proceedings of the 39th
  International Conference on Machine Learning}}
  \emph{(\bibinfo{series}{Proceedings of Machine Learning Research},
  Vol.~\bibinfo{volume}{162})}, \bibfield{editor}{\bibinfo{person}{Kamalika
  Chaudhuri}, \bibinfo{person}{Stefanie Jegelka}, \bibinfo{person}{Le~Song},
  \bibinfo{person}{Csaba Szepesvari}, \bibinfo{person}{Gang Niu}, {and}
  \bibinfo{person}{Sivan Sabato}} (Eds.). \bibinfo{publisher}{PMLR},
  \bibinfo{pages}{2206--2240}.
\newblock
\urldef\tempurl%
\url{https://proceedings.mlr.press/v162/borgeaud22a.html}
\showURL{%
\tempurl}


\bibitem[Buzzega et~al\mbox{.}(2020)]%
        {buzzega2020der}
\bibfield{author}{\bibinfo{person}{Pietro Buzzega}, \bibinfo{person}{Matteo
  Boschini}, \bibinfo{person}{Angelo Porrello}, \bibinfo{person}{Davide Abati},
  {and} \bibinfo{person}{SIMONE CALDERARA}.} \bibinfo{year}{2020}\natexlab{}.
\newblock \showarticletitle{Dark Experience for General Continual Learning: a
  Strong, Simple Baseline}. In \bibinfo{booktitle}{\emph{Advances in Neural
  Information Processing Systems}},
  \bibfield{editor}{\bibinfo{person}{H.~Larochelle},
  \bibinfo{person}{M.~Ranzato}, \bibinfo{person}{R.~Hadsell},
  \bibinfo{person}{M.F. Balcan}, {and} \bibinfo{person}{H.~Lin}} (Eds.),
  Vol.~\bibinfo{volume}{33}. \bibinfo{publisher}{Curran Associates, Inc.},
  \bibinfo{pages}{15920--15930}.
\newblock
\urldef\tempurl%
\url{https://proceedings.neurips.cc/paper_files/paper/2020/file/b704ea2c39778f07c617f6b7ce480e9e-Paper.pdf}
\showURL{%
\tempurl}


\bibitem[Cai et~al\mbox{.}(2025)]%
        {cai2025moc}
\bibfield{author}{\bibinfo{person}{Weilin Cai}, \bibinfo{person}{Le Qin}, {and}
  \bibinfo{person}{Jiayi Huang}.} \bibinfo{year}{2025}\natexlab{}.
\newblock \showarticletitle{MoC-System: Efficient Fault Tolerance for Sparse
  Mixture-of-Experts Model Training}. In \bibinfo{booktitle}{\emph{Proceedings
  of the 30th ACM International Conference on Architectural Support for
  Programming Languages and Operating Systems, Volume 2}} (Rotterdam,
  Netherlands) \emph{(\bibinfo{series}{ASPLOS '25})}.
  \bibinfo{publisher}{Association for Computing Machinery},
  \bibinfo{address}{New York, NY, USA}, \bibinfo{pages}{655–671}.
\newblock
\showISBNx{9798400710797}
\href{https://doi.org/10.1145/3676641.3716006}{doi:\nolinkurl{10.1145/3676641.3716006}}


\bibitem[Cao et~al\mbox{.}(2025)]%
        {cao2025moelightning}
\bibfield{author}{\bibinfo{person}{Shiyi Cao}, \bibinfo{person}{Shu Liu},
  \bibinfo{person}{Tyler Griggs}, \bibinfo{person}{Peter Schafhalter},
  \bibinfo{person}{Xiaoxuan Liu}, \bibinfo{person}{Ying Sheng},
  \bibinfo{person}{Joseph~E. Gonzalez}, \bibinfo{person}{Matei Zaharia}, {and}
  \bibinfo{person}{Ion Stoica}.} \bibinfo{year}{2025}\natexlab{}.
\newblock \showarticletitle{MoE-Lightning: High-Throughput MoE Inference on
  Memory-constrained GPUs}. In \bibinfo{booktitle}{\emph{Proceedings of the
  30th ACM International Conference on Architectural Support for Programming
  Languages and Operating Systems, Volume 1}} (Rotterdam, Netherlands)
  \emph{(\bibinfo{series}{ASPLOS '25})}. \bibinfo{publisher}{Association for
  Computing Machinery}, \bibinfo{address}{New York, NY, USA},
  \bibinfo{pages}{715–730}.
\newblock
\showISBNx{9798400706981}
\href{https://doi.org/10.1145/3669940.3707267}{doi:\nolinkurl{10.1145/3669940.3707267}}


\bibitem[Cao et~al\mbox{.}(2020)]%
        {cao2020polardb}
\bibfield{author}{\bibinfo{person}{Wei Cao}, \bibinfo{person}{Yang Liu},
  \bibinfo{person}{Zhushi Cheng}, \bibinfo{person}{Ning Zheng},
  \bibinfo{person}{Wei Li}, \bibinfo{person}{Wenjie Wu},
  \bibinfo{person}{Linqiang Ouyang}, \bibinfo{person}{Peng Wang},
  \bibinfo{person}{Yijing Wang}, \bibinfo{person}{Ray Kuan},
  \bibinfo{person}{Zhenjun Liu}, \bibinfo{person}{Feng Zhu}, {and}
  \bibinfo{person}{Tong Zhang}.} \bibinfo{year}{2020}\natexlab{}.
\newblock \showarticletitle{{POLARDB} Meets Computational Storage: Efficiently
  Support Analytical Workloads in {Cloud-Native} Relational Database}. In
  \bibinfo{booktitle}{\emph{18th USENIX Conference on File and Storage
  Technologies (FAST 20)}}. \bibinfo{publisher}{USENIX Association},
  \bibinfo{address}{Santa Clara, CA}, \bibinfo{pages}{29--41}.
\newblock
\showISBNx{978-1-939133-12-0}
\urldef\tempurl%
\url{https://www.usenix.org/conference/fast20/presentation/cao-wei}
\showURL{%
\tempurl}


\bibitem[Carbone et~al\mbox{.}(2015a)]%
        {carbone2015asynchronous}
\bibfield{author}{\bibinfo{person}{Paris Carbone}, \bibinfo{person}{Gyula
  F{\'o}ra}, \bibinfo{person}{Stephan Ewen}, \bibinfo{person}{Seif Haridi},
  {and} \bibinfo{person}{Kostas Tzoumas}.} \bibinfo{year}{2015}\natexlab{a}.
\newblock \showarticletitle{Lightweight Asynchronous Snapshots for Distributed
  Dataflows}.
\newblock  (\bibinfo{year}{2015}).
\newblock
\href{https://doi.org/10.48550/arXiv.1506.08603}{doi:\nolinkurl{10.48550/arXiv.1506.08603}}


\bibitem[Carbone et~al\mbox{.}(2015b)]%
        {carbone2015flink}
\bibfield{author}{\bibinfo{person}{Paris Carbone}, \bibinfo{person}{Asterios
  Katsifodimos}, \bibinfo{person}{Stephan Ewen}, \bibinfo{person}{Volker
  Markl}, \bibinfo{person}{Seif Haridi}, {and} \bibinfo{person}{Kostas
  Tzoumas}.} \bibinfo{year}{2015}\natexlab{b}.
\newblock \showarticletitle{Apache flink: Stream and batch processing in a
  single engine}.
\newblock \bibinfo{journal}{\emph{The Bulletin of the Technical Committee on
  Data Engineering}} \bibinfo{volume}{38}, \bibinfo{number}{4}
  (\bibinfo{year}{2015}).
\newblock


\bibitem[Castro~Fernandez et~al\mbox{.}(2013)]%
        {fernandez2013osm}
\bibfield{author}{\bibinfo{person}{Raul Castro~Fernandez},
  \bibinfo{person}{Matteo Migliavacca}, \bibinfo{person}{Evangelia
  Kalyvianaki}, {and} \bibinfo{person}{Peter Pietzuch}.}
  \bibinfo{year}{2013}\natexlab{}.
\newblock \showarticletitle{Integrating scale out and fault tolerance in stream
  processing using operator state management}. In
  \bibinfo{booktitle}{\emph{Proceedings of the 2013 ACM SIGMOD International
  Conference on Management of Data}} (New York, New York, USA)
  \emph{(\bibinfo{series}{SIGMOD '13})}. \bibinfo{publisher}{Association for
  Computing Machinery}, \bibinfo{address}{New York, NY, USA},
  \bibinfo{pages}{725–736}.
\newblock
\showISBNx{9781450320375}
\href{https://doi.org/10.1145/2463676.2465282}{doi:\nolinkurl{10.1145/2463676.2465282}}


\bibitem[Chambers et~al\mbox{.}(2010)]%
        {chambers2010flumejava}
\bibfield{author}{\bibinfo{person}{Craig Chambers}, \bibinfo{person}{Ashish
  Raniwala}, \bibinfo{person}{Frances Perry}, \bibinfo{person}{Stephen Adams},
  \bibinfo{person}{Robert~R. Henry}, \bibinfo{person}{Robert Bradshaw}, {and}
  \bibinfo{person}{Nathan Weizenbaum}.} \bibinfo{year}{2010}\natexlab{}.
\newblock \showarticletitle{FlumeJava: easy, efficient data-parallel
  pipelines}. In \bibinfo{booktitle}{\emph{Proceedings of the 31st ACM SIGPLAN
  Conference on Programming Language Design and Implementation}} (Toronto,
  Ontario, Canada) \emph{(\bibinfo{series}{PLDI '10})}.
  \bibinfo{publisher}{Association for Computing Machinery},
  \bibinfo{address}{New York, NY, USA}, \bibinfo{pages}{363–375}.
\newblock
\showISBNx{9781450300193}
\href{https://doi.org/10.1145/1806596.1806638}{doi:\nolinkurl{10.1145/1806596.1806638}}


\bibitem[Chandramouli et~al\mbox{.}(2014)]%
        {chandramouli2014trill}
\bibfield{author}{\bibinfo{person}{Badrish Chandramouli},
  \bibinfo{person}{Jonathan Goldstein}, \bibinfo{person}{Mike Barnett},
  \bibinfo{person}{Robert DeLine}, \bibinfo{person}{Danyel Fisher},
  \bibinfo{person}{John~C. Platt}, \bibinfo{person}{James~F. Terwilliger},
  {and} \bibinfo{person}{John Wernsing}.} \bibinfo{year}{2014}\natexlab{}.
\newblock \showarticletitle{Trill: a high-performance incremental query
  processor for diverse analytics}.
\newblock \bibinfo{journal}{\emph{Proc. VLDB Endow.}} \bibinfo{volume}{8},
  \bibinfo{number}{4} (\bibinfo{date}{Dec.} \bibinfo{year}{2014}),
  \bibinfo{pages}{401–412}.
\newblock
\showISSN{2150-8097}
\href{https://doi.org/10.14778/2735496.2735503}{doi:\nolinkurl{10.14778/2735496.2735503}}


\bibitem[Chandy and Lamport(1985)]%
        {chandy1985snapshot}
\bibfield{author}{\bibinfo{person}{K.~Mani Chandy} {and}
  \bibinfo{person}{Leslie Lamport}.} \bibinfo{year}{1985}\natexlab{}.
\newblock \showarticletitle{Distributed snapshots: determining global states of
  distributed systems}.
\newblock \bibinfo{journal}{\emph{ACM Trans. Comput. Syst.}}
  \bibinfo{volume}{3}, \bibinfo{number}{1} (\bibinfo{date}{Feb.}
  \bibinfo{year}{1985}), \bibinfo{pages}{63–75}.
\newblock
\showISSN{0734-2071}
\href{https://doi.org/10.1145/214451.214456}{doi:\nolinkurl{10.1145/214451.214456}}


\bibitem[Chang et~al\mbox{.}(2024)]%
        {chang2024gmt}
\bibfield{author}{\bibinfo{person}{Chia-Hao Chang}, \bibinfo{person}{Jihoon
  Han}, \bibinfo{person}{Anand Sivasubramaniam}, \bibinfo{person}{Vikram~Sharma
  Mailthody}, \bibinfo{person}{Zaid Qureshi}, {and} \bibinfo{person}{Wen mei
  Hwu}.} \bibinfo{year}{2024}\natexlab{}.
\newblock \showarticletitle{GMT: GPU Orchestrated Memory Tiering for the Big
  Data Era}. In \bibinfo{booktitle}{\emph{Proceedings of the 29th ACM
  International Conference on Architectural Support for Programming Languages
  and Operating Systems, Volume 3, ASPLOS 2024, La Jolla, CA, USA, 27 April
  2024- 1 May 2024}}, \bibfield{editor}{\bibinfo{person}{Rajiv~Gupta 0001},
  \bibinfo{person}{Nael~B. Abu-Ghazaleh}, \bibinfo{person}{Madan Musuvathi},
  {and} \bibinfo{person}{Dan Tsafrir}} (Eds.). \bibinfo{publisher}{ACM},
  \bibinfo{pages}{464--478}.
\newblock
\href{https://doi.org/10.1145/3620666.3651353}{doi:\nolinkurl{10.1145/3620666.3651353}}


\bibitem[Chang et~al\mbox{.}(2008)]%
        {chang2008bigtable}
\bibfield{author}{\bibinfo{person}{Fay Chang}, \bibinfo{person}{Jeffrey Dean},
  \bibinfo{person}{Sanjay Ghemawat}, \bibinfo{person}{Wilson~C. Hsieh},
  \bibinfo{person}{Deborah~A. Wallach}, \bibinfo{person}{Mike Burrows},
  \bibinfo{person}{Tushar Chandra}, \bibinfo{person}{Andrew Fikes}, {and}
  \bibinfo{person}{Robert~E. Gruber}.} \bibinfo{year}{2008}\natexlab{}.
\newblock \showarticletitle{Bigtable: A Distributed Storage System for
  Structured Data}.
\newblock \bibinfo{journal}{\emph{ACM Trans. Comput. Syst.}}
  \bibinfo{volume}{26}, \bibinfo{number}{2}, Article \bibinfo{articleno}{4}
  (\bibinfo{date}{June} \bibinfo{year}{2008}), \bibinfo{numpages}{26}~pages.
\newblock
\showISSN{0734-2071}
\href{https://doi.org/10.1145/1365815.1365816}{doi:\nolinkurl{10.1145/1365815.1365816}}


\bibitem[Chaudhry et~al\mbox{.}(2019)]%
        {chaudhry2019agem}
\bibfield{author}{\bibinfo{person}{Arslan Chaudhry},
  \bibinfo{person}{Marc'Aurelio Ranzato}, \bibinfo{person}{Marcus Rohrbach},
  {and} \bibinfo{person}{Mohamed Elhoseiny}.} \bibinfo{year}{2019}\natexlab{}.
\newblock \showarticletitle{{Efficient Lifelong Learning with A-GEM}}. In
  \bibinfo{booktitle}{\emph{International Conference on Learning
  Representations}}.
\newblock
\urldef\tempurl%
\url{https://openreview.net/forum?id=Hkf2_sC5FX}
\showURL{%
\tempurl}


\bibitem[Chen et~al\mbox{.}(2024a)]%
        {chen2024singlestorev}
\bibfield{author}{\bibinfo{person}{Cheng Chen}, \bibinfo{person}{Chenzhe Jin},
  \bibinfo{person}{Yunan Zhang}, \bibinfo{person}{Sasha Podolsky},
  \bibinfo{person}{Chun Wu}, \bibinfo{person}{Szu-Po Wang},
  \bibinfo{person}{Eric Hanson}, \bibinfo{person}{Zhou Sun},
  \bibinfo{person}{Robert Walzer}, {and} \bibinfo{person}{Jianguo Wang}.}
  \bibinfo{year}{2024}\natexlab{a}.
\newblock \showarticletitle{SingleStore-V: An Integrated Vector Database System
  in SingleStore}.
\newblock \bibinfo{journal}{\emph{Proc. VLDB Endow.}} \bibinfo{volume}{17},
  \bibinfo{number}{12} (\bibinfo{date}{Aug.} \bibinfo{year}{2024}),
  \bibinfo{pages}{3772–3785}.
\newblock
\showISSN{2150-8097}
\href{https://doi.org/10.14778/3685800.3685805}{doi:\nolinkurl{10.14778/3685800.3685805}}


\bibitem[Chen et~al\mbox{.}(2025)]%
        {chen2025heteroinfer}
\bibfield{author}{\bibinfo{person}{Le Chen}, \bibinfo{person}{Dahu Feng},
  \bibinfo{person}{Erhu Feng}, \bibinfo{person}{Yingrui Wang},
  \bibinfo{person}{Rong Zhao}, \bibinfo{person}{Yubin Xia},
  \bibinfo{person}{Pinjie Xu}, {and} \bibinfo{person}{Haibo Chen}.}
  \bibinfo{year}{2025}\natexlab{}.
\newblock \showarticletitle{Characterizing Mobile SoC for Accelerating
  Heterogeneous LLM Inference}. In \bibinfo{booktitle}{\emph{Proceedings of the
  ACM SIGOPS 31st Symposium on Operating Systems Principles}} (Lotte Hotel
  World, Seoul, Republic of Korea) \emph{(\bibinfo{series}{SOSP '25})}.
  \bibinfo{publisher}{Association for Computing Machinery},
  \bibinfo{address}{New York, NY, USA}, \bibinfo{pages}{359–374}.
\newblock
\showISBNx{9798400718700}
\href{https://doi.org/10.1145/3731569.3764808}{doi:\nolinkurl{10.1145/3731569.3764808}}


\bibitem[Chen et~al\mbox{.}(2024b)]%
        {chen2024punica}
\bibfield{author}{\bibinfo{person}{Lequn Chen}, \bibinfo{person}{Zihao Ye},
  \bibinfo{person}{Yongji Wu}, \bibinfo{person}{Danyang Zhuo},
  \bibinfo{person}{Luis Ceze}, {and} \bibinfo{person}{Arvind Krishnamurthy}.}
  \bibinfo{year}{2024}\natexlab{b}.
\newblock \showarticletitle{Punica: Multi-Tenant LoRA Serving}. In
  \bibinfo{booktitle}{\emph{Proceedings of Machine Learning and Systems}},
  \bibfield{editor}{\bibinfo{person}{P.~Gibbons},
  \bibinfo{person}{G.~Pekhimenko}, {and} \bibinfo{person}{C.~De Sa}} (Eds.),
  Vol.~\bibinfo{volume}{6}. \bibinfo{pages}{1--13}.
\newblock
\urldef\tempurl%
\url{https://proceedings.mlsys.org/paper_files/paper/2024/file/054de805fcceb78a201f5e9d53c85908-Paper-Conference.pdf}
\showURL{%
\tempurl}


\bibitem[Cho et~al\mbox{.}(2024)]%
        {cho2024coaxial}
\bibfield{author}{\bibinfo{person}{Albert Cho}, \bibinfo{person}{Anish Saxena},
  \bibinfo{person}{Moinuddin Qureshi}, {and} \bibinfo{person}{Alexandros
  Daglis}.} \bibinfo{year}{2024}\natexlab{}.
\newblock \showarticletitle{COAXIAL: A CXL-Centric Memory System for Scalable
  Servers}. In \bibinfo{booktitle}{\emph{Proceedings of the International
  Conference for High Performance Computing, Networking, Storage, and
  Analysis}} (Atlanta, GA, USA) \emph{(\bibinfo{series}{SC '24})}.
  \bibinfo{publisher}{IEEE Press}, Article \bibinfo{articleno}{95},
  \bibinfo{numpages}{15}~pages.
\newblock
\showISBNx{9798350352917}
\href{https://doi.org/10.1109/SC41406.2024.00101}{doi:\nolinkurl{10.1109/SC41406.2024.00101}}


\bibitem[Cong et~al\mbox{.}(2026)]%
        {cong2026mico}
\bibfield{author}{\bibinfo{person}{Peizhuang Cong}, \bibinfo{person}{Tong
  Yang}, \bibinfo{person}{Yuchao Zhang}, \bibinfo{person}{Wendong Wang}, {and}
  \bibinfo{person}{Ke Xu}.} \bibinfo{year}{2026}\natexlab{}.
\newblock \showarticletitle{MICO: efficient query scheduling for multi-cloud
  deployed LLM inference service}.
\newblock \bibinfo{journal}{\emph{Science China Information Sciences}}
  \bibinfo{volume}{69}, \bibinfo{number}{3} (\bibinfo{year}{2026}),
  \bibinfo{pages}{132102}.
\newblock


\bibitem[Corbett et~al\mbox{.}(2013)]%
        {corbett2012spanner}
\bibfield{author}{\bibinfo{person}{James~C. Corbett}, \bibinfo{person}{Jeffrey
  Dean}, \bibinfo{person}{Michael Epstein}, \bibinfo{person}{Andrew Fikes},
  \bibinfo{person}{Christopher Frost}, \bibinfo{person}{J.~J. Furman},
  \bibinfo{person}{Sanjay Ghemawat}, \bibinfo{person}{Andrey Gubarev},
  \bibinfo{person}{Christopher Heiser}, \bibinfo{person}{Peter Hochschild},
  \bibinfo{person}{Wilson Hsieh}, \bibinfo{person}{Sebastian Kanthak},
  \bibinfo{person}{Eugene Kogan}, \bibinfo{person}{Hongyi Li},
  \bibinfo{person}{Alexander Lloyd}, \bibinfo{person}{Sergey Melnik},
  \bibinfo{person}{David Mwaura}, \bibinfo{person}{David Nagle},
  \bibinfo{person}{Sean Quinlan}, \bibinfo{person}{Rajesh Rao},
  \bibinfo{person}{Lindsay Rolig}, \bibinfo{person}{Yasushi Saito},
  \bibinfo{person}{Michal Szymaniak}, \bibinfo{person}{Christopher Taylor},
  \bibinfo{person}{Ruth Wang}, {and} \bibinfo{person}{Dale Woodford}.}
  \bibinfo{year}{2013}\natexlab{}.
\newblock \showarticletitle{Spanner: Google’s Globally Distributed Database}.
\newblock \bibinfo{journal}{\emph{ACM Trans. Comput. Syst.}}
  \bibinfo{volume}{31}, \bibinfo{number}{3}, Article \bibinfo{articleno}{8}
  (\bibinfo{date}{Aug.} \bibinfo{year}{2013}), \bibinfo{numpages}{22}~pages.
\newblock
\showISSN{0734-2071}
\href{https://doi.org/10.1145/2491245}{doi:\nolinkurl{10.1145/2491245}}


\bibitem[Crankshaw et~al\mbox{.}(2020)]%
        {crankshaw2020inferline}
\bibfield{author}{\bibinfo{person}{Daniel Crankshaw}, \bibinfo{person}{Gur-Eyal
  Sela}, \bibinfo{person}{Xiangxi Mo}, \bibinfo{person}{Corey Zumar},
  \bibinfo{person}{Ion Stoica}, \bibinfo{person}{Joseph Gonzalez}, {and}
  \bibinfo{person}{Alexey Tumanov}.} \bibinfo{year}{2020}\natexlab{}.
\newblock \showarticletitle{InferLine: latency-aware provisioning and scaling
  for prediction serving pipelines}. In \bibinfo{booktitle}{\emph{Proceedings
  of the 11th ACM Symposium on Cloud Computing}} (Virtual Event, USA)
  \emph{(\bibinfo{series}{SoCC '20})}. \bibinfo{publisher}{Association for
  Computing Machinery}, \bibinfo{address}{New York, NY, USA},
  \bibinfo{pages}{477–491}.
\newblock
\showISBNx{9781450381376}
\href{https://doi.org/10.1145/3419111.3421285}{doi:\nolinkurl{10.1145/3419111.3421285}}


\bibitem[Crankshaw et~al\mbox{.}(2017)]%
        {crankshaw2017clipper}
\bibfield{author}{\bibinfo{person}{Daniel Crankshaw}, \bibinfo{person}{Xin
  Wang}, \bibinfo{person}{Giulio Zhou}, \bibinfo{person}{Michael~J. Franklin},
  \bibinfo{person}{Joseph~E. Gonzalez}, {and} \bibinfo{person}{Ion Stoica}.}
  \bibinfo{year}{2017}\natexlab{}.
\newblock \showarticletitle{Clipper: a low-latency online prediction serving
  system}. In \bibinfo{booktitle}{\emph{Proceedings of the 14th USENIX
  Conference on Networked Systems Design and Implementation}} (Boston, MA, USA)
  \emph{(\bibinfo{series}{NSDI'17})}. \bibinfo{publisher}{USENIX Association},
  \bibinfo{address}{USA}, \bibinfo{pages}{613–627}.
\newblock
\showISBNx{9781931971379}


\bibitem[Cugola and Margara(2012)]%
        {cugola2012processing}
\bibfield{author}{\bibinfo{person}{Gianpaolo Cugola} {and}
  \bibinfo{person}{Alessandro Margara}.} \bibinfo{year}{2012}\natexlab{}.
\newblock \showarticletitle{Processing flows of information: From data stream
  to complex event processing}.
\newblock \bibinfo{journal}{\emph{ACM Comput. Surv.}} \bibinfo{volume}{44},
  \bibinfo{number}{3}, Article \bibinfo{articleno}{15} (\bibinfo{date}{June}
  \bibinfo{year}{2012}), \bibinfo{numpages}{62}~pages.
\newblock
\showISSN{0360-0300}
\href{https://doi.org/10.1145/2187671.2187677}{doi:\nolinkurl{10.1145/2187671.2187677}}


\bibitem[Dao(2024)]%
        {dao2024flashattention2}
\bibfield{author}{\bibinfo{person}{Tri Dao}.} \bibinfo{year}{2024}\natexlab{}.
\newblock \showarticletitle{Flashattention-2: Faster attention with better
  parallelism and work partitioning}. In
  \bibinfo{booktitle}{\emph{International Conference on Learning
  Representations}}, Vol.~\bibinfo{volume}{2024}.
  \bibinfo{pages}{35549--35562}.
\newblock


\bibitem[De~Lange et~al\mbox{.}(2022)]%
        {delange2021survey}
\bibfield{author}{\bibinfo{person}{Matthias De~Lange}, \bibinfo{person}{Rahaf
  Aljundi}, \bibinfo{person}{Marc Masana}, \bibinfo{person}{Sarah Parisot},
  \bibinfo{person}{Xu Jia}, \bibinfo{person}{Aleš Leonardis},
  \bibinfo{person}{Gregory Slabaugh}, {and} \bibinfo{person}{Tinne
  Tuytelaars}.} \bibinfo{year}{2022}\natexlab{}.
\newblock \showarticletitle{A Continual Learning Survey: Defying Forgetting in
  Classification Tasks}.
\newblock \bibinfo{journal}{\emph{IEEE Transactions on Pattern Analysis and
  Machine Intelligence}} \bibinfo{volume}{44}, \bibinfo{number}{7}
  (\bibinfo{year}{2022}), \bibinfo{pages}{3366--3385}.
\newblock
\href{https://doi.org/10.1109/TPAMI.2021.3057446}{doi:\nolinkurl{10.1109/TPAMI.2021.3057446}}


\bibitem[DeCandia et~al\mbox{.}(2007)]%
        {decandia2007dynamo}
\bibfield{author}{\bibinfo{person}{Giuseppe DeCandia}, \bibinfo{person}{Deniz
  Hastorun}, \bibinfo{person}{Madan Jampani}, \bibinfo{person}{Gunavardhan
  Kakulapati}, \bibinfo{person}{Avinash Lakshman}, \bibinfo{person}{Alex
  Pilchin}, \bibinfo{person}{Swaminathan Sivasubramanian},
  \bibinfo{person}{Peter Vosshall}, {and} \bibinfo{person}{Werner Vogels}.}
  \bibinfo{year}{2007}\natexlab{}.
\newblock \showarticletitle{Dynamo: amazon's highly available key-value store}.
  In \bibinfo{booktitle}{\emph{Proceedings of Twenty-First ACM SIGOPS Symposium
  on Operating Systems Principles}} (Stevenson, Washington, USA)
  \emph{(\bibinfo{series}{SOSP '07})}. \bibinfo{publisher}{Association for
  Computing Machinery}, \bibinfo{address}{New York, NY, USA},
  \bibinfo{pages}{205–220}.
\newblock
\showISBNx{9781595935915}
\href{https://doi.org/10.1145/1294261.1294281}{doi:\nolinkurl{10.1145/1294261.1294281}}


\bibitem[Domingo and Kannan(2021)]%
        {domingo2021pfsck}
\bibfield{author}{\bibinfo{person}{David Domingo} {and}
  \bibinfo{person}{Sudarsun Kannan}.} \bibinfo{year}{2021}\natexlab{}.
\newblock \showarticletitle{{pFSCK}: Accelerating File System Checking and
  Repair for Modern Storage}. In \bibinfo{booktitle}{\emph{19th USENIX
  Conference on File and Storage Technologies (FAST 21)}}.
  \bibinfo{publisher}{USENIX Association}, \bibinfo{pages}{113--126}.
\newblock
\showISBNx{978-1-939133-20-5}
\urldef\tempurl%
\url{https://www.usenix.org/conference/fast21/presentation/domingo}
\showURL{%
\tempurl}


\bibitem[Dong et~al\mbox{.}(2025)]%
        {dong2024toleo}
\bibfield{author}{\bibinfo{person}{Juechu Dong}, \bibinfo{person}{Jonah
  Rosenblum}, {and} \bibinfo{person}{Satish Narayanasamy}.}
  \bibinfo{year}{2025}\natexlab{}.
\newblock \showarticletitle{Toleo: Scaling Freshness to Tera-scale Memory Using
  CXL and PIM}. In \bibinfo{booktitle}{\emph{Proceedings of the 29th ACM
  International Conference on Architectural Support for Programming Languages
  and Operating Systems, Volume 4}} (Hilton La Jolla Torrey Pines, La Jolla,
  CA, USA) \emph{(\bibinfo{series}{ASPLOS '24})}.
  \bibinfo{publisher}{Association for Computing Machinery},
  \bibinfo{address}{New York, NY, USA}, \bibinfo{pages}{313–328}.
\newblock
\showISBNx{9798400703911}
\href{https://doi.org/10.1145/3622781.3674180}{doi:\nolinkurl{10.1145/3622781.3674180}}


\bibitem[Dragojevi\'{c} et~al\mbox{.}(2014)]%
        {dragojevic2014farm}
\bibfield{author}{\bibinfo{person}{Aleksandar Dragojevi\'{c}},
  \bibinfo{person}{Dushyanth Narayanan}, \bibinfo{person}{Orion Hodson}, {and}
  \bibinfo{person}{Miguel Castro}.} \bibinfo{year}{2014}\natexlab{}.
\newblock \showarticletitle{FaRM: fast remote memory}. In
  \bibinfo{booktitle}{\emph{Proceedings of the 11th USENIX Conference on
  Networked Systems Design and Implementation}} (Seattle, WA)
  \emph{(\bibinfo{series}{NSDI'14})}. \bibinfo{publisher}{USENIX Association},
  \bibinfo{address}{USA}, \bibinfo{pages}{401–414}.
\newblock
\showISBNx{9781931971096}


\bibitem[Du et~al\mbox{.}(2025)]%
        {du2025prefillonly}
\bibfield{author}{\bibinfo{person}{Kuntai Du}, \bibinfo{person}{Bowen Wang},
  \bibinfo{person}{Chen Zhang}, \bibinfo{person}{Yiming Cheng},
  \bibinfo{person}{Qing Lan}, \bibinfo{person}{Hejian Sang},
  \bibinfo{person}{Yihua Cheng}, \bibinfo{person}{Jiayi Yao},
  \bibinfo{person}{Xiaoxuan Liu}, \bibinfo{person}{Yifan Qiao},
  \bibinfo{person}{Ion Stoica}, {and} \bibinfo{person}{Junchen Jiang}.}
  \bibinfo{year}{2025}\natexlab{}.
\newblock \showarticletitle{PrefillOnly: An Inference Engine for Prefill-only
  Workloads in Large Language Model Applications}.
\newblock  (\bibinfo{year}{2025}).
\newblock
\href{https://doi.org/10.48550/ARXIV.2505.07203}{doi:\nolinkurl{10.48550/ARXIV.2505.07203}}


\bibitem[Eisenman et~al\mbox{.}(2022)]%
        {eisenman2022checknrun}
\bibfield{author}{\bibinfo{person}{Assaf Eisenman},
  \bibinfo{person}{Kiran~Kumar Matam}, \bibinfo{person}{Steven Ingram},
  \bibinfo{person}{Dheevatsa Mudigere}, \bibinfo{person}{Raghuraman
  Krishnamoorthi}, \bibinfo{person}{Krishnakumar Nair}, \bibinfo{person}{Misha
  Smelyanskiy}, {and} \bibinfo{person}{Murali Annavaram}.}
  \bibinfo{year}{2022}\natexlab{}.
\newblock \showarticletitle{{Check-N-Run}: a Checkpointing System for Training
  Deep Learning Recommendation Models}. In \bibinfo{booktitle}{\emph{19th
  USENIX Symposium on Networked Systems Design and Implementation (NSDI 22)}}.
  \bibinfo{publisher}{USENIX Association}, \bibinfo{address}{Renton, WA},
  \bibinfo{pages}{929--943}.
\newblock
\showISBNx{978-1-939133-27-4}
\urldef\tempurl%
\url{https://www.usenix.org/conference/nsdi22/presentation/eisenman}
\showURL{%
\tempurl}


\bibitem[Fan et~al\mbox{.}(2025)]%
        {fan2025spinfer}
\bibfield{author}{\bibinfo{person}{Ruibo Fan}, \bibinfo{person}{Xiangrui Yu},
  \bibinfo{person}{Peijie Dong}, \bibinfo{person}{Zeyu Li}, \bibinfo{person}{Gu
  Gong}, \bibinfo{person}{Qiang Wang}, \bibinfo{person}{Wei Wang}, {and}
  \bibinfo{person}{Xiaowen Chu}.} \bibinfo{year}{2025}\natexlab{}.
\newblock \showarticletitle{SpInfer: Leveraging Low-Level Sparsity for
  Efficient Large Language Model Inference on GPUs}. In
  \bibinfo{booktitle}{\emph{Proceedings of the Twentieth European Conference on
  Computer Systems}} (Rotterdam, Netherlands) \emph{(\bibinfo{series}{EuroSys
  '25})}. \bibinfo{publisher}{Association for Computing Machinery},
  \bibinfo{address}{New York, NY, USA}, \bibinfo{pages}{243–260}.
\newblock
\showISBNx{9798400711961}
\href{https://doi.org/10.1145/3689031.3717481}{doi:\nolinkurl{10.1145/3689031.3717481}}


\bibitem[Fang et~al\mbox{.}(2025)]%
        {fang2025anda}
\bibfield{author}{\bibinfo{person}{Chao Fang}, \bibinfo{person}{Man Shi},
  \bibinfo{person}{Robin Geens}, \bibinfo{person}{Arne Symons},
  \bibinfo{person}{Zhongfeng Wang}, {and} \bibinfo{person}{Marian Verhelst}.}
  \bibinfo{year}{2025}\natexlab{}.
\newblock \showarticletitle{Anda: Unlocking Efficient LLM Inference with a
  Variable-Length Grouped Activation Data Format}. In
  \bibinfo{booktitle}{\emph{2025 IEEE International Symposium on High
  Performance Computer Architecture (HPCA)}}. \bibinfo{pages}{1467--1481}.
\newblock
\href{https://doi.org/10.1109/HPCA61900.2025.00110}{doi:\nolinkurl{10.1109/HPCA61900.2025.00110}}


\bibitem[Fang et~al\mbox{.}(2026)]%
        {fang2026planetserve}
\bibfield{author}{\bibinfo{person}{Fei Fang}, \bibinfo{person}{Yifan Hua},
  \bibinfo{person}{Shengze Wang}, \bibinfo{person}{Ruilin Zhou},
  \bibinfo{person}{Yi Liu}, \bibinfo{person}{Chen Qian}, {and}
  \bibinfo{person}{Xiaoxue Zhang}.} \bibinfo{year}{2026}\natexlab{}.
\newblock \showarticletitle{{PlanetServe}: A Decentralized, Scalable, and
  {Privacy-Preserving} Overlay for Democratizing Large Language Model Serving}.
  In \bibinfo{booktitle}{\emph{23rd USENIX Symposium on Networked Systems
  Design and Implementation (NSDI 26)}}. \bibinfo{publisher}{USENIX
  Association}, \bibinfo{address}{Renton, WA}, \bibinfo{pages}{2111--2129}.
\newblock
\showISBNx{978-1-939133-54-0}
\urldef\tempurl%
\url{https://www.usenix.org/conference/nsdi26/presentation/fang}
\showURL{%
\tempurl}


\bibitem[Federici et~al\mbox{.}(2025)]%
        {federici2025dip}
\bibfield{author}{\bibinfo{person}{Marco Federici}, \bibinfo{person}{Davide
  Belli}, \bibinfo{person}{Mart Van~Baalen}, \bibinfo{person}{Amir Jalalirad},
  \bibinfo{person}{Andrii Skliar}, \bibinfo{person}{Bence Major},
  \bibinfo{person}{Markus Nagel}, {and} \bibinfo{person}{Paul Whatmough}.}
  \bibinfo{year}{2025}\natexlab{}.
\newblock \showarticletitle{Efficient LLM Inference using Dynamic Input Pruning
  and Cache-Aware Masking}. In \bibinfo{booktitle}{\emph{Proceedings of Machine
  Learning and Systems}}, \bibfield{editor}{\bibinfo{person}{M.~Zaharia},
  \bibinfo{person}{G.~Joshi}, {and} \bibinfo{person}{Y.~Lin}} (Eds.),
  Vol.~\bibinfo{volume}{7}. \bibinfo{publisher}{MLSys}.
\newblock
\urldef\tempurl%
\url{https://proceedings.mlsys.org/paper_files/paper/2025/file/afd6374c7f2839cba22f537f15f4f760-Paper-Conference.pdf}
\showURL{%
\tempurl}


\bibitem[Feng et~al\mbox{.}(2025b)]%
        {windserve2025}
\bibfield{author}{\bibinfo{person}{Jingqi Feng}, \bibinfo{person}{Yukai Huang},
  \bibinfo{person}{Rui Zhang}, \bibinfo{person}{Sicheng Liang},
  \bibinfo{person}{Ming Yan}, {and} \bibinfo{person}{Jie Wu}.}
  \bibinfo{year}{2025}\natexlab{b}.
\newblock \showarticletitle{WindServe: Efficient Phase-Disaggregated LLM
  Serving with Stream-based Dynamic Scheduling}. In
  \bibinfo{booktitle}{\emph{Proceedings of the 52nd Annual International
  Symposium on Computer Architecture}} \emph{(\bibinfo{series}{ISCA '25})}.
  \bibinfo{publisher}{Association for Computing Machinery},
  \bibinfo{address}{New York, NY, USA}, \bibinfo{pages}{1283–1295}.
\newblock
\showISBNx{9798400712616}
\href{https://doi.org/10.1145/3695053.3730999}{doi:\nolinkurl{10.1145/3695053.3730999}}


\bibitem[Feng et~al\mbox{.}(2025a)]%
        {feng2025optimus}
\bibfield{author}{\bibinfo{person}{Weiqi Feng}, \bibinfo{person}{Yangrui Chen},
  \bibinfo{person}{Shaoyu Wang}, \bibinfo{person}{Yanghua Peng},
  \bibinfo{person}{Haibin Lin}, {and} \bibinfo{person}{Minlan Yu}.}
  \bibinfo{year}{2025}\natexlab{a}.
\newblock \showarticletitle{Optimus: accelerating large-scale multi-modal LLM
  training by bubble exploitation}. In \bibinfo{booktitle}{\emph{Proceedings of
  the 2025 USENIX Conference on Usenix Annual Technical Conference}} (Boston,
  MA, USA) \emph{(\bibinfo{series}{USENIX ATC '25})}.
  \bibinfo{publisher}{USENIX Association}, \bibinfo{address}{USA}, Article
  \bibinfo{articleno}{10}, \bibinfo{numpages}{17}~pages.
\newblock
\showISBNx{978-1-939133-48-9}


\bibitem[Fu et~al\mbox{.}(2024b)]%
        {fu2024autocheck}
\bibfield{author}{\bibinfo{person}{Xiang Fu}, \bibinfo{person}{Weiping Zhang},
  \bibinfo{person}{Shiman Meng}, \bibinfo{person}{Xin Huang},
  \bibinfo{person}{Wubiao Xu}, \bibinfo{person}{Luanzheng Guo}, {and}
  \bibinfo{person}{Kento Sato}.} \bibinfo{year}{2024}\natexlab{b}.
\newblock \showarticletitle{AutoCheck: Automatically Identifying Variables for
  Checkpointing by Data Dependency Analysis}. In
  \bibinfo{booktitle}{\emph{Proceedings of the International Conference for
  High Performance Computing, Networking, Storage, and Analysis}} (Atlanta, GA,
  USA) \emph{(\bibinfo{series}{SC '24})}. \bibinfo{publisher}{IEEE Press},
  Article \bibinfo{articleno}{99}, \bibinfo{numpages}{16}~pages.
\newblock
\showISBNx{9798350352917}
\href{https://doi.org/10.1109/SC41406.2024.00105}{doi:\nolinkurl{10.1109/SC41406.2024.00105}}


\bibitem[Fu et~al\mbox{.}(2023)]%
        {fu2024autoscratch}
\bibfield{author}{\bibinfo{person}{Yaosheng Fu}, \bibinfo{person}{Evgeny
  Bolotin}, \bibinfo{person}{Aamer Jaleel}, \bibinfo{person}{Gal Dalal},
  \bibinfo{person}{Shie Mannor}, \bibinfo{person}{Jacob Subag},
  \bibinfo{person}{Noam Korem}, \bibinfo{person}{Michael Behar}, {and}
  \bibinfo{person}{David Nellans}.} \bibinfo{year}{2023}\natexlab{}.
\newblock \showarticletitle{AutoScratch: ML-Optimized Cache Management for
  Inference-Oriented GPUs}. In \bibinfo{booktitle}{\emph{Proceedings of Machine
  Learning and Systems}}, \bibfield{editor}{\bibinfo{person}{D.~Song},
  \bibinfo{person}{M.~Carbin}, {and} \bibinfo{person}{T.~Chen}} (Eds.),
  Vol.~\bibinfo{volume}{5}. \bibinfo{publisher}{Curan},
  \bibinfo{pages}{495--512}.
\newblock
\urldef\tempurl%
\url{https://proceedings.mlsys.org/paper_files/paper/2023/file/9d32b9324a89001520ae456b9e5ec73b-Paper-mlsys2023.pdf}
\showURL{%
\tempurl}


\bibitem[Fu et~al\mbox{.}(2022)]%
        {fu2022sfs}
\bibfield{author}{\bibinfo{person}{Yuqi Fu}, \bibinfo{person}{Li Liu},
  \bibinfo{person}{Haoliang Wang}, \bibinfo{person}{Yue Cheng}, {and}
  \bibinfo{person}{Songqing Chen}.} \bibinfo{year}{2022}\natexlab{}.
\newblock \showarticletitle{SFS: smart OS scheduling for serverless functions}.
  In \bibinfo{booktitle}{\emph{Proceedings of the International Conference on
  High Performance Computing, Networking, Storage and Analysis}} (Dallas,
  Texas) \emph{(\bibinfo{series}{SC '22})}. \bibinfo{publisher}{IEEE Press},
  Article \bibinfo{articleno}{42}, \bibinfo{numpages}{16}~pages.
\newblock
\showISBNx{9784665454445}


\bibitem[Fu et~al\mbox{.}(2024a)]%
        {fu2024serverlessllm}
\bibfield{author}{\bibinfo{person}{Yao Fu}, \bibinfo{person}{Leyang Xue},
  \bibinfo{person}{Yeqi Huang}, \bibinfo{person}{Andrei-Octavian Brabete},
  \bibinfo{person}{Dmitrii Ustiugov}, \bibinfo{person}{Yuvraj Patel}, {and}
  \bibinfo{person}{Luo Mai}.} \bibinfo{year}{2024}\natexlab{a}.
\newblock \showarticletitle{ServerlessLLM: low-latency serverless inference for
  large language models}. In \bibinfo{booktitle}{\emph{Proceedings of the 18th
  USENIX Conference on Operating Systems Design and Implementation}} (Santa
  Clara, CA, USA) \emph{(\bibinfo{series}{OSDI'24})}.
  \bibinfo{publisher}{USENIX Association}, \bibinfo{address}{USA}, Article
  \bibinfo{articleno}{8}, \bibinfo{numpages}{19}~pages.
\newblock
\showISBNx{978-1-939133-40-3}


\bibitem[Gao et~al\mbox{.}(2024)]%
        {gao2025cachedattention}
\bibfield{author}{\bibinfo{person}{Bin Gao}, \bibinfo{person}{Zhuomin He},
  \bibinfo{person}{Puru Sharma}, \bibinfo{person}{Qingxuan Kang},
  \bibinfo{person}{Djordje Jevdjic}, \bibinfo{person}{Junbo Deng},
  \bibinfo{person}{Xingkun Yang}, \bibinfo{person}{Zhou Yu}, {and}
  \bibinfo{person}{Pengfei Zuo}.} \bibinfo{year}{2024}\natexlab{}.
\newblock \showarticletitle{Cost-efficient large language model serving for
  multi-turn conversations with CachedAttention}. In
  \bibinfo{booktitle}{\emph{Proceedings of the 2024 USENIX Conference on Usenix
  Annual Technical Conference}} (Santa Clara, CA, USA)
  \emph{(\bibinfo{series}{USENIX ATC'24})}. \bibinfo{publisher}{USENIX
  Association}, \bibinfo{address}{USA}, Article \bibinfo{articleno}{7},
  \bibinfo{numpages}{16}~pages.
\newblock
\showISBNx{978-1-939133-41-0}


\bibitem[Gao et~al\mbox{.}(2026)]%
        {gao2026grace}
\bibfield{author}{\bibinfo{person}{Hongru Gao}, \bibinfo{person}{Shuhao Zhang},
  \bibinfo{person}{Xiaofei Liao}, {and} \bibinfo{person}{Hai Jin}.}
  \bibinfo{year}{2026}\natexlab{}.
\newblock \showarticletitle{GRACE: Alleviating Reconstruction Cost in Dynamic
  Graph Processing Systems}. In \bibinfo{booktitle}{\emph{Proceedings of the
  IEEE International Conference on Data Engineering}}.
\newblock


\bibitem[Gao et~al\mbox{.}(2025b)]%
        {gao2025hcache}
\bibfield{author}{\bibinfo{person}{Shiwei Gao}, \bibinfo{person}{Youmin Chen},
  {and} \bibinfo{person}{Jiwu Shu}.} \bibinfo{year}{2025}\natexlab{b}.
\newblock \showarticletitle{Fast State Restoration in LLM Serving with HCache}.
  In \bibinfo{booktitle}{\emph{Proceedings of the Twentieth European Conference
  on Computer Systems}} (Rotterdam, Netherlands)
  \emph{(\bibinfo{series}{EuroSys '25})}. \bibinfo{publisher}{Association for
  Computing Machinery}, \bibinfo{address}{New York, NY, USA},
  \bibinfo{pages}{128–143}.
\newblock
\showISBNx{9798400711961}
\href{https://doi.org/10.1145/3689031.3696072}{doi:\nolinkurl{10.1145/3689031.3696072}}


\bibitem[Gao et~al\mbox{.}(2025c)]%
        {gao2025weaver}
\bibfield{author}{\bibinfo{person}{Shiwei Gao}, \bibinfo{person}{Qing Wang},
  \bibinfo{person}{Shaoxun Zeng}, \bibinfo{person}{Youyou Lu}, {and}
  \bibinfo{person}{Jiwu Shu}.} \bibinfo{year}{2025}\natexlab{c}.
\newblock \showarticletitle{Weaver: Efficient {Multi-LLM} Serving with
  Attention Offloading}. In \bibinfo{booktitle}{\emph{2025 USENIX Annual
  Technical Conference (USENIX ATC 25)}}. \bibinfo{publisher}{USENIX
  Association}, \bibinfo{address}{Boston, MA}, \bibinfo{pages}{587--595}.
\newblock
\showISBNx{978-1-939133-48-9}
\urldef\tempurl%
\url{https://www.usenix.org/conference/atc25/presentation/gao}
\showURL{%
\tempurl}


\bibitem[Gao et~al\mbox{.}(2025a)]%
        {gao2025deck}
\bibfield{author}{\bibinfo{person}{Xin Gao}, \bibinfo{person}{Sibasish
  Acharya}, \bibinfo{person}{Sihui Han}, \bibinfo{person}{Yongxiong Ren},
  \bibinfo{person}{Yanli Zhao}, \bibinfo{person}{Liang Luo},
  \bibinfo{person}{Chucheng Wang}, \bibinfo{person}{Pradeep Fernando},
  \bibinfo{person}{Saurabh Mishra}, \bibinfo{person}{Siqi Yan},
  \bibinfo{person}{Yicong Du}, \bibinfo{person}{Elzbieta Krepska},
  \bibinfo{person}{Intaik Park}, \bibinfo{person}{Min Ni},
  \bibinfo{person}{Qunshu Zhang}, {and} \bibinfo{person}{Shen Li}.}
  \bibinfo{year}{2025}\natexlab{a}.
\newblock \showarticletitle{DECK: Experiences on Delta Checkpointing for
  Industrial Recommendation Systems}.
\newblock \bibinfo{journal}{\emph{Proc. VLDB Endow.}} \bibinfo{volume}{18},
  \bibinfo{number}{12} (\bibinfo{date}{Aug.} \bibinfo{year}{2025}),
  \bibinfo{pages}{4978–4990}.
\newblock
\showISSN{2150-8097}
\href{https://doi.org/10.14778/3750601.3750621}{doi:\nolinkurl{10.14778/3750601.3750621}}


\bibitem[Gerogiannis et~al\mbox{.}(2025)]%
        {gerogiannis2025deca}
\bibfield{author}{\bibinfo{person}{Gerasimos Gerogiannis},
  \bibinfo{person}{Stijn Eyerman}, \bibinfo{person}{Evangelos Georganas},
  \bibinfo{person}{Wim Heirman}, {and} \bibinfo{person}{Josep Torrellas}.}
  \bibinfo{year}{2025}\natexlab{}.
\newblock \showarticletitle{DECA: A Near-Core LLM Decompression Accelerator
  Grounded on a 3D Roofline Model}. In \bibinfo{booktitle}{\emph{Proceedings of
  the 58th IEEE/ACM International Symposium on Microarchitecture}}
  \emph{(\bibinfo{series}{MICRO '25})}. \bibinfo{publisher}{Association for
  Computing Machinery}, \bibinfo{address}{New York, NY, USA},
  \bibinfo{pages}{184–200}.
\newblock
\showISBNx{9798400715730}
\href{https://doi.org/10.1145/3725843.3756073}{doi:\nolinkurl{10.1145/3725843.3756073}}


\bibitem[Gim et~al\mbox{.}(2024)]%
        {gim2024promptcache}
\bibfield{author}{\bibinfo{person}{In Gim}, \bibinfo{person}{Guojun Chen},
  \bibinfo{person}{Seung-seob Lee}, \bibinfo{person}{Nikhil Sarda},
  \bibinfo{person}{Anurag Khandelwal}, {and} \bibinfo{person}{Lin Zhong}.}
  \bibinfo{year}{2024}\natexlab{}.
\newblock \showarticletitle{Prompt Cache: Modular Attention Reuse for
  Low-Latency Inference}. In \bibinfo{booktitle}{\emph{Proceedings of Machine
  Learning and Systems}}, \bibfield{editor}{\bibinfo{person}{P.~Gibbons},
  \bibinfo{person}{G.~Pekhimenko}, {and} \bibinfo{person}{C.~De Sa}} (Eds.),
  Vol.~\bibinfo{volume}{6}. \bibinfo{pages}{325--338}.
\newblock
\urldef\tempurl%
\url{https://proceedings.mlsys.org/paper_files/paper/2024/file/a66caa1703fe34705a4368c3014c1966-Paper-Conference.pdf}
\showURL{%
\tempurl}


\bibitem[Gim et~al\mbox{.}(2025)]%
        {gim2025pie}
\bibfield{author}{\bibinfo{person}{In Gim}, \bibinfo{person}{Zhiyao Ma},
  \bibinfo{person}{Seung-seob Lee}, {and} \bibinfo{person}{Lin Zhong}.}
  \bibinfo{year}{2025}\natexlab{}.
\newblock \showarticletitle{Pie: A Programmable Serving System for Emerging LLM
  Applications}. In \bibinfo{booktitle}{\emph{Proceedings of the ACM SIGOPS
  31st Symposium on Operating Systems Principles}} (Lotte Hotel World, Seoul,
  Republic of Korea) \emph{(\bibinfo{series}{SOSP '25})}.
  \bibinfo{publisher}{Association for Computing Machinery},
  \bibinfo{address}{New York, NY, USA}, \bibinfo{pages}{415–430}.
\newblock
\showISBNx{9798400718700}
\href{https://doi.org/10.1145/3731569.3764814}{doi:\nolinkurl{10.1145/3731569.3764814}}


\bibitem[Goiri et~al\mbox{.}(2015)]%
        {goiri2015approxhadoop}
\bibfield{author}{\bibinfo{person}{Inigo Goiri}, \bibinfo{person}{Ricardo
  Bianchini}, \bibinfo{person}{Santosh Nagarakatte}, {and}
  \bibinfo{person}{Thu~D. Nguyen}.} \bibinfo{year}{2015}\natexlab{}.
\newblock \showarticletitle{ApproxHadoop: Bringing Approximations to MapReduce
  Frameworks}. In \bibinfo{booktitle}{\emph{Proceedings of the Twentieth
  International Conference on Architectural Support for Programming Languages
  and Operating Systems}} (Istanbul, Turkey) \emph{(\bibinfo{series}{ASPLOS
  '15})}. \bibinfo{publisher}{Association for Computing Machinery},
  \bibinfo{address}{New York, NY, USA}, \bibinfo{pages}{383–397}.
\newblock
\showISBNx{9781450328357}
\href{https://doi.org/10.1145/2694344.2694351}{doi:\nolinkurl{10.1145/2694344.2694351}}


\bibitem[Gong et~al\mbox{.}(2025)]%
        {gong2025pastfuture}
\bibfield{author}{\bibinfo{person}{Ruihao Gong}, \bibinfo{person}{Shihao Bai},
  \bibinfo{person}{Siyu Wu}, \bibinfo{person}{Yunqian Fan},
  \bibinfo{person}{Zaijun Wang}, \bibinfo{person}{Xiuhong Li},
  \bibinfo{person}{Hailong Yang}, {and} \bibinfo{person}{Xianglong Liu}.}
  \bibinfo{year}{2025}\natexlab{}.
\newblock \showarticletitle{Past-Future Scheduler for LLM Serving under SLA
  Guarantees}. In \bibinfo{booktitle}{\emph{Proceedings of the 30th ACM
  International Conference on Architectural Support for Programming Languages
  and Operating Systems, Volume 2}} (Rotterdam, Netherlands)
  \emph{(\bibinfo{series}{ASPLOS '25})}. \bibinfo{publisher}{Association for
  Computing Machinery}, \bibinfo{address}{New York, NY, USA},
  \bibinfo{pages}{798–813}.
\newblock
\showISBNx{9798400710797}
\href{https://doi.org/10.1145/3676641.3716011}{doi:\nolinkurl{10.1145/3676641.3716011}}


\bibitem[Gu et~al\mbox{.}(2025)]%
        {gu2025cent}
\bibfield{author}{\bibinfo{person}{Yufeng Gu}, \bibinfo{person}{Alireza
  Khadem}, \bibinfo{person}{Sumanth Umesh}, \bibinfo{person}{Ning Liang},
  \bibinfo{person}{Xavier Servot}, \bibinfo{person}{Onur Mutlu},
  \bibinfo{person}{Ravi Iyer}, {and} \bibinfo{person}{Reetuparna Das}.}
  \bibinfo{year}{2025}\natexlab{}.
\newblock \showarticletitle{PIM Is All You Need: A CXL-Enabled GPU-Free System
  for Large Language Model Inference}. In \bibinfo{booktitle}{\emph{Proceedings
  of the 30th ACM International Conference on Architectural Support for
  Programming Languages and Operating Systems, Volume 2}} (Rotterdam,
  Netherlands) \emph{(\bibinfo{series}{ASPLOS '25})}.
  \bibinfo{publisher}{Association for Computing Machinery},
  \bibinfo{address}{New York, NY, USA}, \bibinfo{pages}{862–881}.
\newblock
\showISBNx{9798400710797}
\href{https://doi.org/10.1145/3676641.3716267}{doi:\nolinkurl{10.1145/3676641.3716267}}


\bibitem[Guan et~al\mbox{.}(2025)]%
        {mercury2025rms}
\bibfield{author}{\bibinfo{person}{Yue Guan}, \bibinfo{person}{Xinwei Qiang},
  \bibinfo{person}{Zaifeng Pan}, \bibinfo{person}{Daniels Johnson},
  \bibinfo{person}{Yuanwei Fang}, \bibinfo{person}{Keren Zhou},
  \bibinfo{person}{Yuke Wang}, \bibinfo{person}{Wanlu Li},
  \bibinfo{person}{Yufei Ding}, {and} \bibinfo{person}{Adnan Aziz}.}
  \bibinfo{year}{2025}\natexlab{}.
\newblock \showarticletitle{Mercury: Unlocking Multi-GPU Operator Optimization
  for LLMs via Remote Memory Scheduling}. In
  \bibinfo{booktitle}{\emph{Proceedings of the ACM SIGOPS 31st Symposium on
  Operating Systems Principles}} (Lotte Hotel World, Seoul, Republic of Korea)
  \emph{(\bibinfo{series}{SOSP '25})}. \bibinfo{publisher}{Association for
  Computing Machinery}, \bibinfo{address}{New York, NY, USA},
  \bibinfo{pages}{1046–1061}.
\newblock
\showISBNx{9798400718700}
\href{https://doi.org/10.1145/3731569.3764798}{doi:\nolinkurl{10.1145/3731569.3764798}}


\bibitem[Gujarati et~al\mbox{.}(2020)]%
        {gujarati2020clockwork}
\bibfield{author}{\bibinfo{person}{Arpan Gujarati}, \bibinfo{person}{Reza
  Karimi}, \bibinfo{person}{Safya Alzayat}, \bibinfo{person}{Wei Hao},
  \bibinfo{person}{Antoine Kaufmann}, \bibinfo{person}{Ymir Vigfusson}, {and}
  \bibinfo{person}{Jonathan Mace}.} \bibinfo{year}{2020}\natexlab{}.
\newblock \showarticletitle{Serving DNNs like clockwork: performance
  predictability from the bottom up}. In \bibinfo{booktitle}{\emph{Proceedings
  of the 14th USENIX Conference on Operating Systems Design and
  Implementation}} \emph{(\bibinfo{series}{OSDI'20})}.
  \bibinfo{publisher}{USENIX Association}, \bibinfo{address}{USA}, Article
  \bibinfo{articleno}{25}, \bibinfo{numpages}{20}~pages.
\newblock
\showISBNx{978-1-939133-19-9}


\bibitem[Gulisano et~al\mbox{.}(2012)]%
        {gulisano2012streamcloud}
\bibfield{author}{\bibinfo{person}{Vincenzo Gulisano}, \bibinfo{person}{Ricardo
  Jiménez-Peris}, \bibinfo{person}{Marta Patiño-Martínez},
  \bibinfo{person}{Claudio Soriente}, {and} \bibinfo{person}{Patrick
  Valduriez}.} \bibinfo{year}{2012}\natexlab{}.
\newblock \showarticletitle{StreamCloud: An Elastic and Scalable Data Streaming
  System}.
\newblock \bibinfo{journal}{\emph{IEEE Transactions on Parallel and Distributed
  Systems}} \bibinfo{volume}{23}, \bibinfo{number}{12} (\bibinfo{year}{2012}),
  \bibinfo{pages}{2351--2365}.
\newblock
\href{https://doi.org/10.1109/TPDS.2012.24}{doi:\nolinkurl{10.1109/TPDS.2012.24}}


\bibitem[Guo et~al\mbox{.}(2024)]%
        {guo2024gmlake}
\bibfield{author}{\bibinfo{person}{Cong Guo}, \bibinfo{person}{Rui Zhang},
  \bibinfo{person}{Jiale Xu}, \bibinfo{person}{Jingwen Leng},
  \bibinfo{person}{Zihan Liu}, \bibinfo{person}{Ziyu Huang},
  \bibinfo{person}{Minyi Guo}, \bibinfo{person}{Hao Wu},
  \bibinfo{person}{Shouren Zhao}, \bibinfo{person}{Junping Zhao}, {and}
  \bibinfo{person}{Ke Zhang}.} \bibinfo{year}{2024}\natexlab{}.
\newblock \showarticletitle{GMLake: Efficient and Transparent GPU Memory
  Defragmentation for Large-scale DNN Training with Virtual Memory Stitching}.
  In \bibinfo{booktitle}{\emph{Proceedings of the 29th ACM International
  Conference on Architectural Support for Programming Languages and Operating
  Systems, Volume 2}} (La Jolla, CA, USA) \emph{(\bibinfo{series}{ASPLOS
  '24})}. \bibinfo{publisher}{Association for Computing Machinery},
  \bibinfo{address}{New York, NY, USA}, \bibinfo{pages}{450–466}.
\newblock
\showISBNx{9798400703850}
\href{https://doi.org/10.1145/3620665.3640423}{doi:\nolinkurl{10.1145/3620665.3640423}}


\bibitem[Guo et~al\mbox{.}(2022)]%
        {guo2022manu}
\bibfield{author}{\bibinfo{person}{Rentong Guo}, \bibinfo{person}{Xiaofan
  Luan}, \bibinfo{person}{Long Xiang}, \bibinfo{person}{Xiao Yan},
  \bibinfo{person}{Xiaomeng Yi}, \bibinfo{person}{Jigao Luo},
  \bibinfo{person}{Qianya Cheng}, \bibinfo{person}{Weizhi Xu},
  \bibinfo{person}{Jiarui Luo}, \bibinfo{person}{Frank Liu},
  \bibinfo{person}{Zhenshan Cao}, \bibinfo{person}{Yanliang Qiao},
  \bibinfo{person}{Ting Wang}, \bibinfo{person}{Bo Tang}, {and}
  \bibinfo{person}{Charles Xie}.} \bibinfo{year}{2022}\natexlab{}.
\newblock \showarticletitle{Manu: a cloud native vector database management
  system}.
\newblock \bibinfo{journal}{\emph{Proc. VLDB Endow.}} \bibinfo{volume}{15},
  \bibinfo{number}{12} (\bibinfo{date}{Aug.} \bibinfo{year}{2022}),
  \bibinfo{pages}{3548–3561}.
\newblock
\showISSN{2150-8097}
\href{https://doi.org/10.14778/3554821.3554843}{doi:\nolinkurl{10.14778/3554821.3554843}}


\bibitem[Guo et~al\mbox{.}(2025)]%
        {guo2025gllm}
\bibfield{author}{\bibinfo{person}{Tianyu Guo}, \bibinfo{person}{Xianwei
  Zhang}, \bibinfo{person}{Jiangsu Du}, \bibinfo{person}{Zhiguang Chen},
  \bibinfo{person}{Nong Xiao}, {and} \bibinfo{person}{Yutong Lu}.}
  \bibinfo{year}{2025}\natexlab{}.
\newblock \showarticletitle{gLLM: Global Balanced Pipeline Parallelism Systems
  for Distributed LLMs Serving with Token Throttling}. In
  \bibinfo{booktitle}{\emph{Proceedings of the International Conference for
  High Performance Computing, Networking, Storage and Analysis}}
  \emph{(\bibinfo{series}{SC '25})}. \bibinfo{publisher}{Association for
  Computing Machinery}, \bibinfo{address}{New York, NY, USA},
  \bibinfo{pages}{1725–1741}.
\newblock
\showISBNx{9798400714665}
\href{https://doi.org/10.1145/3712285.3759823}{doi:\nolinkurl{10.1145/3712285.3759823}}


\bibitem[Guo and Joshi(2026)]%
        {guo2026splitzip}
\bibfield{author}{\bibinfo{person}{Yipin Guo} {and} \bibinfo{person}{Siddharth
  Joshi}.} \bibinfo{year}{2026}\natexlab{}.
\newblock \showarticletitle{SplitZip: Ultra Fast Lossless KV Compression for
  Disaggregated LLM Serving}.
\newblock  (\bibinfo{year}{2026}).
\newblock
\href{https://doi.org/10.48550/ARXIV.2605.01708}{doi:\nolinkurl{10.48550/ARXIV.2605.01708}}


\bibitem[Guti\'{e}rrez et~al\mbox{.}(2024)]%
        {wang2024hipporag}
\bibfield{author}{\bibinfo{person}{Bernal~Jim\'{e}nez Guti\'{e}rrez},
  \bibinfo{person}{Yiheng Shu}, \bibinfo{person}{Yu Gu},
  \bibinfo{person}{Michihiro Yasunaga}, {and} \bibinfo{person}{Yu Su}.}
  \bibinfo{year}{2024}\natexlab{}.
\newblock \showarticletitle{HippoRAG: Neurobiologically Inspired Long-Term
  Memory for Large Language Models}. In \bibinfo{booktitle}{\emph{Advances in
  Neural Information Processing Systems}},
  \bibfield{editor}{\bibinfo{person}{A.~Globerson},
  \bibinfo{person}{L.~Mackey}, \bibinfo{person}{D.~Belgrave},
  \bibinfo{person}{A.~Fan}, \bibinfo{person}{U.~Paquet},
  \bibinfo{person}{J.~Tomczak}, {and} \bibinfo{person}{C.~Zhang}} (Eds.),
  Vol.~\bibinfo{volume}{37}. \bibinfo{publisher}{Curran Associates, Inc.},
  \bibinfo{pages}{59532--59569}.
\newblock
\href{https://doi.org/10.52202/079017-1902}{doi:\nolinkurl{10.52202/079017-1902}}


\bibitem[Guu et~al\mbox{.}(2020)]%
        {guu2020realm}
\bibfield{author}{\bibinfo{person}{Kelvin Guu}, \bibinfo{person}{Kenton Lee},
  \bibinfo{person}{Zora Tung}, \bibinfo{person}{Panupong Pasupat}, {and}
  \bibinfo{person}{Ming-Wei Chang}.} \bibinfo{year}{2020}\natexlab{}.
\newblock \showarticletitle{REALM: retrieval-augmented language model
  pre-training}. In \bibinfo{booktitle}{\emph{Proceedings of the 37th
  International Conference on Machine Learning}}
  \emph{(\bibinfo{series}{ICML'20})}. \bibinfo{publisher}{JMLR.org}, Article
  \bibinfo{articleno}{368}, \bibinfo{numpages}{10}~pages.
\newblock


\bibitem[Hayes et~al\mbox{.}(2020)]%
        {hayes2020remind}
\bibfield{author}{\bibinfo{person}{Tyler~L. Hayes}, \bibinfo{person}{Kushal
  Kafle}, \bibinfo{person}{Robik Shrestha}, \bibinfo{person}{Manoj Acharya},
  {and} \bibinfo{person}{Christopher Kanan}.} \bibinfo{year}{2020}\natexlab{}.
\newblock \showarticletitle{REMIND Your Neural Network to Prevent Catastrophic
  Forgetting}. In \bibinfo{booktitle}{\emph{Computer Vision -- ECCV 2020}},
  \bibfield{editor}{\bibinfo{person}{Andrea Vedaldi}, \bibinfo{person}{Horst
  Bischof}, \bibinfo{person}{Thomas Brox}, {and} \bibinfo{person}{Jan-Michael
  Frahm}} (Eds.). \bibinfo{publisher}{Springer International Publishing},
  \bibinfo{address}{Cham}, \bibinfo{pages}{466--483}.
\newblock
\showISBNx{978-3-030-58598-3}


\bibitem[He et~al\mbox{.}(2025a)]%
        {he2025waferllm}
\bibfield{author}{\bibinfo{person}{Congjie He}, \bibinfo{person}{Yeqi Huang},
  \bibinfo{person}{Pei Mu}, \bibinfo{person}{Ziming Miao},
  \bibinfo{person}{Jilong Xue}, \bibinfo{person}{Lingxiao Ma},
  \bibinfo{person}{Fan Yang}, {and} \bibinfo{person}{Luo Mai}.}
  \bibinfo{year}{2025}\natexlab{a}.
\newblock \showarticletitle{WaferLLM: large language model inference at wafer
  scale}. In \bibinfo{booktitle}{\emph{Proceedings of the 19th USENIX
  Conference on Operating Systems Design and Implementation}} (Boston, MA, USA)
  \emph{(\bibinfo{series}{OSDI '25})}. \bibinfo{publisher}{USENIX Association},
  \bibinfo{address}{USA}, Article \bibinfo{articleno}{15},
  \bibinfo{numpages}{17}~pages.
\newblock
\showISBNx{978-1-939133-47-2}


\bibitem[He et~al\mbox{.}(2025b)]%
        {he2025resourcemultiplexing}
\bibfield{author}{\bibinfo{person}{Yongjun He}, \bibinfo{person}{Haofeng Yang},
  \bibinfo{person}{Yao Lu}, \bibinfo{person}{Ana Klimovi\'{c}}, {and}
  \bibinfo{person}{Gustavo Alonso}.} \bibinfo{year}{2025}\natexlab{b}.
\newblock \showarticletitle{Resource multiplexing in tuning and serving large
  language models}. In \bibinfo{booktitle}{\emph{Proceedings of the 2025 USENIX
  Conference on Usenix Annual Technical Conference}} (Boston, MA, USA)
  \emph{(\bibinfo{series}{USENIX ATC '25})}. \bibinfo{publisher}{USENIX
  Association}, \bibinfo{address}{USA}, Article \bibinfo{articleno}{97},
  \bibinfo{numpages}{17}~pages.
\newblock
\showISBNx{978-1-939133-48-9}


\bibitem[Hirzel et~al\mbox{.}(2014)]%
        {hirzel2014catalog}
\bibfield{author}{\bibinfo{person}{Martin Hirzel}, \bibinfo{person}{Robert
  Soul\'{e}}, \bibinfo{person}{Scott Schneider}, \bibinfo{person}{Bu\u{g}ra
  Gedik}, {and} \bibinfo{person}{Robert Grimm}.}
  \bibinfo{year}{2014}\natexlab{}.
\newblock \showarticletitle{A catalog of stream processing optimizations}.
\newblock \bibinfo{journal}{\emph{ACM Comput. Surv.}} \bibinfo{volume}{46},
  \bibinfo{number}{4}, Article \bibinfo{articleno}{46} (\bibinfo{date}{March}
  \bibinfo{year}{2014}), \bibinfo{numpages}{34}~pages.
\newblock
\showISSN{0360-0300}
\href{https://doi.org/10.1145/2528412}{doi:\nolinkurl{10.1145/2528412}}


\bibitem[Hoffmann et~al\mbox{.}(2019)]%
        {nasir2019megaphone}
\bibfield{author}{\bibinfo{person}{Moritz Hoffmann}, \bibinfo{person}{Andrea
  Lattuada}, {and} \bibinfo{person}{Frank McSherry}.}
  \bibinfo{year}{2019}\natexlab{}.
\newblock \showarticletitle{Megaphone: latency-conscious state migration for
  distributed streaming dataflows}.
\newblock \bibinfo{journal}{\emph{Proc. VLDB Endow.}} \bibinfo{volume}{12},
  \bibinfo{number}{9} (\bibinfo{date}{May} \bibinfo{year}{2019}),
  \bibinfo{pages}{1002–1015}.
\newblock
\showISSN{2150-8097}
\href{https://doi.org/10.14778/3329772.3329777}{doi:\nolinkurl{10.14778/3329772.3329777}}


\bibitem[Holmes et~al\mbox{.}(2024)]%
        {holmes2024fastgen}
\bibfield{author}{\bibinfo{person}{Connor Holmes}, \bibinfo{person}{Masahiro
  Tanaka}, \bibinfo{person}{Michael Wyatt}, \bibinfo{person}{Ammar~Ahmad Awan},
  \bibinfo{person}{Jeff Rasley}, \bibinfo{person}{Samyam Rajbhandari},
  \bibinfo{person}{Reza~Yazdani Aminabadi}, \bibinfo{person}{Heyang Qin},
  \bibinfo{person}{Arash Bakhtiari}, \bibinfo{person}{Lev Kurilenko}, {and}
  \bibinfo{person}{Yuxiong He}.} \bibinfo{year}{2024}\natexlab{}.
\newblock \showarticletitle{DeepSpeed-FastGen: High-throughput Text Generation
  for LLMs via MII and DeepSpeed-Inference}.
\newblock  (\bibinfo{year}{2024}).
\newblock
\href{https://doi.org/10.48550/ARXIV.2401.08671}{doi:\nolinkurl{10.48550/ARXIV.2401.08671}}


\bibitem[Hong et~al\mbox{.}(2024)]%
        {hong2024dramcache}
\bibfield{author}{\bibinfo{person}{Jeongmin Hong}, \bibinfo{person}{Sungjun
  Cho}, \bibinfo{person}{Geonwoo Park}, \bibinfo{person}{Wonhyuk Yang},
  \bibinfo{person}{Young-Ho Gong}, {and} \bibinfo{person}{Gwangsun Kim}.}
  \bibinfo{year}{2024}\natexlab{}.
\newblock \showarticletitle{Bandwidth-Effective DRAM Cache for GPU s with
  Storage-Class Memory}. In \bibinfo{booktitle}{\emph{2024 IEEE International
  Symposium on High-Performance Computer Architecture (HPCA)}}.
  \bibinfo{pages}{139--155}.
\newblock
\href{https://doi.org/10.1109/HPCA57654.2024.00021}{doi:\nolinkurl{10.1109/HPCA57654.2024.00021}}


\bibitem[Hong et~al\mbox{.}(2025)]%
        {hong2025sola}
\bibfield{author}{\bibinfo{person}{Ke Hong}, \bibinfo{person}{Xiuhong Li},
  \bibinfo{person}{Lufang Chen}, \bibinfo{person}{Qiuli Mao},
  \bibinfo{person}{Guohao Dai}, \bibinfo{person}{Xuefei Ning},
  \bibinfo{person}{Shengen Yan}, \bibinfo{person}{Yun Liang}, {and}
  \bibinfo{person}{Yu Wang}.} \bibinfo{year}{2025}\natexlab{}.
\newblock \showarticletitle{SOLA: Optimizing SLO Attainment for Large Language
  Model Serving with State-Aware Scheduling}. In
  \bibinfo{booktitle}{\emph{Proceedings of Machine Learning and Systems}},
  \bibfield{editor}{\bibinfo{person}{M.~Zaharia}, \bibinfo{person}{G.~Joshi},
  {and} \bibinfo{person}{Y.~Lin}} (Eds.), Vol.~\bibinfo{volume}{7}.
  \bibinfo{publisher}{MLSys}.
\newblock
\urldef\tempurl%
\url{https://proceedings.mlsys.org/paper_files/paper/2025/file/bc82dbfbfa43232be85b8d9838f49c3e-Paper-Conference.pdf}
\showURL{%
\tempurl}


\bibitem[Hu et~al\mbox{.}(2024)]%
        {hu2024aceso}
\bibfield{author}{\bibinfo{person}{Zhisheng Hu}, \bibinfo{person}{Pengfei Zuo},
  \bibinfo{person}{Yizou Chen}, \bibinfo{person}{Chao Wang},
  \bibinfo{person}{Junliang Hu}, {and} \bibinfo{person}{Ming-Chang Yang}.}
  \bibinfo{year}{2024}\natexlab{}.
\newblock \showarticletitle{Aceso: Achieving Efficient Fault Tolerance in
  Memory-Disaggregated Key-Value Stores}. In
  \bibinfo{booktitle}{\emph{Proceedings of the ACM SIGOPS 30th Symposium on
  Operating Systems Principles}} (Austin, TX, USA) \emph{(\bibinfo{series}{SOSP
  '24})}. \bibinfo{publisher}{Association for Computing Machinery},
  \bibinfo{address}{New York, NY, USA}, \bibinfo{pages}{127–143}.
\newblock
\showISBNx{9798400712517}
\href{https://doi.org/10.1145/3694715.3695951}{doi:\nolinkurl{10.1145/3694715.3695951}}


\bibitem[Huang et~al\mbox{.}(2025c)]%
        {huang2025chimera}
\bibfield{author}{\bibinfo{person}{Chunyue Huang}, \bibinfo{person}{Shuang
  Liu}, \bibinfo{person}{Xinyi Zhang}, \bibinfo{person}{Wenhao Li},
  \bibinfo{person}{Wei Lu}, {and} \bibinfo{person}{Xiaoyong Du}.}
  \bibinfo{year}{2025}\natexlab{c}.
\newblock \showarticletitle{Chimera: Mitigating Ownership Transfers in
  Multi-Primary Shared-Storage Cloud-Native Databases}.
\newblock \bibinfo{journal}{\emph{Proc. VLDB Endow.}} \bibinfo{volume}{18},
  \bibinfo{number}{10} (\bibinfo{date}{June} \bibinfo{year}{2025}),
  \bibinfo{pages}{3368–3381}.
\newblock
\showISSN{2150-8097}
\href{https://doi.org/10.14778/3748191.3748201}{doi:\nolinkurl{10.14778/3748191.3748201}}


\bibitem[Huang et~al\mbox{.}(2025a)]%
        {huang2025tigon}
\bibfield{author}{\bibinfo{person}{Yibo Huang}, \bibinfo{person}{Haowei Chen},
  \bibinfo{person}{Newton Ni}, \bibinfo{person}{Yan Sun},
  \bibinfo{person}{Vijay Chidambaram}, \bibinfo{person}{Dixin Tang}, {and}
  \bibinfo{person}{Emmett Witchel}.} \bibinfo{year}{2025}\natexlab{a}.
\newblock \showarticletitle{Tigon: a distributed database for a CXL pod}. In
  \bibinfo{booktitle}{\emph{Proceedings of the 19th USENIX Conference on
  Operating Systems Design and Implementation}} (Boston, MA, USA)
  \emph{(\bibinfo{series}{OSDI '25})}. \bibinfo{publisher}{USENIX Association},
  \bibinfo{address}{USA}, Article \bibinfo{articleno}{7},
  \bibinfo{numpages}{20}~pages.
\newblock
\showISBNx{978-1-939133-47-2}


\bibitem[Huang et~al\mbox{.}(2025b)]%
        {huang2025obscura}
\bibfield{author}{\bibinfo{person}{Yuzhou Huang}, \bibinfo{person}{Yapeng
  Jiang}, \bibinfo{person}{Zicong Hong}, \bibinfo{person}{Wuhui Chen},
  \bibinfo{person}{Bin Wang}, \bibinfo{person}{Weixi Zhu}, \bibinfo{person}{Yue
  Yu}, {and} \bibinfo{person}{Zibin Zheng}.} \bibinfo{year}{2025}\natexlab{b}.
\newblock \showarticletitle{Obscura: concealing recomputation overhead in
  training of large language models with bubble-filling pipeline
  transformation}. In \bibinfo{booktitle}{\emph{Proceedings of the 2025 USENIX
  Conference on Usenix Annual Technical Conference}} (Boston, MA, USA)
  \emph{(\bibinfo{series}{USENIX ATC '25})}. \bibinfo{publisher}{USENIX
  Association}, \bibinfo{address}{USA}, Article \bibinfo{articleno}{40},
  \bibinfo{numpages}{14}~pages.
\newblock
\showISBNx{978-1-939133-48-9}


\bibitem[Huang et~al\mbox{.}(2025d)]%
        {huang2025flowcheck}
\bibfield{author}{\bibinfo{person}{Zimeng Huang}, \bibinfo{person}{Hao Nie},
  \bibinfo{person}{Haonan Jia}, \bibinfo{person}{Bo Jiang},
  \bibinfo{person}{Junchen Guo}, \bibinfo{person}{Jianyuan Lu},
  \bibinfo{person}{Rong Wen}, \bibinfo{person}{Biao Lyu},
  \bibinfo{person}{Shunmin Zhu}, {and} \bibinfo{person}{Xinbing Wang}.}
  \bibinfo{year}{2025}\natexlab{d}.
\newblock \showarticletitle{FlowCheck: Decoupling Checkpointing and Training of
  Large-Scale Models}. In \bibinfo{booktitle}{\emph{Proceedings of the
  Twentieth European Conference on Computer Systems}} (Rotterdam, Netherlands)
  \emph{(\bibinfo{series}{EuroSys '25})}. \bibinfo{publisher}{Association for
  Computing Machinery}, \bibinfo{address}{New York, NY, USA},
  \bibinfo{pages}{1334–1349}.
\newblock
\showISBNx{9798400711961}
\href{https://doi.org/10.1145/3689031.3696088}{doi:\nolinkurl{10.1145/3689031.3696088}}


\bibitem[Iliakopoulou et~al\mbox{.}(2025)]%
        {iliakopoulou2024chameleon}
\bibfield{author}{\bibinfo{person}{Nikoleta Iliakopoulou},
  \bibinfo{person}{Jovan Stojkovic}, \bibinfo{person}{Chloe Alverti},
  \bibinfo{person}{Tianyin Xu}, \bibinfo{person}{Hubertus Franke}, {and}
  \bibinfo{person}{Josep Torrellas}.} \bibinfo{year}{2025}\natexlab{}.
\newblock \showarticletitle{Chameleon: Adaptive Caching and Scheduling for
  Many-Adapter LLM Inference Environments}. In
  \bibinfo{booktitle}{\emph{Proceedings of the 58th IEEE/ACM International
  Symposium on Microarchitecture}} \emph{(\bibinfo{series}{MICRO '25})}.
  \bibinfo{publisher}{Association for Computing Machinery},
  \bibinfo{address}{New York, NY, USA}, \bibinfo{pages}{217--231}.
\newblock
\showISBNx{9798400715730}
\href{https://doi.org/10.1145/3725843.3756083}{doi:\nolinkurl{10.1145/3725843.3756083}}


\bibitem[Isard et~al\mbox{.}(2007)]%
        {isard2007dryad}
\bibfield{author}{\bibinfo{person}{Michael Isard}, \bibinfo{person}{Mihai
  Budiu}, \bibinfo{person}{Yuan Yu}, \bibinfo{person}{Andrew Birrell}, {and}
  \bibinfo{person}{Dennis Fetterly}.} \bibinfo{year}{2007}\natexlab{}.
\newblock \showarticletitle{Dryad: distributed data-parallel programs from
  sequential building blocks}. In \bibinfo{booktitle}{\emph{Proceedings of the
  2nd ACM SIGOPS/EuroSys European Conference on Computer Systems 2007}}
  (Lisbon, Portugal) \emph{(\bibinfo{series}{EuroSys '07})}.
  \bibinfo{publisher}{Association for Computing Machinery},
  \bibinfo{address}{New York, NY, USA}, \bibinfo{pages}{59–72}.
\newblock
\showISBNx{9781595936363}
\href{https://doi.org/10.1145/1272996.1273005}{doi:\nolinkurl{10.1145/1272996.1273005}}


\bibitem[Izacard et~al\mbox{.}(2023)]%
        {izacard2022atlas}
\bibfield{author}{\bibinfo{person}{Gautier Izacard}, \bibinfo{person}{Patrick
  Lewis}, \bibinfo{person}{Maria Lomeli}, \bibinfo{person}{Lucas Hosseini},
  \bibinfo{person}{Fabio Petroni}, \bibinfo{person}{Timo Schick},
  \bibinfo{person}{Jane Dwivedi-Yu}, \bibinfo{person}{Armand Joulin},
  \bibinfo{person}{Sebastian Riedel}, {and} \bibinfo{person}{Edouard Grave}.}
  \bibinfo{year}{2023}\natexlab{}.
\newblock \showarticletitle{Atlas: few-shot learning with retrieval augmented
  language models}.
\newblock \bibinfo{journal}{\emph{J. Mach. Learn. Res.}} \bibinfo{volume}{24},
  \bibinfo{number}{1}, Article \bibinfo{articleno}{251} (\bibinfo{date}{Jan.}
  \bibinfo{year}{2023}), \bibinfo{numpages}{43}~pages.
\newblock
\showISSN{1532-4435}


\bibitem[Jalalian et~al\mbox{.}(2024)]%
        {jalalian2024extmem}
\bibfield{author}{\bibinfo{person}{Sepehr Jalalian}, \bibinfo{person}{Shaurya
  Patel}, \bibinfo{person}{Milad~Rezaei Hajidehi}, \bibinfo{person}{Margo
  Seltzer}, {and} \bibinfo{person}{Alexandra Fedorova}.}
  \bibinfo{year}{2024}\natexlab{}.
\newblock \showarticletitle{EXTMEM: enabling application-aware virtual memory
  management for data-intensive applications}. In
  \bibinfo{booktitle}{\emph{Proceedings of the 2024 USENIX Conference on Usenix
  Annual Technical Conference}} (Santa Clara, CA, USA)
  \emph{(\bibinfo{series}{USENIX ATC'24})}. \bibinfo{publisher}{USENIX
  Association}, \bibinfo{address}{USA}, Article \bibinfo{articleno}{25},
  \bibinfo{numpages}{12}~pages.
\newblock
\showISBNx{978-1-939133-41-0}


\bibitem[Jang et~al\mbox{.}(2024)]%
        {jang2024smartinfinity}
\bibfield{author}{\bibinfo{person}{Hongsun Jang}, \bibinfo{person}{Jaeyong
  Song}, \bibinfo{person}{Jaewon Jung}, \bibinfo{person}{Jaeyoung Park},
  \bibinfo{person}{Youngsok Kim}, {and} \bibinfo{person}{Jinho Lee}.}
  \bibinfo{year}{2024}\natexlab{}.
\newblock \showarticletitle{Smart-Infinity: Fast Large Language Model Training
  using Near-Storage Processing on a Real System}. In
  \bibinfo{booktitle}{\emph{2024 IEEE International Symposium on
  High-Performance Computer Architecture (HPCA)}}. \bibinfo{pages}{345--360}.
\newblock
\href{https://doi.org/10.1109/HPCA57654.2024.00034}{doi:\nolinkurl{10.1109/HPCA57654.2024.00034}}


\bibitem[Jha et~al\mbox{.}(2025)]%
        {jha2025hycache}
\bibfield{author}{\bibinfo{person}{Keshav~Vinayak Jha}, \bibinfo{person}{Shweta
  Pandey}, \bibinfo{person}{Murali Annavaram}, {and} \bibinfo{person}{Arkaprava
  Basu}.} \bibinfo{year}{2025}\natexlab{}.
\newblock \showarticletitle{HyCache: hybrid caching for accelerating DNN input
  preprocessing pipelines}. In \bibinfo{booktitle}{\emph{Proceedings of the
  2025 USENIX Conference on Usenix Annual Technical Conference}} (Boston, MA,
  USA) \emph{(\bibinfo{series}{USENIX ATC '25})}. \bibinfo{publisher}{USENIX
  Association}, \bibinfo{address}{USA}, Article \bibinfo{articleno}{26},
  \bibinfo{numpages}{16}~pages.
\newblock
\showISBNx{978-1-939133-48-9}


\bibitem[Ji et~al\mbox{.}(2025)]%
        {ji2025cxl}
\bibfield{author}{\bibinfo{person}{Houxiang Ji}, \bibinfo{person}{Yifan Yuan},
  \bibinfo{person}{Yang Zhou}, \bibinfo{person}{Ipoom Jeong},
  \bibinfo{person}{Ren Wang}, \bibinfo{person}{Saksham Agarwal}, {and}
  \bibinfo{person}{Nam~Sung Kim}.} \bibinfo{year}{2025}\natexlab{}.
\newblock \showarticletitle{Re-architecting End-host Networking with CXL:
  Coherence, Memory, and Offloading}. In \bibinfo{booktitle}{\emph{Proceedings
  of the 58th IEEE/ACM International Symposium on Microarchitecture}}
  \emph{(\bibinfo{series}{MICRO '25})}. \bibinfo{publisher}{Association for
  Computing Machinery}, \bibinfo{address}{New York, NY, USA},
  \bibinfo{pages}{1809–1823}.
\newblock
\showISBNx{9798400715730}
\href{https://doi.org/10.1145/3725843.3756102}{doi:\nolinkurl{10.1145/3725843.3756102}}


\bibitem[Jiang and Liu(2025)]%
        {jiang2025shadowviews}
\bibfield{author}{\bibinfo{person}{Sheng Jiang} {and} \bibinfo{person}{Ming
  Liu}.} \bibinfo{year}{2025}\natexlab{}.
\newblock \showarticletitle{Building an Elastic Block Storage over {EBOFs}
  Using Shadow Views}. In \bibinfo{booktitle}{\emph{22nd USENIX Symposium on
  Networked Systems Design and Implementation (NSDI 25)}}.
  \bibinfo{publisher}{USENIX Association}, \bibinfo{address}{Philadelphia, PA},
  \bibinfo{pages}{1137--1153}.
\newblock
\showISBNx{978-1-939133-46-5}
\urldef\tempurl%
\url{https://www.usenix.org/conference/nsdi25/presentation/jiang}
\showURL{%
\tempurl}


\bibitem[Jiang et~al\mbox{.}(2025)]%
        {jiang2025neo}
\bibfield{author}{\bibinfo{person}{Xuanlin Jiang}, \bibinfo{person}{Yang Zhou},
  \bibinfo{person}{Shiyi Cao}, \bibinfo{person}{Ion Stoica}, {and}
  \bibinfo{person}{Minlan Yu}.} \bibinfo{year}{2025}\natexlab{}.
\newblock \showarticletitle{NEO: Saving GPU Memory Crisis with CPU Offloading
  for Online LLM Inference}. In \bibinfo{booktitle}{\emph{Proceedings of
  Machine Learning and Systems}},
  \bibfield{editor}{\bibinfo{person}{M.~Zaharia}, \bibinfo{person}{G.~Joshi},
  {and} \bibinfo{person}{Y.~Lin}} (Eds.), Vol.~\bibinfo{volume}{7}.
  \bibinfo{publisher}{MLSys}.
\newblock
\urldef\tempurl%
\url{https://proceedings.mlsys.org/paper_files/paper/2025/file/66a026c0d17040889b50f0dfa650e5e0-Paper-Conference.pdf}
\showURL{%
\tempurl}


\bibitem[JIANG et~al\mbox{.}(2025)]%
        {jiang2025thunderserve}
\bibfield{author}{\bibinfo{person}{YOUHE JIANG}, \bibinfo{person}{Fangcheng
  Fu}, \bibinfo{person}{Xiaozhe Yao}, \bibinfo{person}{Taiyi Wang},
  \bibinfo{person}{Bin CUI}, \bibinfo{person}{Ana Klimovic}, {and}
  \bibinfo{person}{Eiko Yoneki}.} \bibinfo{year}{2025}\natexlab{}.
\newblock \showarticletitle{ThunderServe: High-performance and Cost-efficient
  LLM Serving in Cloud Environments}. In \bibinfo{booktitle}{\emph{Proceedings
  of Machine Learning and Systems}},
  \bibfield{editor}{\bibinfo{person}{M.~Zaharia}, \bibinfo{person}{G.~Joshi},
  {and} \bibinfo{person}{Y.~Lin}} (Eds.), Vol.~\bibinfo{volume}{7}.
  \bibinfo{publisher}{MLSys}.
\newblock
\urldef\tempurl%
\url{https://proceedings.mlsys.org/paper_files/paper/2025/file/c2a0e26dd9ee7d57e92bb1c24b39659a-Paper-Conference.pdf}
\showURL{%
\tempurl}


\bibitem[Jiang et~al\mbox{.}(2024)]%
        {jiang2024megascale}
\bibfield{author}{\bibinfo{person}{Ziheng Jiang}, \bibinfo{person}{Haibin Lin},
  \bibinfo{person}{Yinmin Zhong}, \bibinfo{person}{Qi Huang},
  \bibinfo{person}{Yangrui Chen}, \bibinfo{person}{Zhi Zhang},
  \bibinfo{person}{Yanghua Peng}, \bibinfo{person}{Xiang Li},
  \bibinfo{person}{Cong Xie}, \bibinfo{person}{Shibiao Nong},
  \bibinfo{person}{Yulu Jia}, \bibinfo{person}{Sun He},
  \bibinfo{person}{Hongmin Chen}, \bibinfo{person}{Zhihao Bai},
  \bibinfo{person}{Qi Hou}, \bibinfo{person}{Shipeng Yan},
  \bibinfo{person}{Ding Zhou}, \bibinfo{person}{Yiyao Sheng},
  \bibinfo{person}{Zhuo Jiang}, \bibinfo{person}{Haohan Xu},
  \bibinfo{person}{Haoran Wei}, \bibinfo{person}{Zhang Zhang},
  \bibinfo{person}{Pengfei Nie}, \bibinfo{person}{Leqi Zou},
  \bibinfo{person}{Sida Zhao}, \bibinfo{person}{Liang Xiang},
  \bibinfo{person}{Zherui Liu}, \bibinfo{person}{Zhe Li},
  \bibinfo{person}{Xiaoying Jia}, \bibinfo{person}{Jianxi Ye},
  \bibinfo{person}{Xin Jin}, {and} \bibinfo{person}{Xin Liu}.}
  \bibinfo{year}{2024}\natexlab{}.
\newblock \showarticletitle{MegaScale: scaling large language model training to
  more than 10,000 GPUs}. In \bibinfo{booktitle}{\emph{Proceedings of the 21st
  USENIX Symposium on Networked Systems Design and Implementation}} (Santa
  Clara, CA, USA) \emph{(\bibinfo{series}{NSDI'24})}.
  \bibinfo{publisher}{USENIX Association}, \bibinfo{address}{USA}, Article
  \bibinfo{articleno}{41}, \bibinfo{numpages}{16}~pages.
\newblock
\showISBNx{978-1-939133-39-7}


\bibitem[Jin et~al\mbox{.}(2024)]%
        {jin2024concealing}
\bibfield{author}{\bibinfo{person}{Sian Jin}, \bibinfo{person}{Sheng Di},
  \bibinfo{person}{Fr\'{e}d\'{e}ric Vivien}, \bibinfo{person}{Daoce Wang},
  \bibinfo{person}{Yves Robert}, \bibinfo{person}{Dingwen Tao}, {and}
  \bibinfo{person}{Franck Cappello}.} \bibinfo{year}{2024}\natexlab{}.
\newblock \showarticletitle{Concealing Compression-accelerated I/O for HPC
  Applications through In Situ Task Scheduling}. In
  \bibinfo{booktitle}{\emph{Proceedings of the Nineteenth European Conference
  on Computer Systems}} (Athens, Greece) \emph{(\bibinfo{series}{EuroSys
  '24})}. \bibinfo{publisher}{Association for Computing Machinery},
  \bibinfo{address}{New York, NY, USA}, \bibinfo{pages}{981–998}.
\newblock
\showISBNx{9798400704376}
\href{https://doi.org/10.1145/3627703.3629573}{doi:\nolinkurl{10.1145/3627703.3629573}}


\bibitem[Joo et~al\mbox{.}(2025)]%
        {joo2025coruscant}
\bibfield{author}{\bibinfo{person}{Donghyeon Joo}, \bibinfo{person}{Helya
  Hosseini}, \bibinfo{person}{Ramyad Hadidi}, {and} \bibinfo{person}{Bahar
  Asgari}.} \bibinfo{year}{2025}\natexlab{}.
\newblock \showarticletitle{Coruscant: Co-Designing GPU Kernel and Sparse
  Tensor Core to Advocate Unstructured Sparsity in Efficient LLM Inference}. In
  \bibinfo{booktitle}{\emph{Proceedings of the 58th IEEE/ACM International
  Symposium on Microarchitecture}} \emph{(\bibinfo{series}{MICRO '25})}.
  \bibinfo{publisher}{Association for Computing Machinery},
  \bibinfo{address}{New York, NY, USA}, \bibinfo{pages}{232–245}.
\newblock
\showISBNx{9798400715730}
\href{https://doi.org/10.1145/3725843.3756065}{doi:\nolinkurl{10.1145/3725843.3756065}}


\bibitem[Jégou et~al\mbox{.}(2011)]%
        {jegou2011pq}
\bibfield{author}{\bibinfo{person}{Herve Jégou}, \bibinfo{person}{Matthijs
  Douze}, {and} \bibinfo{person}{Cordelia Schmid}.}
  \bibinfo{year}{2011}\natexlab{}.
\newblock \showarticletitle{Product Quantization for Nearest Neighbor Search}.
\newblock \bibinfo{journal}{\emph{IEEE Transactions on Pattern Analysis and
  Machine Intelligence}} \bibinfo{volume}{33}, \bibinfo{number}{1}
  (\bibinfo{year}{2011}), \bibinfo{pages}{117--128}.
\newblock
\href{https://doi.org/10.1109/TPAMI.2010.57}{doi:\nolinkurl{10.1109/TPAMI.2010.57}}


\bibitem[Kallman et~al\mbox{.}(2008)]%
        {kallman2008hstore}
\bibfield{author}{\bibinfo{person}{Robert Kallman}, \bibinfo{person}{Hideaki
  Kimura}, \bibinfo{person}{Jonathan Natkins}, \bibinfo{person}{Andrew Pavlo},
  \bibinfo{person}{Alexander Rasin}, \bibinfo{person}{Stanley Zdonik},
  \bibinfo{person}{Evan P.~C. Jones}, \bibinfo{person}{Samuel Madden},
  \bibinfo{person}{Michael Stonebraker}, \bibinfo{person}{Yang Zhang},
  \bibinfo{person}{John Hugg}, {and} \bibinfo{person}{Daniel~J. Abadi}.}
  \bibinfo{year}{2008}\natexlab{}.
\newblock \showarticletitle{H-store: a high-performance, distributed main
  memory transaction processing system}.
\newblock \bibinfo{journal}{\emph{Proc. VLDB Endow.}} \bibinfo{volume}{1},
  \bibinfo{number}{2} (\bibinfo{date}{Aug.} \bibinfo{year}{2008}),
  \bibinfo{pages}{1496–1499}.
\newblock
\showISSN{2150-8097}
\href{https://doi.org/10.14778/1454159.1454211}{doi:\nolinkurl{10.14778/1454159.1454211}}


\bibitem[Kamath et~al\mbox{.}(2025)]%
        {kamath2025podattention}
\bibfield{author}{\bibinfo{person}{Aditya~K. Kamath}, \bibinfo{person}{Ramya
  Prabhu}, \bibinfo{person}{Jayashree Mohan}, \bibinfo{person}{Simon Peter},
  \bibinfo{person}{Ramachandran Ramjee}, {and} \bibinfo{person}{Ashish
  Panwar}.} \bibinfo{year}{2025}\natexlab{}.
\newblock \showarticletitle{POD-Attention: Unlocking Full Prefill-Decode
  Overlap for Faster LLM Inference}. In \bibinfo{booktitle}{\emph{Proceedings
  of the 30th ACM International Conference on Architectural Support for
  Programming Languages and Operating Systems, Volume 2}} (Rotterdam,
  Netherlands) \emph{(\bibinfo{series}{ASPLOS '25})}.
  \bibinfo{publisher}{Association for Computing Machinery},
  \bibinfo{address}{New York, NY, USA}, \bibinfo{pages}{897–912}.
\newblock
\showISBNx{9798400710797}
\href{https://doi.org/10.1145/3676641.3715996}{doi:\nolinkurl{10.1145/3676641.3715996}}


\bibitem[Kang et~al\mbox{.}(2025)]%
        {kang2025turboattention}
\bibfield{author}{\bibinfo{person}{Hao Kang}, \bibinfo{person}{Srikant
  Bharadwaj}, \bibinfo{person}{James Hensman}, \bibinfo{person}{Tushar
  Krishna}, \bibinfo{person}{Victor R\"{u}hle}, {and} \bibinfo{person}{Saravan
  Rajmohan}.} \bibinfo{year}{2025}\natexlab{}.
\newblock \showarticletitle{TurboAttention: Efficient attention approximation
  for high throughputs llm}. In \bibinfo{booktitle}{\emph{Proceedings of
  Machine Learning and Systems}},
  \bibfield{editor}{\bibinfo{person}{M.~Zaharia}, \bibinfo{person}{G.~Joshi},
  {and} \bibinfo{person}{Y.~Lin}} (Eds.), Vol.~\bibinfo{volume}{7}.
  \bibinfo{publisher}{MLSys}.
\newblock
\urldef\tempurl%
\url{https://proceedings.mlsys.org/paper_files/paper/2025/file/f4f55846501f3336f293fd8b6de10770-Paper-Conference.pdf}
\showURL{%
\tempurl}


\bibitem[Karpukhin et~al\mbox{.}(2020)]%
        {karpukhin2020dpr}
\bibfield{author}{\bibinfo{person}{Vladimir Karpukhin}, \bibinfo{person}{Barlas
  Oguz}, \bibinfo{person}{Sewon Min}, \bibinfo{person}{Patrick Lewis},
  \bibinfo{person}{Ledell Wu}, \bibinfo{person}{Sergey Edunov},
  \bibinfo{person}{Danqi Chen}, {and} \bibinfo{person}{Wen-tau Yih}.}
  \bibinfo{year}{2020}\natexlab{}.
\newblock \showarticletitle{Dense Passage Retrieval for Open-Domain Question
  Answering}. In \bibinfo{booktitle}{\emph{Proceedings of the 2020 Conference
  on Empirical Methods in Natural Language Processing (EMNLP)}},
  \bibfield{editor}{\bibinfo{person}{Bonnie Webber}, \bibinfo{person}{Trevor
  Cohn}, \bibinfo{person}{Yulan He}, {and} \bibinfo{person}{Yang Liu}} (Eds.).
  \bibinfo{publisher}{Association for Computational Linguistics},
  \bibinfo{address}{Online}, \bibinfo{pages}{6769--6781}.
\newblock
\href{https://doi.org/10.18653/v1/2020.emnlp-main.550}{doi:\nolinkurl{10.18653/v1/2020.emnlp-main.550}}


\bibitem[Khan et~al\mbox{.}(2025)]%
        {khan2025flstore}
\bibfield{author}{\bibinfo{person}{Ahmad~Faraz Khan}, \bibinfo{person}{Samuel
  Fountain}, \bibinfo{person}{Ahmed~M. Abdelmoniem}, \bibinfo{person}{Ali~R.
  Butt}, {and} \bibinfo{person}{Ali Anwar}.} \bibinfo{year}{2025}\natexlab{}.
\newblock \showarticletitle{FLStore: Efficient Federated Learning Storage for
  non-training workloads}. In \bibinfo{booktitle}{\emph{Proceedings of Machine
  Learning and Systems}}, \bibfield{editor}{\bibinfo{person}{M.~Zaharia},
  \bibinfo{person}{G.~Joshi}, {and} \bibinfo{person}{Y.~Lin}} (Eds.),
  Vol.~\bibinfo{volume}{7}. \bibinfo{publisher}{MLSys}.
\newblock
\urldef\tempurl%
\url{https://proceedings.mlsys.org/paper_files/paper/2025/file/f37347375d8b54e3203e5d24aeb6c58c-Paper-Conference.pdf}
\showURL{%
\tempurl}


\bibitem[Khattab and Zaharia(2020)]%
        {khattab2020colbert}
\bibfield{author}{\bibinfo{person}{Omar Khattab} {and} \bibinfo{person}{Matei
  Zaharia}.} \bibinfo{year}{2020}\natexlab{}.
\newblock \showarticletitle{ColBERT: Efficient and Effective Passage Search via
  Contextualized Late Interaction over BERT}. In
  \bibinfo{booktitle}{\emph{Proceedings of the 43rd International ACM SIGIR
  Conference on Research and Development in Information Retrieval}} (Virtual
  Event, China) \emph{(\bibinfo{series}{SIGIR '20})}.
  \bibinfo{publisher}{Association for Computing Machinery},
  \bibinfo{address}{New York, NY, USA}, \bibinfo{pages}{39–48}.
\newblock
\showISBNx{9781450380164}
\href{https://doi.org/10.1145/3397271.3401075}{doi:\nolinkurl{10.1145/3397271.3401075}}


\bibitem[Kim et~al\mbox{.}(2025c)]%
        {kim2025lia}
\bibfield{author}{\bibinfo{person}{Hyungyo Kim}, \bibinfo{person}{Nachuan
  Wang}, \bibinfo{person}{Qirong Xia}, \bibinfo{person}{Jinghan Huang},
  \bibinfo{person}{Amir Yazdanbakhsh}, {and} \bibinfo{person}{Nam~Sung Kim}.}
  \bibinfo{year}{2025}\natexlab{c}.
\newblock \showarticletitle{LIA: A Single-GPU LLM Inference Acceleration with
  Cooperative AMX-Enabled CPU-GPU Computation and CXL Offloading}. In
  \bibinfo{booktitle}{\emph{Proceedings of the 52nd Annual International
  Symposium on Computer Architecture}} \emph{(\bibinfo{series}{ISCA '25})}.
  \bibinfo{publisher}{Association for Computing Machinery},
  \bibinfo{address}{New York, NY, USA}, \bibinfo{pages}{544–558}.
\newblock
\showISBNx{9798400712616}
\href{https://doi.org/10.1145/3695053.3731092}{doi:\nolinkurl{10.1145/3695053.3731092}}


\bibitem[Kim et~al\mbox{.}(2025a)]%
        {kim2025oaken}
\bibfield{author}{\bibinfo{person}{Minsu Kim}, \bibinfo{person}{Seongmin Hong},
  \bibinfo{person}{RyeoWook Ko}, \bibinfo{person}{Soongyu Choi},
  \bibinfo{person}{Hunjong Lee}, \bibinfo{person}{Junsoo Kim},
  \bibinfo{person}{Joo-Young Kim}, {and} \bibinfo{person}{Jongse Park}.}
  \bibinfo{year}{2025}\natexlab{a}.
\newblock \showarticletitle{Oaken: Fast and Efficient LLM Serving with
  Online-Offline Hybrid KV Cache Quantization}. In
  \bibinfo{booktitle}{\emph{Proceedings of the 52nd Annual International
  Symposium on Computer Architecture}} \emph{(\bibinfo{series}{ISCA '25})}.
  \bibinfo{publisher}{Association for Computing Machinery},
  \bibinfo{address}{New York, NY, USA}, \bibinfo{pages}{482–497}.
\newblock
\showISBNx{9798400712616}
\href{https://doi.org/10.1145/3695053.3731019}{doi:\nolinkurl{10.1145/3695053.3731019}}


\bibitem[Kim et~al\mbox{.}(2025b)]%
        {kim2025pimba}
\bibfield{author}{\bibinfo{person}{Wonung Kim}, \bibinfo{person}{Yubin Lee},
  \bibinfo{person}{Yoonsung Kim}, \bibinfo{person}{Jinwoo Hwang},
  \bibinfo{person}{Seongryong Oh}, \bibinfo{person}{Jiyong Jung},
  \bibinfo{person}{Aziz Huseynov}, \bibinfo{person}{Woong~Gyu Park},
  \bibinfo{person}{Chang~Hyun Park}, \bibinfo{person}{Divya Mahajan}, {and}
  \bibinfo{person}{Jongse Park}.} \bibinfo{year}{2025}\natexlab{b}.
\newblock \showarticletitle{Pimba: A Processing-in-Memory Acceleration for
  Post-Transformer Large Language Model Serving}. In
  \bibinfo{booktitle}{\emph{Proceedings of the 58th IEEE/ACM International
  Symposium on Microarchitecture}} \emph{(\bibinfo{series}{MICRO '25})}.
  \bibinfo{publisher}{Association for Computing Machinery},
  \bibinfo{address}{New York, NY, USA}, \bibinfo{pages}{292–307}.
\newblock
\showISBNx{9798400715730}
\href{https://doi.org/10.1145/3725843.3756121}{doi:\nolinkurl{10.1145/3725843.3756121}}


\bibitem[Kirkpatrick et~al\mbox{.}(2017)]%
        {kirkpatrick2017ewc}
\bibfield{author}{\bibinfo{person}{James Kirkpatrick}, \bibinfo{person}{Razvan
  Pascanu}, \bibinfo{person}{Neil Rabinowitz}, \bibinfo{person}{Joel Veness},
  \bibinfo{person}{Guillaume Desjardins}, \bibinfo{person}{Andrei~A. Rusu},
  \bibinfo{person}{Kieran Milan}, \bibinfo{person}{John Quan},
  \bibinfo{person}{Tiago Ramalho}, \bibinfo{person}{Agnieszka
  Grabska-Barwinska}, \bibinfo{person}{Demis Hassabis},
  \bibinfo{person}{Claudia Clopath}, \bibinfo{person}{Dharshan Kumaran}, {and}
  \bibinfo{person}{Raia Hadsell}.} \bibinfo{year}{2017}\natexlab{}.
\newblock \showarticletitle{Overcoming catastrophic forgetting in neural
  networks}.
\newblock \bibinfo{journal}{\emph{Proceedings of the National Academy of
  Sciences}} \bibinfo{volume}{114}, \bibinfo{number}{13}
  (\bibinfo{year}{2017}), \bibinfo{pages}{3521--3526}.
\newblock
\showeprint{https://www.pnas.org/doi/pdf/10.1073/pnas.1611835114}
\href{https://doi.org/10.1073/pnas.1611835114}{doi:\nolinkurl{10.1073/pnas.1611835114}}


\bibitem[Krishnan et~al\mbox{.}(2016)]%
        {chen2018incapprox}
\bibfield{author}{\bibinfo{person}{Dhanya~R. Krishnan}, \bibinfo{person}{Do~Le
  Quoc}, \bibinfo{person}{Pramod Bhatotia}, \bibinfo{person}{Christof Fetzer},
  {and} \bibinfo{person}{Rodrigo Rodrigues}.} \bibinfo{year}{2016}\natexlab{}.
\newblock \showarticletitle{IncApprox: A Data Analytics System for Incremental
  Approximate Computing}. In \bibinfo{booktitle}{\emph{Proceedings of the 25th
  International Conference on World Wide Web}} (Montr\'{e}al, Qu\'{e}bec,
  Canada) \emph{(\bibinfo{series}{WWW '16})}. \bibinfo{publisher}{International
  World Wide Web Conferences Steering Committee}, \bibinfo{address}{Republic
  and Canton of Geneva, CHE}, \bibinfo{pages}{1133–1144}.
\newblock
\showISBNx{9781450341431}
\href{https://doi.org/10.1145/2872427.2883026}{doi:\nolinkurl{10.1145/2872427.2883026}}


\bibitem[Kumar and Sivathanu(2020)]%
        {kumar2020quiver}
\bibfield{author}{\bibinfo{person}{Abhishek~Vijaya Kumar} {and}
  \bibinfo{person}{Muthian Sivathanu}.} \bibinfo{year}{2020}\natexlab{}.
\newblock \showarticletitle{Quiver: an informed storage cache for deep
  learning}. In \bibinfo{booktitle}{\emph{Proceedings of the 18th USENIX
  Conference on File and Storage Technologies}} (Santa Clara, CA, USA)
  \emph{(\bibinfo{series}{FAST'20})}. \bibinfo{publisher}{USENIX Association},
  \bibinfo{address}{USA}, \bibinfo{pages}{283–296}.
\newblock
\showISBNx{9781939133120}


\bibitem[Kwon et~al\mbox{.}(2020)]%
        {kwon2020fvm}
\bibfield{author}{\bibinfo{person}{Dongup Kwon}, \bibinfo{person}{Junehyuk
  Boo}, \bibinfo{person}{Dongryeong Kim}, {and} \bibinfo{person}{Jangwoo Kim}.}
  \bibinfo{year}{2020}\natexlab{}.
\newblock \showarticletitle{FVM: FPGA-assisted virtual device emulation for
  fast, scalable, and flexible storage virtualization}. In
  \bibinfo{booktitle}{\emph{Proceedings of the 14th USENIX Conference on
  Operating Systems Design and Implementation}}
  \emph{(\bibinfo{series}{OSDI'20})}. \bibinfo{publisher}{USENIX Association},
  \bibinfo{address}{USA}, Article \bibinfo{articleno}{54},
  \bibinfo{numpages}{17}~pages.
\newblock
\showISBNx{978-1-939133-19-9}


\bibitem[Kwon et~al\mbox{.}(2023)]%
        {kwon2023pagedattention}
\bibfield{author}{\bibinfo{person}{Woosuk Kwon}, \bibinfo{person}{Zhuohan Li},
  \bibinfo{person}{Siyuan Zhuang}, \bibinfo{person}{Ying Sheng},
  \bibinfo{person}{Lianmin Zheng}, \bibinfo{person}{Cody~Hao Yu},
  \bibinfo{person}{Joseph Gonzalez}, \bibinfo{person}{Hao Zhang}, {and}
  \bibinfo{person}{Ion Stoica}.} \bibinfo{year}{2023}\natexlab{}.
\newblock \showarticletitle{Efficient Memory Management for Large Language
  Model Serving with PagedAttention}. In \bibinfo{booktitle}{\emph{Proceedings
  of the 29th Symposium on Operating Systems Principles}} (Koblenz, Germany)
  \emph{(\bibinfo{series}{SOSP '23})}. \bibinfo{publisher}{Association for
  Computing Machinery}, \bibinfo{address}{New York, NY, USA},
  \bibinfo{pages}{611–626}.
\newblock
\showISBNx{9798400702297}
\href{https://doi.org/10.1145/3600006.3613165}{doi:\nolinkurl{10.1145/3600006.3613165}}


\bibitem[Lakshman and Malik(2010)]%
        {lakshman2010cassandra}
\bibfield{author}{\bibinfo{person}{Avinash Lakshman} {and}
  \bibinfo{person}{Prashant Malik}.} \bibinfo{year}{2010}\natexlab{}.
\newblock \showarticletitle{Cassandra: a decentralized structured storage
  system}.
\newblock \bibinfo{journal}{\emph{SIGOPS Oper. Syst. Rev.}}
  \bibinfo{volume}{44}, \bibinfo{number}{2} (\bibinfo{date}{April}
  \bibinfo{year}{2010}), \bibinfo{pages}{35–40}.
\newblock
\showISSN{0163-5980}
\href{https://doi.org/10.1145/1773912.1773922}{doi:\nolinkurl{10.1145/1773912.1773922}}


\bibitem[Lee et~al\mbox{.}(2025a)]%
        {kim2025paise}
\bibfield{author}{\bibinfo{person}{Hyojung Lee}, \bibinfo{person}{Daehyeon
  Baek}, \bibinfo{person}{Jimyoung Son}, \bibinfo{person}{Jieun Choi},
  \bibinfo{person}{Kihyo Moon}, {and} \bibinfo{person}{Minsung Jang}.}
  \bibinfo{year}{2025}\natexlab{a}.
\newblock \showarticletitle{PAISE: PIM-Accelerated Inference Scheduling Engine
  for Transformer-based LLM}. In \bibinfo{booktitle}{\emph{2025 IEEE
  International Symposium on High Performance Computer Architecture (HPCA)}}.
  \bibinfo{pages}{1707--1719}.
\newblock
\href{https://doi.org/10.1109/HPCA61900.2025.00126}{doi:\nolinkurl{10.1109/HPCA61900.2025.00126}}


\bibitem[Lee et~al\mbox{.}(2025b)]%
        {lee2025tiertune}
\bibfield{author}{\bibinfo{person}{Hwanjun Lee}, \bibinfo{person}{Minho Kim},
  \bibinfo{person}{Yeji Jung}, \bibinfo{person}{Seonmu Oh},
  \bibinfo{person}{Ki-Dong Kang}, \bibinfo{person}{Seunghak Lee}, {and}
  \bibinfo{person}{Daehoon Kim}.} \bibinfo{year}{2025}\natexlab{b}.
\newblock \showarticletitle{Beyond Page Migration: Enhancing Tiered Memory
  Performance via Integrated Last-Level Cache Management and Page Migration}.
  In \bibinfo{booktitle}{\emph{Proceedings of the 58th IEEE/ACM International
  Symposium on Microarchitecture}} \emph{(\bibinfo{series}{MICRO '25})}.
  \bibinfo{publisher}{Association for Computing Machinery},
  \bibinfo{address}{New York, NY, USA}, \bibinfo{pages}{1763–1776}.
\newblock
\showISBNx{9798400715730}
\href{https://doi.org/10.1145/3725843.3756063}{doi:\nolinkurl{10.1145/3725843.3756063}}


\bibitem[Lee et~al\mbox{.}(2024)]%
        {infinigen2024}
\bibfield{author}{\bibinfo{person}{Wonbeom Lee}, \bibinfo{person}{Jungi Lee},
  \bibinfo{person}{Junghwan Seo}, {and} \bibinfo{person}{Jaewoong Sim}.}
  \bibinfo{year}{2024}\natexlab{}.
\newblock \showarticletitle{InfiniGen: efficient generative inference of large
  language models with dynamic KV cache management}. In
  \bibinfo{booktitle}{\emph{Proceedings of the 18th USENIX Conference on
  Operating Systems Design and Implementation}} (Santa Clara, CA, USA)
  \emph{(\bibinfo{series}{OSDI'24})}. \bibinfo{publisher}{USENIX Association},
  \bibinfo{address}{USA}, Article \bibinfo{articleno}{9},
  \bibinfo{numpages}{18}~pages.
\newblock
\showISBNx{978-1-939133-40-3}


\bibitem[Lewis et~al\mbox{.}(2020)]%
        {lewis2020rag}
\bibfield{author}{\bibinfo{person}{Patrick Lewis}, \bibinfo{person}{Ethan
  Perez}, \bibinfo{person}{Aleksandra Piktus}, \bibinfo{person}{Fabio Petroni},
  \bibinfo{person}{Vladimir Karpukhin}, \bibinfo{person}{Naman Goyal},
  \bibinfo{person}{Heinrich K\"{u}ttler}, \bibinfo{person}{Mike Lewis},
  \bibinfo{person}{Wen-tau Yih}, \bibinfo{person}{Tim Rockt\"{a}schel},
  \bibinfo{person}{Sebastian Riedel}, {and} \bibinfo{person}{Douwe Kiela}.}
  \bibinfo{year}{2020}\natexlab{}.
\newblock \showarticletitle{Retrieval-Augmented Generation for
  Knowledge-Intensive NLP Tasks}. In \bibinfo{booktitle}{\emph{Advances in
  Neural Information Processing Systems}},
  \bibfield{editor}{\bibinfo{person}{H.~Larochelle},
  \bibinfo{person}{M.~Ranzato}, \bibinfo{person}{R.~Hadsell},
  \bibinfo{person}{M.F. Balcan}, {and} \bibinfo{person}{H.~Lin}} (Eds.),
  Vol.~\bibinfo{volume}{33}. \bibinfo{publisher}{Curran Associates, Inc.},
  \bibinfo{pages}{9459--9474}.
\newblock
\urldef\tempurl%
\url{https://proceedings.neurips.cc/paper_files/paper/2020/file/6b493230205f780e1bc26945df7481e5-Paper.pdf}
\showURL{%
\tempurl}


\bibitem[Li et~al\mbox{.}(2026a)]%
        {li2026streamfp}
\bibfield{author}{\bibinfo{person}{Changwu Li}, \bibinfo{person}{Tongjun Shi},
  \bibinfo{person}{Shuhao Zhang}, \bibinfo{person}{Binbin Chen},
  \bibinfo{person}{Bingsheng He}, \bibinfo{person}{Xiaofei Liao}, {and}
  \bibinfo{person}{Hai Jin}.} \bibinfo{year}{2026}\natexlab{a}.
\newblock \showarticletitle{StreamFP: Fingerprint-guided Data Selection for
  Efficient Stream Learning}. In \bibinfo{booktitle}{\emph{Proceedings of the
  ACM Web Conference 2026}} (United Arab Emirates) \emph{(\bibinfo{series}{WWW
  '26})}. \bibinfo{publisher}{Association for Computing Machinery},
  \bibinfo{address}{New York, NY, USA}, \bibinfo{pages}{7474–7484}.
\newblock
\showISBNx{9798400723070}
\href{https://doi.org/10.1145/3774904.3792584}{doi:\nolinkurl{10.1145/3774904.3792584}}


\bibitem[Li et~al\mbox{.}(2020)]%
        {li2020pegasus}
\bibfield{author}{\bibinfo{person}{Jialin Li}, \bibinfo{person}{Jacob Nelson},
  \bibinfo{person}{Ellis Michael}, \bibinfo{person}{Xin Jin}, {and}
  \bibinfo{person}{Dan R.~K. Ports}.} \bibinfo{year}{2020}\natexlab{}.
\newblock \showarticletitle{Pegasus: Tolerating skewed workloads in distributed
  storage with in-network coherence directories}. In
  \bibinfo{booktitle}{\emph{Proceedings of the 14th USENIX Conference on
  Operating Systems Design and Implementation}}
  \emph{(\bibinfo{series}{OSDI'20})}. \bibinfo{publisher}{USENIX Association},
  \bibinfo{address}{USA}, Article \bibinfo{articleno}{22},
  \bibinfo{numpages}{20}~pages.
\newblock
\showISBNx{978-1-939133-19-9}


\bibitem[Li et~al\mbox{.}(2025b)]%
        {li2025toppings}
\bibfield{author}{\bibinfo{person}{Suyi Li}, \bibinfo{person}{Hanfeng Lu},
  \bibinfo{person}{Tianyuan Wu}, \bibinfo{person}{Minchen Yu},
  \bibinfo{person}{Qizhen Weng}, \bibinfo{person}{Xusheng Chen},
  \bibinfo{person}{Yizhou Shan}, \bibinfo{person}{Binhang Yuan}, {and}
  \bibinfo{person}{Wei Wang}.} \bibinfo{year}{2025}\natexlab{b}.
\newblock \showarticletitle{TOPPINGS: CPU-assisted, rank-aware adapter serving
  for LLM inference}. In \bibinfo{booktitle}{\emph{Proceedings of the 2025
  USENIX Conference on Usenix Annual Technical Conference}} (Boston, MA, USA)
  \emph{(\bibinfo{series}{USENIX ATC '25})}. \bibinfo{publisher}{USENIX
  Association}, \bibinfo{address}{USA}, Article \bibinfo{articleno}{37},
  \bibinfo{numpages}{17}~pages.
\newblock
\showISBNx{978-1-939133-48-9}


\bibitem[Li et~al\mbox{.}(2026b)]%
        {li2025global}
\bibfield{author}{\bibinfo{person}{Yading Li}, \bibinfo{person}{Dandan Song},
  \bibinfo{person}{Yuhang Tian}, \bibinfo{person}{Hao Wang},
  \bibinfo{person}{Changzhi Zhou}, {and} \bibinfo{person}{Shuhao Zhang}.}
  \bibinfo{year}{2026}\natexlab{b}.
\newblock \showarticletitle{A Framework of Knowledge Graph-Enhanced Large
  Language Model Based on Global Planning}.
\newblock \bibinfo{journal}{\emph{IEEE Transactions on Knowledge and Data
  Engineering}} \bibinfo{volume}{38}, \bibinfo{number}{2}
  (\bibinfo{year}{2026}), \bibinfo{pages}{736--748}.
\newblock
\href{https://doi.org/10.1109/TKDE.2025.3639599}{doi:\nolinkurl{10.1109/TKDE.2025.3639599}}


\bibitem[Li et~al\mbox{.}(2024)]%
        {li2024keldar}
\bibfield{author}{\bibinfo{person}{Yading Li}, \bibinfo{person}{Dandan Song},
  \bibinfo{person}{Changzhi Zhou}, \bibinfo{person}{Yuhang Tian},
  \bibinfo{person}{Hao Wang}, \bibinfo{person}{Ziyi Yang}, {and}
  \bibinfo{person}{Shuhao Zhang}.} \bibinfo{year}{2024}\natexlab{}.
\newblock \showarticletitle{A Framework of Knowledge Graph-Enhanced Large
  Language Model Based on Question Decomposition and Atomic Retrieval}. In
  \bibinfo{booktitle}{\emph{Findings of the Association for Computational
  Linguistics: EMNLP 2024}}, \bibfield{editor}{\bibinfo{person}{Yaser
  Al-Onaizan}, \bibinfo{person}{Mohit Bansal}, {and} \bibinfo{person}{Yun-Nung
  Chen}} (Eds.). \bibinfo{publisher}{Association for Computational
  Linguistics}, \bibinfo{address}{Miami, Florida, USA},
  \bibinfo{pages}{11472--11485}.
\newblock
\href{https://doi.org/10.18653/v1/2024.findings-emnlp.670}{doi:\nolinkurl{10.18653/v1/2024.findings-emnlp.670}}


\bibitem[Li et~al\mbox{.}(2025a)]%
        {liu2025slimpipe}
\bibfield{author}{\bibinfo{person}{Zhouyang Li}, \bibinfo{person}{Yuliang Liu},
  \bibinfo{person}{Wei Zhang}, \bibinfo{person}{Tailing Yuan},
  \bibinfo{person}{Bin Chen}, {and} \bibinfo{person}{Chengru Song}.}
  \bibinfo{year}{2025}\natexlab{a}.
\newblock \showarticletitle{SlimPipe: Memory-Thrifty and Efficient Pipeline
  Parallelism for Long-Context LLM Training}. In
  \bibinfo{booktitle}{\emph{Proceedings of the International Conference for
  High Performance Computing, Networking, Storage and Analysis}}
  \emph{(\bibinfo{series}{SC '25})}. \bibinfo{publisher}{Association for
  Computing Machinery}, \bibinfo{address}{New York, NY, USA},
  \bibinfo{pages}{1409–1428}.
\newblock
\showISBNx{9798400714665}
\href{https://doi.org/10.1145/3712285.3759855}{doi:\nolinkurl{10.1145/3712285.3759855}}


\bibitem[Li and Zhang(2024)]%
        {li2024streamcache}
\bibfield{author}{\bibinfo{person}{Zhiyue Li} {and} \bibinfo{person}{Guangyan
  Zhang}.} \bibinfo{year}{2024}\natexlab{}.
\newblock \showarticletitle{StreamCache: revisiting page cache for file
  scanning on fast storage devices}. In \bibinfo{booktitle}{\emph{Proceedings
  of the 2024 USENIX Conference on Usenix Annual Technical Conference}} (Santa
  Clara, CA, USA) \emph{(\bibinfo{series}{USENIX ATC'24})}.
  \bibinfo{publisher}{USENIX Association}, \bibinfo{address}{USA}, Article
  \bibinfo{articleno}{68}, \bibinfo{numpages}{16}~pages.
\newblock
\showISBNx{978-1-939133-41-0}


\bibitem[Li et~al\mbox{.}(2023)]%
        {li2023alpaserve}
\bibfield{author}{\bibinfo{person}{Zhuohan Li}, \bibinfo{person}{Lianmin
  Zheng}, \bibinfo{person}{Yinmin Zhong}, \bibinfo{person}{Vincent Liu},
  \bibinfo{person}{Ying Sheng}, \bibinfo{person}{Xin Jin},
  \bibinfo{person}{Yanping Huang}, \bibinfo{person}{Zhifeng Chen},
  \bibinfo{person}{Hao Zhang}, \bibinfo{person}{Joseph~E. Gonzalez}, {and}
  \bibinfo{person}{Ion Stoica}.} \bibinfo{year}{2023}\natexlab{}.
\newblock \showarticletitle{{AlpaServe}: Statistical Multiplexing with Model
  Parallelism for Deep Learning Serving}. In \bibinfo{booktitle}{\emph{17th
  USENIX Symposium on Operating Systems Design and Implementation (OSDI 23)}}.
  \bibinfo{publisher}{USENIX Association}, \bibinfo{address}{Boston, MA},
  \bibinfo{pages}{663--679}.
\newblock
\showISBNx{978-1-939133-34-2}
\urldef\tempurl%
\url{https://www.usenix.org/conference/osdi23/presentation/li-zhouhan}
\showURL{%
\tempurl}


\bibitem[Liakopoulos et~al\mbox{.}(2025)]%
        {liakopoulos2025maveriq}
\bibfield{author}{\bibinfo{person}{Dimitrios Liakopoulos},
  \bibinfo{person}{Prasoon Sinha}, \bibinfo{person}{Tianrui Hu},
  \bibinfo{person}{Myungjin Lee}, {and} \bibinfo{person}{Neeraja~J.
  Yadwadkar}.} \bibinfo{year}{2025}\natexlab{}.
\newblock \showarticletitle{MaverIQ: Fingerprint-Guided Extrapolation and
  Fragmentation-Aware Layering for Intent-Based LLM Serving}. In
  \bibinfo{booktitle}{\emph{Proceedings of the International Conference for
  High Performance Computing, Networking, Storage and Analysis}}
  \emph{(\bibinfo{series}{SC '25})}. \bibinfo{publisher}{Association for
  Computing Machinery}, \bibinfo{address}{New York, NY, USA},
  \bibinfo{pages}{1676–1696}.
\newblock
\showISBNx{9798400714665}
\href{https://doi.org/10.1145/3712285.3759867}{doi:\nolinkurl{10.1145/3712285.3759867}}


\bibitem[Lian et~al\mbox{.}(2025)]%
        {lian2025universal}
\bibfield{author}{\bibinfo{person}{Xinyu Lian}, \bibinfo{person}{Sam~Ade
  Jacobs}, \bibinfo{person}{Lev Kurilenko}, \bibinfo{person}{Masahiro Tanaka},
  \bibinfo{person}{Stas Bekman}, \bibinfo{person}{Olatunji Ruwase}, {and}
  \bibinfo{person}{Minjia Zhang}.} \bibinfo{year}{2025}\natexlab{}.
\newblock \showarticletitle{Universal checkpointing: a flexible and efficient
  distributed checkpointing system for large-scale DNN training with
  reconfigurable parallelism}. In \bibinfo{booktitle}{\emph{Proceedings of the
  2025 USENIX Conference on Usenix Annual Technical Conference}} (Boston, MA,
  USA) \emph{(\bibinfo{series}{USENIX ATC '25})}. \bibinfo{publisher}{USENIX
  Association}, \bibinfo{address}{USA}, Article \bibinfo{articleno}{90},
  \bibinfo{numpages}{16}~pages.
\newblock
\showISBNx{978-1-939133-48-9}


\bibitem[Lim et~al\mbox{.}(2021)]%
        {zico21}
\bibfield{author}{\bibinfo{person}{Gangmuk Lim}, \bibinfo{person}{Jeongseob
  Ahn}, \bibinfo{person}{Wencong Xiao}, \bibinfo{person}{Youngjin Kwon}, {and}
  \bibinfo{person}{Myeongjae Jeon}.} \bibinfo{year}{2021}\natexlab{}.
\newblock \showarticletitle{Zico: Efficient {GPU} Memory Sharing for Concurrent
  {DNN} Training}. In \bibinfo{booktitle}{\emph{2021 USENIX Annual Technical
  Conference (USENIX ATC 21)}}. \bibinfo{publisher}{USENIX Association},
  \bibinfo{pages}{161--175}.
\newblock
\showISBNx{978-1-939133-23-6}
\urldef\tempurl%
\url{https://www.usenix.org/conference/atc21/presentation/lim}
\showURL{%
\tempurl}


\bibitem[Lin et~al\mbox{.}(2024)]%
        {lin2024parrot}
\bibfield{author}{\bibinfo{person}{Chaofan Lin}, \bibinfo{person}{Zhenhua Han},
  \bibinfo{person}{Chengruidong Zhang}, \bibinfo{person}{Yuqing Yang},
  \bibinfo{person}{Fan Yang}, \bibinfo{person}{Chen Chen}, {and}
  \bibinfo{person}{Lili Qiu}.} \bibinfo{year}{2024}\natexlab{}.
\newblock \showarticletitle{Parrot: efficient serving of LLM-based applications
  with semantic variable}. In \bibinfo{booktitle}{\emph{Proceedings of the 18th
  USENIX Conference on Operating Systems Design and Implementation}} (Santa
  Clara, CA, USA) \emph{(\bibinfo{series}{OSDI'24})}.
  \bibinfo{publisher}{USENIX Association}, \bibinfo{address}{USA}, Article
  \bibinfo{articleno}{50}, \bibinfo{numpages}{17}~pages.
\newblock
\showISBNx{978-1-939133-40-3}


\bibitem[Lin et~al\mbox{.}(2025a)]%
        {lin2025whatifstragglers}
\bibfield{author}{\bibinfo{person}{Jinkun Lin}, \bibinfo{person}{Ziheng Jiang},
  \bibinfo{person}{Zuquan Song}, \bibinfo{person}{Sida Zhao},
  \bibinfo{person}{Menghan Yu}, \bibinfo{person}{Zhanghan Wang},
  \bibinfo{person}{Chenyuan Wang}, \bibinfo{person}{Zuocheng Shi},
  \bibinfo{person}{Xiang Shi}, \bibinfo{person}{Wei Jia},
  \bibinfo{person}{Zherui Liu}, \bibinfo{person}{Shuguang Wang},
  \bibinfo{person}{Haibin Lin}, \bibinfo{person}{Xin Liu},
  \bibinfo{person}{Aurojit Panda}, {and} \bibinfo{person}{Jinyang Li}.}
  \bibinfo{year}{2025}\natexlab{a}.
\newblock \showarticletitle{Understanding stragglers in large model training
  using what-if analysis}. In \bibinfo{booktitle}{\emph{Proceedings of the 19th
  USENIX Conference on Operating Systems Design and Implementation}} (Boston,
  MA, USA) \emph{(\bibinfo{series}{OSDI '25})}. \bibinfo{publisher}{USENIX
  Association}, \bibinfo{address}{USA}, Article \bibinfo{articleno}{27},
  \bibinfo{numpages}{16}~pages.
\newblock
\showISBNx{978-1-939133-47-2}


\bibitem[Lin et~al\mbox{.}(2025b)]%
        {lin2025weipipe}
\bibfield{author}{\bibinfo{person}{Junfeng Lin}, \bibinfo{person}{Ziming Liu},
  \bibinfo{person}{Yang You}, \bibinfo{person}{Jun Wang},
  \bibinfo{person}{Weihao Zhang}, {and} \bibinfo{person}{Rong Zhao}.}
  \bibinfo{year}{2025}\natexlab{b}.
\newblock \showarticletitle{WeiPipe: Weight Pipeline Parallelism for
  Communication-Effective Long-Context Large Model Training}. In
  \bibinfo{booktitle}{\emph{Proceedings of the 30th ACM SIGPLAN Annual
  Symposium on Principles and Practice of Parallel Programming}} (Las Vegas,
  NV, USA) \emph{(\bibinfo{series}{PPoPP '25})}.
  \bibinfo{publisher}{Association for Computing Machinery},
  \bibinfo{address}{New York, NY, USA}, \bibinfo{pages}{225–238}.
\newblock
\showISBNx{9798400714436}
\href{https://doi.org/10.1145/3710848.3710869}{doi:\nolinkurl{10.1145/3710848.3710869}}


\bibitem[Lin et~al\mbox{.}(2025c)]%
        {lin2025qserve}
\bibfield{author}{\bibinfo{person}{Yujun Lin}, \bibinfo{person}{Haotian Tang},
  \bibinfo{person}{Shang Yang}, \bibinfo{person}{Zhekai Zhang},
  \bibinfo{person}{Guangxuan Xiao}, \bibinfo{person}{Chuang Gan}, {and}
  \bibinfo{person}{Song Han}.} \bibinfo{year}{2025}\natexlab{c}.
\newblock \showarticletitle{QServe:W4A8KV4 Quantization and System Co-design
  for Efficient LLM Serving}. In \bibinfo{booktitle}{\emph{Proceedings of
  Machine Learning and Systems}},
  \bibfield{editor}{\bibinfo{person}{M.~Zaharia}, \bibinfo{person}{G.~Joshi},
  {and} \bibinfo{person}{Y.~Lin}} (Eds.), Vol.~\bibinfo{volume}{7}.
  \bibinfo{publisher}{MLSys}.
\newblock
\urldef\tempurl%
\url{https://proceedings.mlsys.org/paper_files/paper/2025/file/fbe2b2f74a2ece8070d8fb073717bda6-Paper-Conference.pdf}
\showURL{%
\tempurl}


\bibitem[Liu et~al\mbox{.}(2025d)]%
        {liu2025heterrag}
\bibfield{author}{\bibinfo{person}{Chaoqiang Liu}, \bibinfo{person}{Haifeng
  Liu}, \bibinfo{person}{Dan Chen}, \bibinfo{person}{Yu Huang},
  \bibinfo{person}{Yi Zhang}, \bibinfo{person}{Wenjing Xiao},
  \bibinfo{person}{Xiaofei Liao}, {and} \bibinfo{person}{Hai Jin}.}
  \bibinfo{year}{2025}\natexlab{d}.
\newblock \showarticletitle{HeterRAG: Heterogeneous Processing-in-Memory
  Acceleration for Retrieval-augmented Generation}. In
  \bibinfo{booktitle}{\emph{Proceedings of the 52nd Annual International
  Symposium on Computer Architecture}} \emph{(\bibinfo{series}{ISCA '25})}.
  \bibinfo{publisher}{Association for Computing Machinery},
  \bibinfo{address}{New York, NY, USA}, \bibinfo{pages}{884–898}.
\newblock
\showISBNx{9798400712616}
\href{https://doi.org/10.1145/3695053.3731089}{doi:\nolinkurl{10.1145/3695053.3731089}}


\bibitem[Liu et~al\mbox{.}(2025a)]%
        {liu2025beyondhotness}
\bibfield{author}{\bibinfo{person}{Jinshu Liu}, \bibinfo{person}{Hamid Hadian},
  \bibinfo{person}{Hanchen Xu}, {and} \bibinfo{person}{Huaicheng Li}.}
  \bibinfo{year}{2025}\natexlab{a}.
\newblock \showarticletitle{Tiered memory management beyond hotness}. In
  \bibinfo{booktitle}{\emph{Proceedings of the 19th USENIX Conference on
  Operating Systems Design and Implementation}} (Boston, MA, USA)
  \emph{(\bibinfo{series}{OSDI '25})}. \bibinfo{publisher}{USENIX Association},
  \bibinfo{address}{USA}, Article \bibinfo{articleno}{40},
  \bibinfo{numpages}{17}~pages.
\newblock
\showISBNx{978-1-939133-47-2}


\bibitem[Liu et~al\mbox{.}(2026c)]%
        {liu2026sage}
\bibfield{author}{\bibinfo{person}{Jun Liu}, \bibinfo{person}{Peilin Liu},
  \bibinfo{person}{Ruicheng Zhang}, \bibinfo{person}{Zhang Senlei},
  \bibinfo{person}{Yanbo Chen}, \bibinfo{person}{Ziao Wang},
  \bibinfo{person}{Jinyun Yang}, \bibinfo{person}{mingqi wang},
  \bibinfo{person}{Shuhao Zhang}, \bibinfo{person}{Xiaofei Liao}, {and}
  \bibinfo{person}{Hai Jin}.} \bibinfo{year}{2026}\natexlab{c}.
\newblock \showarticletitle{{SAGE}: A Dataflow-Native Framework for Modular,
  Controllable, and Transparent {LLM}-Augmented Reasoning}. In
  \bibinfo{booktitle}{\emph{Forty-third International Conference on Machine
  Learning}}.
\newblock
\urldef\tempurl%
\url{https://openreview.net/forum?id=TXcFJdT7at}
\showURL{%
\tempurl}


\bibitem[Liu et~al\mbox{.}(2025f)]%
        {liu2025hermes}
\bibfield{author}{\bibinfo{person}{Lian Liu}, \bibinfo{person}{Shixin Zhao},
  \bibinfo{person}{Bing Li}, \bibinfo{person}{Haimeng Ren},
  \bibinfo{person}{Zhaohui Xu}, \bibinfo{person}{Mengdi Wang},
  \bibinfo{person}{Xiaowei Li}, \bibinfo{person}{Yinhe Han}, {and}
  \bibinfo{person}{Ying Wang}.} \bibinfo{year}{2025}\natexlab{f}.
\newblock \showarticletitle{Make LLM Inference Affordable to Everyone:
  Augmenting GPU Memory with NDP-DIMM}. In \bibinfo{booktitle}{\emph{2025 IEEE
  International Symposium on High Performance Computer Architecture (HPCA)}}.
  \bibinfo{pages}{1751--1765}.
\newblock
\href{https://doi.org/10.1109/HPCA61900.2025.00129}{doi:\nolinkurl{10.1109/HPCA61900.2025.00129}}


\bibitem[Liu et~al\mbox{.}(2025b)]%
        {liu2025mell}
\bibfield{author}{\bibinfo{person}{Qianli Liu}, \bibinfo{person}{Zicong Hong},
  \bibinfo{person}{Peng Li}, \bibinfo{person}{Fahao Chen}, {and}
  \bibinfo{person}{Song Guo}.} \bibinfo{year}{2025}\natexlab{b}.
\newblock \showarticletitle{Mell: Memory-Efficient Large Language Model Serving
  via Multi-GPU KV Cache Management}. In \bibinfo{booktitle}{\emph{IEEE INFOCOM
  2025 - IEEE Conference on Computer Communications}}. \bibinfo{pages}{1--10}.
\newblock
\href{https://doi.org/10.1109/INFOCOM55648.2025.11044533}{doi:\nolinkurl{10.1109/INFOCOM55648.2025.11044533}}


\bibitem[Liu et~al\mbox{.}(2025c)]%
        {liu2025mario}
\bibfield{author}{\bibinfo{person}{Weijian Liu}, \bibinfo{person}{Mingzhen Li},
  \bibinfo{person}{Guangming Tan}, {and} \bibinfo{person}{Weile Jia}.}
  \bibinfo{year}{2025}\natexlab{c}.
\newblock \showarticletitle{Mario: Near Zero-cost Activation Checkpointing in
  Pipeline Parallelism}. In \bibinfo{booktitle}{\emph{Proceedings of the 30th
  ACM SIGPLAN Annual Symposium on Principles and Practice of Parallel
  Programming}} (Las Vegas, NV, USA) \emph{(\bibinfo{series}{PPoPP '25})}.
  \bibinfo{publisher}{Association for Computing Machinery},
  \bibinfo{address}{New York, NY, USA}, \bibinfo{pages}{197–211}.
\newblock
\showISBNx{9798400714436}
\href{https://doi.org/10.1145/3710848.3710878}{doi:\nolinkurl{10.1145/3710848.3710878}}


\bibitem[Liu et~al\mbox{.}(2026a)]%
        {cacheslide2025}
\bibfield{author}{\bibinfo{person}{Yang Liu}, \bibinfo{person}{Yunfei Gu},
  \bibinfo{person}{Liqiang Zhang}, \bibinfo{person}{Chentao Wu},
  \bibinfo{person}{Guangtao Xue}, \bibinfo{person}{Jie Li},
  \bibinfo{person}{Minyi Guo}, \bibinfo{person}{Junhao Hu}, {and}
  \bibinfo{person}{Jie Meng}.} \bibinfo{year}{2026}\natexlab{a}.
\newblock \showarticletitle{{CacheSlide}: Unlocking Cross {Position-Aware} {KV}
  Cache Reuse for Accelerating {LLM} Serving}. In
  \bibinfo{booktitle}{\emph{24th USENIX Conference on File and Storage
  Technologies (FAST 26)}}. \bibinfo{publisher}{USENIX Association},
  \bibinfo{address}{Santa Clara, CA}, \bibinfo{pages}{83--99}.
\newblock
\showISBNx{978-1-939133-53-3}
\urldef\tempurl%
\url{https://www.usenix.org/conference/fast26/presentation/liu-yang}
\showURL{%
\tempurl}


\bibitem[Liu et~al\mbox{.}(2026b)]%
        {liu2026droidspeak}
\bibfield{author}{\bibinfo{person}{Yuhan Liu}, \bibinfo{person}{Yuyang Huang},
  \bibinfo{person}{Jiayi Yao}, \bibinfo{person}{Shaoting Feng},
  \bibinfo{person}{Zhuohan Gu}, \bibinfo{person}{Kuntai Du},
  \bibinfo{person}{Hanchen Li}, \bibinfo{person}{Yihua Cheng},
  \bibinfo{person}{Junchen Jiang}, \bibinfo{person}{Shan Lu},
  \bibinfo{person}{Madan Musuvathi}, {and} \bibinfo{person}{Esha Choukse}.}
  \bibinfo{year}{2026}\natexlab{b}.
\newblock \showarticletitle{{DroidSpeak}: {KV} Cache Sharing Across Fine-tuned
  Model Variants}. In \bibinfo{booktitle}{\emph{23rd USENIX Symposium on
  Networked Systems Design and Implementation (NSDI 26)}}.
  \bibinfo{publisher}{USENIX Association}, \bibinfo{address}{Renton, WA},
  \bibinfo{pages}{319--338}.
\newblock
\showISBNx{978-1-939133-54-0}
\urldef\tempurl%
\url{https://www.usenix.org/conference/nsdi26/presentation/liu-yuhan}
\showURL{%
\tempurl}


\bibitem[Liu et~al\mbox{.}(2025e)]%
        {liu2025vqllm}
\bibfield{author}{\bibinfo{person}{Zihan Liu}, \bibinfo{person}{Xinhao Luo},
  \bibinfo{person}{Junxian Guo}, \bibinfo{person}{Wentao Ni},
  \bibinfo{person}{Yangjie Zhou}, \bibinfo{person}{Yue Guan},
  \bibinfo{person}{Cong Guo}, \bibinfo{person}{Weihao Cui}, \bibinfo{person}{Yu
  Feng}, \bibinfo{person}{Minyi Guo}, \bibinfo{person}{Yuhao Zhu},
  \bibinfo{person}{Minjia Zhang}, \bibinfo{person}{Chen Jin}, {and}
  \bibinfo{person}{Jingwen Leng}.} \bibinfo{year}{2025}\natexlab{e}.
\newblock \showarticletitle{VQ-LLM: High-performance Code Generation for Vector
  Quantization Augmented LLM Inference}. In \bibinfo{booktitle}{\emph{2025 IEEE
  International Symposium on High Performance Computer Architecture (HPCA)}}.
  \bibinfo{pages}{1496--1509}.
\newblock
\href{https://doi.org/10.1109/HPCA61900.2025.00112}{doi:\nolinkurl{10.1109/HPCA61900.2025.00112}}


\bibitem[Lopez-Paz and Ranzato(2017)]%
        {lopez2017gem}
\bibfield{author}{\bibinfo{person}{David Lopez-Paz} {and}
  \bibinfo{person}{Marc'Aurelio Ranzato}.} \bibinfo{year}{2017}\natexlab{}.
\newblock \showarticletitle{Gradient episodic memory for continual learning}.
  In \bibinfo{booktitle}{\emph{Proceedings of the 31st International Conference
  on Neural Information Processing Systems}} (Long Beach, California, USA)
  \emph{(\bibinfo{series}{NIPS'17})}. \bibinfo{publisher}{Curran Associates
  Inc.}, \bibinfo{address}{Red Hook, NY, USA}, \bibinfo{pages}{6470–6479}.
\newblock
\showISBNx{9781510860964}


\bibitem[Lou et~al\mbox{.}(2026)]%
        {lou2026hydraserve}
\bibfield{author}{\bibinfo{person}{Chiheng Lou}, \bibinfo{person}{Sheng Qi},
  \bibinfo{person}{Chao Jin}, \bibinfo{person}{Dapeng Nie},
  \bibinfo{person}{Haoran Yang}, \bibinfo{person}{Yu Ding},
  \bibinfo{person}{Xuanzhe Liu}, {and} \bibinfo{person}{Xin Jin}.}
  \bibinfo{year}{2026}\natexlab{}.
\newblock \showarticletitle{{HydraServe}: Minimizing Cold Start Latency for
  Serverless {LLM} Serving in Public Clouds}. In \bibinfo{booktitle}{\emph{23rd
  USENIX Symposium on Networked Systems Design and Implementation (NSDI 26)}}.
  \bibinfo{publisher}{USENIX Association}, \bibinfo{address}{Renton, WA},
  \bibinfo{pages}{415--430}.
\newblock
\showISBNx{978-1-939133-54-0}
\urldef\tempurl%
\url{https://www.usenix.org/conference/nsdi26/presentation/lou}
\showURL{%
\tempurl}


\bibitem[Luo and Carey(2019)]%
        {luo2020lsm}
\bibfield{author}{\bibinfo{person}{Chen Luo} {and} \bibinfo{person}{Michael~J.
  Carey}.} \bibinfo{year}{2019}\natexlab{}.
\newblock \showarticletitle{LSM-based storage techniques: a survey}.
\newblock \bibinfo{journal}{\emph{The VLDB Journal}} \bibinfo{volume}{29},
  \bibinfo{number}{1} (\bibinfo{year}{2019}), \bibinfo{pages}{393–418}.
\newblock
\showISSN{0949-877X}
\href{https://doi.org/10.1007/s00778-019-00555-y}{doi:\nolinkurl{10.1007/s00778-019-00555-y}}


\bibitem[Mahgoub et~al\mbox{.}(2021)]%
        {mahgoub2021sonic}
\bibfield{author}{\bibinfo{person}{Ashraf Mahgoub}, \bibinfo{person}{Karthick
  Shankar}, \bibinfo{person}{Subrata Mitra}, \bibinfo{person}{Ana Klimovic},
  \bibinfo{person}{Somali Chaterji}, {and} \bibinfo{person}{Saurabh Bagchi}.}
  \bibinfo{year}{2021}\natexlab{}.
\newblock \showarticletitle{{SONIC}: Application-aware Data Passing for Chained
  Serverless Applications}. In \bibinfo{booktitle}{\emph{2021 USENIX Annual
  Technical Conference (USENIX ATC 21)}}. \bibinfo{publisher}{USENIX
  Association}, \bibinfo{pages}{285--301}.
\newblock
\showISBNx{978-1-939133-23-6}
\urldef\tempurl%
\url{https://www.usenix.org/conference/atc21/presentation/mahgoub}
\showURL{%
\tempurl}


\bibitem[Mahgoub et~al\mbox{.}(2022)]%
        {mahgoub2022orion}
\bibfield{author}{\bibinfo{person}{Ashraf Mahgoub},
  \bibinfo{person}{Edgardo~Barsallo Yi}, \bibinfo{person}{Karthick Shankar},
  \bibinfo{person}{Sameh Elnikety}, \bibinfo{person}{Somali Chaterji}, {and}
  \bibinfo{person}{Saurabh Bagchi}.} \bibinfo{year}{2022}\natexlab{}.
\newblock \showarticletitle{{ORION} and the Three Rights: Sizing, Bundling, and
  Prewarming for Serverless {DAGs}}. In \bibinfo{booktitle}{\emph{16th USENIX
  Symposium on Operating Systems Design and Implementation (OSDI 22)}}.
  \bibinfo{publisher}{USENIX Association}, \bibinfo{address}{Carlsbad, CA},
  \bibinfo{pages}{303--320}.
\newblock
\showISBNx{978-1-939133-28-1}
\urldef\tempurl%
\url{https://www.usenix.org/conference/osdi22/presentation/mahgoub}
\showURL{%
\tempurl}


\bibitem[Malkov and Yashunin(2020)]%
        {malkov2020hnsw}
\bibfield{author}{\bibinfo{person}{Yu~A. Malkov} {and} \bibinfo{person}{D.~A.
  Yashunin}.} \bibinfo{year}{2020}\natexlab{}.
\newblock \showarticletitle{Efficient and Robust Approximate Nearest Neighbor
  Search Using Hierarchical Navigable Small World Graphs}.
\newblock \bibinfo{journal}{\emph{IEEE Transactions on Pattern Analysis and
  Machine Intelligence}} \bibinfo{volume}{42}, \bibinfo{number}{4}
  (\bibinfo{year}{2020}), \bibinfo{pages}{824--836}.
\newblock
\href{https://doi.org/10.1109/TPAMI.2018.2889473}{doi:\nolinkurl{10.1109/TPAMI.2018.2889473}}


\bibitem[Mao et~al\mbox{.}(2025)]%
        {mao2025spacker}
\bibfield{author}{\bibinfo{person}{Yancan Mao}, \bibinfo{person}{Shuhao Zhang},
  {and} \bibinfo{person}{Richard T.~B. Ma}.} \bibinfo{year}{2025}\natexlab{}.
\newblock \showarticletitle{Spacker: Unified State Migration for Distributed
  Streaming}. In \bibinfo{booktitle}{\emph{2025 IEEE 45th International
  Conference on Distributed Computing Systems (ICDCS)}}.
  \bibinfo{pages}{1000--1010}.
\newblock
\href{https://doi.org/10.1109/ICDCS63083.2025.00101}{doi:\nolinkurl{10.1109/ICDCS63083.2025.00101}}


\bibitem[Mao et~al\mbox{.}(2023)]%
        {mao2023morphstream}
\bibfield{author}{\bibinfo{person}{Yancan Mao}, \bibinfo{person}{Jianjun Zhao},
  \bibinfo{person}{Shuhao Zhang}, \bibinfo{person}{Haikun Liu}, {and}
  \bibinfo{person}{Volker Markl}.} \bibinfo{year}{2023}\natexlab{}.
\newblock \showarticletitle{MorphStream: Adaptive Scheduling for Scalable
  Transactional Stream Processing on Multicores}.
\newblock \bibinfo{journal}{\emph{Proc. ACM Manag. Data}} \bibinfo{volume}{1},
  \bibinfo{number}{1}, Article \bibinfo{articleno}{59} (\bibinfo{date}{May}
  \bibinfo{year}{2023}), \bibinfo{numpages}{26}~pages.
\newblock
\href{https://doi.org/10.1145/3588913}{doi:\nolinkurl{10.1145/3588913}}


\bibitem[Maurya et~al\mbox{.}(2025)]%
        {maurya2025mlpoffload}
\bibfield{author}{\bibinfo{person}{Avinash~Kumar Maurya},
  \bibinfo{person}{M.~Mustafa Rafique}, \bibinfo{person}{Franck Cappello},
  {and} \bibinfo{person}{Bogdan Nicolae}.} \bibinfo{year}{2025}\natexlab{}.
\newblock \showarticletitle{MLP-Offload: Multi-Level, Multi-Path Offloading for
  LLM Pre-training to Break the GPU Memory Wall}. In
  \bibinfo{booktitle}{\emph{Proceedings of the International Conference for
  High Performance Computing, Networking, Storage and Analysis}}
  \emph{(\bibinfo{series}{SC '25})}. \bibinfo{publisher}{Association for
  Computing Machinery}, \bibinfo{address}{New York, NY, USA},
  \bibinfo{pages}{1381–1394}.
\newblock
\showISBNx{9798400714665}
\href{https://doi.org/10.1145/3712285.3759864}{doi:\nolinkurl{10.1145/3712285.3759864}}


\bibitem[McSherry et~al\mbox{.}(2013)]%
        {murray2013differential}
\bibfield{author}{\bibinfo{person}{Frank McSherry},
  \bibinfo{person}{Derek~Gordon Murray}, \bibinfo{person}{Rebecca Isaacs},
  {and} \bibinfo{person}{Michael Isard}.} \bibinfo{year}{2013}\natexlab{}.
\newblock \showarticletitle{Differential dataflow.}. In
  \bibinfo{booktitle}{\emph{Proceedings of the Sixth Biennial Conference on
  Innovative Data Systems Research (CIDR)}}.
\newblock
\urldef\tempurl%
\url{https://www.cidrdb.org/cidr2013/Papers/CIDR13_Paper111.pdf}
\showURL{%
\tempurl}


\bibitem[Miao et~al\mbox{.}(2024)]%
        {miao2024specinfer}
\bibfield{author}{\bibinfo{person}{Xupeng Miao}, \bibinfo{person}{Gabriele
  Oliaro}, \bibinfo{person}{Zhihao Zhang}, \bibinfo{person}{Xinhao Cheng},
  \bibinfo{person}{Zeyu Wang}, \bibinfo{person}{Zhengxin Zhang},
  \bibinfo{person}{Rae Ying~Yee Wong}, \bibinfo{person}{Alan Zhu},
  \bibinfo{person}{Lijie Yang}, \bibinfo{person}{Xiaoxiang Shi},
  \bibinfo{person}{Chunan Shi}, \bibinfo{person}{Zhuoming Chen},
  \bibinfo{person}{Daiyaan Arfeen}, \bibinfo{person}{Reyna Abhyankar}, {and}
  \bibinfo{person}{Zhihao Jia}.} \bibinfo{year}{2024}\natexlab{}.
\newblock \showarticletitle{SpecInfer: Accelerating Large Language Model
  Serving with Tree-based Speculative Inference and Verification}. In
  \bibinfo{booktitle}{\emph{Proceedings of the 29th ACM International
  Conference on Architectural Support for Programming Languages and Operating
  Systems, Volume 3}} (La Jolla, CA, USA) \emph{(\bibinfo{series}{ASPLOS
  '24})}. \bibinfo{publisher}{Association for Computing Machinery},
  \bibinfo{address}{New York, NY, USA}, \bibinfo{pages}{932–949}.
\newblock
\showISBNx{9798400703867}
\href{https://doi.org/10.1145/3620666.3651335}{doi:\nolinkurl{10.1145/3620666.3651335}}


\bibitem[Migliavacca et~al\mbox{.}(2010)]%
        {fernandez2010seep}
\bibfield{author}{\bibinfo{person}{Matteo Migliavacca}, \bibinfo{person}{David
  Eyers}, \bibinfo{person}{Jean Bacon}, \bibinfo{person}{Yiannis Papagiannis},
  \bibinfo{person}{Brian Shand}, {and} \bibinfo{person}{Peter Pietzuch}.}
  \bibinfo{year}{2010}\natexlab{}.
\newblock \showarticletitle{SEEP: scalable and elastic event processing}. In
  \bibinfo{booktitle}{\emph{Middleware '10 Posters and Demos Track}}
  (Bangalore, India) \emph{(\bibinfo{series}{Middleware Posters '10})}.
  \bibinfo{publisher}{Association for Computing Machinery},
  \bibinfo{address}{New York, NY, USA}, Article \bibinfo{articleno}{4},
  \bibinfo{numpages}{2}~pages.
\newblock
\showISBNx{9781450306010}
\href{https://doi.org/10.1145/1930028.1930032}{doi:\nolinkurl{10.1145/1930028.1930032}}


\bibitem[Mo et~al\mbox{.}(2025)]%
        {mo2025hetis}
\bibfield{author}{\bibinfo{person}{Zizhao Mo}, \bibinfo{person}{Jianxiong
  Liao}, \bibinfo{person}{Huanle Xu}, \bibinfo{person}{Zhi Zhou}, {and}
  \bibinfo{person}{ChengZhong Xu}.} \bibinfo{year}{2025}\natexlab{}.
\newblock \showarticletitle{Hetis: Serving LLMs in Heterogeneous GPU Clusters
  with Fine-grained and Dynamic Parallelism}. In
  \bibinfo{booktitle}{\emph{Proceedings of the International Conference for
  High Performance Computing, Networking, Storage and Analysis}}
  \emph{(\bibinfo{series}{SC '25})}. \bibinfo{publisher}{Association for
  Computing Machinery}, \bibinfo{address}{New York, NY, USA},
  \bibinfo{pages}{1710–1724}.
\newblock
\showISBNx{9798400714665}
\href{https://doi.org/10.1145/3712285.3759784}{doi:\nolinkurl{10.1145/3712285.3759784}}


\bibitem[Mohan et~al\mbox{.}(2021)]%
        {mohan2021checkfreq}
\bibfield{author}{\bibinfo{person}{Jayashree Mohan}, \bibinfo{person}{Amar
  Phanishayee}, {and} \bibinfo{person}{Vijay Chidambaram}.}
  \bibinfo{year}{2021}\natexlab{}.
\newblock \showarticletitle{{CheckFreq}: Frequent, {Fine-Grained} {DNN}
  Checkpointing}. In \bibinfo{booktitle}{\emph{19th USENIX Conference on File
  and Storage Technologies (FAST 21)}}. \bibinfo{publisher}{USENIX
  Association}, \bibinfo{pages}{203--216}.
\newblock
\showISBNx{978-1-939133-20-5}
\urldef\tempurl%
\url{https://www.usenix.org/conference/fast21/presentation/mohan}
\showURL{%
\tempurl}


\bibitem[Murray et~al\mbox{.}(2013)]%
        {murray2013naiad}
\bibfield{author}{\bibinfo{person}{Derek~G. Murray}, \bibinfo{person}{Frank
  McSherry}, \bibinfo{person}{Rebecca Isaacs}, \bibinfo{person}{Michael Isard},
  \bibinfo{person}{Paul Barham}, {and} \bibinfo{person}{Mart\'{\i}n Abadi}.}
  \bibinfo{year}{2013}\natexlab{}.
\newblock \showarticletitle{Naiad: a timely dataflow system}. In
  \bibinfo{booktitle}{\emph{Proceedings of the Twenty-Fourth ACM Symposium on
  Operating Systems Principles}} (Farminton, Pennsylvania)
  \emph{(\bibinfo{series}{SOSP '13})}. \bibinfo{publisher}{Association for
  Computing Machinery}, \bibinfo{address}{New York, NY, USA},
  \bibinfo{pages}{439–455}.
\newblock
\showISBNx{9781450323888}
\href{https://doi.org/10.1145/2517349.2522738}{doi:\nolinkurl{10.1145/2517349.2522738}}


\bibitem[Murray et~al\mbox{.}(2011)]%
        {murray2011ciel}
\bibfield{author}{\bibinfo{person}{Derek~G. Murray}, \bibinfo{person}{Malte
  Schwarzkopf}, \bibinfo{person}{Christopher Smowton}, \bibinfo{person}{Steven
  Smith}, \bibinfo{person}{Anil Madhavapeddy}, {and} \bibinfo{person}{Steven
  Hand}.} \bibinfo{year}{2011}\natexlab{}.
\newblock \showarticletitle{CIEL: a universal execution engine for distributed
  data-flow computing}. In \bibinfo{booktitle}{\emph{Proceedings of the 8th
  USENIX Conference on Networked Systems Design and Implementation}} (Boston,
  MA) \emph{(\bibinfo{series}{NSDI'11})}. \bibinfo{publisher}{USENIX
  Association}, \bibinfo{address}{USA}, \bibinfo{pages}{113–126}.
\newblock


\bibitem[Na et~al\mbox{.}(2025)]%
        {na2025flexinfer}
\bibfield{author}{\bibinfo{person}{Seonjin Na}, \bibinfo{person}{Geonhwa
  Jeong}, \bibinfo{person}{Byung~Hoon Ahn}, \bibinfo{person}{Aaron Jezghani},
  \bibinfo{person}{Jeffrey Young}, \bibinfo{person}{Christopher~J. Hughes},
  \bibinfo{person}{Tushar Krishna}, {and} \bibinfo{person}{Hyesoon Kim}.}
  \bibinfo{year}{2025}\natexlab{}.
\newblock \showarticletitle{FlexInfer: Flexible LLM Inference with CPU
  Computations}. In \bibinfo{booktitle}{\emph{Proceedings of Machine Learning
  and Systems}}, \bibfield{editor}{\bibinfo{person}{M.~Zaharia},
  \bibinfo{person}{G.~Joshi}, {and} \bibinfo{person}{Y.~Lin}} (Eds.),
  Vol.~\bibinfo{volume}{7}. \bibinfo{publisher}{MLSys}.
\newblock
\urldef\tempurl%
\url{https://proceedings.mlsys.org/paper_files/paper/2025/file/698cfaf72a208aef2e78bcac55b74328-Paper-Conference.pdf}
\showURL{%
\tempurl}


\bibitem[Nayak et~al\mbox{.}(2024)]%
        {nayak2024fusemax}
\bibfield{author}{\bibinfo{person}{Nandeeka Nayak}, \bibinfo{person}{Xinrui
  Wu}, \bibinfo{person}{Toluwanimi~O. Odemuyiwa}, \bibinfo{person}{Michael
  Pellauer}, \bibinfo{person}{Joel~S. Emer}, {and}
  \bibinfo{person}{Christopher~W. Fletcher}.} \bibinfo{year}{2024}\natexlab{}.
\newblock \showarticletitle{FuseMax: Leveraging Extended Einsums to Optimize
  Attention Accelerator Design}. In \bibinfo{booktitle}{\emph{2024 57th
  IEEE/ACM International Symposium on Microarchitecture (MICRO)}}.
  \bibinfo{pages}{1458--1473}.
\newblock
\href{https://doi.org/10.1109/MICRO61859.2024.00107}{doi:\nolinkurl{10.1109/MICRO61859.2024.00107}}


\bibitem[Nian et~al\mbox{.}(2026)]%
        {nian2026cacheflow}
\bibfield{author}{\bibinfo{person}{Sean Nian}, \bibinfo{person}{Jiahao Fang},
  \bibinfo{person}{Qilong Feng}, \bibinfo{person}{Zhiyu Wu}, {and}
  \bibinfo{person}{Fan Lai}.} \bibinfo{year}{2026}\natexlab{}.
\newblock \showarticletitle{CacheFlow: Efficient LLM Serving with 3D-Parallel
  KV Cache Restoration}.
\newblock  (\bibinfo{year}{2026}).
\newblock
\href{https://doi.org/10.48550/ARXIV.2604.25080}{doi:\nolinkurl{10.48550/ARXIV.2604.25080}}


\bibitem[Oh et~al\mbox{.}(2024)]%
        {oh2024exegpt}
\bibfield{author}{\bibinfo{person}{Hyungjun Oh}, \bibinfo{person}{Kihong Kim},
  \bibinfo{person}{Jaemin Kim}, \bibinfo{person}{Sungkyun Kim},
  \bibinfo{person}{Junyeol Lee}, \bibinfo{person}{Du-seong Chang}, {and}
  \bibinfo{person}{Jiwon Seo}.} \bibinfo{year}{2024}\natexlab{}.
\newblock \showarticletitle{ExeGPT: Constraint-Aware Resource Scheduling for
  LLM Inference}. In \bibinfo{booktitle}{\emph{Proceedings of the 29th ACM
  International Conference on Architectural Support for Programming Languages
  and Operating Systems, Volume 2}} (La Jolla, CA, USA)
  \emph{(\bibinfo{series}{ASPLOS '24})}. \bibinfo{publisher}{Association for
  Computing Machinery}, \bibinfo{address}{New York, NY, USA},
  \bibinfo{pages}{369–384}.
\newblock
\showISBNx{9798400703850}
\href{https://doi.org/10.1145/3620665.3640383}{doi:\nolinkurl{10.1145/3620665.3640383}}


\bibitem[Ongaro and Ousterhout(2014)]%
        {ongaro2014raft}
\bibfield{author}{\bibinfo{person}{Diego Ongaro} {and} \bibinfo{person}{John
  Ousterhout}.} \bibinfo{year}{2014}\natexlab{}.
\newblock \showarticletitle{In search of an understandable consensus
  algorithm}. In \bibinfo{booktitle}{\emph{Proceedings of the 2014 USENIX
  Conference on USENIX Annual Technical Conference}} (Philadelphia, PA)
  \emph{(\bibinfo{series}{USENIX ATC'14})}. \bibinfo{publisher}{USENIX
  Association}, \bibinfo{address}{USA}, \bibinfo{pages}{305–320}.
\newblock
\showISBNx{9781931971102}


\bibitem[Ousterhout et~al\mbox{.}(2010)]%
        {ousterhout2011ramcloud}
\bibfield{author}{\bibinfo{person}{John Ousterhout}, \bibinfo{person}{Parag
  Agrawal}, \bibinfo{person}{David Erickson}, \bibinfo{person}{Christos
  Kozyrakis}, \bibinfo{person}{Jacob Leverich}, \bibinfo{person}{David
  Mazi\`{e}res}, \bibinfo{person}{Subhasish Mitra}, \bibinfo{person}{Aravind
  Narayanan}, \bibinfo{person}{Guru Parulkar}, \bibinfo{person}{Mendel
  Rosenblum}, \bibinfo{person}{Stephen~M. Rumble}, \bibinfo{person}{Eric
  Stratmann}, {and} \bibinfo{person}{Ryan Stutsman}.}
  \bibinfo{year}{2010}\natexlab{}.
\newblock \showarticletitle{The case for RAMClouds: scalable high-performance
  storage entirely in DRAM}.
\newblock \bibinfo{journal}{\emph{SIGOPS Oper. Syst. Rev.}}
  \bibinfo{volume}{43}, \bibinfo{number}{4} (\bibinfo{date}{Jan.}
  \bibinfo{year}{2010}), \bibinfo{pages}{92–105}.
\newblock
\showISSN{0163-5980}
\href{https://doi.org/10.1145/1713254.1713276}{doi:\nolinkurl{10.1145/1713254.1713276}}


\bibitem[Pan et~al\mbox{.}(2025d)]%
        {pan2025marconi}
\bibfield{author}{\bibinfo{person}{Rui Pan}, \bibinfo{person}{Zhuang Wang},
  \bibinfo{person}{Zhen Jia}, \bibinfo{person}{Can Karakus},
  \bibinfo{person}{Luca Zancato}, \bibinfo{person}{Tri Dao},
  \bibinfo{person}{Yida Wang}, {and} \bibinfo{person}{Ravi Netravali}.}
  \bibinfo{year}{2025}\natexlab{d}.
\newblock \showarticletitle{Marconi: Prefix Caching for the Era of Hybrid
  LLMs}. In \bibinfo{booktitle}{\emph{Proceedings of Machine Learning and
  Systems}}, \bibfield{editor}{\bibinfo{person}{M.~Zaharia},
  \bibinfo{person}{G.~Joshi}, {and} \bibinfo{person}{Y.~Lin}} (Eds.),
  Vol.~\bibinfo{volume}{7}. \bibinfo{publisher}{MLSys}.
\newblock
\urldef\tempurl%
\url{https://proceedings.mlsys.org/paper_files/paper/2025/file/7c180af017258d239bac6248d1eb26ac-Paper-Conference.pdf}
\showURL{%
\tempurl}


\bibitem[Pan et~al\mbox{.}(2025b)]%
        {pan2025instattention}
\bibfield{author}{\bibinfo{person}{Xiurui Pan}, \bibinfo{person}{Endian Li},
  \bibinfo{person}{Qiao Li}, \bibinfo{person}{Shengwen Liang},
  \bibinfo{person}{Yizhou Shan}, \bibinfo{person}{Ke Zhou},
  \bibinfo{person}{Yingwei Luo}, \bibinfo{person}{Xiaolin Wang}, {and}
  \bibinfo{person}{Jie Zhang}.} \bibinfo{year}{2025}\natexlab{b}.
\newblock \showarticletitle{InstAttention: In-Storage Attention Offloading for
  Cost-Effective Long-Context LLM Inference}. In \bibinfo{booktitle}{\emph{2025
  IEEE International Symposium on High Performance Computer Architecture
  (HPCA)}}. \bibinfo{pages}{1510--1525}.
\newblock
\href{https://doi.org/10.1109/HPCA61900.2025.00113}{doi:\nolinkurl{10.1109/HPCA61900.2025.00113}}


\bibitem[Pan et~al\mbox{.}(2025c)]%
        {fsmoe25}
\bibfield{author}{\bibinfo{person}{Xinglin Pan}, \bibinfo{person}{Wenxiang
  Lin}, \bibinfo{person}{Lin Zhang}, \bibinfo{person}{Shaohuai Shi},
  \bibinfo{person}{Zhenheng Tang}, \bibinfo{person}{Rui Wang},
  \bibinfo{person}{Bo Li}, {and} \bibinfo{person}{Xiaowen Chu}.}
  \bibinfo{year}{2025}\natexlab{c}.
\newblock \showarticletitle{FSMoE: A Flexible and Scalable Training System for
  Sparse Mixture-of-Experts Models}. In \bibinfo{booktitle}{\emph{Proceedings
  of the 30th ACM International Conference on Architectural Support for
  Programming Languages and Operating Systems, Volume 1}} (Rotterdam,
  Netherlands) \emph{(\bibinfo{series}{ASPLOS '25})}.
  \bibinfo{publisher}{Association for Computing Machinery},
  \bibinfo{address}{New York, NY, USA}, \bibinfo{pages}{524–539}.
\newblock
\showISBNx{9798400706981}
\href{https://doi.org/10.1145/3669940.3707272}{doi:\nolinkurl{10.1145/3669940.3707272}}


\bibitem[Pan et~al\mbox{.}(2025a)]%
        {pan2025fasttree}
\bibfield{author}{\bibinfo{person}{Zaifeng Pan}, \bibinfo{person}{Yitong Ding},
  \bibinfo{person}{Yue Guan}, \bibinfo{person}{Zheng Wang},
  \bibinfo{person}{Zhongkai Yu}, \bibinfo{person}{Xulong Tang},
  \bibinfo{person}{Yida Wang}, {and} \bibinfo{person}{Yufei Ding}.}
  \bibinfo{year}{2025}\natexlab{a}.
\newblock \showarticletitle{FastTree: Optimizing Attention Kernel and Runtime
  for Tree-Structured LLM Inference}. In \bibinfo{booktitle}{\emph{Proceedings
  of Machine Learning and Systems}}.
\newblock


\bibitem[Park et~al\mbox{.}(2025b)]%
        {park2025clap}
\bibfield{author}{\bibinfo{person}{Junhyeok Park}, \bibinfo{person}{Sungbin
  Jang}, \bibinfo{person}{Osang Kwon}, \bibinfo{person}{Yongho Lee}, {and}
  \bibinfo{person}{Seokin Hong}.} \bibinfo{year}{2025}\natexlab{b}.
\newblock \showarticletitle{Leveraging Chiplet-Locality for Efficient Memory
  Mapping in Multi-Chip Module GPUs}. In \bibinfo{booktitle}{\emph{Proceedings
  of the 58th IEEE/ACM International Symposium on Microarchitecture}}
  \emph{(\bibinfo{series}{MICRO '25})}. \bibinfo{publisher}{Association for
  Computing Machinery}, \bibinfo{address}{New York, NY, USA},
  \bibinfo{pages}{1040–1057}.
\newblock
\showISBNx{9798400715730}
\href{https://doi.org/10.1145/3725843.3756090}{doi:\nolinkurl{10.1145/3725843.3756090}}


\bibitem[Park et~al\mbox{.}(2025a)]%
        {park2025decdec}
\bibfield{author}{\bibinfo{person}{Yeonhong Park}, \bibinfo{person}{Jake Hyun},
  \bibinfo{person}{Hojoon Kim}, {and} \bibinfo{person}{Jae~W. Lee}.}
  \bibinfo{year}{2025}\natexlab{a}.
\newblock \showarticletitle{{DecDEC}: A Systems Approach to Advancing {Low-Bit}
  {LLM} Quantization}. In \bibinfo{booktitle}{\emph{19th USENIX Symposium on
  Operating Systems Design and Implementation (OSDI 25)}}.
  \bibinfo{publisher}{USENIX Association}, \bibinfo{address}{Boston, MA},
  \bibinfo{pages}{803--819}.
\newblock
\showISBNx{978-1-939133-47-2}
\urldef\tempurl%
\url{https://www.usenix.org/conference/osdi25/presentation/park-yeonhong}
\showURL{%
\tempurl}


\bibitem[Patel et~al\mbox{.}(2024)]%
        {patel2024splitwise}
\bibfield{author}{\bibinfo{person}{Pratyush Patel}, \bibinfo{person}{Esha
  Choukse}, \bibinfo{person}{Chaojie Zhang}, \bibinfo{person}{Aashaka Shah},
  \bibinfo{person}{Íñigo Goiri}, \bibinfo{person}{Saeed Maleki}, {and}
  \bibinfo{person}{Ricardo Bianchini}.} \bibinfo{year}{2024}\natexlab{}.
\newblock \showarticletitle{Splitwise: Efficient Generative LLM Inference Using
  Phase Splitting}. In \bibinfo{booktitle}{\emph{2024 ACM/IEEE 51st Annual
  International Symposium on Computer Architecture (ISCA)}}.
  \bibinfo{pages}{118--132}.
\newblock
\href{https://doi.org/10.1109/ISCA59077.2024.00019}{doi:\nolinkurl{10.1109/ISCA59077.2024.00019}}


\bibitem[Peng and Dabek(2010)]%
        {peng2010percolator}
\bibfield{author}{\bibinfo{person}{Daniel Peng} {and} \bibinfo{person}{Frank
  Dabek}.} \bibinfo{year}{2010}\natexlab{}.
\newblock \showarticletitle{Large-scale Incremental Processing Using
  Distributed Transactions and Notifications}. In \bibinfo{booktitle}{\emph{9th
  USENIX Symposium on Operating Systems Design and Implementation (OSDI 10)}}.
  \bibinfo{publisher}{USENIX Association}, \bibinfo{address}{Vancouver, BC}.
\newblock
\urldef\tempurl%
\url{https://www.usenix.org/conference/osdi10/large-scale-incremental-processing-using-distributed-transactions-and}
\showURL{%
\tempurl}


\bibitem[Peng et~al\mbox{.}(2024)]%
        {peng2024scalacache}
\bibfield{author}{\bibinfo{person}{Li Peng}, \bibinfo{person}{Yuda An},
  \bibinfo{person}{You Zhou}, \bibinfo{person}{Chenxi Wang},
  \bibinfo{person}{Qiao Li}, \bibinfo{person}{Chuanning Cheng}, {and}
  \bibinfo{person}{Jie Zhang}.} \bibinfo{year}{2024}\natexlab{}.
\newblock \showarticletitle{ScalaCache: scalable user-space page cache
  management with software-hardware coordination}. In
  \bibinfo{booktitle}{\emph{Proceedings of the 2024 USENIX Conference on Usenix
  Annual Technical Conference}} (Santa Clara, CA, USA)
  \emph{(\bibinfo{series}{USENIX ATC'24})}. \bibinfo{publisher}{USENIX
  Association}, \bibinfo{address}{USA}, Article \bibinfo{articleno}{72},
  \bibinfo{numpages}{18}~pages.
\newblock
\showISBNx{978-1-939133-41-0}


\bibitem[Prabhu et~al\mbox{.}(2020)]%
        {prabhu2020gdumb}
\bibfield{author}{\bibinfo{person}{Ameya Prabhu}, \bibinfo{person}{Philip H.~S.
  Torr}, {and} \bibinfo{person}{Puneet~K. Dokania}.}
  \bibinfo{year}{2020}\natexlab{}.
\newblock \showarticletitle{GDumb: A Simple Approach that Questions Our
  Progress in Continual Learning}. In \bibinfo{booktitle}{\emph{Computer Vision
  -- ECCV 2020}}, \bibfield{editor}{\bibinfo{person}{Andrea Vedaldi},
  \bibinfo{person}{Horst Bischof}, \bibinfo{person}{Thomas Brox}, {and}
  \bibinfo{person}{Jan-Michael Frahm}} (Eds.). \bibinfo{publisher}{Springer
  International Publishing}, \bibinfo{address}{Cham},
  \bibinfo{pages}{524--540}.
\newblock
\showISBNx{978-3-030-58536-5}


\bibitem[Prabhu et~al\mbox{.}(2025)]%
        {prabhu2025vattention}
\bibfield{author}{\bibinfo{person}{Ramya Prabhu}, \bibinfo{person}{Ajay Nayak},
  \bibinfo{person}{Jayashree Mohan}, \bibinfo{person}{Ramachandran Ramjee},
  {and} \bibinfo{person}{Ashish Panwar}.} \bibinfo{year}{2025}\natexlab{}.
\newblock \showarticletitle{vAttention: Dynamic Memory Management for Serving
  LLMs without PagedAttention}. In \bibinfo{booktitle}{\emph{Proceedings of the
  30th ACM International Conference on Architectural Support for Programming
  Languages and Operating Systems, Volume 1}} (Rotterdam, Netherlands)
  \emph{(\bibinfo{series}{ASPLOS '25})}. \bibinfo{publisher}{Association for
  Computing Machinery}, \bibinfo{address}{New York, NY, USA},
  \bibinfo{pages}{1133–1150}.
\newblock
\showISBNx{9798400706981}
\href{https://doi.org/10.1145/3669940.3707256}{doi:\nolinkurl{10.1145/3669940.3707256}}


\bibitem[Qin et~al\mbox{.}(2026)]%
        {qin2026prfaas}
\bibfield{author}{\bibinfo{person}{Ruoyu Qin}, \bibinfo{person}{Weiran He},
  \bibinfo{person}{Yaoyu Wang}, \bibinfo{person}{Zheming Li},
  \bibinfo{person}{Xinran Xu}, \bibinfo{person}{Yongwei Wu},
  \bibinfo{person}{Weimin Zheng}, {and} \bibinfo{person}{Mingxing Zhang}.}
  \bibinfo{year}{2026}\natexlab{}.
\newblock \showarticletitle{Prefill-as-a-Service: KVCache of Next-Generation
  Models Could Go Cross-Datacenter}.
\newblock  (\bibinfo{year}{2026}).
\newblock
\href{https://doi.org/10.48550/ARXIV.2604.15039}{doi:\nolinkurl{10.48550/ARXIV.2604.15039}}


\bibitem[Qin et~al\mbox{.}(2025)]%
        {qin2025mooncake}
\bibfield{author}{\bibinfo{person}{Ruoyu Qin}, \bibinfo{person}{Zheming Li},
  \bibinfo{person}{Weiran He}, \bibinfo{person}{Jialei Cui},
  \bibinfo{person}{Heyi Tang}, \bibinfo{person}{Feng Ren},
  \bibinfo{person}{Teng Ma}, \bibinfo{person}{Shangming Cai},
  \bibinfo{person}{Yineng Zhang}, \bibinfo{person}{Mingxing Zhang},
  \bibinfo{person}{Yongwei Wu}, \bibinfo{person}{Weimin Zheng}, {and}
  \bibinfo{person}{Xinran Xu}.} \bibinfo{year}{2025}\natexlab{}.
\newblock \showarticletitle{Mooncake: A KVCache-centric Disaggregated
  Architecture for LLM Serving}.
\newblock \bibinfo{journal}{\emph{ACM Trans. Storage}} (\bibinfo{date}{Nov.}
  \bibinfo{year}{2025}).
\newblock
\showISSN{1553-3077}
\href{https://doi.org/10.1145/3773772}{doi:\nolinkurl{10.1145/3773772}}
\newblock
\shownote{Just Accepted}.


\bibitem[Qiu et~al\mbox{.}(2025)]%
        {qiu2025hotrap}
\bibfield{author}{\bibinfo{person}{Jiansheng Qiu}, \bibinfo{person}{Fangzhou
  Yuan}, \bibinfo{person}{Mingyu Gao}, {and} \bibinfo{person}{Huanchen Zhang}.}
  \bibinfo{year}{2025}\natexlab{}.
\newblock \showarticletitle{HotRAP: hot record retention and promotion for
  LSM-trees with tiered storage}. In \bibinfo{booktitle}{\emph{Proceedings of
  the 2025 USENIX Conference on Usenix Annual Technical Conference}} (Boston,
  MA, USA) \emph{(\bibinfo{series}{USENIX ATC '25})}.
  \bibinfo{publisher}{USENIX Association}, \bibinfo{address}{USA}, Article
  \bibinfo{articleno}{30}, \bibinfo{numpages}{15}~pages.
\newblock
\showISBNx{978-1-939133-48-9}


\bibitem[Quinn et~al\mbox{.}(2025)]%
        {quinn2025accelerating}
\bibfield{author}{\bibinfo{person}{Derrick Quinn}, \bibinfo{person}{Mohammad
  Nouri}, \bibinfo{person}{Neel Patel}, \bibinfo{person}{John Salihu},
  \bibinfo{person}{Alireza Salemi}, \bibinfo{person}{Sukhan Lee},
  \bibinfo{person}{Hamed Zamani}, {and} \bibinfo{person}{Mohammad Alian}.}
  \bibinfo{year}{2025}\natexlab{}.
\newblock \showarticletitle{Accelerating Retrieval-Augmented Generation}. In
  \bibinfo{booktitle}{\emph{Proceedings of the 30th ACM International
  Conference on Architectural Support for Programming Languages and Operating
  Systems, Volume 1}} (Rotterdam, Netherlands) \emph{(\bibinfo{series}{ASPLOS
  '25})}. \bibinfo{publisher}{Association for Computing Machinery},
  \bibinfo{address}{New York, NY, USA}, \bibinfo{pages}{15–32}.
\newblock
\showISBNx{9798400706981}
\href{https://doi.org/10.1145/3669940.3707264}{doi:\nolinkurl{10.1145/3669940.3707264}}


\bibitem[Quoc et~al\mbox{.}(2018)]%
        {wang2018approxjoin}
\bibfield{author}{\bibinfo{person}{Do~Le Quoc}, \bibinfo{person}{Istemi~Ekin
  Akkus}, \bibinfo{person}{Pramod Bhatotia}, \bibinfo{person}{Spyros Blanas},
  \bibinfo{person}{Ruichuan Chen}, \bibinfo{person}{Christof Fetzer}, {and}
  \bibinfo{person}{Thorsten Strufe}.} \bibinfo{year}{2018}\natexlab{}.
\newblock \showarticletitle{ApproxJoin: Approximate Distributed Joins}. In
  \bibinfo{booktitle}{\emph{Proceedings of the ACM Symposium on Cloud
  Computing}} (Carlsbad, CA, USA) \emph{(\bibinfo{series}{SoCC '18})}.
  \bibinfo{publisher}{Association for Computing Machinery},
  \bibinfo{address}{New York, NY, USA}, \bibinfo{pages}{426–438}.
\newblock
\showISBNx{9781450360111}
\href{https://doi.org/10.1145/3267809.3267834}{doi:\nolinkurl{10.1145/3267809.3267834}}


\bibitem[Quoc et~al\mbox{.}(2017)]%
        {hu2018streamapprox}
\bibfield{author}{\bibinfo{person}{Do~Le Quoc}, \bibinfo{person}{Ruichuan
  Chen}, \bibinfo{person}{Pramod Bhatotia}, \bibinfo{person}{Christof Fetzer},
  \bibinfo{person}{Volker Hilt}, {and} \bibinfo{person}{Thorsten Strufe}.}
  \bibinfo{year}{2017}\natexlab{}.
\newblock \showarticletitle{StreamApprox: approximate computing for stream
  analytics}. In \bibinfo{booktitle}{\emph{Proceedings of the 18th
  ACM/IFIP/USENIX Middleware Conference}} (Las Vegas, Nevada)
  \emph{(\bibinfo{series}{Middleware '17})}. \bibinfo{publisher}{Association
  for Computing Machinery}, \bibinfo{address}{New York, NY, USA},
  \bibinfo{pages}{185–197}.
\newblock
\showISBNx{9781450347204}
\href{https://doi.org/10.1145/3135974.3135989}{doi:\nolinkurl{10.1145/3135974.3135989}}


\bibitem[Raybuck et~al\mbox{.}(2021)]%
        {raybuck2021hemem}
\bibfield{author}{\bibinfo{person}{Amanda Raybuck}, \bibinfo{person}{Tim
  Stamler}, \bibinfo{person}{Wei Zhang}, \bibinfo{person}{Mattan Erez}, {and}
  \bibinfo{person}{Simon Peter}.} \bibinfo{year}{2021}\natexlab{}.
\newblock \showarticletitle{HeMem: Scalable Tiered Memory Management for Big
  Data Applications and Real NVM}. In \bibinfo{booktitle}{\emph{Proceedings of
  the ACM SIGOPS 28th Symposium on Operating Systems Principles}} (Virtual
  Event, Germany) \emph{(\bibinfo{series}{SOSP '21})}.
  \bibinfo{publisher}{Association for Computing Machinery},
  \bibinfo{address}{New York, NY, USA}, \bibinfo{pages}{392–407}.
\newblock
\showISBNx{9781450387095}
\href{https://doi.org/10.1145/3477132.3483550}{doi:\nolinkurl{10.1145/3477132.3483550}}


\bibitem[Rebuffi et~al\mbox{.}(2017)]%
        {rebuffi2017icarl}
\bibfield{author}{\bibinfo{person}{Sylvestre-Alvise Rebuffi},
  \bibinfo{person}{Alexander Kolesnikov}, \bibinfo{person}{Georg Sperl}, {and}
  \bibinfo{person}{Christoph~H. Lampert}.} \bibinfo{year}{2017}\natexlab{}.
\newblock \showarticletitle{iCaRL: Incremental Classifier and Representation
  Learning}. In \bibinfo{booktitle}{\emph{2017 IEEE Conference on Computer
  Vision and Pattern Recognition (CVPR)}}. \bibinfo{publisher}{IEEE},
  \bibinfo{pages}{5533–5542}.
\newblock
\href{https://doi.org/10.1109/cvpr.2017.587}{doi:\nolinkurl{10.1109/cvpr.2017.587}}


\bibitem[Ren et~al\mbox{.}(2020)]%
        {ren2020hmg}
\bibfield{author}{\bibinfo{person}{Xiaowei Ren}, \bibinfo{person}{Daniel
  Lustig}, \bibinfo{person}{Evgeny Bolotin}, \bibinfo{person}{Aamer Jaleel},
  \bibinfo{person}{Oreste Villa}, {and} \bibinfo{person}{David Nellans}.}
  \bibinfo{year}{2020}\natexlab{}.
\newblock \showarticletitle{HMG: Extending Cache Coherence Protocols Across
  Modern Hierarchical Multi-GPU Systems}. In \bibinfo{booktitle}{\emph{2020
  IEEE International Symposium on High Performance Computer Architecture
  (HPCA)}}. \bibinfo{pages}{582--595}.
\newblock
\href{https://doi.org/10.1109/HPCA47549.2020.00054}{doi:\nolinkurl{10.1109/HPCA47549.2020.00054}}


\bibitem[Rolnick et~al\mbox{.}(2019)]%
        {rolnick2019experience}
\bibfield{author}{\bibinfo{person}{David Rolnick}, \bibinfo{person}{Arun
  Ahuja}, \bibinfo{person}{Jonathan Schwarz}, \bibinfo{person}{Timothy
  Lillicrap}, {and} \bibinfo{person}{Gregory Wayne}.}
  \bibinfo{year}{2019}\natexlab{}.
\newblock \showarticletitle{Experience Replay for Continual Learning}. In
  \bibinfo{booktitle}{\emph{Advances in Neural Information Processing
  Systems}}, \bibfield{editor}{\bibinfo{person}{H.~Wallach},
  \bibinfo{person}{H.~Larochelle}, \bibinfo{person}{A.~Beygelzimer},
  \bibinfo{person}{F.~d\textquotesingle Alch\'{e}-Buc},
  \bibinfo{person}{E.~Fox}, {and} \bibinfo{person}{R.~Garnett}} (Eds.),
  Vol.~\bibinfo{volume}{32}. \bibinfo{publisher}{Curran Associates, Inc.}
\newblock
\urldef\tempurl%
\url{https://proceedings.neurips.cc/paper_files/paper/2019/file/fa7cdfad1a5aaf8370ebeda47a1ff1c3-Paper.pdf}
\showURL{%
\tempurl}


\bibitem[Romero et~al\mbox{.}(2021)]%
        {romero2021infaas}
\bibfield{author}{\bibinfo{person}{Francisco Romero}, \bibinfo{person}{Qian
  Li}, \bibinfo{person}{Neeraja~J. Yadwadkar}, {and} \bibinfo{person}{Christos
  Kozyrakis}.} \bibinfo{year}{2021}\natexlab{}.
\newblock \showarticletitle{{INFaaS}: Automated Model-less Inference Serving}.
  In \bibinfo{booktitle}{\emph{2021 USENIX Annual Technical Conference (USENIX
  ATC 21)}}. \bibinfo{publisher}{USENIX Association},
  \bibinfo{pages}{397--411}.
\newblock
\showISBNx{978-1-939133-23-6}
\urldef\tempurl%
\url{https://www.usenix.org/conference/atc21/presentation/romero}
\showURL{%
\tempurl}


\bibitem[Ruan et~al\mbox{.}(2026)]%
        {ruan2026libra}
\bibfield{author}{\bibinfo{person}{Chaoyi Ruan}, \bibinfo{person}{Yinhe Chen},
  \bibinfo{person}{Dongqi Tian}, \bibinfo{person}{Yandong Shi},
  \bibinfo{person}{Yongji Wu}, \bibinfo{person}{Jialin Li}, {and}
  \bibinfo{person}{Cheng Li}.} \bibinfo{year}{2026}\natexlab{}.
\newblock \showarticletitle{Libra: Flexible Request Partitioning and Scheduling
  for Serving Unbalanced and Dynamic {LLM} Workloads}. In
  \bibinfo{booktitle}{\emph{23rd USENIX Symposium on Networked Systems Design
  and Implementation (NSDI 26)}}. \bibinfo{publisher}{USENIX Association},
  \bibinfo{address}{Renton, WA}, \bibinfo{pages}{1243--1258}.
\newblock
\showISBNx{978-1-939133-54-0}
\urldef\tempurl%
\url{https://www.usenix.org/conference/nsdi26/presentation/ruan-libra}
\showURL{%
\tempurl}


\bibitem[Sanovar et~al\mbox{.}(2025)]%
        {sanovar2024leanattention}
\bibfield{author}{\bibinfo{person}{Rya Sanovar}, \bibinfo{person}{Srikant
  Bharadwaj}, \bibinfo{person}{Ren\'{e}e St.~Amant}, \bibinfo{person}{Victor
  R\"{u}hle}, {and} \bibinfo{person}{Saravan Rajmohan}.}
  \bibinfo{year}{2025}\natexlab{}.
\newblock \showarticletitle{LeanAttention: Hardware-Aware Scalable Attention
  Mechanism for the Decode-Phase of Transformers}. In
  \bibinfo{booktitle}{\emph{Proceedings of Machine Learning and Systems}},
  \bibfield{editor}{\bibinfo{person}{M.~Zaharia}, \bibinfo{person}{G.~Joshi},
  {and} \bibinfo{person}{Y.~Lin}} (Eds.), Vol.~\bibinfo{volume}{7}.
  \bibinfo{publisher}{MLSys}.
\newblock
\urldef\tempurl%
\url{https://proceedings.mlsys.org/paper_files/paper/2025/file/16ec6494e9b5a4138de7238761d715b4-Paper-Conference.pdf}
\showURL{%
\tempurl}


\bibitem[Sarthi et~al\mbox{.}(2024)]%
        {sarthi2024raptor}
\bibfield{author}{\bibinfo{person}{Parth Sarthi}, \bibinfo{person}{Salman
  Abdullah}, \bibinfo{person}{Aditi Tuli}, \bibinfo{person}{Shubh Khanna},
  \bibinfo{person}{Anna Goldie}, {and} \bibinfo{person}{Christopher Manning}.}
  \bibinfo{year}{2024}\natexlab{}.
\newblock \showarticletitle{RAPTOR: Recursive Abstractive Processing for
  Tree-Organized Retrieval}. In \bibinfo{booktitle}{\emph{International
  Conference on Learning Representations}},
  \bibfield{editor}{\bibinfo{person}{B.~Kim}, \bibinfo{person}{Y.~Yue},
  \bibinfo{person}{S.~Chaudhuri}, \bibinfo{person}{K.~Fragkiadaki},
  \bibinfo{person}{M.~Khan}, {and} \bibinfo{person}{Y.~Sun}} (Eds.),
  Vol.~\bibinfo{volume}{2024}. \bibinfo{pages}{32628--32649}.
\newblock
\urldef\tempurl%
\url{https://proceedings.iclr.cc/paper_files/paper/2024/file/8a2acd174940dbca361a6398a4f9df91-Paper-Conference.pdf}
\showURL{%
\tempurl}


\bibitem[Schneider(1990)]%
        {schneider1990state}
\bibfield{author}{\bibinfo{person}{Fred~B. Schneider}.}
  \bibinfo{year}{1990}\natexlab{}.
\newblock \showarticletitle{Implementing fault-tolerant services using the
  state machine approach: a tutorial}.
\newblock \bibinfo{journal}{\emph{ACM Comput. Surv.}} \bibinfo{volume}{22},
  \bibinfo{number}{4} (\bibinfo{date}{Dec.} \bibinfo{year}{1990}),
  \bibinfo{pages}{299–319}.
\newblock
\showISSN{0360-0300}
\href{https://doi.org/10.1145/98163.98167}{doi:\nolinkurl{10.1145/98163.98167}}


\bibitem[Shapiro et~al\mbox{.}(2011)]%
        {shapiro2011crdt}
\bibfield{author}{\bibinfo{person}{Marc Shapiro}, \bibinfo{person}{Nuno
  Pregui{\c{c}}a}, \bibinfo{person}{Carlos Baquero}, {and}
  \bibinfo{person}{Marek Zawirski}.} \bibinfo{year}{2011}\natexlab{}.
\newblock \showarticletitle{Conflict-Free Replicated Data Types}. In
  \bibinfo{booktitle}{\emph{Stabilization, Safety, and Security of Distributed
  Systems}}, \bibfield{editor}{\bibinfo{person}{Xavier D{\'e}fago},
  \bibinfo{person}{Franck Petit}, {and} \bibinfo{person}{Vincent Villain}}
  (Eds.). \bibinfo{publisher}{Springer Berlin Heidelberg},
  \bibinfo{address}{Berlin, Heidelberg}, \bibinfo{pages}{386--400}.
\newblock
\showISBNx{978-3-642-24550-3}


\bibitem[Shen et~al\mbox{.}(2019)]%
        {shen2019nexus}
\bibfield{author}{\bibinfo{person}{Haichen Shen}, \bibinfo{person}{Lequn Chen},
  \bibinfo{person}{Yuchen Jin}, \bibinfo{person}{Liangyu Zhao},
  \bibinfo{person}{Bingyu Kong}, \bibinfo{person}{Matthai Philipose},
  \bibinfo{person}{Arvind Krishnamurthy}, {and} \bibinfo{person}{Ravi
  Sundaram}.} \bibinfo{year}{2019}\natexlab{}.
\newblock \showarticletitle{Nexus: a GPU cluster engine for accelerating
  DNN-based video analysis}. In \bibinfo{booktitle}{\emph{Proceedings of the
  27th ACM Symposium on Operating Systems Principles}} (Huntsville, Ontario,
  Canada) \emph{(\bibinfo{series}{SOSP '19})}. \bibinfo{publisher}{Association
  for Computing Machinery}, \bibinfo{address}{New York, NY, USA},
  \bibinfo{pages}{322–337}.
\newblock
\showISBNx{9781450368735}
\href{https://doi.org/10.1145/3341301.3359658}{doi:\nolinkurl{10.1145/3341301.3359658}}


\bibitem[Sheng et~al\mbox{.}(2023a)]%
        {sheng2023slora}
\bibfield{author}{\bibinfo{person}{Ying Sheng}, \bibinfo{person}{Shiyi Cao},
  \bibinfo{person}{Dacheng Li}, \bibinfo{person}{Coleman Hooper},
  \bibinfo{person}{Nicholas Lee}, \bibinfo{person}{Shuo Yang},
  \bibinfo{person}{Christopher Chou}, \bibinfo{person}{Banghua Zhu},
  \bibinfo{person}{Lianmin Zheng}, \bibinfo{person}{Kurt Keutzer},
  \bibinfo{person}{Joseph~E. Gonzalez}, {and} \bibinfo{person}{Ion Stoica}.}
  \bibinfo{year}{2023}\natexlab{a}.
\newblock \showarticletitle{S-LoRA: Serving Thousands of Concurrent LoRA
  Adapters}.
\newblock  (\bibinfo{year}{2023}).
\newblock
\href{https://doi.org/10.48550/ARXIV.2311.03285}{doi:\nolinkurl{10.48550/ARXIV.2311.03285}}


\bibitem[Sheng et~al\mbox{.}(2024)]%
        {sheng2024fairnessserving}
\bibfield{author}{\bibinfo{person}{Ying Sheng}, \bibinfo{person}{Shiyi Cao},
  \bibinfo{person}{Dacheng Li}, \bibinfo{person}{Banghua Zhu},
  \bibinfo{person}{Zhuohan Li}, \bibinfo{person}{Danyang Zhuo},
  \bibinfo{person}{Joseph~E. Gonzalez}, {and} \bibinfo{person}{Ion Stoica}.}
  \bibinfo{year}{2024}\natexlab{}.
\newblock \showarticletitle{Fairness in serving large language models}. In
  \bibinfo{booktitle}{\emph{Proceedings of the 18th USENIX Conference on
  Operating Systems Design and Implementation}} (Santa Clara, CA, USA)
  \emph{(\bibinfo{series}{OSDI'24})}. \bibinfo{publisher}{USENIX Association},
  \bibinfo{address}{USA}, Article \bibinfo{articleno}{52},
  \bibinfo{numpages}{24}~pages.
\newblock
\showISBNx{978-1-939133-40-3}


\bibitem[Sheng et~al\mbox{.}(2023b)]%
        {sheng2023flexgen}
\bibfield{author}{\bibinfo{person}{Ying Sheng}, \bibinfo{person}{Lianmin
  Zheng}, \bibinfo{person}{Binhang Yuan}, \bibinfo{person}{Zhuohan Li},
  \bibinfo{person}{Max Ryabinin}, \bibinfo{person}{Beidi Chen},
  \bibinfo{person}{Percy Liang}, \bibinfo{person}{Christopher R\'{e}},
  \bibinfo{person}{Ion Stoica}, {and} \bibinfo{person}{Ce Zhang}.}
  \bibinfo{year}{2023}\natexlab{b}.
\newblock \showarticletitle{FlexGen: high-throughput generative inference of
  large language models with a single GPU}. In
  \bibinfo{booktitle}{\emph{Proceedings of the 40th International Conference on
  Machine Learning}} (Honolulu, Hawaii, USA)
  \emph{(\bibinfo{series}{ICML'23})}. \bibinfo{publisher}{JMLR.org}, Article
  \bibinfo{articleno}{1288}, \bibinfo{numpages}{23}~pages.
\newblock


\bibitem[Singh et~al\mbox{.}(2021)]%
        {singh2021freshdiskann}
\bibfield{author}{\bibinfo{person}{Aditi Singh}, \bibinfo{person}{Suhas~Jayaram
  Subramanya}, \bibinfo{person}{Ravishankar Krishnaswamy}, {and}
  \bibinfo{person}{Harsha~Vardhan Simhadri}.} \bibinfo{year}{2021}\natexlab{}.
\newblock \showarticletitle{FreshDiskANN: A Fast and Accurate Graph-Based ANN
  Index for Streaming Similarity Search}.
\newblock \bibinfo{journal}{\emph{arXiv preprint arXiv:2105.09613}}
  (\bibinfo{year}{2021}).
\newblock
\href{https://doi.org/10.48550/arXiv.2105.09613}{doi:\nolinkurl{10.48550/arXiv.2105.09613}}


\bibitem[Song et~al\mbox{.}(2024)]%
        {song2024powerinfer}
\bibfield{author}{\bibinfo{person}{Yixin Song}, \bibinfo{person}{Zeyu Mi},
  \bibinfo{person}{Haotong Xie}, {and} \bibinfo{person}{Haibo Chen}.}
  \bibinfo{year}{2024}\natexlab{}.
\newblock \showarticletitle{PowerInfer: Fast Large Language Model Serving with
  a Consumer-grade GPU}. In \bibinfo{booktitle}{\emph{Proceedings of the ACM
  SIGOPS 30th Symposium on Operating Systems Principles}} (Austin, TX, USA)
  \emph{(\bibinfo{series}{SOSP '24})}. \bibinfo{publisher}{Association for
  Computing Machinery}, \bibinfo{address}{New York, NY, USA},
  \bibinfo{pages}{590–606}.
\newblock
\showISBNx{9798400712517}
\href{https://doi.org/10.1145/3694715.3695964}{doi:\nolinkurl{10.1145/3694715.3695964}}


\bibitem[Su et~al\mbox{.}(2025)]%
        {su2025seesaw}
\bibfield{author}{\bibinfo{person}{Qidong Su}, \bibinfo{person}{Wei Zhao},
  \bibinfo{person}{Xin Li}, \bibinfo{person}{Muralidhar Andoorveedu},
  \bibinfo{person}{Chenhao Jiang}, \bibinfo{person}{Zhanda Zhu},
  \bibinfo{person}{Kevin Song}, \bibinfo{person}{Christina Giannoula}, {and}
  \bibinfo{person}{Gennady Pekhimenko}.} \bibinfo{year}{2025}\natexlab{}.
\newblock \showarticletitle{Seesaw: High-throughput LLM Inference via Model
  Re-sharding}. In \bibinfo{booktitle}{\emph{Proceedings of Machine Learning
  and Systems}}, \bibfield{editor}{\bibinfo{person}{M.~Zaharia},
  \bibinfo{person}{G.~Joshi}, {and} \bibinfo{person}{Y.~Lin}} (Eds.),
  Vol.~\bibinfo{volume}{7}. \bibinfo{publisher}{MLSys}.
\newblock
\urldef\tempurl%
\url{https://proceedings.mlsys.org/paper_files/paper/2025/file/cbc4ab80cd77aa0eb87da062fbcddb46-Paper-Conference.pdf}
\showURL{%
\tempurl}


\bibitem[Sun et~al\mbox{.}(2024)]%
        {sun2024llumnix}
\bibfield{author}{\bibinfo{person}{Biao Sun}, \bibinfo{person}{Ziming Huang},
  \bibinfo{person}{Hanyu Zhao}, \bibinfo{person}{Wencong Xiao},
  \bibinfo{person}{Xinyi Zhang}, \bibinfo{person}{Yong Li}, {and}
  \bibinfo{person}{Wei Lin}.} \bibinfo{year}{2024}\natexlab{}.
\newblock \showarticletitle{Llumnix: dynamic scheduling for large language
  model serving}. In \bibinfo{booktitle}{\emph{Proceedings of the 18th USENIX
  Conference on Operating Systems Design and Implementation}} (Santa Clara, CA,
  USA) \emph{(\bibinfo{series}{OSDI'24})}. \bibinfo{publisher}{USENIX
  Association}, \bibinfo{address}{USA}, Article \bibinfo{articleno}{10},
  \bibinfo{numpages}{19}~pages.
\newblock
\showISBNx{978-1-939133-40-3}


\bibitem[Sun et~al\mbox{.}(2025)]%
        {sun2025mepipe}
\bibfield{author}{\bibinfo{person}{Zhenbo Sun}, \bibinfo{person}{Shengqi Chen},
  \bibinfo{person}{Yuanwei Wang}, \bibinfo{person}{Jian Sha},
  \bibinfo{person}{Guanyu Feng}, {and} \bibinfo{person}{Wenguang Chen}.}
  \bibinfo{year}{2025}\natexlab{}.
\newblock \showarticletitle{MEPipe: Democratizing LLM Training with
  Memory-Efficient Slice-Level Pipeline Scheduling on Cost-Effective
  Accelerators}. In \bibinfo{booktitle}{\emph{Proceedings of the Twentieth
  European Conference on Computer Systems}} (Rotterdam, Netherlands)
  \emph{(\bibinfo{series}{EuroSys '25})}. \bibinfo{publisher}{Association for
  Computing Machinery}, \bibinfo{address}{New York, NY, USA},
  \bibinfo{pages}{1263–1278}.
\newblock
\showISBNx{9798400711961}
\href{https://doi.org/10.1145/3689031.3717469}{doi:\nolinkurl{10.1145/3689031.3717469}}


\bibitem[Tabatabai et~al\mbox{.}(2024)]%
        {tabatabai2024fbmm}
\bibfield{author}{\bibinfo{person}{Bijan Tabatabai}, \bibinfo{person}{James
  Sorenson}, {and} \bibinfo{person}{Michael~M. Swift}.}
  \bibinfo{year}{2024}\natexlab{}.
\newblock \showarticletitle{FBMM: making memory management extensible with
  filesystems}. In \bibinfo{booktitle}{\emph{Proceedings of the 2024 USENIX
  Conference on Usenix Annual Technical Conference}} (Santa Clara, CA, USA)
  \emph{(\bibinfo{series}{USENIX ATC'24})}. \bibinfo{publisher}{USENIX
  Association}, \bibinfo{address}{USA}, Article \bibinfo{articleno}{48},
  \bibinfo{numpages}{14}~pages.
\newblock
\showISBNx{978-1-939133-41-0}


\bibitem[Tang et~al\mbox{.}(2016)]%
        {zhang2016elastic}
\bibfield{author}{\bibinfo{person}{Shanjiang Tang}, \bibinfo{person}{BingSheng
  He}, \bibinfo{person}{Shuhao Zhang}, {and} \bibinfo{person}{Zhaojie Niu}.}
  \bibinfo{year}{2016}\natexlab{}.
\newblock \showarticletitle{Elastic Multi-resource Fairness: Balancing Fairness
  and Efficiency in Coupled CPU-GPU Architectures}. In
  \bibinfo{booktitle}{\emph{SC '16: Proceedings of the International Conference
  for High Performance Computing, Networking, Storage and Analysis}}.
  \bibinfo{pages}{875--886}.
\newblock
\href{https://doi.org/10.1109/SC.2016.74}{doi:\nolinkurl{10.1109/SC.2016.74}}


\bibitem[Tang et~al\mbox{.}(2024)]%
        {tang2025freesam}
\bibfield{author}{\bibinfo{person}{Xilin Tang}, \bibinfo{person}{Feng Zhang},
  \bibinfo{person}{Shuhao Zhang}, \bibinfo{person}{Yani Liu},
  \bibinfo{person}{Bingsheng He}, {and} \bibinfo{person}{Xiaoyong Du}.}
  \bibinfo{year}{2024}\natexlab{}.
\newblock \showarticletitle{Enabling Adaptive Sampling for Intra-Window Join:
  Simultaneously Optimizing Quantity and Quality}.
\newblock \bibinfo{journal}{\emph{Proc. ACM Manag. Data}} \bibinfo{volume}{2},
  \bibinfo{number}{4}, Article \bibinfo{articleno}{198} (\bibinfo{date}{Sept.}
  \bibinfo{year}{2024}), \bibinfo{numpages}{31}~pages.
\newblock
\href{https://doi.org/10.1145/3677134}{doi:\nolinkurl{10.1145/3677134}}


\bibitem[Thomson et~al\mbox{.}(2012)]%
        {thomson2012calvin}
\bibfield{author}{\bibinfo{person}{Alexander Thomson},
  \bibinfo{person}{Thaddeus Diamond}, \bibinfo{person}{Shu-Chun Weng},
  \bibinfo{person}{Kun Ren}, \bibinfo{person}{Philip Shao}, {and}
  \bibinfo{person}{Daniel~J. Abadi}.} \bibinfo{year}{2012}\natexlab{}.
\newblock \showarticletitle{Calvin: fast distributed transactions for
  partitioned database systems}. In \bibinfo{booktitle}{\emph{Proceedings of
  the 2012 ACM SIGMOD International Conference on Management of Data}}
  (Scottsdale, Arizona, USA) \emph{(\bibinfo{series}{SIGMOD '12})}.
  \bibinfo{publisher}{Association for Computing Machinery},
  \bibinfo{address}{New York, NY, USA}, \bibinfo{pages}{1–12}.
\newblock
\showISBNx{9781450312479}
\href{https://doi.org/10.1145/2213836.2213838}{doi:\nolinkurl{10.1145/2213836.2213838}}


\bibitem[Tian et~al\mbox{.}(2025)]%
        {tian2025clone}
\bibfield{author}{\bibinfo{person}{Chunlin Tian}, \bibinfo{person}{Xinpeng
  Qin}, \bibinfo{person}{Kahou Tam}, \bibinfo{person}{Li Li},
  \bibinfo{person}{Zijian Wang}, \bibinfo{person}{Yuanzhe Zhao},
  \bibinfo{person}{Minglei Zhang}, {and} \bibinfo{person}{Chengzhong Xu}.}
  \bibinfo{year}{2025}\natexlab{}.
\newblock \showarticletitle{CLONE: customizing LLMs for efficient latency-aware
  inference at the edge}. In \bibinfo{booktitle}{\emph{Proceedings of the 2025
  USENIX Conference on Usenix Annual Technical Conference}} (Boston, MA, USA)
  \emph{(\bibinfo{series}{USENIX ATC '25})}. \bibinfo{publisher}{USENIX
  Association}, \bibinfo{address}{USA}, Article \bibinfo{articleno}{34},
  \bibinfo{numpages}{23}~pages.
\newblock
\showISBNx{978-1-939133-48-9}


\bibitem[Vijaya~Kumar et~al\mbox{.}(2025)]%
        {kumar2025aqa}
\bibfield{author}{\bibinfo{person}{Abhishek Vijaya~Kumar},
  \bibinfo{person}{Gianni Antichi}, {and} \bibinfo{person}{Rachee Singh}.}
  \bibinfo{year}{2025}\natexlab{}.
\newblock \showarticletitle{Aqua: Network-Accelerated Memory Offloading for
  LLMs in Scale-Up GPU Domains}. In \bibinfo{booktitle}{\emph{Proceedings of
  the 30th ACM International Conference on Architectural Support for
  Programming Languages and Operating Systems, Volume 2}} (Rotterdam,
  Netherlands) \emph{(\bibinfo{series}{ASPLOS '25})}.
  \bibinfo{publisher}{Association for Computing Machinery},
  \bibinfo{address}{New York, NY, USA}, \bibinfo{pages}{48–62}.
\newblock
\showISBNx{9798400710797}
\href{https://doi.org/10.1145/3676641.3715983}{doi:\nolinkurl{10.1145/3676641.3715983}}


\bibitem[Vuppalapati and Agarwal(2024)]%
        {vuppalapati2024colloid}
\bibfield{author}{\bibinfo{person}{Midhul Vuppalapati} {and}
  \bibinfo{person}{Rachit Agarwal}.} \bibinfo{year}{2024}\natexlab{}.
\newblock \showarticletitle{Tiered Memory Management: Access Latency is the
  Key!}. In \bibinfo{booktitle}{\emph{Proceedings of the ACM SIGOPS 30th
  Symposium on Operating Systems Principles}} (Austin, TX, USA)
  \emph{(\bibinfo{series}{SOSP '24})}. \bibinfo{publisher}{Association for
  Computing Machinery}, \bibinfo{address}{New York, NY, USA},
  \bibinfo{pages}{79–94}.
\newblock
\showISBNx{9798400712517}
\href{https://doi.org/10.1145/3694715.3695968}{doi:\nolinkurl{10.1145/3694715.3695968}}


\bibitem[Vuppalapati et~al\mbox{.}(2020)]%
        {vuppalapati2020snowflake}
\bibfield{author}{\bibinfo{person}{Midhul Vuppalapati}, \bibinfo{person}{Justin
  Miron}, \bibinfo{person}{Rachit Agarwal}, \bibinfo{person}{Dan Truong},
  \bibinfo{person}{Ashish Motivala}, {and} \bibinfo{person}{Thierry Cruanes}.}
  \bibinfo{year}{2020}\natexlab{}.
\newblock \showarticletitle{Building an elastic query engine on disaggregated
  storage}. In \bibinfo{booktitle}{\emph{Proceedings of the 17th Usenix
  Conference on Networked Systems Design and Implementation}} (Santa Clara, CA,
  USA) \emph{(\bibinfo{series}{NSDI'20})}. \bibinfo{publisher}{USENIX
  Association}, \bibinfo{address}{USA}, \bibinfo{pages}{449–462}.
\newblock
\showISBNx{9781939133137}


\bibitem[Wan et~al\mbox{.}(2025)]%
        {wan2025bytecheckpoint}
\bibfield{author}{\bibinfo{person}{Borui Wan}, \bibinfo{person}{Mingji Han},
  \bibinfo{person}{Yiyao Sheng}, \bibinfo{person}{Yanghua Peng},
  \bibinfo{person}{Haibin Lin}, \bibinfo{person}{Mofan Zhang},
  \bibinfo{person}{Zhichao Lai}, \bibinfo{person}{Menghan Yu},
  \bibinfo{person}{Junda Zhang}, \bibinfo{person}{Zuquan Song},
  \bibinfo{person}{Xin Liu}, {and} \bibinfo{person}{Chuan Wu}.}
  \bibinfo{year}{2025}\natexlab{}.
\newblock \showarticletitle{{ByteCheckpoint}: A Unified Checkpointing System
  for Large Foundation Model Development}. In \bibinfo{booktitle}{\emph{22nd
  USENIX Symposium on Networked Systems Design and Implementation (NSDI 25)}}.
  \bibinfo{publisher}{USENIX Association}, \bibinfo{address}{Philadelphia, PA},
  \bibinfo{pages}{559--578}.
\newblock
\showISBNx{978-1-939133-46-5}
\urldef\tempurl%
\url{https://www.usenix.org/conference/nsdi25/presentation/wan-borui}
\showURL{%
\tempurl}


\bibitem[Wang et~al\mbox{.}(2025b)]%
        {wang2025kvcachewild}
\bibfield{author}{\bibinfo{person}{Jiahao Wang}, \bibinfo{person}{Jinbo Han},
  \bibinfo{person}{Xingda Wei}, \bibinfo{person}{Sijie Shen},
  \bibinfo{person}{Dingyan Zhang}, \bibinfo{person}{Chenguang Fang},
  \bibinfo{person}{Rong Chen}, \bibinfo{person}{Wenyuan Yu}, {and}
  \bibinfo{person}{Haibo Chen}.} \bibinfo{year}{2025}\natexlab{b}.
\newblock \showarticletitle{KVCache cache in the wild: characterizing and
  optimizing KVCache cache at a large cloud provider}. In
  \bibinfo{booktitle}{\emph{Proceedings of the 2025 USENIX Conference on Usenix
  Annual Technical Conference}} (Boston, MA, USA)
  \emph{(\bibinfo{series}{USENIX ATC '25})}. \bibinfo{publisher}{USENIX
  Association}, \bibinfo{address}{USA}, Article \bibinfo{articleno}{28},
  \bibinfo{numpages}{18}~pages.
\newblock
\showISBNx{978-1-939133-48-9}


\bibitem[Wang et~al\mbox{.}(2025d)]%
        {wang2025sirius}
\bibfield{author}{\bibinfo{person}{Jiali Wang}, \bibinfo{person}{Yankui Wang},
  \bibinfo{person}{Mingcong Han}, {and} \bibinfo{person}{Rong Chen}.}
  \bibinfo{year}{2025}\natexlab{d}.
\newblock \showarticletitle{Colocating {ML} Inference and Training with Fast
  {GPU} Memory Handover}. In \bibinfo{booktitle}{\emph{2025 USENIX Annual
  Technical Conference (USENIX ATC 25)}}. \bibinfo{publisher}{USENIX
  Association}, \bibinfo{address}{Boston, MA}, \bibinfo{pages}{1657--1675}.
\newblock
\showISBNx{978-1-939133-48-9}
\urldef\tempurl%
\url{https://www.usenix.org/conference/atc25/presentation/wang-jiali}
\showURL{%
\tempurl}


\bibitem[Wang et~al\mbox{.}(2021)]%
        {wang2021milvus}
\bibfield{author}{\bibinfo{person}{Jianguo Wang}, \bibinfo{person}{Xiaomeng
  Yi}, \bibinfo{person}{Rentong Guo}, \bibinfo{person}{Hai Jin},
  \bibinfo{person}{Peng Xu}, \bibinfo{person}{Shengjun Li},
  \bibinfo{person}{Xiangyu Wang}, \bibinfo{person}{Xiangzhou Guo},
  \bibinfo{person}{Chengming Li}, \bibinfo{person}{Xiaohai Xu},
  \bibinfo{person}{Kun Yu}, \bibinfo{person}{Yuxing Yuan},
  \bibinfo{person}{Yinghao Zou}, \bibinfo{person}{Jiquan Long},
  \bibinfo{person}{Yudong Cai}, \bibinfo{person}{Zhenxiang Li},
  \bibinfo{person}{Zhifeng Zhang}, \bibinfo{person}{Yihua Mo},
  \bibinfo{person}{Jun Gu}, \bibinfo{person}{Ruiyi Jiang}, \bibinfo{person}{Yi
  Wei}, {and} \bibinfo{person}{Charles Xie}.} \bibinfo{year}{2021}\natexlab{}.
\newblock \showarticletitle{Milvus: A Purpose-Built Vector Data Management
  System}. In \bibinfo{booktitle}{\emph{Proceedings of the 2021 International
  Conference on Management of Data}} (Virtual Event, China)
  \emph{(\bibinfo{series}{SIGMOD '21})}. \bibinfo{publisher}{Association for
  Computing Machinery}, \bibinfo{address}{New York, NY, USA},
  \bibinfo{pages}{2614–2627}.
\newblock
\showISBNx{9781450383431}
\href{https://doi.org/10.1145/3448016.3457550}{doi:\nolinkurl{10.1145/3448016.3457550}}


\bibitem[Wang et~al\mbox{.}(2018)]%
        {wang2018superneurons}
\bibfield{author}{\bibinfo{person}{Linnan Wang}, \bibinfo{person}{Jinmian Ye},
  \bibinfo{person}{Yiyang Zhao}, \bibinfo{person}{Wei Wu}, \bibinfo{person}{Ang
  Li}, \bibinfo{person}{Shuaiwen~Leon Song}, \bibinfo{person}{Zenglin Xu},
  {and} \bibinfo{person}{Tim Kraska}.} \bibinfo{year}{2018}\natexlab{}.
\newblock \showarticletitle{Superneurons: dynamic GPU memory management for
  training deep neural networks}. In \bibinfo{booktitle}{\emph{Proceedings of
  the 23rd ACM SIGPLAN Symposium on Principles and Practice of Parallel
  Programming}} (Vienna, Austria) \emph{(\bibinfo{series}{PPoPP '18})}.
  \bibinfo{publisher}{Association for Computing Machinery},
  \bibinfo{address}{New York, NY, USA}, \bibinfo{pages}{41–53}.
\newblock
\showISBNx{9781450349826}
\href{https://doi.org/10.1145/3178487.3178491}{doi:\nolinkurl{10.1145/3178487.3178491}}


\bibitem[Wang et~al\mbox{.}(2026)]%
        {wang2026candor}
\bibfield{author}{\bibinfo{person}{Mingqi Wang}, \bibinfo{person}{Junyao Dong},
  \bibinfo{person}{Zhuoyan Wu}, \bibinfo{person}{Jun Liu},
  \bibinfo{person}{Ruicheng Zhang}, \bibinfo{person}{Jianjun Zhao},
  \bibinfo{person}{Ruipeng Wan}, \bibinfo{person}{Xinyan Lei},
  \bibinfo{person}{Shuhao Zhang}, \bibinfo{person}{Bolong Zheng},
  \bibinfo{person}{Haikun Liu}, \bibinfo{person}{Xiaofei Liao}, {and}
  \bibinfo{person}{Hai Jin}.} \bibinfo{year}{2026}\natexlab{}.
\newblock \showarticletitle{CANDOR-Bench: Benchmarking In-Memory Continuous
  ANNS under Dynamic Open-World Streams [Experiments \& Analysis]}.
\newblock \bibinfo{journal}{\emph{Proc. ACM Manag. Data}} \bibinfo{volume}{4},
  \bibinfo{number}{1}, Article \bibinfo{articleno}{16} (\bibinfo{date}{April}
  \bibinfo{year}{2026}), \bibinfo{numpages}{27}~pages.
\newblock
\href{https://doi.org/10.1145/3786630}{doi:\nolinkurl{10.1145/3786630}}


\bibitem[Wang et~al\mbox{.}(2025c)]%
        {wang2025finemem}
\bibfield{author}{\bibinfo{person}{Xiaoyang Wang}, \bibinfo{person}{Yongkun
  Li}, \bibinfo{person}{Kan Wu}, \bibinfo{person}{Wenzhe Zhu},
  \bibinfo{person}{Yuqi Li}, {and} \bibinfo{person}{Yinlong Xu}.}
  \bibinfo{year}{2025}\natexlab{c}.
\newblock \showarticletitle{FineMem: breaking the allocation overhead vs.
  memory waste dilemma in fine-grained disaggregated memory management}. In
  \bibinfo{booktitle}{\emph{Proceedings of the 19th USENIX Conference on
  Operating Systems Design and Implementation}} (Boston, MA, USA)
  \emph{(\bibinfo{series}{OSDI '25})}. \bibinfo{publisher}{USENIX Association},
  \bibinfo{address}{USA}, Article \bibinfo{articleno}{4},
  \bibinfo{numpages}{18}~pages.
\newblock
\showISBNx{978-1-939133-47-2}


\bibitem[Wang et~al\mbox{.}(2023)]%
        {wang2023dscstudy}
\bibfield{author}{\bibinfo{person}{Xin Wang}, \bibinfo{person}{Zhengru Wang},
  \bibinfo{person}{Zhenyu Wu}, \bibinfo{person}{Shuhao Zhang},
  \bibinfo{person}{Xuanhua Shi}, {and} \bibinfo{person}{Li Lu}.}
  \bibinfo{year}{2023}\natexlab{}.
\newblock \showarticletitle{Data Stream Clustering: An In-depth Empirical
  Study}.
\newblock \bibinfo{journal}{\emph{Proc. ACM Manag. Data}} \bibinfo{volume}{1},
  \bibinfo{number}{2}, Article \bibinfo{articleno}{162} (\bibinfo{date}{June}
  \bibinfo{year}{2023}), \bibinfo{numpages}{26}~pages.
\newblock
\href{https://doi.org/10.1145/3589307}{doi:\nolinkurl{10.1145/3589307}}


\bibitem[Wang et~al\mbox{.}(2025a)]%
        {wang2025wlbllm}
\bibfield{author}{\bibinfo{person}{Zheng Wang}, \bibinfo{person}{Anna Cai},
  \bibinfo{person}{Xinfeng Xie}, \bibinfo{person}{Zaifeng Pan},
  \bibinfo{person}{Yue Guan}, \bibinfo{person}{Weiwei Chu},
  \bibinfo{person}{Jie Wang}, \bibinfo{person}{Shikai Li},
  \bibinfo{person}{Jianyu Huang}, \bibinfo{person}{Chris Cai},
  \bibinfo{person}{Yuchen Hao}, {and} \bibinfo{person}{Yufei Ding}.}
  \bibinfo{year}{2025}\natexlab{a}.
\newblock \showarticletitle{WLB-LLM: workload-balanced 4D parallelism for large
  language model training}. In \bibinfo{booktitle}{\emph{Proceedings of the
  19th USENIX Conference on Operating Systems Design and Implementation}}
  (Boston, MA, USA) \emph{(\bibinfo{series}{OSDI '25})}.
  \bibinfo{publisher}{USENIX Association}, \bibinfo{address}{USA}, Article
  \bibinfo{articleno}{43}, \bibinfo{numpages}{17}~pages.
\newblock
\showISBNx{978-1-939133-47-2}


\bibitem[Wang et~al\mbox{.}(2024)]%
        {wang2024mostream}
\bibfield{author}{\bibinfo{person}{Zhengru Wang}, \bibinfo{person}{Xin Wang},
  {and} \bibinfo{person}{Shuhao Zhang}.} \bibinfo{year}{2024}\natexlab{}.
\newblock \showarticletitle{MOStream: A Modular and Self-Optimizing Data Stream
  Clustering Algorithm}. In \bibinfo{booktitle}{\emph{2024 IEEE International
  Conference on Data Mining (ICDM)}}. \bibinfo{pages}{500--509}.
\newblock
\href{https://doi.org/10.1109/ICDM59182.2024.00057}{doi:\nolinkurl{10.1109/ICDM59182.2024.00057}}


\bibitem[Wei et~al\mbox{.}(2025)]%
        {wei2025phoenixos}
\bibfield{author}{\bibinfo{person}{Xingda Wei}, \bibinfo{person}{Zhuobin
  Huang}, \bibinfo{person}{Tianle Sun}, \bibinfo{person}{Yingyi Hao},
  \bibinfo{person}{Rong Chen}, \bibinfo{person}{Mingcong Han},
  \bibinfo{person}{Jinyu Gu}, {and} \bibinfo{person}{Haibo Chen}.}
  \bibinfo{year}{2025}\natexlab{}.
\newblock \showarticletitle{PhoenixOS: Concurrent OS-level GPU Checkpoint and
  Restore with Validated Speculation}. In \bibinfo{booktitle}{\emph{Proceedings
  of the ACM SIGOPS 31st Symposium on Operating Systems Principles}} (Lotte
  Hotel World, Seoul, Republic of Korea) \emph{(\bibinfo{series}{SOSP '25})}.
  \bibinfo{publisher}{Association for Computing Machinery},
  \bibinfo{address}{New York, NY, USA}, \bibinfo{pages}{996–1013}.
\newblock
\showISBNx{9798400718700}
\href{https://doi.org/10.1145/3731569.3764813}{doi:\nolinkurl{10.1145/3731569.3764813}}


\bibitem[Wei et~al\mbox{.}(2015)]%
        {ren2015drtmh}
\bibfield{author}{\bibinfo{person}{Xingda Wei}, \bibinfo{person}{Jiaxin Shi},
  \bibinfo{person}{Yanzhe Chen}, \bibinfo{person}{Rong Chen}, {and}
  \bibinfo{person}{Haibo Chen}.} \bibinfo{year}{2015}\natexlab{}.
\newblock \showarticletitle{Fast in-memory transaction processing using RDMA
  and HTM}. In \bibinfo{booktitle}{\emph{Proceedings of the 25th Symposium on
  Operating Systems Principles}} (Monterey, California)
  \emph{(\bibinfo{series}{SOSP '15})}. \bibinfo{publisher}{Association for
  Computing Machinery}, \bibinfo{address}{New York, NY, USA},
  \bibinfo{pages}{87–104}.
\newblock
\showISBNx{9781450338349}
\href{https://doi.org/10.1145/2815400.2815419}{doi:\nolinkurl{10.1145/2815400.2815419}}


\bibitem[Wu et~al\mbox{.}(2024a)]%
        {loongserve2024}
\bibfield{author}{\bibinfo{person}{Bingyang Wu}, \bibinfo{person}{Shengyu Liu},
  \bibinfo{person}{Yinmin Zhong}, \bibinfo{person}{Peng Sun},
  \bibinfo{person}{Xuanzhe Liu}, {and} \bibinfo{person}{Xin Jin}.}
  \bibinfo{year}{2024}\natexlab{a}.
\newblock \showarticletitle{LoongServe: Efficiently Serving Long-Context Large
  Language Models with Elastic Sequence Parallelism}. In
  \bibinfo{booktitle}{\emph{Proceedings of the ACM SIGOPS 30th Symposium on
  Operating Systems Principles}} (Austin, TX, USA) \emph{(\bibinfo{series}{SOSP
  '24})}. \bibinfo{publisher}{Association for Computing Machinery},
  \bibinfo{address}{New York, NY, USA}, \bibinfo{pages}{640–654}.
\newblock
\showISBNx{9798400712517}
\href{https://doi.org/10.1145/3694715.3695948}{doi:\nolinkurl{10.1145/3694715.3695948}}


\bibitem[Wu et~al\mbox{.}(2026)]%
        {wu2026fastserve}
\bibfield{author}{\bibinfo{person}{Bingyang Wu}, \bibinfo{person}{Yinmin
  Zhong}, \bibinfo{person}{Zili Zhang}, \bibinfo{person}{Shengyu Liu},
  \bibinfo{person}{Fangyue Liu}, \bibinfo{person}{Yuanhang Sun},
  \bibinfo{person}{Gang Huang}, \bibinfo{person}{Xuanzhe Liu}, {and}
  \bibinfo{person}{Xin Jin}.} \bibinfo{year}{2026}\natexlab{}.
\newblock \showarticletitle{{FastServe}: {Iteration-Level} Preemptive
  Scheduling for Large Language Model Inference}. In
  \bibinfo{booktitle}{\emph{23rd USENIX Symposium on Networked Systems Design
  and Implementation (NSDI 26)}}. \bibinfo{publisher}{USENIX Association},
  \bibinfo{address}{Renton, WA}, \bibinfo{pages}{57--74}.
\newblock
\showISBNx{978-1-939133-54-0}
\urldef\tempurl%
\url{https://www.usenix.org/conference/nsdi26/presentation/wu-bingyang}
\showURL{%
\tempurl}


\bibitem[Wu et~al\mbox{.}(2024b)]%
        {wu2024dlora}
\bibfield{author}{\bibinfo{person}{Bingyang Wu}, \bibinfo{person}{Ruidong Zhu},
  \bibinfo{person}{Zili Zhang}, \bibinfo{person}{Peng Sun},
  \bibinfo{person}{Xuanzhe Liu}, {and} \bibinfo{person}{Xin Jin}.}
  \bibinfo{year}{2024}\natexlab{b}.
\newblock \showarticletitle{dLoRA: dynamically orchestrating requests and
  adapters for LoRA LLM serving}. In \bibinfo{booktitle}{\emph{Proceedings of
  the 18th USENIX Conference on Operating Systems Design and Implementation}}
  (Santa Clara, CA, USA) \emph{(\bibinfo{series}{OSDI'24})}.
  \bibinfo{publisher}{USENIX Association}, \bibinfo{address}{USA}, Article
  \bibinfo{articleno}{49}, \bibinfo{numpages}{17}~pages.
\newblock
\showISBNx{978-1-939133-40-3}


\bibitem[Wu et~al\mbox{.}(2023)]%
        {wu2023sentistream}
\bibfield{author}{\bibinfo{person}{Yuhao Wu}, \bibinfo{person}{Karthick
  Sharma}, \bibinfo{person}{Chun Seah}, {and} \bibinfo{person}{Shuhao Zhang}.}
  \bibinfo{year}{2023}\natexlab{}.
\newblock \showarticletitle{{S}enti{S}tream: A Co-Training Framework for
  Adaptive Online Sentiment Analysis in Evolving Data Streams}. In
  \bibinfo{booktitle}{\emph{Proceedings of the 2023 Conference on Empirical
  Methods in Natural Language Processing}},
  \bibfield{editor}{\bibinfo{person}{Houda Bouamor}, \bibinfo{person}{Juan
  Pino}, {and} \bibinfo{person}{Kalika Bali}} (Eds.).
  \bibinfo{publisher}{Association for Computational Linguistics},
  \bibinfo{address}{Singapore}, \bibinfo{pages}{6198--6212}.
\newblock
\href{https://doi.org/10.18653/v1/2023.emnlp-main.380}{doi:\nolinkurl{10.18653/v1/2023.emnlp-main.380}}


\bibitem[Xia et~al\mbox{.}(2024)]%
        {xia2024quantllm}
\bibfield{author}{\bibinfo{person}{Haojun Xia}, \bibinfo{person}{Zhen Zheng},
  \bibinfo{person}{Xiaoxia Wu}, \bibinfo{person}{Shiyang Chen},
  \bibinfo{person}{Zhewei Yao}, \bibinfo{person}{Stephen Youn},
  \bibinfo{person}{Arash Bakhtiari}, \bibinfo{person}{Michael Wyatt},
  \bibinfo{person}{Donglin Zhuang}, \bibinfo{person}{Zhongzhu Zhou},
  \bibinfo{person}{Olatunji Ruwase}, \bibinfo{person}{Yuxiong He}, {and}
  \bibinfo{person}{Shuaiwen~Leon Song}.} \bibinfo{year}{2024}\natexlab{}.
\newblock \showarticletitle{Quant-LLM: accelerating the serving of large
  language models via FP6-centric algorithm-system co-design on modern GPUs}.
  In \bibinfo{booktitle}{\emph{Proceedings of the 2024 USENIX Conference on
  Usenix Annual Technical Conference}} (Santa Clara, CA, USA)
  \emph{(\bibinfo{series}{USENIX ATC'24})}. \bibinfo{publisher}{USENIX
  Association}, \bibinfo{address}{USA}, Article \bibinfo{articleno}{43},
  \bibinfo{numpages}{15}~pages.
\newblock
\showISBNx{978-1-939133-41-0}


\bibitem[Xiang et~al\mbox{.}(2025)]%
        {xiang2025aegaeon}
\bibfield{author}{\bibinfo{person}{Yuxing Xiang}, \bibinfo{person}{Xue Li},
  \bibinfo{person}{Kun Qian}, \bibinfo{person}{Yufan Yang},
  \bibinfo{person}{Diwen Zhu}, \bibinfo{person}{Wenyuan Yu},
  \bibinfo{person}{Ennan Zhai}, \bibinfo{person}{Xuanzhe Liu},
  \bibinfo{person}{Xin Jin}, {and} \bibinfo{person}{Jingren Zhou}.}
  \bibinfo{year}{2025}\natexlab{}.
\newblock \showarticletitle{Aegaeon: Effective GPU Pooling for Concurrent LLM
  Serving on the Market}. In \bibinfo{booktitle}{\emph{Proceedings of the ACM
  SIGOPS 31st Symposium on Operating Systems Principles}} (Lotte Hotel World,
  Seoul, Republic of Korea) \emph{(\bibinfo{series}{SOSP '25})}.
  \bibinfo{publisher}{Association for Computing Machinery},
  \bibinfo{address}{New York, NY, USA}, \bibinfo{pages}{1030–1045}.
\newblock
\showISBNx{9798400718700}
\href{https://doi.org/10.1145/3731569.3764815}{doi:\nolinkurl{10.1145/3731569.3764815}}


\bibitem[Xiong et~al\mbox{.}(2024)]%
        {xiong2024layerkv}
\bibfield{author}{\bibinfo{person}{Yi Xiong}, \bibinfo{person}{Hao Wu},
  \bibinfo{person}{Changxu Shao}, \bibinfo{person}{Ziqing Wang},
  \bibinfo{person}{Rui Zhang}, \bibinfo{person}{Yuhong Guo},
  \bibinfo{person}{Junping Zhao}, \bibinfo{person}{Ke Zhang}, {and}
  \bibinfo{person}{Zhenxuan Pan}.} \bibinfo{year}{2024}\natexlab{}.
\newblock \showarticletitle{LayerKV: Optimizing Large Language Model Serving
  with Layer-wise KV Cache Management}.
\newblock  (\bibinfo{year}{2024}).
\newblock
\href{https://doi.org/10.48550/ARXIV.2410.00428}{doi:\nolinkurl{10.48550/ARXIV.2410.00428}}


\bibitem[Xu et~al\mbox{.}(2025)]%
        {xu2025place}
\bibfield{author}{\bibinfo{person}{Haike Xu}, \bibinfo{person}{Magdalen~Dobson
  Manohar}, \bibinfo{person}{Philip~A. Bernstein}, \bibinfo{person}{Badrish
  Chandramouli}, \bibinfo{person}{Richard Wen}, {and}
  \bibinfo{person}{Harsha~Vardhan Simhadri}.} \bibinfo{year}{2025}\natexlab{}.
\newblock \showarticletitle{In-Place Updates of a Graph Index for Streaming
  Approximate Nearest Neighbor Search}.
\newblock  (\bibinfo{year}{2025}).
\newblock
\href{https://doi.org/10.48550/ARXIV.2502.13826}{doi:\nolinkurl{10.48550/ARXIV.2502.13826}}


\bibitem[Xu et~al\mbox{.}(2023)]%
        {xu2023spfresh}
\bibfield{author}{\bibinfo{person}{Yuming Xu}, \bibinfo{person}{Hengyu Liang},
  \bibinfo{person}{Jin Li}, \bibinfo{person}{Shuotao Xu}, \bibinfo{person}{Qi
  Chen}, \bibinfo{person}{Qianxi Zhang}, \bibinfo{person}{Cheng Li},
  \bibinfo{person}{Ziyue Yang}, \bibinfo{person}{Fan Yang},
  \bibinfo{person}{Yuqing Yang}, \bibinfo{person}{Peng Cheng}, {and}
  \bibinfo{person}{Mao Yang}.} \bibinfo{year}{2023}\natexlab{}.
\newblock \showarticletitle{SPFresh: Incremental In-Place Update for
  Billion-Scale Vector Search}. In \bibinfo{booktitle}{\emph{Proceedings of the
  29th Symposium on Operating Systems Principles}} (Koblenz, Germany)
  \emph{(\bibinfo{series}{SOSP '23})}. \bibinfo{publisher}{Association for
  Computing Machinery}, \bibinfo{address}{New York, NY, USA},
  \bibinfo{pages}{545–561}.
\newblock
\showISBNx{9798400702297}
\href{https://doi.org/10.1145/3600006.3613166}{doi:\nolinkurl{10.1145/3600006.3613166}}


\bibitem[Xu et~al\mbox{.}(2024)]%
        {xu2024leap}
\bibfield{author}{\bibinfo{person}{Yanchao Xu}, \bibinfo{person}{Dongxiang
  Zhang}, \bibinfo{person}{Shuhao Zhang}, \bibinfo{person}{Sai Wu},
  \bibinfo{person}{Zexu Feng}, {and} \bibinfo{person}{Gang Chen}.}
  \bibinfo{year}{2024}\natexlab{}.
\newblock \showarticletitle{Predictive and Near-Optimal Sampling for View
  Materialization in Video Databases}.
\newblock \bibinfo{journal}{\emph{Proc. ACM Manag. Data}} \bibinfo{volume}{2},
  \bibinfo{number}{1}, Article \bibinfo{articleno}{19} (\bibinfo{date}{March}
  \bibinfo{year}{2024}), \bibinfo{numpages}{27}~pages.
\newblock
\href{https://doi.org/10.1145/3639274}{doi:\nolinkurl{10.1145/3639274}}


\bibitem[Yang et~al\mbox{.}(2025)]%
        {yang2025lserve}
\bibfield{author}{\bibinfo{person}{Shang Yang}, \bibinfo{person}{Junxian Guo},
  \bibinfo{person}{Haotian Tang}, \bibinfo{person}{Qinghao Hu},
  \bibinfo{person}{Guangxuan Xiao}, \bibinfo{person}{Jiaming Tang},
  \bibinfo{person}{Yujun Lin}, \bibinfo{person}{Zhijian Liu},
  \bibinfo{person}{Yao Lu}, {and} \bibinfo{person}{Song Han}.}
  \bibinfo{year}{2025}\natexlab{}.
\newblock \showarticletitle{LServe: Efficient Long-sequence LLM Serving with
  Unified Sparse Attention}. In \bibinfo{booktitle}{\emph{Proceedings of
  Machine Learning and Systems}},
  \bibfield{editor}{\bibinfo{person}{M.~Zaharia}, \bibinfo{person}{G.~Joshi},
  {and} \bibinfo{person}{Y.~Lin}} (Eds.), Vol.~\bibinfo{volume}{7}.
  \bibinfo{publisher}{MLSys}.
\newblock
\urldef\tempurl%
\url{https://proceedings.mlsys.org/paper_files/paper/2025/file/cc8c6b9d89f7a898a29f58869b238e46-Paper-Conference.pdf}
\showURL{%
\tempurl}


\bibitem[Yao et~al\mbox{.}(2025)]%
        {yao2025cacheblend}
\bibfield{author}{\bibinfo{person}{Jiayi Yao}, \bibinfo{person}{Hanchen Li},
  \bibinfo{person}{Yuhan Liu}, \bibinfo{person}{Siddhant Ray},
  \bibinfo{person}{Yihua Cheng}, \bibinfo{person}{Qizheng Zhang},
  \bibinfo{person}{Kuntai Du}, \bibinfo{person}{Shan Lu}, {and}
  \bibinfo{person}{Junchen Jiang}.} \bibinfo{year}{2025}\natexlab{}.
\newblock \showarticletitle{CacheBlend: Fast Large Language Model Serving for
  RAG with Cached Knowledge Fusion}. In \bibinfo{booktitle}{\emph{Proceedings
  of the Twentieth European Conference on Computer Systems}} (Rotterdam,
  Netherlands) \emph{(\bibinfo{series}{EuroSys '25})}.
  \bibinfo{publisher}{Association for Computing Machinery},
  \bibinfo{address}{New York, NY, USA}, \bibinfo{pages}{94–109}.
\newblock
\showISBNx{9798400711961}
\href{https://doi.org/10.1145/3689031.3696098}{doi:\nolinkurl{10.1145/3689031.3696098}}


\bibitem[Ye et~al\mbox{.}(2025)]%
        {ye2025flashinfer}
\bibfield{author}{\bibinfo{person}{Zihao Ye}, \bibinfo{person}{Lequn Chen},
  \bibinfo{person}{Ruihang Lai}, \bibinfo{person}{Wuwei Lin},
  \bibinfo{person}{Yineng Zhang}, \bibinfo{person}{Stephanie Wang},
  \bibinfo{person}{Tianqi Chen}, \bibinfo{person}{Baris Kasikci},
  \bibinfo{person}{Vinod Grover}, \bibinfo{person}{Arvind Krishnamurthy}, {and}
  \bibinfo{person}{Luis Ceze}.} \bibinfo{year}{2025}\natexlab{}.
\newblock \showarticletitle{FlashInfer: Efficient and Customizable Attention
  Engine for LLM Inference Serving}. In \bibinfo{booktitle}{\emph{Proceedings
  of Machine Learning and Systems}},
  \bibfield{editor}{\bibinfo{person}{M.~Zaharia}, \bibinfo{person}{G.~Joshi},
  {and} \bibinfo{person}{Y.~Lin}} (Eds.), Vol.~\bibinfo{volume}{7}.
  \bibinfo{publisher}{MLSys}.
\newblock
\urldef\tempurl%
\url{https://proceedings.mlsys.org/paper_files/paper/2025/file/dbf02b21d77409a2db30e56866a8ab3a-Paper-Conference.pdf}
\showURL{%
\tempurl}


\bibitem[Yu et~al\mbox{.}(2022)]%
        {yu2022orca}
\bibfield{author}{\bibinfo{person}{Gyeong-In Yu}, \bibinfo{person}{Joo~Seong
  Jeong}, \bibinfo{person}{Geon-Woo Kim}, \bibinfo{person}{Soojeong Kim}, {and}
  \bibinfo{person}{Byung-Gon Chun}.} \bibinfo{year}{2022}\natexlab{}.
\newblock \showarticletitle{Orca: A Distributed Serving System for
  {Transformer-Based} Generative Models}. In \bibinfo{booktitle}{\emph{16th
  USENIX Symposium on Operating Systems Design and Implementation (OSDI 22)}}.
  \bibinfo{publisher}{USENIX Association}, \bibinfo{address}{Carlsbad, CA},
  \bibinfo{pages}{521--538}.
\newblock
\showISBNx{978-1-939133-28-1}
\urldef\tempurl%
\url{https://www.usenix.org/conference/osdi22/presentation/yu}
\showURL{%
\tempurl}


\bibitem[Yu et~al\mbox{.}(2025b)]%
        {yu2025pensieve}
\bibfield{author}{\bibinfo{person}{Lingfan Yu}, \bibinfo{person}{Jinkun Lin},
  {and} \bibinfo{person}{Jinyang Li}.} \bibinfo{year}{2025}\natexlab{b}.
\newblock \showarticletitle{Stateful Large Language Model Serving with
  Pensieve}. In \bibinfo{booktitle}{\emph{Proceedings of the Twentieth European
  Conference on Computer Systems}} (Rotterdam, Netherlands)
  \emph{(\bibinfo{series}{EuroSys '25})}. \bibinfo{publisher}{Association for
  Computing Machinery}, \bibinfo{address}{New York, NY, USA},
  \bibinfo{pages}{144–158}.
\newblock
\showISBNx{9798400711961}
\href{https://doi.org/10.1145/3689031.3696086}{doi:\nolinkurl{10.1145/3689031.3696086}}


\bibitem[Yu et~al\mbox{.}(2025c)]%
        {yu2025prism}
\bibfield{author}{\bibinfo{person}{Shan Yu}, \bibinfo{person}{Yifan Qiao},
  \bibinfo{person}{Mingyuan Ma}, \bibinfo{person}{Yangmin Li},
  \bibinfo{person}{Shuo Yang}, \bibinfo{person}{Xinyuan Tong},
  \bibinfo{person}{Yang Wang}, \bibinfo{person}{Zhiqiang Xie},
  \bibinfo{person}{Yuwei An}, \bibinfo{person}{Shiyi Cao}, \bibinfo{person}{Ke
  Bao}, \bibinfo{person}{Deepak Vij}, \bibinfo{person}{Xiaoning Ding},
  \bibinfo{person}{Yichen Wang}, \bibinfo{person}{Qingda Lu},
  \bibinfo{person}{Zhong Wang}, \bibinfo{person}{Gao Gao},
  \bibinfo{person}{Harry Xu}, \bibinfo{person}{Junyi Shu},
  \bibinfo{person}{Jiarong Xing}, {and} \bibinfo{person}{Ying Sheng}.}
  \bibinfo{year}{2025}\natexlab{c}.
\newblock \showarticletitle{Prism: Cost-Efficient Multi-LLM Serving via GPU
  Memory Ballooning}.
\newblock  (\bibinfo{year}{2025}).
\newblock
\href{https://doi.org/10.48550/ARXIV.2505.04021}{doi:\nolinkurl{10.48550/ARXIV.2505.04021}}


\bibitem[Yu et~al\mbox{.}(2016)]%
        {yu2016tictoc}
\bibfield{author}{\bibinfo{person}{Xiangyao Yu}, \bibinfo{person}{Andrew
  Pavlo}, \bibinfo{person}{Daniel Sanchez}, {and} \bibinfo{person}{Srinivas
  Devadas}.} \bibinfo{year}{2016}\natexlab{}.
\newblock \showarticletitle{TicToc: Time Traveling Optimistic Concurrency
  Control}. In \bibinfo{booktitle}{\emph{Proceedings of the 2016 International
  Conference on Management of Data}} (San Francisco, California, USA)
  \emph{(\bibinfo{series}{SIGMOD '16})}. \bibinfo{publisher}{Association for
  Computing Machinery}, \bibinfo{address}{New York, NY, USA},
  \bibinfo{pages}{1629–1642}.
\newblock
\showISBNx{9781450335317}
\href{https://doi.org/10.1145/2882903.2882935}{doi:\nolinkurl{10.1145/2882903.2882935}}


\bibitem[Yu et~al\mbox{.}(2025a)]%
        {yu2025iccache}
\bibfield{author}{\bibinfo{person}{Yifan Yu}, \bibinfo{person}{Yu Gan},
  \bibinfo{person}{Nikhil Sarda}, \bibinfo{person}{Lillian Tsai},
  \bibinfo{person}{Jiaming Shen}, \bibinfo{person}{Yanqi Zhou},
  \bibinfo{person}{Arvind Krishnamurthy}, \bibinfo{person}{Fan Lai},
  \bibinfo{person}{Hank Levy}, {and} \bibinfo{person}{David Culler}.}
  \bibinfo{year}{2025}\natexlab{a}.
\newblock \showarticletitle{IC-Cache: Efficient Large Language Model Serving
  via In-context Caching}. In \bibinfo{booktitle}{\emph{Proceedings of the ACM
  SIGOPS 31st Symposium on Operating Systems Principles}} (Lotte Hotel World,
  Seoul, Republic of Korea) \emph{(\bibinfo{series}{SOSP '25})}.
  \bibinfo{publisher}{Association for Computing Machinery},
  \bibinfo{address}{New York, NY, USA}, \bibinfo{pages}{375–398}.
\newblock
\showISBNx{9798400718700}
\href{https://doi.org/10.1145/3731569.3764829}{doi:\nolinkurl{10.1145/3731569.3764829}}


\bibitem[Yue et~al\mbox{.}(2025)]%
        {yue2025marco}
\bibfield{author}{\bibinfo{person}{Ziyang Yue}, \bibinfo{person}{Bolong Zheng},
  \bibinfo{person}{Ling Xu}, \bibinfo{person}{Kanru Xu},
  \bibinfo{person}{Shuhao Zhang}, \bibinfo{person}{Yajuan Du},
  \bibinfo{person}{Yunjun Gao}, \bibinfo{person}{Xiaofang Zhou}, {and}
  \bibinfo{person}{Christian~S. Jensen}.} \bibinfo{year}{2025}\natexlab{}.
\newblock \showarticletitle{Select Edges Wisely: Monotonic Path Aware Graph
  Layout Optimization for Disk-Based ANN Search}.
\newblock \bibinfo{journal}{\emph{Proc. VLDB Endow.}} \bibinfo{volume}{18},
  \bibinfo{number}{11} (\bibinfo{date}{July} \bibinfo{year}{2025}),
  \bibinfo{pages}{4337–4349}.
\newblock
\showISSN{2150-8097}
\href{https://doi.org/10.14778/3749646.3749697}{doi:\nolinkurl{10.14778/3749646.3749697}}


\bibitem[Yun et~al\mbox{.}(2024)]%
        {yun2024duplex}
\bibfield{author}{\bibinfo{person}{Sungmin Yun}, \bibinfo{person}{Kwanhee
  Kyung}, \bibinfo{person}{Juhwan Cho}, \bibinfo{person}{Jaewan Choi},
  \bibinfo{person}{Jongmin Kim}, \bibinfo{person}{Byeongho Kim},
  \bibinfo{person}{Sukhan Lee}, \bibinfo{person}{Kyomin Sohn}, {and}
  \bibinfo{person}{Jung~Ho Ahn}.} \bibinfo{year}{2024}\natexlab{}.
\newblock \showarticletitle{Duplex: A Device for Large Language Models with
  Mixture of Experts, Grouped Query Attention, and Continuous Batching}. In
  \bibinfo{booktitle}{\emph{Proceedings of the 2024 57th IEEE/ACM International
  Symposium on Microarchitecture}} (Austin, TX, USA)
  \emph{(\bibinfo{series}{MICRO '24})}. \bibinfo{publisher}{IEEE Press},
  \bibinfo{pages}{1429–1443}.
\newblock
\href{https://doi.org/10.1109/MICRO61859.2024.00105}{doi:\nolinkurl{10.1109/MICRO61859.2024.00105}}


\bibitem[Zaharia et~al\mbox{.}(2013)]%
        {zaharia2013dstreams}
\bibfield{author}{\bibinfo{person}{Matei Zaharia}, \bibinfo{person}{Tathagata
  Das}, \bibinfo{person}{Haoyuan Li}, \bibinfo{person}{Timothy Hunter},
  \bibinfo{person}{Scott Shenker}, {and} \bibinfo{person}{Ion Stoica}.}
  \bibinfo{year}{2013}\natexlab{}.
\newblock \showarticletitle{Discretized streams: fault-tolerant streaming
  computation at scale}. In \bibinfo{booktitle}{\emph{Proceedings of the
  Twenty-Fourth ACM Symposium on Operating Systems Principles}} (Farminton,
  Pennsylvania) \emph{(\bibinfo{series}{SOSP '13})}.
  \bibinfo{publisher}{Association for Computing Machinery},
  \bibinfo{address}{New York, NY, USA}, \bibinfo{pages}{423–438}.
\newblock
\showISBNx{9781450323888}
\href{https://doi.org/10.1145/2517349.2522737}{doi:\nolinkurl{10.1145/2517349.2522737}}


\bibitem[Zamanian et~al\mbox{.}(2017)]%
        {zamanian2017endmyth}
\bibfield{author}{\bibinfo{person}{Erfan Zamanian}, \bibinfo{person}{Carsten
  Binnig}, \bibinfo{person}{Tim Harris}, {and} \bibinfo{person}{Tim Kraska}.}
  \bibinfo{year}{2017}\natexlab{}.
\newblock \showarticletitle{The end of a myth: distributed transactions can
  scale}.
\newblock \bibinfo{journal}{\emph{Proc. VLDB Endow.}} \bibinfo{volume}{10},
  \bibinfo{number}{6} (\bibinfo{date}{Feb.} \bibinfo{year}{2017}),
  \bibinfo{pages}{685–696}.
\newblock
\showISSN{2150-8097}
\href{https://doi.org/10.14778/3055330.3055335}{doi:\nolinkurl{10.14778/3055330.3055335}}


\bibitem[Zeng et~al\mbox{.}(2024a)]%
        {zeng2024libamm}
\bibfield{author}{\bibinfo{person}{Xianzhi Zeng}, \bibinfo{person}{Wenchao
  Jiang}, {and} \bibinfo{person}{Shuhao Zhang}.}
  \bibinfo{year}{2024}\natexlab{a}.
\newblock \showarticletitle{LibAMM: Empirical Insights into Approximate
  Computing for Accelerating Matrix Multiplication}. In
  \bibinfo{booktitle}{\emph{Advances in Neural Information Processing
  Systems}}, \bibfield{editor}{\bibinfo{person}{A.~Globerson},
  \bibinfo{person}{L.~Mackey}, \bibinfo{person}{D.~Belgrave},
  \bibinfo{person}{A.~Fan}, \bibinfo{person}{U.~Paquet},
  \bibinfo{person}{J.~Tomczak}, {and} \bibinfo{person}{C.~Zhang}} (Eds.),
  Vol.~\bibinfo{volume}{37}. \bibinfo{publisher}{Curran Associates, Inc.},
  \bibinfo{pages}{60517--60530}.
\newblock
\href{https://doi.org/10.52202/079017-1935}{doi:\nolinkurl{10.52202/079017-1935}}


\bibitem[Zeng and Zhang(2023)]%
        {zeng2023cstreamicde}
\bibfield{author}{\bibinfo{person}{Xianzhi Zeng} {and} \bibinfo{person}{Shuhao
  Zhang}.} \bibinfo{year}{2023}\natexlab{}.
\newblock \showarticletitle{Parallelizing Stream Compression for IoT
  Applications on Asymmetric Multicores}. In \bibinfo{booktitle}{\emph{2023
  IEEE 39th International Conference on Data Engineering (ICDE)}}.
  \bibinfo{pages}{950--964}.
\newblock
\href{https://doi.org/10.1109/ICDE55515.2023.00078}{doi:\nolinkurl{10.1109/ICDE55515.2023.00078}}


\bibitem[Zeng and Zhang(2024)]%
        {zeng2024cstream}
\bibfield{author}{\bibinfo{person}{Xianzhi Zeng} {and} \bibinfo{person}{Shuhao
  Zhang}.} \bibinfo{year}{2024}\natexlab{}.
\newblock \showarticletitle{CStream: Parallel Data Stream Compression on
  Multicore Edge Devices}.
\newblock \bibinfo{journal}{\emph{IEEE Transactions on Knowledge and Data
  Engineering}} \bibinfo{volume}{36}, \bibinfo{number}{11}
  (\bibinfo{year}{2024}), \bibinfo{pages}{5889--5904}.
\newblock
\href{https://doi.org/10.1109/TKDE.2024.3386862}{doi:\nolinkurl{10.1109/TKDE.2024.3386862}}


\bibitem[Zeng et~al\mbox{.}(2024b)]%
        {zeng2024pecj}
\bibfield{author}{\bibinfo{person}{Xianzhi Zeng}, \bibinfo{person}{Shuhao
  Zhang}, \bibinfo{person}{Hongbin Zhong}, \bibinfo{person}{Hao Zhang},
  \bibinfo{person}{Mian Lu}, \bibinfo{person}{Zhao Zheng}, {and}
  \bibinfo{person}{Yuqiang Chen}.} \bibinfo{year}{2024}\natexlab{b}.
\newblock \showarticletitle{PECJ: Stream Window Join on Disorder Data Streams
  with Proactive Error Compensation}.
\newblock \bibinfo{journal}{\emph{Proc. ACM Manag. Data}} \bibinfo{volume}{2},
  \bibinfo{number}{1}, Article \bibinfo{articleno}{13} (\bibinfo{date}{March}
  \bibinfo{year}{2024}), \bibinfo{numpages}{24}~pages.
\newblock
\href{https://doi.org/10.1145/3639268}{doi:\nolinkurl{10.1145/3639268}}


\bibitem[Zhan et~al\mbox{.}(2025)]%
        {zhan2025assyllm}
\bibfield{author}{\bibinfo{person}{Shichen Zhan}, \bibinfo{person}{Li Li},
  {and} \bibinfo{person}{Chengzhong Xu}.} \bibinfo{year}{2025}\natexlab{}.
\newblock \showarticletitle{AssyLLM: efficient federated fine-tuning of LLMs
  via assembling pre-trained blocks}. In \bibinfo{booktitle}{\emph{Proceedings
  of the 2025 USENIX Conference on Usenix Annual Technical Conference}}
  (Boston, MA, USA) \emph{(\bibinfo{series}{USENIX ATC '25})}.
  \bibinfo{publisher}{USENIX Association}, \bibinfo{address}{USA}, Article
  \bibinfo{articleno}{99}, \bibinfo{numpages}{15}~pages.
\newblock
\showISBNx{978-1-939133-48-9}


\bibitem[Zhang et~al\mbox{.}(2025a)]%
        {zhang2025jenga}
\bibfield{author}{\bibinfo{person}{Chen Zhang}, \bibinfo{person}{Kuntai Du},
  \bibinfo{person}{Shu Liu}, \bibinfo{person}{Woosuk Kwon},
  \bibinfo{person}{Xiangxi Mo}, \bibinfo{person}{Yufeng Wang},
  \bibinfo{person}{Xiaoxuan Liu}, \bibinfo{person}{Kaichao You},
  \bibinfo{person}{Zhuohan Li}, \bibinfo{person}{Mingsheng Long},
  \bibinfo{person}{Jidong Zhai}, \bibinfo{person}{Joseph Gonzalez}, {and}
  \bibinfo{person}{Ion Stoica}.} \bibinfo{year}{2025}\natexlab{a}.
\newblock \showarticletitle{Jenga: Effective Memory Management for Serving LLM
  with Heterogeneity}. In \bibinfo{booktitle}{\emph{Proceedings of the ACM
  SIGOPS 31st Symposium on Operating Systems Principles}} (Lotte Hotel World,
  Seoul, Republic of Korea) \emph{(\bibinfo{series}{SOSP '25})}.
  \bibinfo{publisher}{Association for Computing Machinery},
  \bibinfo{address}{New York, NY, USA}, \bibinfo{pages}{446–461}.
\newblock
\showISBNx{9798400718700}
\href{https://doi.org/10.1145/3731569.3764823}{doi:\nolinkurl{10.1145/3731569.3764823}}


\bibitem[Zhang et~al\mbox{.}(2020b)]%
        {zhang2020finestream}
\bibfield{author}{\bibinfo{person}{Feng Zhang}, \bibinfo{person}{Lin Yang},
  \bibinfo{person}{Shuhao Zhang}, \bibinfo{person}{Bingsheng He},
  \bibinfo{person}{Wei Lu}, {and} \bibinfo{person}{Xiaoyong Du}.}
  \bibinfo{year}{2020}\natexlab{b}.
\newblock \showarticletitle{FineStream: fine-grained window-based stream
  processing on CPU-GPU integrated architectures}. In
  \bibinfo{booktitle}{\emph{Proceedings of the 2020 USENIX Conference on Usenix
  Annual Technical Conference}} \emph{(\bibinfo{series}{USENIX ATC'20})}.
  \bibinfo{publisher}{USENIX Association}, \bibinfo{address}{USA}, Article
  \bibinfo{articleno}{43}, \bibinfo{numpages}{15}~pages.
\newblock
\showISBNx{978-1-939133-14-4}


\bibitem[Zhang et~al\mbox{.}(2017c)]%
        {zhang2016corun}
\bibfield{author}{\bibinfo{person}{Feng Zhang}, \bibinfo{person}{Jidong Zhai},
  \bibinfo{person}{Bingsheng He}, \bibinfo{person}{Shuhao Zhang}, {and}
  \bibinfo{person}{Wenguang Chen}.} \bibinfo{year}{2017}\natexlab{c}.
\newblock \showarticletitle{Understanding Co-Running Behaviors on Integrated
  CPU/GPU Architectures}.
\newblock \bibinfo{journal}{\emph{IEEE Transactions on Parallel and Distributed
  Systems}} \bibinfo{volume}{28}, \bibinfo{number}{3} (\bibinfo{year}{2017}),
  \bibinfo{pages}{905--918}.
\newblock
\href{https://doi.org/10.1109/TPDS.2016.2586074}{doi:\nolinkurl{10.1109/TPDS.2016.2586074}}


\bibitem[Zhang et~al\mbox{.}(2021b)]%
        {zhang2021caerus}
\bibfield{author}{\bibinfo{person}{Hong Zhang}, \bibinfo{person}{Yupeng Tang},
  \bibinfo{person}{Anurag Khandelwal}, \bibinfo{person}{Jingrong Chen}, {and}
  \bibinfo{person}{Ion Stoica}.} \bibinfo{year}{2021}\natexlab{b}.
\newblock \showarticletitle{Caerus: {NIMBLE} Task Scheduling for Serverless
  Analytics}. In \bibinfo{booktitle}{\emph{18th USENIX Symposium on Networked
  Systems Design and Implementation (NSDI 21)}}. \bibinfo{publisher}{USENIX
  Association}, \bibinfo{pages}{653--669}.
\newblock
\showISBNx{978-1-939133-21-2}
\urldef\tempurl%
\url{https://www.usenix.org/conference/nsdi21/presentation/zhang-hong}
\showURL{%
\tempurl}


\bibitem[Zhang et~al\mbox{.}(2023b)]%
        {zhang2023openmldb}
\bibfield{author}{\bibinfo{person}{Hao Zhang}, \bibinfo{person}{Xianzhi Zeng},
  \bibinfo{person}{Shuhao Zhang}, \bibinfo{person}{Xinyi Liu},
  \bibinfo{person}{Mian Lu}, {and} \bibinfo{person}{Zhao Zheng}.}
  \bibinfo{year}{2023}\natexlab{b}.
\newblock \showarticletitle{Scalable Online Interval Join on Modern Multicore
  Processors in OpenMLDB}. In \bibinfo{booktitle}{\emph{2023 IEEE 39th
  International Conference on Data Engineering (ICDE)}}.
  \bibinfo{pages}{3031--3042}.
\newblock
\href{https://doi.org/10.1109/ICDE55515.2023.00232}{doi:\nolinkurl{10.1109/ICDE55515.2023.00232}}


\bibitem[Zhang et~al\mbox{.}(2023d)]%
        {zhang2023g10}
\bibfield{author}{\bibinfo{person}{Haoyang Zhang}, \bibinfo{person}{Yirui
  Zhou}, \bibinfo{person}{Yuqi Xue}, \bibinfo{person}{Yiqi Liu}, {and}
  \bibinfo{person}{Jian Huang}.} \bibinfo{year}{2023}\natexlab{d}.
\newblock \showarticletitle{G10: Enabling An Efficient Unified GPU Memory and
  Storage Architecture with Smart Tensor Migrations}. In
  \bibinfo{booktitle}{\emph{Proceedings of the 56th Annual IEEE/ACM
  International Symposium on Microarchitecture}} (Toronto, ON, Canada)
  \emph{(\bibinfo{series}{MICRO '23})}. \bibinfo{publisher}{Association for
  Computing Machinery}, \bibinfo{address}{New York, NY, USA},
  \bibinfo{pages}{395–410}.
\newblock
\showISBNx{9798400703294}
\href{https://doi.org/10.1145/3613424.3614309}{doi:\nolinkurl{10.1145/3613424.3614309}}


\bibitem[Zhang et~al\mbox{.}(2025b)]%
        {zhang2025qfactory}
\bibfield{author}{\bibinfo{person}{Qihao Zhang}, \bibinfo{person}{Mingshu
  Zhai}, \bibinfo{person}{Rui Sun}, {and} \bibinfo{person}{Jidong Zhai}.}
  \bibinfo{year}{2025}\natexlab{b}.
\newblock \showarticletitle{QFactory: accelerating quantized large language
  model serving with Qtile graphs}. In \bibinfo{booktitle}{\emph{Proceedings of
  the 2025 USENIX Conference on Usenix Annual Technical Conference}} (Boston,
  MA, USA) \emph{(\bibinfo{series}{USENIX ATC '25})}.
  \bibinfo{publisher}{USENIX Association}, \bibinfo{address}{USA}, Article
  \bibinfo{articleno}{38}, \bibinfo{numpages}{16}~pages.
\newblock
\showISBNx{978-1-939133-48-9}


\bibitem[Zhang et~al\mbox{.}(2026a)]%
        {zhang2026neuromem}
\bibfield{author}{\bibinfo{person}{Ruicheng Zhang}, \bibinfo{person}{Xinyi Li},
  \bibinfo{person}{Tianyi Xu}, \bibinfo{person}{Shuhao Zhang},
  \bibinfo{person}{Xiaofei Liao}, {and} \bibinfo{person}{Hai Jin}.}
  \bibinfo{year}{2026}\natexlab{a}.
\newblock \showarticletitle{Neuromem: A Granular Decomposition of the Streaming
  Lifecycle in External Memory for {LLM}s}. In
  \bibinfo{booktitle}{\emph{Forty-third International Conference on Machine
  Learning}}.
\newblock
\urldef\tempurl%
\url{https://openreview.net/forum?id=mO7DgwFFVe}
\showURL{%
\tempurl}


\bibitem[Zhang et~al\mbox{.}(2017a)]%
        {zhang2017revisiting}
\bibfield{author}{\bibinfo{person}{Shuhao Zhang}, \bibinfo{person}{Bingsheng
  He}, \bibinfo{person}{Daniel Dahlmeier}, \bibinfo{person}{Amelie~Chi Zhou},
  {and} \bibinfo{person}{Thomas Heinze}.} \bibinfo{year}{2017}\natexlab{a}.
\newblock \showarticletitle{Revisiting the Design of Data Stream Processing
  Systems on Multi-Core Processors}. In \bibinfo{booktitle}{\emph{2017 IEEE
  33rd International Conference on Data Engineering (ICDE)}}.
  \bibinfo{pages}{659--670}.
\newblock
\href{https://doi.org/10.1109/ICDE.2017.119}{doi:\nolinkurl{10.1109/ICDE.2017.119}}


\bibitem[Zhang et~al\mbox{.}(2019)]%
        {zhang2019briskstream}
\bibfield{author}{\bibinfo{person}{Shuhao Zhang}, \bibinfo{person}{Jiong He},
  \bibinfo{person}{Amelie~Chi Zhou}, {and} \bibinfo{person}{Bingsheng He}.}
  \bibinfo{year}{2019}\natexlab{}.
\newblock \showarticletitle{BriskStream: Scaling Data Stream Processing on
  Shared-Memory Multicore Architectures}. In
  \bibinfo{booktitle}{\emph{Proceedings of the 2019 International Conference on
  Management of Data}} (Amsterdam, Netherlands) \emph{(\bibinfo{series}{SIGMOD
  '19})}. \bibinfo{publisher}{Association for Computing Machinery},
  \bibinfo{address}{New York, NY, USA}, \bibinfo{pages}{705–722}.
\newblock
\showISBNx{9781450356435}
\href{https://doi.org/10.1145/3299869.3300067}{doi:\nolinkurl{10.1145/3299869.3300067}}


\bibitem[Zhang et~al\mbox{.}(2021a)]%
        {zhang2021intrawindowjoin}
\bibfield{author}{\bibinfo{person}{Shuhao Zhang}, \bibinfo{person}{Yancan Mao},
  \bibinfo{person}{Jiong He}, \bibinfo{person}{Philipp~M. Grulich},
  \bibinfo{person}{Steffen Zeuch}, \bibinfo{person}{Bingsheng He},
  \bibinfo{person}{Richard T.~B. Ma}, {and} \bibinfo{person}{Volker Markl}.}
  \bibinfo{year}{2021}\natexlab{a}.
\newblock \showarticletitle{Parallelizing Intra-Window Join on Multicores: An
  Experimental Study}. In \bibinfo{booktitle}{\emph{Proceedings of the 2021
  International Conference on Management of Data}} (Virtual Event, China)
  \emph{(\bibinfo{series}{SIGMOD '21})}. \bibinfo{publisher}{Association for
  Computing Machinery}, \bibinfo{address}{New York, NY, USA},
  \bibinfo{pages}{2089–2101}.
\newblock
\showISBNx{9781450383431}
\href{https://doi.org/10.1145/3448016.3452793}{doi:\nolinkurl{10.1145/3448016.3452793}}


\bibitem[Zhang et~al\mbox{.}(2026b)]%
        {zhang2026flowrag}
\bibfield{author}{\bibinfo{person}{Senlei Zhang}, \bibinfo{person}{Tongjun
  Shi}, \bibinfo{person}{Dandan Song}, \bibinfo{person}{Luan Zhang},
  \bibinfo{person}{Shuhao Zhang}, \bibinfo{person}{Xiaofei Liao}, {and}
  \bibinfo{person}{Hai Jin}.} \bibinfo{year}{2026}\natexlab{b}.
\newblock \showarticletitle{FlowRAG: Continual Learning for Dynamic Retriever
  in Retrieval-Augmented Generation}. In \bibinfo{booktitle}{\emph{Proceedings
  of the ACM Web Conference 2026}} (United Arab Emirates)
  \emph{(\bibinfo{series}{WWW '26})}. \bibinfo{publisher}{Association for
  Computing Machinery}, \bibinfo{address}{New York, NY, USA},
  \bibinfo{pages}{2160–2170}.
\newblock
\showISBNx{9798400723070}
\href{https://doi.org/10.1145/3774904.3792361}{doi:\nolinkurl{10.1145/3774904.3792361}}


\bibitem[Zhang et~al\mbox{.}(2023a)]%
        {zhang2024survey}
\bibfield{author}{\bibinfo{person}{Shuhao Zhang}, \bibinfo{person}{Juan Soto},
  {and} \bibinfo{person}{Volker Markl}.} \bibinfo{year}{2023}\natexlab{a}.
\newblock \showarticletitle{A survey on transactional stream processing}.
\newblock \bibinfo{journal}{\emph{The VLDB Journal}} \bibinfo{volume}{33},
  \bibinfo{number}{2} (\bibinfo{year}{2023}), \bibinfo{pages}{451–479}.
\newblock
\showISSN{0949-877X}
\href{https://doi.org/10.1007/s00778-023-00814-z}{doi:\nolinkurl{10.1007/s00778-023-00814-z}}


\bibitem[Zhang et~al\mbox{.}(2017b)]%
        {zhang2017mqo}
\bibfield{author}{\bibinfo{person}{Shuhao Zhang}, \bibinfo{person}{Hoang~Tam
  Vo}, \bibinfo{person}{Daniel Dahlmeier}, {and} \bibinfo{person}{Bingsheng
  He}.} \bibinfo{year}{2017}\natexlab{b}.
\newblock \showarticletitle{Multi-Query Optimization for Complex Event
  Processing in SAP ESP}. In \bibinfo{booktitle}{\emph{2017 IEEE 33rd
  International Conference on Data Engineering (ICDE)}}.
  \bibinfo{pages}{1213--1224}.
\newblock
\href{https://doi.org/10.1109/ICDE.2017.166}{doi:\nolinkurl{10.1109/ICDE.2017.166}}


\bibitem[Zhang et~al\mbox{.}(2020a)]%
        {zhang2020concurrent}
\bibfield{author}{\bibinfo{person}{Shuhao Zhang}, \bibinfo{person}{Yingjun Wu},
  \bibinfo{person}{Feng Zhang}, {and} \bibinfo{person}{Bingsheng He}.}
  \bibinfo{year}{2020}\natexlab{a}.
\newblock \showarticletitle{Towards Concurrent Stateful Stream Processing on
  Multicore Processors}. In \bibinfo{booktitle}{\emph{2020 IEEE 36th
  International Conference on Data Engineering (ICDE)}}.
  \bibinfo{pages}{1537--1548}.
\newblock
\href{https://doi.org/10.1109/ICDE48307.2020.00136}{doi:\nolinkurl{10.1109/ICDE48307.2020.00136}}


\bibitem[Zhang et~al\mbox{.}(2020c)]%
        {zhang2020hardware}
\bibfield{author}{\bibinfo{person}{Shuhao Zhang}, \bibinfo{person}{Feng Zhang},
  \bibinfo{person}{Yingjun Wu}, \bibinfo{person}{Bingsheng He}, {and}
  \bibinfo{person}{Paul Johns}.} \bibinfo{year}{2020}\natexlab{c}.
\newblock \showarticletitle{Hardware-Conscious Stream Processing: A Survey}.
\newblock \bibinfo{journal}{\emph{SIGMOD Rec.}} \bibinfo{volume}{48},
  \bibinfo{number}{4} (\bibinfo{date}{Feb.} \bibinfo{year}{2020}),
  \bibinfo{pages}{18–29}.
\newblock
\showISSN{0163-5808}
\href{https://doi.org/10.1145/3385658.3385662}{doi:\nolinkurl{10.1145/3385658.3385662}}


\bibitem[Zhang et~al\mbox{.}(2026c)]%
        {zhang2026jitserve}
\bibfield{author}{\bibinfo{person}{Wei Zhang}, \bibinfo{person}{Zhiyu Wu},
  \bibinfo{person}{Yi Mu}, \bibinfo{person}{Rui Ning}, \bibinfo{person}{Banruo
  Liu}, \bibinfo{person}{Nikhil Sarda}, \bibinfo{person}{Myungjin Lee}, {and}
  \bibinfo{person}{Fan Lai}.} \bibinfo{year}{2026}\natexlab{c}.
\newblock \showarticletitle{{JITServe}: {SLO-aware} {LLM} Serving with
  Imprecise Request Information}. In \bibinfo{booktitle}{\emph{23rd USENIX
  Symposium on Networked Systems Design and Implementation (NSDI 26)}}.
  \bibinfo{publisher}{USENIX Association}, \bibinfo{address}{Renton, WA},
  \bibinfo{pages}{825--848}.
\newblock
\showISBNx{978-1-939133-54-0}
\urldef\tempurl%
\url{https://www.usenix.org/conference/nsdi26/presentation/zhang-wei}
\showURL{%
\tempurl}


\bibitem[Zhang et~al\mbox{.}(2023c)]%
        {zhang2023compressstreamdb}
\bibfield{author}{\bibinfo{person}{Yu Zhang}, \bibinfo{person}{Feng Zhang},
  \bibinfo{person}{Hourun Li}, \bibinfo{person}{Shuhao Zhang}, {and}
  \bibinfo{person}{Xiaoyong Du}.} \bibinfo{year}{2023}\natexlab{c}.
\newblock \showarticletitle{CompressStreamDB: Fine-Grained Adaptive Stream
  Processing without Decompression}. In \bibinfo{booktitle}{\emph{2023 IEEE
  39th International Conference on Data Engineering (ICDE)}}.
  \bibinfo{pages}{408--422}.
\newblock
\href{https://doi.org/10.1109/ICDE55515.2023.00038}{doi:\nolinkurl{10.1109/ICDE55515.2023.00038}}


\bibitem[Zhang et~al\mbox{.}(2024c)]%
        {yu2025adaptivecompression}
\bibfield{author}{\bibinfo{person}{Yu Zhang}, \bibinfo{person}{Feng Zhang},
  \bibinfo{person}{Hourun Li}, \bibinfo{person}{Shuhao Zhang},
  \bibinfo{person}{Xiaoguang Guo}, \bibinfo{person}{Yuxing Chen},
  \bibinfo{person}{Anqun Pan}, {and} \bibinfo{person}{Xiaoyong Du}.}
  \bibinfo{year}{2024}\natexlab{c}.
\newblock \showarticletitle{Data-Aware Adaptive Compression for Stream
  Processing}.
\newblock \bibinfo{journal}{\emph{IEEE Transactions on Knowledge and Data
  Engineering}} \bibinfo{volume}{36}, \bibinfo{number}{9}
  (\bibinfo{year}{2024}), \bibinfo{pages}{4531--4549}.
\newblock
\href{https://doi.org/10.1109/TKDE.2024.3377710}{doi:\nolinkurl{10.1109/TKDE.2024.3377710}}


\bibitem[Zhang et~al\mbox{.}(2024b)]%
        {zhang2024rummy}
\bibfield{author}{\bibinfo{person}{Zili Zhang}, \bibinfo{person}{Fangyue Liu},
  \bibinfo{person}{Gang Huang}, \bibinfo{person}{Xuanzhe Liu}, {and}
  \bibinfo{person}{Xin Jin}.} \bibinfo{year}{2024}\natexlab{b}.
\newblock \showarticletitle{Fast vector query processing for large datasets
  beyond GPU memory with reordered pipelining}. In
  \bibinfo{booktitle}{\emph{Proceedings of the 21st USENIX Symposium on
  Networked Systems Design and Implementation}} (Santa Clara, CA, USA)
  \emph{(\bibinfo{series}{NSDI'24})}. \bibinfo{publisher}{USENIX Association},
  \bibinfo{address}{USA}, Article \bibinfo{articleno}{2},
  \bibinfo{numpages}{18}~pages.
\newblock
\showISBNx{978-1-939133-39-7}


\bibitem[Zhang et~al\mbox{.}(2024a)]%
        {dong2024qhitter}
\bibfield{author}{\bibinfo{person}{Zhenyu Zhang}, \bibinfo{person}{Shiwei Liu},
  \bibinfo{person}{Runjin Chen}, \bibinfo{person}{Bhavya Kailkhura},
  \bibinfo{person}{Beidi Chen}, {and} \bibinfo{person}{Zhangyang Wang}.}
  \bibinfo{year}{2024}\natexlab{a}.
\newblock \showarticletitle{Q-Hitter: A Better Token Oracle for Efficient LLM
  Inference via Sparse-Quantized KV Cache}. In
  \bibinfo{booktitle}{\emph{Proceedings of Machine Learning and Systems}},
  \bibfield{editor}{\bibinfo{person}{P.~Gibbons},
  \bibinfo{person}{G.~Pekhimenko}, {and} \bibinfo{person}{C.~De Sa}} (Eds.),
  Vol.~\bibinfo{volume}{6}. \bibinfo{pages}{381--394}.
\newblock
\urldef\tempurl%
\url{https://proceedings.mlsys.org/paper_files/paper/2024/file/bbb7506579431a85861a05fff048d3e1-Paper-Conference.pdf}
\showURL{%
\tempurl}


\bibitem[Zhao et~al\mbox{.}(2024b)]%
        {zhao2024recovery}
\bibfield{author}{\bibinfo{person}{Jianjun Zhao}, \bibinfo{person}{Haikun Liu},
  \bibinfo{person}{Shuhao Zhang}, \bibinfo{person}{Zhuohui Duan},
  \bibinfo{person}{Xiaofei Liao}, \bibinfo{person}{Hai Jin}, {and}
  \bibinfo{person}{Yu Zhang}.} \bibinfo{year}{2024}\natexlab{b}.
\newblock \showarticletitle{Fast Parallel Recovery for Transactional Stream
  Processing on Multicores}. In \bibinfo{booktitle}{\emph{2024 IEEE 40th
  International Conference on Data Engineering (ICDE)}}.
  \bibinfo{pages}{1478--1491}.
\newblock
\href{https://doi.org/10.1109/ICDE60146.2024.00122}{doi:\nolinkurl{10.1109/ICDE60146.2024.00122}}


\bibitem[Zhao et~al\mbox{.}(2025a)]%
        {zhao2025rtsfaas}
\bibfield{author}{\bibinfo{person}{Jianjun Zhao}, \bibinfo{person}{Haikun Liu},
  \bibinfo{person}{Shuhao Zhang}, \bibinfo{person}{Haodi Lu},
  \bibinfo{person}{Yancan Mao}, \bibinfo{person}{Zhuohui Duan},
  \bibinfo{person}{Xiaofei Liao}, {and} \bibinfo{person}{Hai Jin}.}
  \bibinfo{year}{2025}\natexlab{a}.
\newblock \showarticletitle{Towards high-performance transactional stateful
  serverless workflows with affinity-aware leasing}. In
  \bibinfo{booktitle}{\emph{Proceedings of the 2025 USENIX Conference on Usenix
  Annual Technical Conference}} (Boston, MA, USA)
  \emph{(\bibinfo{series}{USENIX ATC '25})}. \bibinfo{publisher}{USENIX
  Association}, \bibinfo{address}{USA}, Article \bibinfo{articleno}{91},
  \bibinfo{numpages}{17}~pages.
\newblock
\showISBNx{978-1-939133-48-9}


\bibitem[Zhao et~al\mbox{.}(2025b)]%
        {zhao2025morphstream}
\bibfield{author}{\bibinfo{person}{Jianjun Zhao}, \bibinfo{person}{Yancan Mao},
  \bibinfo{person}{Zhonghao Yang}, \bibinfo{person}{Haikun Liu}, {and}
  \bibinfo{person}{Shuhao Zhang}.} \bibinfo{year}{2025}\natexlab{b}.
\newblock \showarticletitle{Scalable Transactional Stream Processing on
  Multicore Processors}.
\newblock \bibinfo{journal}{\emph{IEEE Transactions on Knowledge and Data
  Engineering}} \bibinfo{volume}{37}, \bibinfo{number}{7}
  (\bibinfo{year}{2025}), \bibinfo{pages}{4254--4269}.
\newblock
\href{https://doi.org/10.1109/TKDE.2025.3556741}{doi:\nolinkurl{10.1109/TKDE.2025.3556741}}


\bibitem[Zhao et~al\mbox{.}(2026)]%
        {zhao2025equilibria}
\bibfield{author}{\bibinfo{person}{Kaiyang Zhao}, \bibinfo{person}{Neha
  Gholkar}, \bibinfo{person}{Hasan Maruf}, \bibinfo{person}{Abhishek Dhanotia},
  \bibinfo{person}{Johannes Weiner}, \bibinfo{person}{Gregory Price},
  \bibinfo{person}{Ning Sun}, \bibinfo{person}{Bhavya Dwivedi},
  \bibinfo{person}{Stuart Clark}, {and} \bibinfo{person}{Dimitrios Skarlatos}.}
  \bibinfo{year}{2026}\natexlab{}.
\newblock \showarticletitle{Equilibria: Fair Multi-Tenant CXL Memory Tiering At
  Scale}.
\newblock \bibinfo{journal}{\emph{arXiv preprint arXiv:2602.08800}}
  (\bibinfo{year}{2026}).
\newblock
\href{https://doi.org/10.48550/arXiv.2602.08800}{doi:\nolinkurl{10.48550/arXiv.2602.08800}}


\bibitem[Zhao et~al\mbox{.}(2024a)]%
        {zhao2024atom}
\bibfield{author}{\bibinfo{person}{Yilong Zhao}, \bibinfo{person}{Chien-Yu
  Lin}, \bibinfo{person}{Kan Zhu}, \bibinfo{person}{Zihao Ye},
  \bibinfo{person}{Lequn Chen}, \bibinfo{person}{Size Zheng},
  \bibinfo{person}{Luis Ceze}, \bibinfo{person}{Arvind Krishnamurthy},
  \bibinfo{person}{Tianqi Chen}, {and} \bibinfo{person}{Baris Kasikci}.}
  \bibinfo{year}{2024}\natexlab{a}.
\newblock \showarticletitle{Atom: Low-Bit Quantization for Efficient and
  Accurate LLM Serving}. In \bibinfo{booktitle}{\emph{Proceedings of Machine
  Learning and Systems}}, \bibfield{editor}{\bibinfo{person}{P.~Gibbons},
  \bibinfo{person}{G.~Pekhimenko}, {and} \bibinfo{person}{C.~De Sa}} (Eds.),
  Vol.~\bibinfo{volume}{6}. \bibinfo{pages}{196--209}.
\newblock
\urldef\tempurl%
\url{https://proceedings.mlsys.org/paper_files/paper/2024/file/5edb57c05c81d04beb716ef1d542fe9e-Paper-Conference.pdf}
\showURL{%
\tempurl}


\bibitem[Zhao and Wang(2024)]%
        {zhao2024alise}
\bibfield{author}{\bibinfo{person}{Youpeng Zhao} {and} \bibinfo{person}{Jun
  Wang}.} \bibinfo{year}{2024}\natexlab{}.
\newblock \showarticletitle{ALISE: Accelerating Large Language Model Serving
  with Speculative Scheduling}. In \bibinfo{booktitle}{\emph{2024 ACM/IEEE
  International Conference On Computer Aided Design (ICCAD)}}.
  \bibinfo{pages}{1--9}.
\newblock
\href{https://doi.org/10.1145/3676536.3676659}{doi:\nolinkurl{10.1145/3676536.3676659}}


\bibitem[Zheng et~al\mbox{.}(2024b)]%
        {zheng2023sglang}
\bibfield{author}{\bibinfo{person}{Lianmin Zheng}, \bibinfo{person}{Liangsheng
  Yin}, \bibinfo{person}{Zhiqiang Xie}, \bibinfo{person}{Chuyue Sun},
  \bibinfo{person}{Jeff Huang}, \bibinfo{person}{Cody~Hao Yu},
  \bibinfo{person}{Shiyi Cao}, \bibinfo{person}{Christos Kozyrakis},
  \bibinfo{person}{Ion Stoica}, \bibinfo{person}{Joseph~E. Gonzalez},
  \bibinfo{person}{Clark Barrett}, {and} \bibinfo{person}{Ying Sheng}.}
  \bibinfo{year}{2024}\natexlab{b}.
\newblock \showarticletitle{SGLang: Efficient Execution of Structured Language
  Model Programs}. In \bibinfo{booktitle}{\emph{Advances in Neural Information
  Processing Systems}}, \bibfield{editor}{\bibinfo{person}{A.~Globerson},
  \bibinfo{person}{L.~Mackey}, \bibinfo{person}{D.~Belgrave},
  \bibinfo{person}{A.~Fan}, \bibinfo{person}{U.~Paquet},
  \bibinfo{person}{J.~Tomczak}, {and} \bibinfo{person}{C.~Zhang}} (Eds.),
  Vol.~\bibinfo{volume}{37}. \bibinfo{publisher}{Curran Associates, Inc.},
  \bibinfo{pages}{62557--62583}.
\newblock
\href{https://doi.org/10.52202/079017-2000}{doi:\nolinkurl{10.52202/079017-2000}}


\bibitem[Zheng et~al\mbox{.}(2024a)]%
        {zheng2024vmcu}
\bibfield{author}{\bibinfo{person}{Size Zheng}, \bibinfo{person}{Renze Chen},
  \bibinfo{person}{Meng Li}, \bibinfo{person}{Zihao Ye}, \bibinfo{person}{Luis
  Ceze}, {and} \bibinfo{person}{Yun Liang}.} \bibinfo{year}{2024}\natexlab{a}.
\newblock \showarticletitle{vMCU: Coordinated Memory Management and Kernel
  Optimization for DNN Inference on MCUs}. In
  \bibinfo{booktitle}{\emph{Proceedings of Machine Learning and Systems}},
  \bibfield{editor}{\bibinfo{person}{P.~Gibbons},
  \bibinfo{person}{G.~Pekhimenko}, {and} \bibinfo{person}{C.~De Sa}} (Eds.),
  Vol.~\bibinfo{volume}{6}. \bibinfo{pages}{452--464}.
\newblock
\urldef\tempurl%
\url{https://proceedings.mlsys.org/paper_files/paper/2024/file/d5a655b8b373737b4f2aea8f78e5e754-Paper-Conference.pdf}
\showURL{%
\tempurl}


\bibitem[Zheng et~al\mbox{.}(2025)]%
        {zheng2025save}
\bibfield{author}{\bibinfo{person}{Wenxin Zheng}, \bibinfo{person}{Bin Xu},
  \bibinfo{person}{Jinyu Gu}, {and} \bibinfo{person}{Haibo Chen}.}
  \bibinfo{year}{2025}\natexlab{}.
\newblock \showarticletitle{SAVE: software-implemented fault tolerance for
  model inference against GPU memory bit flips}. In
  \bibinfo{booktitle}{\emph{Proceedings of the 2025 USENIX Conference on Usenix
  Annual Technical Conference}} (Boston, MA, USA)
  \emph{(\bibinfo{series}{USENIX ATC '25})}. \bibinfo{publisher}{USENIX
  Association}, \bibinfo{address}{USA}, Article \bibinfo{articleno}{94},
  \bibinfo{numpages}{20}~pages.
\newblock
\showISBNx{978-1-939133-48-9}


\bibitem[Zhong et~al\mbox{.}(2024a)]%
        {zhong2024memstrata}
\bibfield{author}{\bibinfo{person}{Yuhong Zhong}, \bibinfo{person}{Daniel~S.
  Berger}, \bibinfo{person}{Carl Waldspurger}, \bibinfo{person}{Ryan Wee},
  \bibinfo{person}{Ishwar Agarwal}, \bibinfo{person}{Rajat Agarwal},
  \bibinfo{person}{Frank Hady}, \bibinfo{person}{Karthik Kumar},
  \bibinfo{person}{Mark~D. Hill}, \bibinfo{person}{Mosharaf Chowdhury}, {and}
  \bibinfo{person}{Asaf Cidon}.} \bibinfo{year}{2024}\natexlab{a}.
\newblock \showarticletitle{Managing Memory Tiers with {CXL} in Virtualized
  Environments}. In \bibinfo{booktitle}{\emph{18th USENIX Symposium on
  Operating Systems Design and Implementation (OSDI 24)}}.
  \bibinfo{publisher}{USENIX Association}, \bibinfo{address}{Santa Clara, CA},
  \bibinfo{pages}{37--56}.
\newblock
\showISBNx{978-1-939133-40-3}
\urldef\tempurl%
\url{https://www.usenix.org/conference/osdi24/presentation/zhong-yuhong}
\showURL{%
\tempurl}


\bibitem[Zhong et~al\mbox{.}(2024b)]%
        {zhong2024distserve}
\bibfield{author}{\bibinfo{person}{Yinmin Zhong}, \bibinfo{person}{Shengyu
  Liu}, \bibinfo{person}{Junda Chen}, \bibinfo{person}{Jianbo Hu},
  \bibinfo{person}{Yibo Zhu}, \bibinfo{person}{Xuanzhe Liu},
  \bibinfo{person}{Xin Jin}, {and} \bibinfo{person}{Hao Zhang}.}
  \bibinfo{year}{2024}\natexlab{b}.
\newblock \showarticletitle{DistServe: disaggregating prefill and decoding for
  goodput-optimized large language model serving}. In
  \bibinfo{booktitle}{\emph{Proceedings of the 18th USENIX Conference on
  Operating Systems Design and Implementation}} (Santa Clara, CA, USA)
  \emph{(\bibinfo{series}{OSDI'24})}. \bibinfo{publisher}{USENIX Association},
  \bibinfo{address}{USA}, Article \bibinfo{articleno}{11},
  \bibinfo{numpages}{18}~pages.
\newblock
\showISBNx{978-1-939133-40-3}


\bibitem[Zhong et~al\mbox{.}(2024c)]%
        {zhong2024unimem}
\bibfield{author}{\bibinfo{person}{Yijie Zhong}, \bibinfo{person}{Minqiang
  Zhou}, \bibinfo{person}{Zhirong Shen}, {and} \bibinfo{person}{Jiwu Shu}.}
  \bibinfo{year}{2024}\natexlab{c}.
\newblock \showarticletitle{UniMem: redesigning disaggregated memory within a
  unified local-remote memory hierarchy}. In
  \bibinfo{booktitle}{\emph{Proceedings of the 2024 USENIX Conference on Usenix
  Annual Technical Conference}} (Santa Clara, CA, USA)
  \emph{(\bibinfo{series}{USENIX ATC'24})}. \bibinfo{publisher}{USENIX
  Association}, \bibinfo{address}{USA}, Article \bibinfo{articleno}{29},
  \bibinfo{numpages}{15}~pages.
\newblock
\showISBNx{978-1-939133-41-0}


\bibitem[Zhou et~al\mbox{.}(2025)]%
        {zhou2025ferret}
\bibfield{author}{\bibinfo{person}{Yuhao Zhou}, \bibinfo{person}{Yuxin Tian},
  \bibinfo{person}{Jindi Lv}, \bibinfo{person}{Mingjia Shi},
  \bibinfo{person}{Yuanxi Li}, \bibinfo{person}{Qing Ye},
  \bibinfo{person}{Shuhao Zhang}, {and} \bibinfo{person}{Jiancheng Lv}.}
  \bibinfo{year}{2025}\natexlab{}.
\newblock \showarticletitle{Ferret: An Efficient Online Continual Learning
  Framework under Varying Memory Constraints}. In
  \bibinfo{booktitle}{\emph{2025 IEEE/CVF Conference on Computer Vision and
  Pattern Recognition (CVPR)}}. \bibinfo{pages}{4850--4861}.
\newblock
\href{https://doi.org/10.1109/CVPR52734.2025.00457}{doi:\nolinkurl{10.1109/CVPR52734.2025.00457}}


\bibitem[Zhou et~al\mbox{.}(2024)]%
        {zhou2024neomem}
\bibfield{author}{\bibinfo{person}{Zhe Zhou}, \bibinfo{person}{Yiqi Chen},
  \bibinfo{person}{Tao Zhang}, \bibinfo{person}{Yang Wang},
  \bibinfo{person}{Ran Shu}, \bibinfo{person}{Shuotao Xu},
  \bibinfo{person}{Peng Cheng}, \bibinfo{person}{Lei Qu},
  \bibinfo{person}{Yongqiang Xiong}, \bibinfo{person}{Jie Zhang}, {and}
  \bibinfo{person}{Guangyu Sun}.} \bibinfo{year}{2024}\natexlab{}.
\newblock \showarticletitle{NeoMem: Hardware/Software Co-Design for CXL-Native
  Memory Tiering}. In \bibinfo{booktitle}{\emph{Proceedings of the 2024 57th
  IEEE/ACM International Symposium on Microarchitecture}} (Austin, TX, USA)
  \emph{(\bibinfo{series}{MICRO '24})}. \bibinfo{publisher}{IEEE Press},
  \bibinfo{pages}{1518–1531}.
\newblock
\href{https://doi.org/10.1109/MICRO61859.2024.00111}{doi:\nolinkurl{10.1109/MICRO61859.2024.00111}}


\bibitem[Zhu et~al\mbox{.}(2025)]%
        {zhu2024sampleattention}
\bibfield{author}{\bibinfo{person}{Qianchao Zhu}, \bibinfo{person}{Jiangfei
  Duan}, \bibinfo{person}{Chang Chen}, \bibinfo{person}{Siran Liu},
  \bibinfo{person}{Xiuhong Li}, \bibinfo{person}{Guanyu Feng},
  \bibinfo{person}{Xin Lv}, \bibinfo{person}{Xiao Chuanfu},
  \bibinfo{person}{Dahua Lin}, {and} \bibinfo{person}{Chao Yang}.}
  \bibinfo{year}{2025}\natexlab{}.
\newblock \showarticletitle{SampleAttention: Near-Lossless Acceleration of Long
  Context LLM Inference with Adaptive Structured Sparse Attention}. In
  \bibinfo{booktitle}{\emph{Proceedings of Machine Learning and Systems}},
  \bibfield{editor}{\bibinfo{person}{M.~Zaharia}, \bibinfo{person}{G.~Joshi},
  {and} \bibinfo{person}{Y.~Lin}} (Eds.), Vol.~\bibinfo{volume}{7}.
  \bibinfo{publisher}{MLSys}.
\newblock
\urldef\tempurl%
\url{https://proceedings.mlsys.org/paper_files/paper/2025/file/2d04d97593c8c33d415337f408ed0e1b-Paper-Conference.pdf}
\showURL{%
\tempurl}


\bibitem[Zuo et~al\mbox{.}(2021)]%
        {zuo2021race}
\bibfield{author}{\bibinfo{person}{Pengfei Zuo}, \bibinfo{person}{Jiazhao Sun},
  \bibinfo{person}{Liu Yang}, \bibinfo{person}{Shuangwu Zhang}, {and}
  \bibinfo{person}{Yu Hua}.} \bibinfo{year}{2021}\natexlab{}.
\newblock \showarticletitle{One-sided {RDMA-Conscious} Extendible Hashing for
  Disaggregated Memory}. In \bibinfo{booktitle}{\emph{2021 USENIX Annual
  Technical Conference (USENIX ATC 21)}}. \bibinfo{publisher}{USENIX
  Association}, \bibinfo{pages}{15--29}.
\newblock
\showISBNx{978-1-939133-23-6}
\urldef\tempurl%
\url{https://www.usenix.org/conference/atc21/presentation/zuo}
\showURL{%
\tempurl}


\end{thebibliography}

\end{document}